\documentclass[12pt]{article}
\usepackage{graphicx}
\usepackage{amsmath}
\usepackage{amssymb}
\usepackage{amsbsy}
\usepackage{bm}   
\usepackage{framed}  

\usepackage{hyperref}
\hypersetup{linktocpage}

\hypersetup{
      colorlinks=true,
       linkcolor=blue,
}
\numberwithin{equation}{section}
\newcommand{\bit}{\begin{itemize}}
\newcommand{\eit}{\end{itemize}}

\def\benu{\begin{enumerate}}
\def\eenu{\end{enumerate}}
\def\noi{\noindent}
\def\btab{\begin{tabbing}}
\def\etab{\end{tabbing}}

\def\bit{\begin{itemize}}
\def\eit{\end{itemize}}
\def\beq{\begin{equation}}
\def\eeq{\end{equation}}
\def\bec{\begin{center}}
\def\eec{\end{center}}
\def\btable{\begin{tabular}}
\def\etable{\end{tabular}}
\def\beqr{\begin{eqnarray}}
\def\eeqr{\end{eqnarray}}
\def\rarw{\rightarrow}
\def\Rarw{\Rightarrow}

\def\om{\omega}
\def\gm{\gamma}
\def\Gm{\Gamma}

\def\eps{\epsilon}

\def\al{\alpha}
\def\bt{\beta}

\def\dl{\delta}
\def\Dl{\Delta}
\def\sg{\sigma}
\def\Om{\Omega}
\def\rarw{\rightarrow}
\def\del{\partial}

\def\half{\frac{1}{2}}
\def\qrtr{\frac{1}{4}}

\def\btab{\begin{tabbing}}
\def\etab{\end{tabbing}}
\def\beqrs{\begin{eqnarray*}}
\def\eeqrs{\end{eqnarray*}}
\def\noi{\noindent}
\def\lan{\langle}
\def\ran{\rangle}
\def\bfig{\begin{figure}}
\def\efig{\end{figure}}
\def\fr{\frac}
\def\barr{\begin{array}}
\def\earr{\end{array}}
\def\non{\nonumber}
\def\bibi{\bibitem}

 \fontfamily{ptm}\selectfont
 \usepackage{mathptmx}           

\title{\bf Theory of beam echoes}

\author{Tanaji Sen, FNAL, Batavia, IL 60510}
\date{}

\begin{document}

\maketitle


\begin{abstract}
We develop the theory of beam echoes in circular accelerators under several different conditions. 
We derive detailed expressions for the echo amplitude and pulse width with nonlinear quadrupole and dipole
kicks, first without and then with momentum spread. We use the theory with the linearized dipole and
quadrupole kicks to solve the diffusion equation for different dependencies of the diffusion coefficient on the
action. We then consider the use of multiple quadrupole kicks to increase the maximum echo amplitude.
We have extended these calculations partially to the 2D case and we also have partial results for longitudinal
echoes. 
\end{abstract}

\tableofcontents

\clearpage

\section{Introduction}

The concept of beam echoes was introduced by    Stupakov \cite{Stupakov} and was first measured in the
longitudinal plane at the Fermilab Accumulator \cite{Fermi_AA}. Since then it has been observed at other
accelerators. We list  references on echoes from [5] - [12] in  reverse chronological order.

The lecture notes  by Alex Chao \cite{Chao} inspired me to  write these notes in  the same spirit.
These notes are written mainly for the benefits of researchers and students. The other purpose is to serve as
reference material for journal articles.

The contents in this paper are  arranged in roughly in the order in which the topics were worked on.
A word of caution: be aware of typographical errors especially in the equations, of which  there may be quite a few.

\section{Influence of a dipole kick}

\subsection{Dipole moment}

Here we consider the general case where the dipole kicker is at a phase advance $\Dl\mu$ from the BPM location where the centroid is
located. The changes $(\Dl x, \Dl x')$ in position and slope at the BPM are related to the changes $(0, \theta)$ at the kicker
via the transfer matrix between the two locations
\beq
\left[\begin{array}{c} \Dl x \\ \Dl x' \end{array} \right] = \left( \begin{array}{c c} M_{11} & M_{12} \\ M_{21} & M_{22} \end{array}
\right) \left[\begin{array}{c} 0 \\ \theta \end{array} \right] = \left[\begin{array}{c} M_{12}\theta \\
 \ M_{22}\theta \end{array} \right]
\eeq
The transfer matrix elements from the kicker to the BPM are
\[ M_{12} = \sqrt{\bt \bt_K}\sin\Dl\mu, \; \; \; M_{22} = \sqrt{\frac{\bt_K}{\bt}} (\cos\Dl\mu - \al \sin\Dl\mu ) \]
where $\bt_K$ is the beta function at the kicker and $(\bt, \al)$ are the values at the BPM. In terms of the conjugate variables $x, p$
defined by $p = \bt x' + \al x$, the changes at the BPM location are
\beqrs
\Dl x & = &  \theta \sqrt{\bt \bt_K}\sin\Dl\mu   \\
\Dl p & = & \bt \Dl x' + \al\Dl x = \bt \sqrt{\frac{\bt_K}{\bt}} (\cos\Dl\mu - \al \sin\Dl\mu )\theta + \al 
\sqrt{\bt \bt_K}\sin\Dl\mu \theta  = \theta \sqrt{\bt \bt_K} \cos\Dl\mu 
\eeqrs
In terms of these variables, the dipole kick, the action $J = (x^2 + p^2)/(2\bt)$ changes to 
\[ J_1 = \frac{1}{2\bt}[(x - \Dl x)^2 + (p - \Dl p)^2] = J - \sqrt{\frac{\bt_K}{\bt}}\theta(\sin\Dl\mu x + \cos\Dl\mu p) 
 + \half \bt_K\theta^2 
\]
If the density distribution before the kick was $\psi_0(J)$, after the kick it is $\psi_1(J_1)$.
We use the transformation to action angle variables $J,\phi$
\[ x = \sqrt{2\bt J} \cos\phi, \;\;\;\; p = -\sqrt{2\bt J}\sin\phi \]
to obtain immediately after the kick
\[ \psi_1(J,\phi) = \psi_0(J + \sqrt{2\bt_K J}\theta \sin(\phi - \Dl\mu) + \half \bt_K\theta^2)  \]
At time $t$ after the kick, the distribution evolves to
\[ \psi_2(J,\phi,t) = \psi_1(J,\phi - \om(J)t ) \]
The dipole moment at time $t$ is 
\[  \lan x \ran(t) = \int dJ \int d\phi \sqrt{2\bt J}\cos\phi \psi_2(J,\phi,t) \]
Taking the initial distribution to be
\beq
\psi_0(J) = \frac{1}{2\pi J_0}\exp[-\frac{J}{J_0}]
\eeq
where $J_0= \eps$ is equal to the initial emittance. 
Hence
\beq
\lan x\ran (t)  =  \frac{1}{2\pi J_0}\exp[\half \bt_K\theta^2] \int dJ\sqrt{2\bt J} \int d\phi \exp[-\frac{1}{J_0}(J + \sqrt{2\bt_K J}\theta \sin(\phi - \om(J) t - \Dl\mu) )]
\eeq
The $\phi$ integration is done using
\[ \int d\phi \cos\phi \exp[- a \sin(\phi - b)] = 2\pi I_1(a) \sin b \]
To to the $J$ integration we assume a linear dependence on $J$
\[ \om(J) = \om_0 + \om' J \]
where $\om'$ is a constant. Hence
\[ \lan x\ran (t)  =  \frac{1}{J_0}\exp[\half \bt_K\theta^2] \int dJ \sqrt{2\bt J} \exp[-\frac{J}{J_0}] 
I_1(\sqrt{2\bt_K J}\theta) \sin(\om_0 t + \om' J t  + \Dl\mu) )]
\]
Changing variable to $a = \sqrt{2\bt J}$, we have
\[  \lan x\ran (t)  =  \frac{1}{\bt J_0}\exp[\half \bt_K\theta^2] \int da \; a^2 I_1(\sqrt{\frac{\bt_K}{\bt}}\frac{\theta}{J_0}a)
\exp[-\frac{a^2}{2\bt J_0}] {\rm Im}[\exp[i(\om_0 t + \Dl\mu)]\exp[i \frac{\om' t}{2\bt} a^2]] 
\]
Using
\[ \int_0^{\infty} da\; a^2 \exp[-A a^2] I_1(B a) = \frac{B}{4 A^2} \exp[\frac{B^2}{4A}]
\]
Substituting
\[ A =  \frac{1}{2\bt J_0}( 1 - i \om' J_0 t) = \frac{1}{2\bt J_0}( 1 - i \Theta), B = \sqrt{\frac{\bt_K}{\bt}}\frac{\theta}{J_0}, \; \;
\Theta = \om' J_0 t
\]
we have
\[ \frac{B}{4 A^2} = \sqrt{\frac{\bt_K}{\bt}}\frac{\bt^2 \theta J_0}{(1 - i\Theta)^2}, \;\;\;
\frac{B^2}{4A} = \frac{\bt_K \theta^2}{2J_0} \frac{i\Theta}{(1 - i\Theta)} 
\]
Hence
\beq
\lan x\ran (t)  = \theta  \sqrt{\bt_K \bt} {\rm Im}\left[ \frac{e^{i(\om_0 t + \Dl\mu)}}{(1 - i\Theta)^2}
\exp[\frac{\bt_K\theta^2}{2J_0} \frac{i\Theta}{(1 - i\Theta)}] \right]  \nonumber 
\eeq
Let
\beq
 \eta = \om_0 t + \Dl\mu + \frac{\bt_K\theta^2}{2J_0} \frac{\Theta}{(1 + \Theta^2)} 
\eeq
We have to evaluate
\beqrs
 {\rm Im}\left[ \exp[-\frac{\bt_K\theta^2}{2J_0} \frac{\Theta^2}{1+\Theta^2} \frac{ \exp[i\eta]}{(1 - i\Theta)^2} \right]
 & = & \exp[-\frac{\bt_K\theta^2}{2J_0} \frac{\Theta^2}{1+\Theta^2}]\frac{1}{(1+\Theta^2)^2}
 {\rm Im}\left[ (1 + 2i\Theta - \Theta^2) e^{i\eta} \right] \\
& = & \exp[-\frac{\bt_K\theta^2}{2J_0} \frac{\Theta^2}{1+\Theta^2}]\frac{1}{(1+\Theta^2)^2}
\left[(1 - \Theta^2)\sin\eta + 2\Theta \cos\eta \right]
\eeqrs
Let 
\[ \sin \nu - \frac{2\Theta}{(1+\Theta^2)}, \;\; \Rarw \cos\nu = \sqrt{1-\sin^2\nu} = \frac{1 - \Theta^2}{1+\Theta^2} \]
Hence
\[ \left[\frac{(1 - \Theta^2)}{(1+\Theta^2)}\sin\eta + \frac{2\Theta}{(1+\Theta^2)}\cos\eta \right] = \sin(\eta + \nu) \]
where 
\[ \tan \nu = \frac{2\Theta}{1- \Theta^2} \]
Hence
\beq
\lan x\ran (t)  =\frac{ \theta  \sqrt{\bt_K \bt}}{(1+\Theta^2)}\exp[-\frac{\bt_K\theta^2}{2J_0} \frac{\Theta^2}{1+\Theta^2}]
\sin(\eta + \nu) 
\eeq
The amplitude of the dipole kick is
\beq
\lan x\ran^{amp} (t)  =\frac{ \theta  \sqrt{\bt_K \bt}}{(1+\Theta^2)}\exp[-\frac{\bt_K\theta^2}{2J_0} \frac{\Theta^2}{1+\Theta^2}]
\eeq
This is independent of the phase advance $\Dl\mu$ from the kicker to the BPM.

If the dipole kicker and BPM are at the same location, $\bt_K = \bt$ and $\Dl\mu = 0$.

\subsection{Emittance Growth}

Here we consider the time evolution of the second order moments. 

At time $t$ following the dipole kick, the distribution is 
\beq
\psi_2(J,\phi, t) = \psi_0(J+\theta\sqrt{2\bt J}\sin(\phi - \om(J)t) + \half\bt \theta^2)
\eeq
Hence
\beq
\lan x^2 \ran  =  \int_0^{\infty}dJ \int_0^{2\pi} d\phi 2\bt J\cos^2 \phi \psi_2(J,\phi,t) 
\eeq
With $\psi_0 = (1/2\pi J_0)\exp[-J/J_0]$, introducing $z = J/J_0$, then
\[
\psi_2(z,\phi,t) = \frac{1}{2\pi J_0}\exp[-z]\exp[-\theta \sqrt{2\bt_k J_0 z}\sin(\phi-\om(z)t)]
\exp[-\frac{\bt\theta^2}{2 J_0}]
\]
and
\beq
 \lan x^2 \ran = \frac{\bt J_0 }{2\pi }\exp[-\frac{\bt\theta^2}{2 J_0}]\int dz d\phi \; z 
\exp[-(z+\theta \sqrt{2\bt_k J_0 z}\sin(\phi-\om(z)t))](1 + \cos 2\phi)
\eeq
There are two integrations over $\phi$ for which we use the integration results
\beqrs
\int_0^{2\phi}d\phi \; \exp[-a \sin(\phi + b)] & = & 2\pi I_0(a) \\
\int_0^{2\phi}d\phi \; \exp[-a \sin(\phi + b)]  \cos2\phi & = & -2\pi I_2(a)\cos 2b
\eeqrs
Hence
\beq
 \lan x^2 \ran = \bt J_0 \exp[-\frac{\bt\theta^2}{2 J_0}]\int dz \; z \exp[-z]
\left(I_0(\theta\sqrt{\frac{2\bt_K z}{2 J_0}}) - I_2(\theta\sqrt{\frac{2\bt_K z}{2 J_0}})\cos 2\om(z)t \right)
\eeq
Writing $\cos 2\om(z)t = \cos 2[\om_{\bt} + \om'J_0 z]t = {\rm Re}[e^{i2\om_{\bt}t} e^{i2\om' J_0 t z}]$, we use these results to do the
integration over $z$,
\beqrs
\int dz \;\; z \exp[-z] I_0(a\sqrt{z}) & = & (1 + \frac{1}{4}a^2)\exp[\frac{a^2}{4}] \\
\int dz \;\; z \exp[-b z] I_2(a\sqrt{z}) & = & \frac{a^2}{4b^3}\exp[\frac{a^2}{4b}]
\eeqrs
Here $a = \theta\sqrt{2\bt_K/J_0}$ and $b = 1 - i 2\om' J_0t$. Define the dimensionless variables
\beq
 \Theta_2 = 2\om' J_0 t, \;\;\; a_K = \frac{\bt_K\theta^2}{2J_0}, \;\;\;  \Psi_2 = 2 \om_{\bt} t + \frac{a_K}{1+\Theta_2^2}
\eeq
We have
 \[ \lan x^2 \ran = \bt J_0 \exp[-a_K]\left[ (1 + a_K)\exp[a_K] - a_K{\rm Re}
\left(\frac{e^{i\om_{\bt}t}}{(1 - i\Theta_2)^3}\exp[\frac{a_K}{1 - i\Theta_2}]\right)\right]
\]
We evaluate the second term separately
\beqrs
{\rm Re}\left(\frac{e^{i\om_{\bt}t}}{(1 - i\Theta_2)^3}\exp[\frac{a_K}{1 - i\Theta_2}]\right) & = & \frac{1}{(1+\Theta_2^2)^3}
{\rm Re}\left[ (1+i\Theta_2)^3\exp[i(\om_{\bt}t + \frac{a_K(1 + i\Theta_2)}{1+\Theta_2^2})] \right] \\
&  = & \frac{1}{(1+\Theta_2^2)^3} \exp[\frac{a_K}{1+\Theta_2^2}]
{\rm Re}\left[ (1+i\Theta_2)^3 \exp[i\Psi_2]\right] \\
&  = & \frac{1}{(1+\Theta_2^2)^3} \exp[\frac{a_K}{1+\Theta_2^2}]
\left[ (1 - 3\Theta_2^2)\cos\Psi_2 - (3\Theta_2 - \Theta_2^3)\sin\Psi_2 \right]
\eeqrs
Using $ (1 - 3\Theta_2^2)^2 + (3\Theta_2 - \Theta_2^3)^2 = (1 + \Theta_2^2)^3$, we cam write
\[ (1 - 3\Theta_2^2)\cos\Psi_2 - (3\Theta_2 - \Theta_2^3)\sin\Psi_2 = (1 + \Theta_2^2)^{3/2}\cos(\Psi_2 + \nu)\]
where 
\[ \cos\nu = \frac{(1 - 3\Theta_2^2)}{(1 + \Theta_2^2)^{3/2}}, \;\;\; \sin\nu = \frac{3\Theta_2 - \Theta_2^3}{(1 + \Theta_2^2)^{3/2}}\;\;\;
\tan\nu = \frac{3\Theta_2 - \Theta_2^3}{1 - 3\Theta_2^2} \]
Let $\al = \tan\Theta_2$, then
\[ \tan \nu = \frac{3\tan\al - \tan^3\al}{1 - 3\tan^2\al} = \tan 3\al, \;\;\; \nu = 3\al = 3\tan^{-1}\Theta_2 \]
Hence
\beqr
 \lan x^2 \ran & = & \bt J_0 \exp[-a_K]\left[ (1 +a_K) \exp[a_K] - a_K\frac{1}{(1+\Theta_2^2)^{3/2}} \exp[\frac{a_K}{1+\Theta_2^2}]
\cos(\Psi_2 + 3\tan^{-1}\Theta_2) \right] \nonumber \\
& = & \bt J_0 \left[ (1 + a_K) - \frac{a_K}{(1+\Theta_2^2)^{3/2}}\exp[-\frac{a_K \Theta_2^2}{1+\Theta_2^2}] \cos(\Psi_2 + 3\tan^{-1}\Theta_2) \right] \nonumber \\
& = & \bt J_0 + \half \bt \bt_K \theta^2  
- \frac{\bt \bt_K \theta^2}{2(1+\Theta_2^2)^{3/2}} \exp[-\frac{\bt_K\theta^2\Theta_2^2}{2J_0(1+\Theta_2^2)}]
 \cos(\Psi_2 + 3\tan^{-1}\Theta_2)  
\eeqr
Define
\beqr
\Sigma_0 & = & \bt J_0 , \;\;\; \Dl \Sigma = \half \bt \bt_K \theta^2, \\
A_2 & = & \frac{\bt \bt_K \theta^2}{2(1+\Theta_2^2)^{3/2}} \exp[-\frac{\bt_K\theta^2\Theta_2^2}{2J_0(1+\Theta_2^2)}]
\eeqr
Then
\beq
\lan x^2 \ran = \Sigma_0 + \Dl \Sigma - A_2 \cos(\Psi_2 + 3\tan^{-1}\Theta_2)
\eeq
The first term corresponds to the initial emittance emittance, the remaining terms represent the change. At long times, the last term will decay exponentially, so asymptotically at long times, the change is
\[  \lim_{t\rarw \infty} \Dl \lan x^2 \ran =  \half \bt \bt_K \theta^2 \]

The rms emittance is found from
\beqr
\eps &  = & [ \lan x^2 \ran \lan (x')^2 \ran - (\lan x x' \ran)^2 ] ^{1/2} \nonumber \\
     & = & \frac{1}{\bt}[\lan x^2 \ran \lan (p - \al x)^2 \ran - (\lan x(p - \al x)\ran)^2]^{1/2}
\nonumber \\
     & = & \frac{1}{\bt}[\lan x^2 \ran \lan p^2 \ran - (\lan x p \ran)^2]^{1/2}
\eeqr

We have
\beqr
\lan p^2 \ran & = & 2\bt \int_0^{\infty}dJ \int_0^{2\pi}d\phi  J \sin^2 \phi \psi_2(J,\phi,t)
\nonumber \\
& = & \Sigma_0 + \Dl \Sigma + A_2 \cos(\Psi_2 + 3\tan^{-1}\Theta_2)
\eeqr
while
\beqr
\lan x p \ran & = & 2\bt \int_0^{\infty}dJ \int_0^{2\pi}d\phi  J \sin \phi \cos\phi \psi_2(J,\phi,t) \nonumber \\
& = & A_2 \sin(\Psi_2 + 3\tan^{-1}\Theta_2)
\eeqr
To arrive at this result, we use the integration
\[
\int_0^{2\phi}d\phi \; \exp[-a \sin(\phi + b)]  \sin 2\phi  =  2\pi I_2(a)\sin 2b 
\]
Hence
 \[ \lan x p \ran = \bt J_0 a_K{\rm IP}\left(\frac{e^{i\om_{\bt}t}}{(1 - i\Theta_2)^3}\exp[\frac{a_K}{1 - i\Theta_2}]\right)
\]
Since we now take the imaginary part, the cosine function in $\lan x^2 \ran$ is replaced by the sine function. 

Hence the time dependent rms emittance is
\beq
\eps = \frac{1}{\bt}[ (\Sigma_0 + \Dl \Sigma)^2 - A_2(t)^2 ]^{1/2}
\eeq
At large times we expect $A_2 \rarw 0$, hence
\beq
\lim_{t\rarw \infty}\eps = \frac{1}{\bt}(\Sigma_0 + \Dl\Sigma) = J_0 + \half\bt \theta^2
\eeq

\clearpage
\section{Evolution of the tune spread after dipole kick}

The rms tune spread $\lan \sg(\nu) \ran$ is given by
\beq
\sg(\nu) = [ \lan (\Dl \nu)^2 \ran - (\lan \Dl\nu\ran)^2 ]^{1/2}  \label{eq: rms_dnu}
\eeq
Assuming as above that the tune spread as a function of $J$ is $\Dl\nu(J) = \nu' J$, the mean tune spread is
\beqr
\lan \Dl\nu \ran & = & \int dJ \int d\phi \Dl\nu(J) \psi_2(J,\phi,t) = \nu'  \int dJ \int d\phi J \psi_2(J,\phi,t) \nonumber \\
& = &  \frac{\nu'}{2\pi \eps_0}\exp[- \frac{\bt_K\theta^2}{2\eps_0}] \int dJ \; J e^{-J/\eps_0} \int d\phi \exp[-\frac{1}{\eps_0}\sqrt{2\bt_K J}\theta \sin(\phi - \om(J) t) ]
\eeqr
To integrate over $\phi$, we use
\beq
\int d\phi \exp[-a \sin(\phi + b)] = 2 \pi I_0(a)
\eeq
Hence 
\beq
\lan \Dl\nu \ran = \frac{\nu'}{\eps_0}\exp[- \frac{\bt_K\theta^2}{2\eps_0}] \int dJ \; J e^{-J/\eps_0} I_0(\frac{\sqrt{2\bt_K J}\theta}{\eps_0}  )
\eeq
This is independent of time. Introducing $z = J/\eps_0$ and using
\[ \int dz z \exp[-z] I_0(a \sqrt{z}) = (1 + \frac{a^2}{4}) \exp[\frac{a^2}{4}]
\]
Here with $a = \sqrt{2 \bt_K/\eps_0} \theta$, we have
\beq
\lan \Dl\nu \ran = \nu'\eps_0\exp[\half \bt_K\theta^2](1 + \frac{\beta_K \theta^2}{2 \eps_0})\exp[\frac{\bt_K\theta^2}{2\eps_0}]
=  \nu'\eps_0 (1 + \frac{\beta_K \theta^2}{2 \eps_0})
\eeq
And
\beq
\lan ( \Dl\nu )^2\ran = \frac{(\nu')^2}{\eps_0}\exp[- \frac{\bt_K\theta^2}{2\eps_0}] \int dJ \; J^2 e^{-J/\eps_0} I_0(\frac{\sqrt{2\bt_K J}\theta}{\eps_0}  )
\eeq
Using
\[ \int dz z^2 \exp[-z] I_0(a \sqrt{z}) = (2 + a^2 + (\frac{a^2}{4})^2) \exp[\frac{a^2}{4}]
\]
Hence
\beq
\lan ( \Dl\nu )^2\ran = (\nu' \eps_0)^2 [ 2 + 2 \frac{\bt_K \theta^2}{\eps_0} + (\frac{\beta_K \theta^2}{2 \eps_0})^2 ]
\eeq
Hence, the rms tune spread in the presence of the dipole kick is 
\beqr
\sg(\nu) & = & \nu' \eps_0 [  2 + 2 \frac{\bt_K \theta^2}{\eps_0} + (\frac{\beta_K \theta^2}{2 \eps_0})^2 - 
(1 + \frac{\beta_K \theta^2}{2 \eps_0})^2 ]^{1/2} \nonumber \\
& = & \nu' \eps_0 [ 1 + \frac{\bt_K \theta^2}{\eps_0} ]^{1/2}   \label{eq: rmsdnu}
\eeqr

\clearpage

\section{Higher order echoes following a dipole and quad kicks}

The $n$th order moment following these kicks is 
\beqr
\lan x^n(t) \ran & = & \int dJ \int d\phi (2\bt J)^{n/2}\cos^n\phi \psi_5(J,\phi,t) \\
\psi_5(J,\phi, t) & =  & \bt_k \theta \sqrt{\frac{2}{\bt}}\om'(J)\tau J^{3/2}\psi_0'
\sin [2( \phi - \om(J)(t-\tau))]\cos[\phi - \om(J)t]
\eeqr

\noi Second order moment echo 
\beq
\lan x^2(t) \ran = \bt_k \theta \sqrt{\frac{2}{\bt}}\tau 2\bt \int dJ \; \om'(J) J J^{3/2}\psi_0'
\int d\phi \cos^2 \phi \sin [2( \phi - \om(J)(t-\tau))]\cos[\phi - \om(J)t]
\eeq
The integration over $\phi$ is
\[ I = \int d\phi \cos^2 \phi \sin [2( \phi - \om(J)(t-\tau))]\cos[\phi - \om(J)t] \equiv 0 \]
because the integral decomposes into a sum of integrals of the form $\int d\phi \sin m\phi$, 
$m\ne 0$; all of which vanish. Note that any integration over a product such as
$[\sin,\cos] (m_1\phi + a_1) [\sin,\cos] (m_2\phi + a_2)...$ etc where the trig function can be
either sine or cosine always vanishes if the sum $m_1 + m_2 + ...= {\rm odd}$. 

Hence there is no echo in the second order moment.

\noi Echo in the third order moment
\beq
\lan x^3(t) \ran = \bt_k \theta \sqrt{\frac{2}{\bt}}\tau (2\bt)^{3/2} \int dJ \; \om'(J) J^{3/2}
 J^{3/2}\psi_0'
\int d\phi \cos^3 \phi \sin [2( \phi - \om(J)(t-\tau))]\cos[\phi - \om(J)t]
\eeq
The $\phi$ integration gives
\beqrs 
\int d\phi \cos^3 \phi \sin [2( \phi - \om(J)(t-\tau))]\cos[\phi - \om(J)t] & = &  -\fr{\pi}{8}
[ 3\sin \om(J)(t-2 \tau) + \sin \om(J)(3t-2\tau) ]
\eeqrs
Let $\om(J) = \om_{\bt} + \om' J$, then
\[
\lan x^3(t) \ran = -\fr{\pi}{2}\bt_k \bt \theta \om' \tau  \int dJ \;  J^3 \psi_0'
 [ 3\sin \om(J)(t-2 \tau) + \sin \om(J)(3t-2\tau) ]
\]
Define the phase variables
\beq
\Phi = \om_{\bt}(t - 2\tau),\;\;\; \xi_1 = \om'(t-2\tau)J_0, \;\;\; \Phi_3 = \om_{\bt}(3t - 2\tau), \;\;\;
\xi_3 = \om'(3 t-2\tau)J_0
\eeq
Substituting $\psi_0'= -(1/2\pi J_0^2)\exp[-J/J_0]$, we can write
\[
\lan x^3(t) \ran = \fr{1}{4 J_0^2}\bt_k \bt \theta \om' \tau  \int dJ \;  J^3 
{\rm Im}\left( 3 \exp[i\Phi] \exp[i \xi_1 J/J_0] + \exp[i\Phi_3] \exp[i \xi_3 J/J_0]\right)
\]
Introducing the integration variable $z = J/J_0$, we have two integrals of the form
\[ \int dz  \; z^3 \exp[-(1 - i b)z] = \fr{6}{(1- i b)^4} \]
Hence
\beq
\lan x^3(t) \ran = \fr{3}{2 }\bt_k \bt \theta \om' \tau J_0^2
{\rm Im}\left( 3 \fr{\exp[i\Phi]}{(1 - i\xi_1)^4} + \fr{\exp[i\Phi_3]}{(1 - i\xi_3)^4} \right)
\eeq
Let
\beqrs
1 + i\xi_1 & = & (1 + \xi_1^2)^{1/2}\exp[i \Theta_1], \;\;\; \Theta_1 = {\rm Arctan}[\xi_1] \\
1 + i\xi_3 & = & (1 + \xi_3^2)^{1/2}\exp[i \Theta_3], \;\;\; \Theta_1 = {\rm Arctan}[\xi_3] 
\eeqrs
Then
\[ \fr{1}{(1 - i\xi_1)^4} = \frac{(1 + i\xi_1)^4}{(1 + \xi_1^2)^4} = \fr{1}{(1 + \xi_1^2)^2}
\exp[4 i \Theta_1], \;\;\;\; \fr{1}{(1 - i\xi_3)^4} = \fr{1}{(1 + \xi_3^2)^2}\exp[4 i \Theta_3]
\]
Hence
\beq
\lan x^3(t) \ran = \fr{3}{2 }\bt_k \bt \theta \om' \tau J_0^2
\left( \fr{3}{(1 + \xi_1^2)^2} \sin (\Phi + 4\Theta_1) + \fr{}{(1 + \xi_3^2)^2} \sin (\Phi_3 + 
4\Theta_3)  \right)
\eeq
The first term has a maximum at $\xi_1 = 0$ or at $t  2 \tau$, the same time as the echo in 
$\lan x \ran$. The amplitude of this echo at $t= 2\tau$ is (ignoring the contribution from the 
second term) is
\beq
\lan x^3(t=2\tau) \ran^{amp} = \fr{9}{2 }\bt_k \bt \theta \om' \tau J_0^2 
\eeq
while the second term has a maximum at $\xi_3 = 0$ or at the earlier time $t = 2\tau/3$, its
amplitude being (again ignoring the contribution of the 1st term)
\beq
\lan x^3(t= \fr{2}{3}\tau) \ran^{amp} = \fr{3}{2 }\bt_k \bt \theta \om' \tau J_0^2 
\eeq
This echo at the earlier time has an amplitude one-third of the amplitude at the later time
$t= 2\tau$. 

Since the echo in $\lan x\ran$ at $t=2\tau$ has an amplitude 
$\lan x(2\tau)\ran^{amp} = \bt_K \theta \om' \tau J_0$, we can write
\beq
\lan x^3(t=2\tau) \ran^{amp} = \fr{9}{2 } \bt J_0 \lan x(2\tau)\ran^{amp} =
\fr{9}{2 } \sg_x^2 \lan x(2\tau)\ran^{amp}
\eeq
where $\sg_x$ is the initial beam size at the BPM. 

\noi {\sf Conjecture 1}: With an odd moment echo $\lan x^n\ran$, $n$ is odd, there is an echo
at this time
\[ T_n = \frac{2}{n}\tau \]

\noi {\sf Conjecture 2}: With an odd moment echo, there will be $n- 2$ echoes if $n > 1$. For
example, with $n=5$, there will be 3 echoes. 

\noi {\sf Above conjectures are wrong}: This follows from the integrations
\beqrs
\int d\phi \cos^5 \phi \sin [2( \phi - \om(J)(t-\tau))]\cos[\phi - \om(J)t] & = &  -\fr{5\pi}{2^5}
[ 2\sin \om(J)(t-2 \tau) + \sin \om(J)(3t-2\tau) ] \\
2\int d\phi \cos^7 \phi \sin [2( \phi - \om(J)(t-\tau))]\cos[\phi - \om(J)t] & = &  -\fr{7\pi}{2^7}
[ 5\sin \om(J)(t-2 \tau) + 3\sin \om(J)(3t-2\tau) ]
\eeqrs
Hence the higher order echoes are only at times $t= (2/3)\tau, 2\tau$.

\clearpage

\section{Decoherence functional, tune spread and all that}

In \cite{Stup_Park_Shil}, the rms tune spread is related to a decoherence functional.

First, using the density distribution function $\rho(J)$ is used to define a tune distribution function
$f(\Dl \nu)$ such that $f(\Dl \nu) \Dl\nu$ is the probability for a particle to have the tune shift $\Dl\nu$
within the range $d \Dl\nu$. This is defined via the tune shift function $\Dl \nu(J)$ as 
\beq
f(\Dl\nu) = \int dJ \rho(J) \dl(\Dl\nu - \Dl\nu(J)) 
\eeq
This distribution function can be used to calculate mean values as
\beq
\lan \Dl\nu \ran \equiv \int f(\Dl\nu) \Dl \nu \; d\Dl\nu = \int dJ  \rho(J) \Dl\nu(J)
\eeq
The decoherence functional is defined as the normalized time dependent centroid as 
\beq
K(t) = \frac{1}{\sqrt{\beta_K \beta}\theta} \lan x(t) \ran   \label{eq: centroid_K}
\eeq
where $\theta$ is the dipole kick. This can be written in units of turns $N$ with $t = N T_{rev}$.

{\em In the limit of an infinitesimally small dipole kick}, it was shown in \cite{Stup_Chao_93} that the decoherence
functional can be written in the form 
\beq
K(N) = \int_{-\infty}^{\infty}\sin[2\pi N(\nu_0 + \Dl \nu)] f(\Dl \nu) d\nu = {\rm Im}\left[e^{2\pi  i N \nu_0}
\int e^{2\pi  i N \Dl \nu} f(\Dl \nu) d\Dl \nu \right]   \label{eq: Kdef}
\eeq
This has the property that $K(N=0) = 1$ (because $f(\Dl\nu)$ is normalized to unity). This is not a property of
the definition of $K(N)$ in Eq.(\ref{eq: centroid_K}) above. More on this discrepancy below. 

The amplitude of the centroid oscillations is then given by the modulus of the complex integral
\beq
K_{amp}(N) = | \int e^{2\pi  i N \Dl \nu} f(\Dl \nu) d\Dl \nu |
\eeq
With this form of $K_{amp}(N)$, it follows that the tune spread $\sg(\nu)$ defined in Eq. (\ref{eq: rms_dnu}) 
can be found from
\beq
\sg(\nu) = \frac{1}{2\pi}\sqrt{- \frac{d^2}{DNA^2}K_{amp}(N)|_{N=0}}
\eeq
Comment: This is not obvious; if it is true, it must use the fact that $K_{amp}$ is the absolute value.
\beq
K_{amp}(N) = [ (\int \cos(2\pi N \Dl\nu)f(\Dl \nu) d\Dl \nu)^2 + (\int \sin(2\pi N \Dl\nu)f(\Dl \nu) d\Dl \nu)^2 ]^{1/2}
\eeq

Writing $\rho(J) = \frac{1}{2\pi \eps_0}\exp[- J/\eps_0]$, and $\Dl \nu(J)= \nu' J$, we have
\beq
f(\Dl\nu) = \frac{1}{2\pi \eps_0}\int dJ \exp[-J/ESP_0] \dl(\Dl \nu - \nu' J)  
 = \frac{1}{2\pi}\int dz \exp[-z] \dl(\Dl\nu - \nu' \eps_0 z)
\eeq
and therefore
\beqr
K_{amp}(N) & = & | \int dz \;  \exp[-( 1 - 2\pi  i N \nu' \eps_0) z ] | = 
| \frac{1}{1- 2\pi i N \nu'\eps_0}(1 - \lim_{z\rarw \infty}\exp[-z]\exp[ 2\pi  i N \nu' \eps_0 z] ) | \non \\
& = & \frac{1}{(1 + (2\pi N \nu'\eps_0)^2)}|1 + 2\pi i N \nu'\eps_0| = \frac{1}{(1 + (2\pi N \nu'\eps_0)^2)^{1/2}} 
 \label{eq: Kamp1}
\eeqr
This form for the centroid is {\em wrong}, as can be seen by comparing this with the exact form
\cite{Chao}, Eq. (14)
\beqr
\frac{\lan x(N)\ran_{amp}}{\sqrt{\bt \bt_K}\theta} & = & \sqrt{\frac{\bt_K}{\bt}}
\frac{1}{1 + (2\pi  \nu' \eps_0 N)^2}\exp[ -\frac{\bt_K \theta^2}{2 \eps_0} 
\frac{(2\pi \nu' \eps_0 N)^2}{1 + (2\pi \nu' \eps_0  N)^2}] \\
& = & \sqrt{\frac{\bt_K}{\bt}}\frac{1}{1 + (2\pi  \nu' \eps_0  N)^2}  + O(\theta)
\eeqr
Even dropping terms of $O(\theta)$  in the RHS and putting $\bt_K = \bt$, this does not reduce to the form of
$K_{amp}(N)$. 
Instead, the correct relation appears to be 
\beq
\frac{\lan x(N)\ran_{amp}}{\sqrt{\bt \bt_K}\theta} = |K_{amp}(N)|^2  + O(\theta)
\eeq

The correct expression above for the centroid comes from
\beqr
\lan x(t) \ran & = & \sqrt{2\bt \eps_0} \exp[-\frac{\bt_k\theta^2}{2 \eps_0}] {\rm Im}\left\{ 
e^{i \om_{\bt}t} \int dz \; \sqrt{z} \exp[-\{ 1 - i \om'\eps_0 t  \}z]  I_{1}(\frac{\sqrt{2}\bt_K\theta}{\sqrt{\bt\eps_0}}
\sqrt{z})  \right\} \\
 & \approx & \bt_K \theta {\rm Im}\left\{ e^{i \om_{\bt}t} \int dz \;  z  \exp[-\{ 1 - i \om'\eps_0 t  \}z]   \right\} 
 + O(\theta^2)
\eeqr
In the second line, we approximated $I_1(x) \approx x/2  + O(x^3)$

The problem is in the definition of $K(N)$. Comparing the last equation above with Eq.(\ref{eq: Kamp1}), we see
that $K_{amp}$ and hence $K$ have a missing factor of $z$ in the integrand. This is clear from the definition
of $K(N)$ in Eq.(\ref{eq: Kdef}) which has no information about the transverse amplitude of particles. 
Therefore even for infinitesimally small kicks the following holds
\beq
\frac{1}{\sqrt{\beta_K \beta}\theta} \lan x(N) \ran  \ne  \int_{-\infty}^{\infty}\sin[2\pi N(\nu_0 + \Dl \nu)] f(\Dl \nu) d\nu 
\eeq
The RHS is only an average over the phase but without averaging over the amplitude. 

While we can use the definitions of $K(N)$ in Eq.(\ref{eq: Kdef}) and $K_{amp}$ in Eq.(\ref{eq: Kamp1}), 
it is not clear how they are related to the decoherence of the centroid. 

Definitions of the decoherence time $N_{decoh}$ in \cite{Stup_Park_Shil}
\bit
\item At time $N_{decoh}$, $K_{amp}N_{decoh} = 0.5 K_{amp}(0) = 0.5$. From this definition and 
Eq.(\ref{eq: Kamp1}), it follows that 
\beq 
N_{decoh, 1} = \frac{\sqrt{3}}{2\pi \nu'\eps_0} = \frac{0.276}{\nu'\eps_0}
\eeq

\item Definition from feedback theory for emittance growth: The behavior of $K(N)$ is relevant in this case,
since a strong feedback system must damp beam oscillations quicker than the decoherence time. This
implies that only the initial stage of the decoherence process leads to residual emittance growth. The claim is \\
{\em Because $K_{amp}(N)$ has a quadratic dependence in the limit $N \rarw 0$} \\
an adequate definition is
\beq
N_{decoh} \equiv \frac{1}{\sqrt{- K_{amp}''|_{N=0}}} = \frac{1}{2\pi \sg(\nu)}
\eeq
Reason behind the claim: It may be true on general grounds that $K(N)$ starts with maximum value of 1 at $N=0$
and decreases thereafter. So $K_{amp}'(N = 0) = 0$ and $K_{amp}(N > 0) < K_{amp}(0)$. Why is this true?

In this case 
\[ K_{amp}(N) = 1 - \half (\frac{N}{N_{decoh}})^2 , \;\;\;\; N \ll N_{decoh}
\]
Using Eq.(\ref{eq: Kamp1}) again, it follows that 
\[ K_{amp}'(N) = - \frac{(2\pi \nu')^2 N}{[1 + (2\pi\nu' N)^2]}, \;\;\;  K_{amp}'(0) = 0, \; K_{amp}'(N) < 0 
\]
and from the 2nd derivative,
\beq
N_{decoh, 2} = \frac{1}{2\pi \nu'\eps_0} = \frac{0.159}{\nu'\eps_0}
\eeq

\item Definition from noise: Another definition comes from the effects of noise without feedback. Assume
that at each turn, the beam receives uncorrelated kicks of amplitude $a_m$ at turn $m$. Then the average
displacement after $N$ turns is 
\beq
\Dl x_c = \sum_{m=0}^N K(N - m) a_m
\eeq
Question: How does this follow from the definition of $K(m)$?  Note that this uses the oscillating decoherence
function $K(m)$ and not just the amplitude of this function. 

The averaged (over noise) squared displacement is (assuming uncorrelated kicks so that 
$\lan a_m a_n \ran = \lan a^2 \ran \dl_{mn}$ )
\beq
\lan (\Dl x_c)^2 \ran = \sum_{n=0}^N\sum_{m=0}^N K(N - m) K(N - n) \lan a_n a_m\ran =
\lan a^2 \ran \sum_{m=0}^N K^2(N-m) = \lan a^2 \ran \sum_{m=0}^N K^2(m)
\eeq
{\em Definition of decoherence time $N_{decoh}$:}
\beq
\lan (\Dl x_c)^2 \ran \equiv \half N_{decoh} \lan a^2 \ran
\eeq
Comment: This definition does not make sense. The average squared displacement grows with time due to
white noise, but the expression above assumes it becomes constant. Inconsistent definition.

Continuing with the above, we have
\beq
N_{decoh, 3} = 2 \sum_{m=0}^{\infty}K^2(m) \approx  \int_0^{\infty} K_{amp}^2(m) dm
\eeq
where in the last approximation it was assumed that $K(n)$ oscillates rapidly at the betatron frequency, so
that its squared average is half the squared average of the amplitude. 
Again using Eq.(\ref{eq: Kamp1}), we have
\beq
N_{decoh, 3} = \frac{1}{4 \nu' \eps_0} = \frac{0.25}{\nu'\eps_0}
\eeq

\item Hierarchy of decoherence times
\[ N_{decoh, 1} > N_{decoh, 3} > N_{decoh, 2} \]

\eit

{\sf Question: How would the definition of $K(N)$ be generalized for arbitrary kick amplitudes?}

The basic definition of $K(N)$ in Eq.(\ref{eq: Kdef}) is a Sine transform of the tune distribution function. 
The amplitude function $K_{amp}(N)$ is the same as the amplitude of the Fourier transform. 

\subsection{Decoherence time from the centroid evolution}

From the exact expression for the amplitude of the centroid
\beq
\lan x(N)\ran_{amp} =  \bt_K \theta 
\frac{1}{1 + (2\pi  \nu' \eps_0 N)^2}\exp[ -\frac{\bt_K \theta^2}{2 \eps_0} 
\frac{(2\pi \nu' \eps_0 N)^2}{1 + (2\pi \nu' \eps_0  N)^2} ]   \label{eq: cent_amp}
\eeq
we can find approximately the time at which the centroid amplitude falls to $1/e$ of its initial value.
Define the parameter
\beq
b^2 = \frac{\bt_K \theta^2}{2 \eps_0} = \half \frac{\bt}{\bt_K} a^2, \;\;\; a = \frac{\bt_K \theta}{\sg_0},
\;\;\; \sg_0 = \sqrt{\bt \eps_0}
\eeq
Here $a$ is the dipole kick amplitude relative to the rms beam size $\sg_0$. Setting the amplitude of the
exponential factor in Eq.(\ref{eq: cent_amp}) to -1 yields the approximate decoherence turn number
\beqr
N_{decoh}^{(1)}  & \approx & \frac{1}{\sqrt{b^2 - 1}} \frac{1}{2\pi \nu' \eps_0}   \nonumber \\
  & \approx  &   \frac{1}{2\pi \nu' \eps_0}  \frac{1}{\sqrt{\frac{\bt_K \theta^2}{2 \eps_0} - 1}}   
\label{eq: Ndecoh1}
\eeqr
This form can be written in terms of the rms tune spread after the dipole kick, given in Eq.(\ref{eq: rmsdnu}).
\[ b^2 = \half (\frac{\sg(\nu)}{\nu'\eps_0} - 1) \]
Hence, we also have
\beq
N_{decoh}^{(1)}  \approx \frac{1}{2\pi \nu' \eps_0}  \sqrt{\frac{2}{\frac{\sg(\nu)}{\nu'\eps_0} - 3}}
\eeq

At this time, the ratio of the centroid amplitude to its initial value is
\beq
\frac{\lan x(N_{decoh})\ran_{amp}}{\lan x(0) \ran_{amp}} = (1 - \frac{1}{b^2}) e^{-1} 
\eeq
The RHS approaches $1/e$ for $b \gg 1$. This analysis above assumes that $b > 1$, 

The completely general equation to be solved is 
\[ \frac{1}{1 + \Theta^2} \exp[ - b^2 \frac{\Theta^2}{1 + \Theta^2}]  = \exp[-1]  \]
where we defined $\Theta = 2\pi \nu' \eps_0 N$. This equation can be rewritten as
\beq
\ln(1 + \Theta^2) + b^2\frac{\Theta^2}{1 + \Theta^2} = 1
\eeq
When $b^2 \gg 1$, we have the solution in Eq.(\ref{eq: Ndecoh1}). 

In the opposite limit when $b^2 \ll 1$, i.e. for very weak kicks, if we drop the 2nd term, then in this limit
we have
\beq
N_{decoh}^{(2)}  \approx \sqrt{e - 1} \frac{1}{2\pi \nu' \eps_0} , \;\;\;\;  b^2 \ll 1
\eeq
Since we have 
\[  N_{decoh}^{(2)} >  N_{decoh}^{(1)} \]
this analysis shows that the decoherence time decreases as the dipole kick increases. 

The intermediate case when $b^2 \sim O{1}$ needs more work.

\clearpage

\section{Decoherence and echoes in 1D with chromatic tune spread}

Now the betatron tune depends on the synchrotron motion via the chromaticity $\chi$
\beq
\om(J, \dl) = \om_{\bt}[1 + \chi \dl(t)] + \om' J = \om_{\bt}[1 + \chi \hat{\dl} \cos(\om_s t + \phi_{s0})] + \om' J
\eeq
where $\hat{\dl}$ is the amplitude of the relative energy deviation for the particle and $\phi_{s0}$ is
the initial synchrotron phase of the particle. Note that the chromaticity $\chi$ is defined here as
\[ \chi \dl = \fr{\om(J=0, \dl) - \om_{\bt}}{\om_{\bt}} = \fr{\nu(J=0, \dl) - \nu_{\bt}}{\nu_{\bt}} \]
i.e. the chromaticity is scaled by the nominal tune $\nu_{\bt}$. Another definition of the linear chromaticity is to
define it as the 1st term in a Taylor series expansion of $\nu(\dl)$ as
\[ \nu(\dl) = \nu_{\bt} + \xi \dl + O(\dl^2) \]
Hence $\chi = \xi/\nu_{\bt}$. 

Alternatively, it may be better to consider the complete longitudinal phase space $(\hat{z}, \hat{\dl})$. In this 
case, assuming linear motion
\beq
\left[ \barr{c}\hat{z}(t) \\ \hat{\dl}(t) \earr \right] = \left[ \barr{cc}\cos \om_s t & \sin \om_s t \\ - \sin\om_s t & \cos\om_s t 
\earr\right]
\left[ \barr{c}\hat{z}_0 \\ \hat{\dl}_0 \earr \right]
\eeq
For this to be valid, $\hat{z}, \hat{\dl}$ must have the same dimensions. \\
 One way would be to scale the physical quantities $(z, \dl)$ by their rms values as 
\[ \hat{z} = \fr{z}{\sg_z},  \;\;\;  \hat{\dl} = \fr{\dl}{\sg_{\dl}}  \]

So the betatron tune changes with time as
\beq
\om(J,t) = \om_{\bt}[1 + \chi\sg_{\dl}\{ \hat{\dl}_0 \cos\om_s t - \hat{z}_0 \sin\om_s t\}] + \om' J
\eeq
In this case, the betatron phase advances with time as
\beq
\Dl \phi(t) = \int_0^t dt'\; \om(J,t') = (\om_{\bt}+\om' J)t + \chi \sg_{\dl}\fr{\om_{\bt}}{\om_s}[\hat{\dl}_0\sin\om_s t
 - \hat{z}_0(1 - \cos\om_s t)]
 \equiv \Dl\phi_{\bt}(t) + \Dl \phi_s(t)
\eeq
where $(\Dl \phi_{\bt}(t), \Dl\phi_s(t)$ denote the betatron and synchrotron motion contributions to the phase 
change, and $\Dl\phi(0) = 0$. 

If the initial transverse distribution is $\psi_0(J)$, after the dipole kick, the distribution is 
\beq
\psi_1(J,\phi) = \psi_0(x,p - \bt_K\theta) = \psi_0(J + \theta\sqrt{2J/\bt}\sin\phi + \half \bt_K \theta^2)
\eeq
We assume that the change in the betatron motion does not affect the longitudinal motion, so that there are
no changes to the longitudinal distributions. 

At time $t$ after the dipole kick, the distribution changes to
\beq
\psi_2(J,\phi,t) = \psi_1(J, \phi - \Dl\phi(t)) = \psi_0(J + \bt_K\theta\sqrt{2J/\bt}\sin\phi_{-t} + \half \bt_K \theta^2), 
\;\;\; \phi_{-t} = \phi - \Dl\phi(t)
\eeq

The centroid can be found by the usual procedure, except that the average must be done over initial coordinates
in longitudinal phase space
\beq
\lan x(t) \ran = \int dJ \int d\phi \int d\hat{z}_0 \int d\hat{\dl}_0 \psi_2(J,\phi,t) \psi_s(\hat{z}_0, \hat{\dl}_0) \sqrt{2\bt J}\cos\phi
\eeq
where we assumed that the transverse and longitudinal distributions are uncoupled. 

Assume that the longitudinal distributions are Gaussian, so that
\beq
\psi_s(\hat{z}_0, \hat{\dl}_0) = \frac{1}{2\pi  }\exp[-\frac{\hat{z}_0^2}{2} - \fr{\hat{\dl}_0^2}{2}]
\eeq
Substituting for $\psi_2$, we have
\beqr
\lan x(t) \ran &  = &  \int dJ \int d\phi \sqrt{2\bt J}\cos\phi \int d\hat{z}_0 \int d\hat{\dl}_0
 \psi_0(J + \bt_K\theta\sqrt{2J/\bt}\sin\phi_{-t} + \half \bt_K \theta^2) \psi_s(\hat{z}_0, \hat{\dl}_0)  \nonumber \\
& = & \fr{1}{2\pi\eps_0} \fr{1}{2\pi  }\sqrt{2\bt}e^{-\half \bt_K \theta^2}
\int dJ \int d\phi  \sqrt{J}e^{-J/\eps_0} \cos\phi  \nonumber \\
&  & \times \int dz_0 \int d\hat{\dl}_0 \exp[-\frac{\hat{z}_0^2}{2} - \fr{\hat{\dl}_0^2}{2}]
\exp[- \fr{\bt_K\theta\sqrt{2J/\bt}}{\eps_0} \sin( \phi - \Dl \phi_{\bt}(t) - \Dl \phi_s(t))
\eeqr
The $\hat{z}_0, \hat{\dl}_0$ integrations are of  the form 
\beq
I_s(a,b,c) = \int d\hat{z}_0 \int d\hat{\dl}_0 \exp[-\frac{\hat{z}_0^2}{2} - \fr{\hat{\dl}_0^2}{2}]
\exp[- a \sin(\phi_{-,\bt} + b \hat{z}_0 - c \hat{\dl}_0)]
\eeq
where 
\[ a = \fr{\bt_K\theta\sqrt{2J/\bt}}{\eps_0}, \;\;\; b = \chi\sg_{\dl} \fr{\om_{\bt}}{\om_s}(1 - \cos\om_s t), \;\;\;
c = \chi\sg_{\dl} \fr{\om_{\bt}}{\om_s}\sin\om_s t, \;\;\; \phi_{-,\bt} = \phi - \Dl\phi_{\bt}(t)
\]
We use the expansion
\[
e^{- a \sin\theta} = \sum_{n=-\infty}^{\infty}i^n I_n(a) e^{i n \theta}
\]
We obtain
\beqr
I_s(a,b,c) &  = & \sum_{n=-\infty}^{\infty}i^n I_n(a) e^{i n \phi_{-,\bt}}\int d\hat{z}_0 e^{-i n b \hat{z}_0}\exp[-\frac{\hat{z}_0^2}{2}]
\int d\hat{\dl}_0    e^{-i n c \hat{\dl}_0}  \exp[- \fr{\hat{\dl}_0^2}{2}]
\eeqr
We use the integrations
\[ \int_{-\infty}^{\infty} dx\; \exp[-a x^2] \cos[b x] = \sqrt{\fr{\pi}{a}} \exp[- \fr{b^2}{4 a}] , \;\;\;
\int_{-\infty}^{\infty} dx\; \exp[-a x^2] \sin[b x] = 0
\]
Hence
\beq
I_s(a,b,c) =  2\pi    \sum_{n=-\infty}^{\infty}i^n I_n(a) e^{i n \phi_{-,\bt}} 
\exp\left\{-\half[ (n b )^2 + (n c)^2] \right\}
\eeq
We have
\beqrs
\lan x(t) \ran & = & \fr{1}{2\pi \eps_0} \fr{1}{2\pi  }\sqrt{2\bt}e^{-\half \bt_K \theta^2}
\int dJ \int d\phi  \sqrt{J}e^{-J/\eps_0} \cos\phi I_s(a,b,c) \\
& = &  \fr{ \sqrt{2\bt}}{2\pi\eps_0}e^{-\half \bt_K \theta^2}
\int dJ \int d\phi  \sqrt{J}e^{-J/\eps_0} \cos\phi \sum_{n=-\infty}^{\infty}i^n I_n(a) e^{i n (\phi - \Dl\phi_{\bt})}
e^{-[ (n b )^2 + (n c )^2]/2} \\
& = &  \fr{\sqrt{2\bt}}{2\pi \eps_0} e^{-\half \bt_K \theta^2}
\int dJ \; \sqrt{J}e^{-J/\eps_0}  \sum_{n=-\infty}^{\infty}i^n I_n(a) e^{- i n \Dl\phi_{\bt})} e^{-[ (n b )^2 + (n c )^2]/2} \\
& & \times {\rm Re}\left[ \int d\phi \exp\{i (n+1)\phi \} \right] \\
& = &  \fr{1}{\eps_0} \sqrt{2\bt}e^{-\half \bt_K \theta^2}e^{-[ b^2 + c^2]/2}
{\rm Re}\left[ \int dJ \; \sqrt{J}e^{-J/\eps_0} i^{-1} I_{-1}(a)  e^{ i \Dl\phi_{\bt}}   \right] 
\eeqrs
Using
\[ {\rm Re}[ -i f(z)] = {\rm Im}[f(z)], \;\;\; I_{-m}(z) = I_m(z) \]
and introducing the scaled variable 
\[ u = J/\eps_0, \;\;\; a = \sqrt{2}\bt_K\theta \sqrt{\fr{u}{\bt \eps_0}} = b_2 \sqrt{u}; \;\;\; b_2 = 
\sqrt{2}\frac{\bt_K\theta}{\sg_0}
\]
we have
\[ \Dl \om_{\bt} = \om_{\bt} t + \om' \eps_0 u t \]
Hence
\beqr
\lan x(t) \ran & = & \sqrt{2\bt\eps_0}e^{-\half \bt_K \theta^2} e^{-\half[ ( b )^2 + ( c )^2] } 
{\rm Im}\left[ e^{i\om_{\bt}t} \int du \; \sqrt{u} I_{1}(b_2\sqrt{u})  e^{ -(1 - i \om'\eps_0 t)u}   \right]  \nonumber \\
& = & \bt_K\theta  e^{-\half[ ( b )^2 + ( c )^2] } 
{\rm Im}\left[ \fr{ e^{i\om_{\bt}t}}{(1 - i\om'\eps_0 t)^2}
\exp[ \fr{\bt_K\theta^2}{2\eps_0}\fr{i \om'\eps_0 t}{(1 - i\om'\eps_0 t)}] \right]   \nonumber \\
& = & \fr{\bt_K\theta}{1+ \Theta^2} e^{-\half[ ( b )^2 + ( c )^2] } 
\exp[ -\frac{\bt_K \theta^2}{2\eps_0}\fr{\Theta^2}{(1 + \Theta^2)}]  \nonumber \\
&  &  \times \sin[\om_{\bt}t + \frac{\bt_K \theta^2}{2\eps_0}\fr{\Theta}{(1 + \Theta^2)} + 2{\rm Arctan}\Theta]
\eeqr
where as before
\[ \Theta(t) = \om'\eps_0 t \]
Substituting for $b$ and $c$, 
\beqrs
 b^2 + c^2 & = & (\chi \sg_{\dl} \fr{\om_{\bt}}{\om_s})^2[ (1 - \cos\om_s t)^2 + \sin^2\om_s t]  \\
&  = & 2(\chi \sg_{\dl} \fr{\om_{\bt}}{\om_s})^2[1 - \cos\om_s t] = [2\chi \sg_{\dl} \fr{\om_{\bt}}{\om_s}\sin(\om_s t/2)]^2
\eeqrs
Hence the amplitude of the decoherence is in the presence of chromaticity
\beqr
\lan x(t) \ran_{amp} & = &  \fr{\bt_K\theta}{1+ \Theta^2} 
\exp\{- 2 (\fr{\chi \sg_{\dl}\om_{\bt}}{\om_s}\sin(\om_s t/2))^2 \} 
\exp[ -\frac{\bt_K \theta^2}{2\eps_0}\fr{\Theta^2}{(1 + \Theta^2)}]  \\
& \equiv & \exp[-\half \al^2] \lan x(t) \ran_{amp}(\chi=0) \\
\al & = & 2 \chi \sg_{\dl}\fr{\om_{\bt}}{\om_s}\sin(\om_s t/2)  = \sqrt{b^2 + c^2}
\eeqr
We see that chromaticity results in a multiplicative exponential factor which oscillates with the synchrotron
frequency. 

Note: The factor of $\al $ defined above is the same as the $\al$ defined in Eq.(7a) of \cite{Meller}, once
we take into account the difference in definition of chromaticities. Here we have defined the chromaticity
$\chi$ via the change of tune amplitude
\beq 
|\Dl\nu_{\bt}| = \nu_{\bt} \chi \dl
\eeq
while Meller et al define it via  $|\Dl \nu_{\bt}| = \chi \dl$.

\subsection{Emittance growth with chromatic tune spread}

The rms emittance is given by
\beq
\eps = \left[ \lan x^2 \ran \lan p^2 \ran - (\lan x p \ran)^2 \right]^{1/2}
\eeq
We have
\beq
\lan x^2 \ran = \int dJ \int d\phi \int d\hat{z}_0 \int d\hat{\dl}_0 \psi_2(J,\phi,t) \psi_s(\hat{z}_0, \hat{\dl}_0)
2\bt J\cos^2\phi
\eeq
The integrations over the longitudinal variables are unchanged, leaving us with
\beqrs
\lan x^2 \ran & = & \fr{1}{4\pi^2 \eps_0} 2\bt  e^{-\half \bt_K \theta^2/\eps_0}
\int dJ \int d\phi  \sqrt{J}e^{-J/\eps_0} \cos^2\phi  \; I_s(a,b,c) \\
& = & \fr{\bt}{2\pi \eps_0}   e^{-\half \bt_K \theta^2/\eps_0}
\int dJ \; J e^{-J/\eps_0}  \sum_{n=-\infty}^{\infty}i^n I_n(a) e^{- i n \Dl\phi_{\bt})} e^{-[ (n b )^2 + (n c )^2]/2} \\
& & \times \left[ \int d\phi \exp\{i n \phi \}[1 + \half(e^{2 i \phi} + e^{-2 i \phi})] \right] \\
& = & \fr{\bt}{\eps_0}   e^{-\half \bt_K \theta^2/ \eps_0}
\int dJ \; J e^{-J/\eps_0}  \sum_{n=-\infty}^{\infty}i^n I_n(a) e^{- i n \Dl\phi_{\bt})} e^{-[ (n b )^2 + (n c )^2]/2} \\
& & \times \left[ \dl_{n,0} + \half( \dl_{n,-2}+  \dl_{n,2}) \right] \\
& = & \bt \eps_0   e^{-\half \bt_K \theta^2/ \eps_0} 
\int du \; u e^u  \left[  I_0(a)  - I_2(a)e^{-2[ b^2 + c^2]} \cos 2\Dl\phi_{\bt} \right] \\
\eeqrs
For the 1st integral we use ($a = b_2 \sqrt{u}$)
\[  \int du \; u \exp[-u] I_0(b_2 \sqrt{u}) = (1 + \fr{b_2^2}{4})\exp[\fr{b_2^2}{4}]  \]
For the 2nd integral, we write it as
\beqrs
\mbox{} & = & \int du \; u e^u   I_2(b_2 \sqrt{u}) {\rm Re}[e^{2i(\om_{\bt}+\om'\eps_0 u)t}] \\
& = & {\rm Re}\left[e^{2i\om_{\bt}t} \int du \; u  I_2(b_2 \sqrt{u}) \exp[- (1 - 2i\om'\eps_0)u] \right] \\
& = & {\rm Re}\left[e^{2i\om_{\bt}t} \fr{b_2^2}{4(1 - 2i\om'\eps_0)^3} \exp[\fr{b_2^2}{4(1 - 2i\om'\eps_0)}]
\right]
  \eeqrs
  Let as before
  \beq
  \Theta = \om_{\bt}t, \; \; \; \Theta_2 = 2 \om' \eps_0 t, \;\;\; \Psi_2 = 2\Theta + \fr{b_2^2 \Theta_2}{4(1 + \Theta_2^2)}
  \eeq
  Evaluating the second term above separately,
  \beqrs
{\rm Re}\left[\fr{e^{2i\Theta}}{(1 - i\Theta_2)^3}\exp[\fr{b_2^2}{4(1 - i\Theta_2)}] \right]
  & = & \frac{1}{(1+ \Theta_2^2)^3}{\rm Re}\left[(1 + i\Theta_2)^3
    \exp[ 2i \Theta + \fr{b_2^2}{4(1 +\Theta_2)^2}(1 + i \Theta_2)] \right] \\
& = & \frac{1}{(1+ \Theta_2^2)^3}\exp[ \fr{b_2^2}{4(1 +\Theta_2^2)}]
{\rm Re}\left[(1 + i\Theta_2)^3 \exp[ i\{2\Theta+\fr{b_2^2}{4(1 +\Theta_2^2)} \Theta_2 \}] \right] \\
& = & \frac{1}{(1+ \Theta_2^2)^3}\exp[ \fr{b_2^2}{4(1 +\Theta_2^2)}]
{\rm Re}\left[(1 + i\Theta_2)^3 \exp[ i \Psi_2] \right] \\
& = & \frac{1}{(1+ \Theta_2^2)^3}\exp[ \fr{b_2^2}{4(1 +\Theta_2^2)}]
\left[(1 - 3\Theta_2^2)\cos\Psi_2 - (3\Theta_2 - \Theta_2^3)\sin\Psi_2 \right] \\
\eeqrs
Using
\[ (1 - 3\Theta_2^2)^2 + (3\Theta_2 - \Theta_2^3)^2 = (1 + \Theta_2^2)^3 \]
we can write
\[ (1 - 3\Theta_2^2)\cos\Psi_2 - (3\Theta_2 - \Theta_2^3)\sin\Psi_2  = (1 + \Theta_2^2)^{3/2} \cos(\Psi_2 + \Dl\Psi)
\]
where
\[ \tan \Dl\Psi = \frac{3\Theta_2 - \Theta_2^3}{1 - 3\Theta_2^2}, \Rarw \Dl\Psi = 3 {\rm Arctan}[\Theta_2]  \]
Combining the two terms from the integration (after using $b^2 + c^2 = \al^2$)
\beq
\lan x^2 \ran = \bt \eps_0  e^{-\half \bt_K \theta^2/ \eps_0} \left[
  (1 + \fr{b_2^2}{4})\exp[\fr{b_2^2}{4}] - e^{-2 \al^2}
  \frac{b_2^2}{4(1+ \Theta_2^2)^3}\exp[ \fr{b_2^2}{4(1 +\Theta_2^2)}]\cos(\Psi_2 + \Dl\Psi) \right]
\eeq
Since $b_2^2/4 = (\bt_K \theta)^2/(2 \bt \eps_0) = \half (\bt_K^2 /\bt) \theta^2/\eps_0 $ reduces to
$\half \bt_K \theta^2/\eps_0$ when $\bt_K \equiv \bt$, I should replace $\half \bt_K \theta^2/\eps_0$ by
$b_2^2/4$.  Hence
\beqrs
\lan x(t)^2 \ran & = &  \bt \eps_0  \left[ 1 + \fr{b_2^2}{4} -
  \frac{e^{-2 \al^2}}{(1+ \Theta_2^2)^3}\exp[ -\fr{b_2^2}{4}(1 - \fr{1}{(1 +\Theta_2^2)})]\cos(\Psi_2 + \Dl\Psi) \right] \\
& = &  \bt \eps_0  \left[ 1 + \fr{b_2^2}{4} -
  \frac{e^{-2 \al^2} b_2^2}{4(1+ \Theta_2^2)^3}\exp[ -\fr{b_2^2}{4}\fr{\Theta_2^2}{(1 +\Theta_2^2)}]\cos(\Psi_2 + \Dl\Psi) \right] \\
& = & \bt \eps_0  \left[ 1 + \fr{b_2^2}{4} - e^{-2 \al^2} A_2 \cos(\Psi_2 + \Dl\Psi) \right]
\eeqrs
where
\beqrs
A_2(t) & =  & \frac{b_2^2}{4(1+ \Theta_2^2)^3}\exp[ -\fr{b_2^2}{4}\fr{\Theta_2^2}{(1 +\Theta_2^2)}] \\
& = & \fr{(\bt_K \theta)^2}{2\bt \eps_0(1+ \Theta_2^2)^3}
\exp[ - \fr{(\bt_K \theta)^2 \Theta_2^2}{2\bt \eps_0(1+ \Theta_2^2)}
  \eeqrs
  At $t=0$,
  \[ \al(0) = 0, \;\;\; \Theta_2(0) = 0, \;\;\; A_2(0) = \frac{b_2^2}{4},  \;\;\; \Psi_2(0) = 0 = \Dl\Psi(0) \]
hence we have $\lan x(0)^2 \ran = \bt \eps_0$ as desired, In the opposite limit of
long times $\lim_{t\rarw \infty}A_2(t) = 0$, hence we have
  \beq
  \lim_{t\rarw \infty} \lan x(t)^2 \ran = \bt \eps_0  \left[ 1 + \fr{(\bt_K\theta)^2}{2\bt \eps_0} \right]
  \eeq
  The chromaticity has no impact on the asymptotic value of  $ \lan x(t)^2 \ran$. 

  The remaining rms values are
\beqrs
\lan p^2 \ran & = &  \int dJ \int d\phi \int d\hat{z}_0 \int d\hat{\dl}_0 \psi_2(J,\phi,t) \psi_s(\hat{z}_0, \hat{\dl}_0)
2\bt J\sin^2\phi \\
& = & \bt \eps_0  \left[ 1 + \fr{b_2^2}{4} + e^{-2 \al^2} A_2 \cos(\Psi_2 + \Dl\Psi) \right]
\eeqrs
At $t = 0$, we have
\[ \lan p(0)^2 \ran = \bt \eps_0  [ 1 + \fr{b_2^2}{2}] = \bt\eps_0 + (\bt_K \theta)^2 \]

Hence
\[ \lan x^2 \ran \lan p^2 \ran = 
(\bt \eps_0  \left[ (1 + \fr{b_2^2}{4})^2 - e^{-4 \al^2}  A_2^2 \cos^2(\Psi_2 + \Dl\Psi) \right]
\]

Similarly,
\beqrs
\lan x p \ran & = &  \int dJ \int d\phi \int d\hat{z}_0 \int d\hat{\dl}_0 \psi_2(J,\phi,t) \psi_s(\hat{z}_0, \hat{\dl}_0)
2\bt J\cos \phi \sin\phi \\
& = & \bt \eps_0 e^{-2 \al^2} \fr{b_2^2}{4}{\rm Im}\left( \fr{e^{2i\Theta}}{(1 - i\Theta_2)^3}\exp[\fr{b_2^2}{4(1 - i\Theta_2)}] \right) \\
& = & \bt \eps_0 e^{-2 \al^2} A_2 \sin(\Psi_2 + \Dl\Psi)
\eeqrs
It follows that the rms emittance is
\beqr
\eps(t) & = &\eps_0  \left[(1 + \fr{b_2^2}{4})^2 -  e^{-4 \al^2} A_2^2 \cos^2(\Psi_2 + \Dl\Psi)
  - e^{-4 \al^2} A_2^2 \sin^2(\Psi_2 + \Dl\Psi) \right]^{1/2} \nonumber \\
& = & \eps_0  \left[(1 + \fr{b_2^2}{4})^2 - e^{-4 \al^2} A_2^2 \right]^{1/2}
\eeqr
At long times, we have
\beq
\lim_{t\rarw \infty}\eps(t) = \eps_0 (1 + \fr{(\bt_K \theta)^2}{2\bt \eps_0})
\eeq
The asymptotic emittance is not affected by the chromaticity.
At intermediate times, the time dependent factor is changed by the multiplicative factor
\[ \exp[- 4 \al(t)^2] = \exp[ - (4 \chi \sg_{\dl}\fr{\om_{\bt}}{\om_s}\sin(\om_s t/2))^2] \]

Time constant for the emittance decay
\beqrs
2\eps \fr{d\eps}{dt} & = &  - \eps_0^2 \left[ 2 e^{-4\al^2} A_2 \fr{d A_2}{dt} - 8\al\fr{d\al}{dt}e^{-4\al^2}A_2^2 \right] \\
& = & - 2\eps_0^2  e^{-4\al^2} A_2 \left[ \fr{d A_2}{dt} - 4\al\fr{d\al}{dt} A_2 \right] \\
\Rarw \fr{1}{\eps} \fr{d\eps}{dt} & = & \fr{1}{ [(1 + \fr{b_2^2}{4})^2 - e^{-4 \al^2} A_2^2 ]  }
e^{-4\al^2} A_2 \left[ \fr{d A_2}{dt} - 4\al\fr{d\al}{dt} A_2 \right]
\eeqrs

\subsection{Echo amplitude with chromatic tune spread}

Without the chromatic tune spread and in the simplified nonlinear dipole theory we had 
for the dipole moment
\beqr
\lan x(t) \ran & = & \frac{\sqrt{2\bt \eps_0}}{2\pi} \exp[-\frac{\bt_k\theta^2}{2 \eps_0}]\int dz \; \sqrt{z} \exp[- z] 
T_{\phi}(z) \\
T_{\phi}(z)  & = &  {\rm Re} \left\{  \int d\phi e^{i\phi} \exp\left[ - a_{\theta}\sqrt{2z} \sin( \phi_{- \Dl\phi} - \tau \om - 
\half q + Q z  \sin 2\phi_{-\Dl \phi} ) \right] \right\}
\eeqr
where before $\Dl \phi = \om(J)t = (\om_{\bt} + \om' \eps_0 u)t \equiv \Dl \om_{\bt} $. To include the chromatic
effects, we replace $\Dl\phi$ in $T_{\phi}$ above by $\Dl\phi_{\bt} + \Dl\phi_s$ where
\beq
\Dl \phi_s(t) = \chi \sg_{\dl} \fr{\om_{\bt}}{\om_s} [\hat{\dl}_0\sin\om_s t - \hat{z}_0(1 - \cos\om_s t)] 
\equiv b \hat{\dl}_0 - c \hat{z}_0
\eeq
and average over $(\hat{\dl}_0, \hat{z}_0)$.

\clearpage

\section{Analysis of echoes with diffusion}

I follow Chao's notation. The coordinates used are $x, p$ and action angle coordinates $J, \phi$ are related as
\beqrs
x = \sqrt{2\bt J}\cos\phi , \; \; \; & \;\;\; p = \al x + \bt x' = - \sqrt{2\bt J} \sin\phi  \\
J = \frac{1}{2\bt}(x^2 + p^2)  ,   &   \tan\phi = -\frac{p}{x} \\
\frac{\del J}{\del x} = \sqrt{\frac{2J}{\bt}}\cos\phi  & \frac{\del J}{\del p} = -\sqrt{\frac{2J}{\bt}}\sin\phi
\eeqrs

The initial distribution is exponential in the action
\beq
\psi_0(J) = \frac{1}{2\pi J_0}\exp[ - \frac{J}{J_0}]
\eeq
where $J_0$ is the average action of the beam distribution and related to the rms emittance $\eps$ by
\[ J_0 = \eps \]
Following the dipole kick by an angle $\theta$, the beam distribution is
\beq
\psi_1(J, \phi) = \psi_0(x, p - \bt \theta) \simeq \psi_0(J) - \bt \theta\psi_0' \frac{\del J}{\del p} + O[(\bt\theta)^2]
\eeq
In the absence of diffusion, the distribution would obey
\[ \psi_2(J,\phi,t) = \psi_1(J, \phi - \om(J)t) \]
In the presence of diffusion, the distribution $\psi_2$ after the dipole kick evolves according to the diffusion equation
\beq
\frac{\del\psi_2}{\del t} = \frac{\del }{\del J}[ D(J) \frac{\del \psi_2}{\del J}]
\eeq
To solve this equation, transform from the independent variables $J, \phi$ to $J, v = \phi - \om(J)t$. The derivatives transform as
\[ \frac{\del}{\del J} \rarw \frac{\del }{\del J} + \frac{\del v}{\del J}\frac{\del}{\del v} = 
\frac{\del }{\del J} -  \om' t \frac{\del}{\del v}, \; \; \; 
 \frac{\del}{\del \phi} \rarw  \frac{\del v}{\del \phi}\frac{\del}{\del v} = \frac{\del}{\del v}
\]
The RHS of the diffusion equation transforms to
\beqrs
\frac{\del }{\del J}[ D(J) \frac{\del \psi_2}{\del J}]  & \rarw & [\frac{\del }{\del J} -  
\om' t \frac{\del}{\del v}] [D(J)(\frac{\del }{\del J} -  \om' t \frac{\del}{\del v})]\psi_2 \\
 & = & \frac{\del }{\del J}[D(J)\frac{\del \psi_2 }{\del J}] -  t\frac{\del }{\del J} 
 [D(J)\om^{'} \frac{\del \psi_2}{\del v}] - 
 \om' t D(J) \frac{\del^2 \psi_2}{\del v \del J} + (\om' t)^2 D(J) \frac{\del^2 \psi_2}{\del v^2}
\eeqrs
Assuming for the moment that $D(J)= D_0$, const. and $\om(J) = \om_0 + \om' J$, the diffusion 
equation simplifies to
\[ 
\frac{\del\psi_2}{\del t} = D_0[\frac{\del^2 \psi_2 }{\del J^2}
- 2\om' t \frac{\del^2 \psi_2 }{\del J\del v} + (\om' t)^2\frac{\del^2 \psi_2 }{\del v^2}]
\]
Assuming that the phase variations in the distribution are more significant than the 
amplitude variations and keeping only the term that increases fastest with time, we have
the simplified diffusion equation
\[
\frac{\del\psi_2}{\del t} = D_0 (\om' t)^2\frac{\del^2 \psi_2 }{\del v^2}
\]
This assumes the following
\[ \frac{\del^2 \psi_2 }{\del J^2}, |2\om' t \frac{\del^2 \psi_2 }{\del J\del v}|
\ll (\om' t)^2\frac{\del^2 \psi_2 }{\del v^2}
\]
We will check whether these conditions are satisfied when we have constructed some solutions.
Under these conditions, the above diffusion equation would also be valid when the diffusion
coefficient is a function of the action, hence we have
\beq
\frac{\del\psi_2}{\del t} = D(J) (\om' t)^2\frac{\del^2 \psi_2 }{\del v^2}
\label{eq: diffeq_simple}
\eeq
The solution to a PDE of the form
\[ \psi_t = A(t) \psi_{vv} \] 
is of the form (by separation of variables)
\[ \psi(t) = \psi(0)\exp[- \int A(t)dt] \sin v \]
Hence the solution to Eq.(\ref{eq: diffeq_simple}) is 
\beq
\psi_2(J,v,t) = \psi_2(J,v,0)\exp[-\frac{1}{3}D(J)(\om')^2 t^3]\sin v
\eeq
where $\psi_2(J,v,0)=\psi_1(J,v,0)=\psi_0(J)+\theta\sqrt{2\bt J}\psi_0'(J)$. Since the
1st term $\psi_0(J)$ will not contribute to the dipole moment, it can be dropped. Hence
\beq
\psi_2(J,v,t) = \theta\sqrt{2\bt J}\psi_0'(J)\exp[-\frac{1}{3}D(J)(\om')^2 t^3]\sin v
\eeq

We now check if the assumptions made in writing the simplified diffusion equation above
are valid for this solution. For simplicity here we assume $D(J)=D_0$
\beqrs
\frac{\del^2 \psi_2}{\del J^2} & = & \theta\sqrt{2\bt}\exp[-\frac{1}{3}D_0\om' t^3] \sin v \\
&  &  \frac{1}{4 J^{3/2}}[4 J\psi_0^{''}+ 4J^2 \psi_0^{'''}-\psi_0^{'}] \\
&  &  - \frac{2D_0}{3 J^{1/2}}[2J\om^{'}\om^{''}\psi_0^{''} + 
\psi_0\om^{'}( J(\om^{''})^2  + \om^{'}(\om^{''}+ J \om^{'''}))]t^3  \\
&  & + \frac{4}{9}J^{1/2} D_0^2 (\om^{'} \om^{''})^2 \psi_0^{'} t^6
\eeqrs
Keeping the dominant $t^6$ term at long time, the condition 
\beq
 \frac{\del^2 \psi_2}{\del J^2} \ll (\om't)^2 \frac{\del^2 \psi_2}{\del v^2} 
\label{eq: 2nd_derivs}
\eeq
requires that
\[ \frac{4}{9} D_0^2 (\om^{'} \om^{''})^2 t^6 \ll (\om^{'} t)^2 \]
We can consider two cases
\bit
\item  $\om^{''} = 0$

In this case, the $t^3$ terms also vanish, and we are left with the condition

\[ 
J^2 \frac{\psi_0^{'''}}{\psi_0^{'}} + J\frac{\psi_0^{''}}{\psi_0^{'}} - \frac{1}{4} 
\ll J^2(\om^{'} t)^2 
\]
Applying this to the initial distribution
\[ \psi_0(J) = \frac{1}{2\pi J_0}\exp[- \frac{J}{J_0}]
\]
requires 
\[ (\frac{J}{J_0})^2 - \frac{J}{J_0} - \frac{1}{4} \ll (\om^{'} t)^2 \]
This implies that at a given time $t$, the approximation makes the diffusion equation valid
in actions utp a value $J$ satisfying this inequality.

\item $\om^{''} \neq 0$

In this case, the analysis is valid for times $t$ and weak diffusion $D_0$ and
detuning so that
\[ \frac{4}{9} D_0^2 (\om^{''})^2 t^4 \ll 1 \]
\eit

{\tt Continuation of echo analysis }

Assuming that these conditions are obeyed
\[ |\frac{\del^2 \psi_2}{\del J^2}|, 2|\om' t \frac{\del^2\psi_2}{\del J \del v}| \ll (\om' t)^2 \frac{\del^2\psi_2}{\del v^2} \]
the distribution obeys
\beq
\frac{\del \psi_2}{\del t} = D(J)(\om' t)^2 \frac{\del^2 \psi_2}{\del v^2}
\eeq
The solution of 
\[ \psi_t = f(t) \psi_{vv} \]
is of the form
\[ \psi(v, t) = \psi(0) \exp[-\int f(t) dt]\sin(v + c) \]
where $\psi(0)$ is the distribution at time $t=0$ and $c$ is an arbitrary constant.  In our case to match the distribution at $t=0$, I will put $c=0$. Then
\beq
\psi_2(J.v,t) = \psi_2(J,v,0)\exp[-\frac{1}{3}D(J) (\om')^2 t^3] \sin v 
\eeq
We have
\[ \psi_2(J,v,0) = \psi_1(J,v,0) = \psi_0(J) + \theta \sqrt{2\bt J}\psi_0^{'}(J)\sin v \]
Since $\psi_0(J)$ does not contribute to the dipole moment, it can be dropped. Thus
\beq
\psi_2(J,v,t) = \theta \sqrt{2\bt J}\psi_0^{'}(J)\exp[-\frac{1}{3}D(J) (\om')^2 t^3] \sin v
\eeq
where $v = \phi - \om t$. Following the quad kick at time $t = \tau$, 
\beqrs
\psi_4(J,\phi) & = & \psi_3(x, p + qx) \approx \psi_3(J,\phi) + qx \frac{\del \psi_3}{\del p} \\
& = & \psi_3(J,\phi) - q\sqrt{2\bt J}\cos\phi [ \sqrt{\frac{2J}{\bt}}\sin\phi 
\frac{\del\psi_3}{\del J} + \frac{1}{\sqrt{2\bt J}}\sin\phi \frac{\del\psi_3}{\del\phi} ]
\eeqrs
Keeping only the dominant term from $\del\phi_3/\del J$, 
\[ \frac{\del \psi_3}{\del J} \simeq - \theta \om' \tau \psi_0'\sqrt{2\bt J}\cos(\phi - \om \tau) \]
Hence, keeping only the term linear in quadrupole strength $q$, 
\[ \psi_4(J,\phi) = \theta q (\om' \tau) \sqrt{2\bt J}\cos\phi \exp[-\frac{1}{3}D(J) (\om')^2 t^3]\sin 2\phi \cos(\phi - \om \tau) \]
At times $t > \tau$, the distribution is given by the diffusion equation
\beq
\frac{\del\psi_5}{\del t} =\frac{\del}{\del J} [ D(J)\frac{\del \psi_5}{\del J} ]
\eeq
Changing the phase variable from $\phi$ to $u = \phi - \om(J) (t - \tau)$
 \[ \frac{\del}{\del J} \rarw \frac{\del}{\del J} - \om'(t - \tau)\frac{\del}{\del u} \]
and the diffusion equation for $\psi_5$ is
\[ \frac{\del}{\del t}\psi_5 = [\frac{\del}{\del J} - \om'(t - \tau)\frac{\del}{\del u}][
D(J) (\frac{\del}{\del J} - \om'(t - \tau)\frac{\del}{\del u})]\psi_5
\]
Under the approximation that $[\om'(t-\tau)]^2 \del^2\psi_5/\del u^2$ is the dominant term on the RHS,
the diffusion equation simplifies to
\beq
\frac{\del}{\del t}\psi_5 = [\om'(t-\tau)]^2 D(J) \frac{\del^2\psi_5}{\del u^2}
\eeq
Assuming that the initial condition satisfies $\psi_5(J,\phi,t=\tau) = \psi_4(J,\phi)$, the solution
can be written as
\beq
\psi_5(J,\phi,t) = \psi_4(J,\phi)\exp[-\frac{1}{3}D(J)(\om')^2 (t - \tau)^3]\sin (u + c)
\eeq
where 
\[
\psi_4(J,\phi) = \theta q (\om' \tau)\sqrt{2\bt J}J\psi_0'\exp[-\frac{1}{3}D(J)(\om')^2 \tau^3]
\sin 2\phi \cos(\phi - \om\tau) 
\]
Writing the product of trigonometric terms as
\[ \sin 2\phi \cos(\phi - \om\tau)  = \half[\sin(3\phi-\om\tau) + \sin(\phi+ \om\tau)]
\]
The first term is a third harmonic and gives rise to a sextupole echo and may have a negligible
impact on the dipole echo. Assuming this is the case, er can keep only the second term above.

Matching the solutions for $\psi_5$ and $\psi_4$ at $t=\tau$ requires 
\[ \sin(u + c)|_{t=\tau} = \sin(\phi + c) = \sin(\phi + \om \tau), \;\;\; c = \om \tau \]

{\tt Consider the consequences of ignoring the sextupole echo term. It has the same weight as the
dipole term but it could average out in the dipole moment. Also $\psi_5$ and $\psi_4$ could not be
matched at $t\tau$. Also the normalization is not preserved, i.e.
\[ \int \psi_4(J,\phi)dJ d\phi \ne \int \psi_5(J,\phi)dJ d\phi \]
}

With the above and $u+c = \phi - \om(t-2\tau)$, we have
\beq
\psi_5(J,\phi,t) = \half \theta q (\om' \tau)\sqrt{2\bt J}J\psi_0'
\exp[-\frac{1}{3}D(J)(\om')^2 ((t-\tau)^3 + \tau^3)]\sin (\phi - \om(t-2\tau))
\eeq
The echo amplitude is 
\beqrs
\lan x \ran & = & \sqrt{2\bt}\int dJ \sqrt{J} \int d\phi \cos\phi \psi_5 \\
 & = & \half\theta q (2\bt \tau) \int dJ \om' J^2 \psi_0'\exp[-\frac{1}{3}D(J)(\om')^2 t_1^3]
\int \cos \phi \sin(\phi - \om(t-2\tau)) \\
 & = & -\pi\bt \theta q \tau\int dJ \om' J^2 \psi_0'\exp[-\frac{1}{3}D(J)(\om')^2 t_1^3]
\sin(\om(t - 2\tau))
\eeqrs
where we defined 
\[ t_1^3 = (t-\tau)^3 + \tau^3 \]

Consider the complexified form of the above integral
\[ I = {\rm Im}[\int dJ \om' J^2 \psi_0'\exp[-\frac{1}{3}D(J)(\om')^2 t_1^3]
\exp[i(\om(t - 2\tau))] 
\]
In Ref.\cite{Stup_Chao_93}, the statement is made that assuming that $\psi_0'$ is monotonic, a Schwarz
inequality can be used to show that the maximum of the integral occurs at $t=2\tau$ where the 
complex exponential factor assumes its largest value of unity. 

In most of the following, we consider the initial distribution and the action dependent 
transverse angular frequency to be of the form
\beq
\psi_0(J)  =  \frac{1}{2\pi J_0}\exp[-\frac{J}{J_0}], \;\;\;\; \om(J) = \om_0 + \om' J
\eeq

\subsection{Constant diffusion coefficient}

\[ D(J) = D_0 \]
The dipole moment at time $t$ is 
\[
\lan x \ran(t) = \frac{\pi \bt \theta q \tau \om'}{2\pi J_0^2} \exp[-\frac{1}{3}D_0(\om')^2 t_1^3]
{\rm Im}[ \exp[i\om_0(t-2\tau)]\int J^2 \exp[-\frac{i\om'(t-2\tau)J}{J_0}] dJ 
\]
Using the result 
\[ \int_0^{\infty} dJ J^2 \exp[-a J] = \frac{2}{a^3} \]
we have
\[
\lan x \ran(t) = \bt\theta q \tau \om' J_0\exp[-\frac{1}{3}D_0(\om')^2 t_1^3]
{\rm Im}[ \frac{\exp[i\Phi]}{(1+i\xi)^3}]
\]
Expanding we have
\[ 
\lan x \ran(t) = \frac{\bt\theta q \tau \om' J_0}{(1+ \xi^2)^{3/2}}
[\xi(3 - \xi^2)\cos\Phi + (1 - 3\xi^2)\sin\Phi]
\exp[-\frac{1}{3}D_0(\om')^2 t_1^3]
\]
At $t=2\tau$, both $\xi and \Phi$ vanish, hence so does $\lan x\ran$. However the echo amplitude
achieves its maximum in the vicinity of $2\tau$. 

The maximum amplitude near $t=2\tau$ is 
\beq
\lan x \ran(2\tau) = \bt\theta q \tau \om' J_0
\exp[-\frac{2}{3}D_0(\om')^2 \tau^3]
\eeq
As a function of the delay $\tau$, the amplitude has a maximum at a delay $\tau_m$ such that
\beq
\tau_m^3 = \frac{1}{2D_0(\om')^2}
\eeq
and the maximum amplitude at this delay is
\beq
\lan x \ran(2\tau_m) = \frac{\bt\theta q J_0}{2D_0 \om'}
\eeq
From a knowledge of both $\tau_m$ and the amplitude at $2\tau_m$, both the diffusion coefficient and the detuning
can be determined. 

\vspace{2em}
\noi \underline{Recap of assumptions}
\benu
\item The dipole kick amplitude is much less than the beam size
\[ \bt \theta \ll \sg \]

\item The quadrupole kick is also small enough for the 1st order Taylor expansion to be valid or that it satisfies 
\[  q \ll 
\]

\item The distribution function obeys the diffusion equation in action $J$ alone
\[ \frac{\del}{\del t}\psi = \frac{\del}{\del J}[D(J)\frac{\del \psi}{\del J}] \]

\item The diffusion equation is solved under the approximations 
\beqrs
\frac{\del}{\del J}[D(J)\frac{\del \psi_2}{\del J}] & \ll & (\om' t)^2 D(J) \frac{\del^2\psi_2}{\del v^2}  \\
| t\frac{\del }{\del J}[D(J) \om' \frac{\del \psi_2}{\del v}] | & \ll &  (\om' t)^2 D(J) \frac{\del^2\psi_2}{\del v^2}
\eeqrs

\item In the complete expression for the distribution $\psi_4$ right after the quad kick is 
\[ \psi_4 = \theta q (\om' \tau) \sqrt{2\bt J}J \psi_0' \exp[-\frac{1}{3}D_0 (\om')^2 \tau^3]\half
[\sin(3\phi - \om\tau) + \sin(\phi + \om\tau) ]
\]
The 3rd harmonic term is dropped in the evolution of the distribution with diffusion following the quad kick because it is a sextupole term which will not contribute to the dipole echo. 

Q: What are the consequences of dropping this term?

\item In the diffusion equation for $\psi_5$, the assumptions  are
\beqrs
\frac{\del}{\del J}[D(J)\frac{\del \psi_5}{\del J}] & \ll & (\om'( t-\tau))^2 D(J) \frac{\del^2\psi_5}{\del v^2}  \\
| (t-\tau)\frac{\del }{\del J}[D(J) \om' \frac{\del \psi_5}{\del v}] | & \ll &  (\om'( t- \tau))^2 D(J) \frac{\del^2\psi_5}{\del v^2}
\eeqrs
Are these inequalities satisfied at time $t = \tau_m, 2\tau_m$?

\item The solution of the diffusion equation 
\beq
 \frac{\del\psi_5}{\del t} = (\om' (t - \tau))^2 D(J) \frac{\del^2 \psi_5}{\del u^2}
\label{eq: diff_psi5}
\eeq
is of the form
\[ \psi_5(J,\phi,t \ge \tau) = A \exp[-\frac{1}{3}D(J) (\om')^2 (t-\tau)^3]\sin(u + c) \]
has to match the solution at $t= \tau$. i.e. 
\[ \psi_5(J,\phi, \tau) = \psi_4(J, \phi) \]
This can be done only by dropping the $\sin(3\phi - \om\tau)$ term in $\psi_4$.

Is it possible to solve the equation Eq. (\ref{eq: diff_psi5}) by use of a Fourier series with more terms?

\eenu

\vspace{2em}

\noi Check inequalities for $\psi_2$

Is this satisfied for $0 \le t \le \tau$
\[ | \frac{\del }{\del J}[D(J) \frac{\del\psi_2}{\del J}] | \ll 
| (\om' t)^2 D(J)\frac{\del^2\psi_2}{\del v^2} |
 \]
where 
\[ \psi_2 = \theta\sqrt{2\bt}(\sqrt{J}\psi_0')\exp[-\frac{1}{3}D(J)(\om')^2 t^3]\sin v \]
For simplicity we assume that $D$ and $\om'$ are constants
\[ D(J)= D_0, \;\;\; \om = \om_0 + \om' J \]
In this case, the inequality can be written as
\[ 
| \frac{J^2}{J_0^2} - \frac{J}{J_0}  - \frac{1}{4} | \ll (\om^{'} t)^2 
\]
The above can be written as 
\[ (\frac{J}{J_0} - \half)^2 - \half \ll  (\om^{'} t)^2  J^2 \]

Is this satisfied at $t = \tau_m =[ (1/(2D_0(\om')^2))]^{1/3}$? The above inequality becomes

\[ \frac{J}{J_0} \ll \half + [\half + J^2 (\frac{\om'}{2D_0})^{2/3}]^{1/2}
\]
{\tt Note: Explore the consequences of this inequality. Does it set limits on $D_0$?}

\vspace{2em}

\noi Diffusion constant in terms of the optimum delay \newline
Writing
\[ \om' = \frac{\mu}{J_0}\om_{rev} = \frac{\mu}{\eps} \om_{rev} \]
where we have used $J_0 = \eps$, 
Then, the constant coefficient $D_0$ is 
\beq
D_0 = \frac{1}{2} (\frac{\eps}{\om_{rev}})^2 \frac{1}{\mu^2 \tau_m^3}
\eeq

\noi Optimum detuning $\mu_m$ \newline
The maximum of the echo amplitude as a function of the detuning occurs when $(\om_m')^2 = (3/4)/(D_0\tau^3)$ or
the diffusion coefficient $D_0$ is 
\beq
D_0 = \frac{3}{16}\frac{1}{\mu_m^2 \tau^3} 
\eeq

\subsubsection{FWHM calculation}

At a time $t = 2\tau + \Dl t_H$, the amplitude falls to half the value at $t = 2\tau$, hence
\[
\frac{\bt \theta q J_0 \om' \tau}{[1 + (\om' J_0 \Dl t_H)^2]^{3/2}} \exp[-\frac{1}{3}D_0(\om')^2[(t+\Dl t_H)^3 + \tau^3]] = \half \bt \theta q J_0 \om' \tau \exp[-\frac{2}{3}D_0(\om')^2 \tau^3] 
\]
Expanding to 1st order in $\Dl t_H/\tau$, so that $ (\tau + \Dl t_H)^3 \simeq \tau^3 + 3 \Dl t_H \tau^2 $, we solve for
$\Dl t_H$ from the equation
\[
\frac{1}{[1 + (\om' J_0 \Dl t_H)^2]^{3/2}} \exp[-D_0 (\om')^2 \tau^2 \Dl t_H] = \half
\]
or equivalently
\[
2^{2/3}\exp[-\frac{2}{3}D_0 (\om')^2 \tau^2 \Dl t_H] = 1 + (\om' J_0 \Dl t_H)^2
\]
Expanding the exponential and keeping only to linear order in $D_0$, we have
\beq
(\om' J_0 )^2 (\Dl t_H)^2 + \frac{2}{3}2^{2/3}D_0 (\om')^2 \tau^2 \Dl t_H - (2^{2/3} - 1) = 0
\eeq
which on assuming that $D_0
^2$ is small enough to satisfy
\[ (2^{2/3}-1) J_0^2 \gg [\frac{2^{2/3}}{3}D_0 \om' \tau^2]^2 \]
we have for the full width $\Dl t_{FWHM} = 2\Dl t_H$
\beq
\Dl t_H = 2 \frac{\sqrt{2^{2/3}-1}}{\om' J_0} - \frac{2^{5/3}}{3}D_0 (\frac{\tau}{J_0})^2 + 
\frac{2^{4/3}}{9\sqrt{2^{2/3}-1}}(\frac{\tau^4 \om'}{J_0^3})  D_0^2 
\eeq
where again, the last term in $D_0^2$ can be dropped.

\subsubsection{Escape time}

This time is given by
\beq
t_{esc} = \int dJ \frac{J}{D(J)} 
\eeq
With $D(J) = D_0$, we have
\beq
t_{esc} = \frac{J_A^2}{2 D_0}
\eeq
where $J_A$ is the action at the aperture.  {\tt Express $J_A$ in terms of the amplitude at the aperture; see Edwards and Syphers}

Clearly, the escape time must be larger than the decoherence time $\tau_D \simeq 1/(\om' J_0) = 1/(\mu \om_{rev})$
or
\beq
\tau_D \ll t_{esc}, \Rarw D_0 \ll \half \om_{rev} \mu J_A^2 
\eeq

\subsubsection{Summary of results with a constant diffusion coefficient}

 The amplitude of the echo at $t \simeq 2\tau$
\beq
\lan x \ran^{echo, max amp} = \bt \theta q J_0 \om' \tau \exp[-\frac{2}{3}D_0 (\om')^2 \tau^3]
\label{eq: echo_amp}
\eeq
assuming that $\om(J) = \om_0 + \om' J$, 
and the initial distribution is $\psi_0(J) = \frac{1}{2\pi J_0} \exp[-J/J_0]$. 
The constant angular frequency coefficient $\om' = (\om_{rev}/\eps)\mu$, where $\mu$ is the
detuning coefficient defined by the action dependent tune
\[ \nu(J) - \nu_0 = \frac{J}{\eps}\mu \]
Here $\eps$ is the rms un-normalized emittance and $\om_{rev}$ is the angular revolution frequency.

At other times $t$, the echo amplitude behaves as
\beqr
\lan x \ran^{echo, amp} & = & \frac{\bt \theta q J_0 \om' \tau}{(1 + \xi^2)^{3/2}} 
\exp[-\frac{1}{3}D_0 (\om')^2 t_1^3] \\
\xi & = & \om' (t - 2\tau) J_0, \;\;\; t_1^3 = (t - \tau)^3 + \tau^3
\label{eq: echo_timeamp}
\eeqr
To find the half width time $\Dl t_H$ at which the amplitude falls to half its maximum, we have
\[ 
\frac{1}{(1 + (\om' J_0 \Dl t_H)^2)^{3/2}} 
\exp[-\frac{1}{3}D_0 (\om')^2 ((\tau + \Dl t_H)^3 + \tau^3)] = \half
\exp[-\frac{2}{3}D_0 (\om')^2 \tau^3]
\]
Keeping terms to first order in $D_0$ results in a quadratic for $\Dl t_H$. Taking the negative root
and defining $\Dl \tau_{FWHM} = 2 |\Dl t_H|$, 
it follows that the FWHM of the echo pulse is
\beq
\Dl\tau_{FWHM} = \frac{2\sqrt{2^{2/3}-1}}{\om' J_0} + \frac{2^{5/3}}{3}D_0(\frac{\tau}{J_0})^2 
 + O(D_0^2)
\label{eq: FWHM_0}
\eeq
The width of the echo pulse vanishes for delays $\tau \ge \tau_{max}$, i.e.
\beqr
\Dl\tau_{FWHM} & \rarw &  0,   \;\;\; {\rm for} \;\;\; \tau \ge \tau_{max}  \nonumber \\
\tau_{max} & = & \left[(\frac{6\sqrt{2^{2/3}-1}}{2^{5/3}})(\frac{J_0}{\om' D_0})\right]^{1/2}
 = \left[(\frac{6\sqrt{2^{2/3}-1}}{2^{5/3}})(\frac{1}{\om_{rev}\mu D_0})\right]^{1/2}\frac{\eps}{2}
\label{eq: tau_max}
\eeqr

One of the conditions for the solution in Eq.(\ref{eq: echo_amp}) to be valid is that the diffusion
must be sufficiently weak as to satisfy
\beq
D_0 \ll D_{max} = 2 \om_{rev} \mu \eps^2
\eeq

If the echo amplitude is measured as a function of the detuning, the maximum amplitude occurs
at a value $\mu = \mu_m$ such that 
\beq
D_0 = \frac{3}{4}(\frac{\eps}{\om_{rev}})^2 \frac{1}{\tau^3 \mu_m^2}
\label{eq: D0_0}
\eeq
This relation determines the diffusion coefficient $D_0$, given the delay $\tau$ and
the optimum detuning $\mu_m$.

\subsection{Polynomial diffusion coefficients}

Consider the diffusion coefficient to be of the form
\beq
D(J) = D_0 + \sum_{n=1} D_n (\frac{J}{J_0})^n
\eeq
The average dipole moment is given by
\beqr
\lan x(t) \ran & = & \int x \psi_5(J, \phi) dJ d\phi \nonumber \\
 & = & - \pi \bt \theta q \om' \tau \exp[ - \frac{1}{3}D_0 (\om')^2 t_1^3] 
{\rm Im}[e^{[i \Phi_0]} \int J^2 \psi_0' \exp[-\frac{1}{3}(\om')^2 t_1^3 \sum_n D_n (J/J_0)^n]
e^{[i\Phi_1 J]} dJ  \nonumber \\
& = & \half \bt \theta q \mu \tau \om_{rev} \exp[ -\frac{1}{3}D_0 (\om')^2 t_1^3] 
{\rm Im}[e^{[i \Phi_0]} \int z^2 \psi_0' \exp[-\frac{1}{3}(\om')^2 t_1^3 \sum_n D_n z^n]
e^{[i\Phi_1 J_0 z]} dz 
\label{eq: dipmoment_n}
\eeqr
where
\[ \Phi_0 = \om_0(t - 2\tau), \;\;\;\; \Phi_1 = \om'(t - 2\tau) \]
and we used 
\[ \psi_0'(J) = -\frac{1}{2\pi J_0^2} \exp[-\frac{J}{J_0}] \]
The general form of the integral is
\beqrs
{\rm Int} & = &  \int dz z^2 \exp[- a z] \exp[-\sum_n b_n z^n] \\
a & = & 1 - i \Phi_1 J_0 = 1 - \mu\om_{rev}(t-2\tau) , \; \; \;  
b_n = \frac{1}{3} D_n(\om')^2 t_1^3 
\eeqrs

\subsection{Constant and linear diffusion coefficient}

Here we assume that the diffusion is of the form
\beq
D(J) = D_0 + D_1(\frac{J}{J_0})
\eeq
where $D_0$ and $D_1$ have the same dimensions.

The integral is 
\[ {\rm Int}(J)  =   \int dJ J^2 \exp[- ( a + b_1)J] = \frac{2}{(a + b_1)^3} \]
where 
\[ a + b_1 = \frac{1}{J_0}[(1 + \frac{1}{3}D_1 (\om')^2 t_1^3) - i\Phi_1 J_0 ] \]

in this case, the dipole moment is given by
\beqr
\lan x \ran (t) & = & \bt \theta q \om' \tau J_0 \exp[-\frac{1}{3}D_0(\om')^2 t_1^3]
 {\rm Im}[\frac{e^{i\Phi}}{(\al - i\xi})^3] \\
 & = & \theta q \om' \tau J_0 \exp[-\frac{1}{3}D_0(\om')^2 t_1^3]
\frac{[ (3 \al^2 - \xi^2)\xi\cos\Phi + (\al^2 - 3 \xi^2)\al \sin\Phi]}{(\al^2 + \xi^2)^3} \\
 t_1^3 & = & (t-\tau)^3 + \tau^3, \;\;\; \Phi = \om_0(t - 2\tau)  \nonumber \\
\al &  = & 1 + \frac{1}{3}D_1(\om')^2 t_1^3, \;\;\; \xi = \om' (t - 2\tau) J_0 = \om_{rev}\mu(t - 2\tau)
\eeqr
The amplitude of this echo is determined by the amplitude of $e^{i\Phi}/(\al - i\xi)^3$. Since
$\Phi$ is a fast varying phase while $\xi$ is slowly varying, it follows that
$ampl[e^{i\Phi}/(\al - i\xi)^3]= ampl[1//(\al - i\xi)^3]= 1/(\al^2 + \xi^2)^{3/2}$.
Hence the amplitude of the echo is
\beq
\lan x \ran (t) ^{amp} = \frac{\bt \theta q \om' \tau J_0  }{(\al^2 + \xi^2)^{3/2}} \exp[-\frac{1}{3}D_0(\om')^2 t_1^3] 
\label{eq: amp_0_1}
\eeq
At the time of the 1st echo, $t=2\tau$, the echo amplitude is
\beqr
\lan x \ran (2\tau) ^{amp} & = & \frac{\bt \theta q \om' \tau J_0  }{\al_1^3} \exp[-\frac{2}{3}D_0(\om')^2 \tau^3]  \\
\al_1 & = & 1 + \frac{2}{3}D_1(\om')^2 \tau^3 
\eeqr

At the time of the 2nd echo, $t=4\tau$, the different parameters are
\beqrs
 t_1^3 & = & 28\tau^3, \;\;\; \Phi_2 = 2\om_0\tau   \\
\al_2 &  = & 1 + \frac{28}{3}D_1(\om')^2 \tau^3, \;\;\; \xi_2 = 2 \om' \tau J_0 
\eeqrs
and the echo amplitude is
\beqr
\lan x \ran (4\tau) ^{amp} & = & \frac{\bt \theta q \om' \tau J_0  }{(\al_2^2+\xi_2^2)^{3/2}} 
\exp[-\frac{28}{3}D_0(\om')^2 \tau^3] 
\eeqr

At the 3rd echo $t=6\tau$, the parameters are 
\beqrs
 t_1^3 & = & 126\tau^3, \;\;\; \Phi_3 = 4\om_0\tau  \\
\al_3 &  = & 1 + 42 D_1(\om')^2 \tau^3, \;\;\; \xi_2 = 4 \om' \tau J_0 
\eeqrs
and the echo amplitude is
\beqr
\lan x \ran (6\tau) ^{amp} & = & \frac{\bt \theta q \om' \tau J_0  }{(\al_3^2+\xi_3^2)^{3/2}} 
\exp[- 42 D_0(\om')^2 \tau^3] 
\eeqr

\subsubsection{Summary of results with constant and linear diffusion coefficient}

\clearpage

\subsubsection{Maxima of the echo amplitude as functions of detuning and delay time}

Using 
\[ \om' = \frac{\om_{rev}}{J_0}\mu = \frac{\om_{rev}}{\eps}\mu  \]
define the following variables of the diffusion coefficients
\beq
d_0 =  \frac{2}{3}D_0 (\frac{\om_{rev}}{\eps})^2, \;\;\; d_1 = \frac{2}{3}D_1 (\frac{\om_{rev}}{\eps})^2
\eeq
Then the echo amplitude at $t=2\tau$ scaled by the dipole and quad kick strengths is
\beq
\frac{\lan x \ran (2\tau) ^{amp}}{\bt \theta q \om_{rev}} = \frac{\mu \tau}{(1 + d_1 \mu^2\tau^3)^3}\exp[- d_0 \mu^2 \tau^3]
\eeq
Since we have two undetermined coefficients $d_0, d_1$, we need two measured variables.

As a function of the time delay, this amplitude has a maximum at a delay $\tau = \tau_m$, such that
the two coefficients can be related as 
\beq
d_1 = \frac{1 - 3 d_0 \mu^2 \tau_m^3}{\mu^2 \tau_m^3( 8 + 3 d_0 \mu^2 \tau_m^3)} 
\label{eq: d1_1}
\eeq
It is understood that $\mu$ is held fixed at $\mu_f$ while finding the optimum delay $\tau_m$ 
Define
\beq
c_{\tau} = \mu_f^2 \tau_m^3 
\eeq
Substituting this into the equation for the relative amplitude, we have using 
$1 + d_1 c_{\tau} = 9/(8 + 3 d_0 c_{\tau})$ for the maximum amplitude obtained at the delay 
$\tau_m$
\beq
\frac{\lan x\ran_{max}(\tau_m)}{\bt \theta} = \om_{rev}q \mu \tau_m[\frac{8 + 3 d_0 c_{\tau}}{9}]^3
\exp[-d_0 c_{\tau}]
\eeq
This equation can be solved for $d_0$ and subsequently $d_1$ can be found. We require that the solution
for $d_0$ obey $ 3 d_0 c_{\tau} \le 1$ in order for $d_1 \ge 0$.

Similarly, as a function of the detuning, the amplitude has a maximum at $\mu = \mu_m$ such that
\beq
d_1 = \frac{1 - 2 d_0 \mu_m^2 \tau^3}{\mu_m^2 \tau^3( 5 + 2 d_0 \mu_m^2 \tau^3)}
\label{eq: d1_2}
\eeq
Here $\tau$ is held fixed at $\tau_f$ while finding the optimum in $\mu$. Define
\beq
c_{\mu} = \mu_m^2 \tau_f^3 
\eeq
Again, substituting back for $d_1$, we can write the maximum relative amplitude at $\mu_m$ as
\beq
\frac{\lan x\ran_{max}(\mu_m)}{\bt \theta} = \om_{rev}q \mu_m \tau_f[\frac{5 + 2 d_0 c_{\mu}}{6}]^3
\exp[-d_0 c_{\mu}]
\eeq
Here $d_1 \ge 0$ requires that the solution for $d_0$ obey $ 2 d_0 c_{\mu} \le 1$.

If both $\mu_m$ and $\tau_m$ are measured, then the diffusion coefficient $d_0$ can be found from
\beq
\frac{1 - 3 d_0 c_{\tau}}{c_{\tau}(8 + 3 d_0 c_{\tau})} = \frac{1 - 2 d_0 c_{\mu}}{c_{\mu}(5 + 2 d_0 c_{\mu})} 
\eeq
which has the solutions
\beq
d_0 = \frac{1}{12c_{\mu}c_{\tau}}\left[ 2c_{\mu}+3 c_{\tau} \pm \sqrt{\frac{(2c_{\mu}-3c_{\tau})(2c_{\mu}^2 + 67c_{\mu}c_{\tau} + 3c_{\tau}^2)}
{c_{\mu}- c_{\tau}}} \right]
\eeq
The negative root is permissible if $d_0 \ge 0$. 
Once $d_0$ is determined, $d_1$ can be determined from either of Equations (\ref{eq: d1_1}) or 
(\ref{eq: d1_2}). Positivity of $d_1$ requires that the above solution obey
\[
d_0 \le \frac{1}{2 c_{\mu}}, \;\;\;\; d_0 \le \frac{1}{3 c_{\tau}}
\]

\noi \underline{Case} $D_1=0$ 

As a function of the delay $\tau$, the maximum amplitude occurs at $d_0 = 1/(3c_{\tau})$ or 
\beq
 D_0 = \frac{1}{2}(\frac{\eps}{\om_{rev}})^2 \frac{1}{\mu^2 \tau_m^3}
\eeq
while the solution for optimum detuning $\mu_m$ gives $d_0 = 1/(2c_{\mu})$ which is the same
as the solution in Eq.(\ref{eq: D0_0}).

\noi \underline{Case} $D_0=0$ 

The optimum delay $\tau_m$ occurs at $d_1 = 1/(8 c_{\tau})$ or
\beq
D_1 = \frac{3}{16}(\frac{\eps}{\om_{rev}})^2 \frac{1}{\mu^2 \tau_m^3}
\eeq
while the optimum detuning $\mu_m$ occurs at $d_1 = 1/(5 c_{\mu})$ or
\beq
D_1 = \frac{3}{10}(\frac{\eps}{\om_{rev}})^2 \frac{1}{\mu_m^2 \tau^3}
\eeq


\subsubsection{Full width at half maximum}

We make the following approximations 
\bit
\item $D_0 (\om')^2 \tau^2 \Dl t_h \ll 1$ so that we can expand to 1st order
\[ \exp[-\frac{2}{3}D_0(\om')^2 \tau^2 \Dl t_h] \simeq 1 - \frac{2}{3}D_0(\om')^2 \tau^2 \Dl t_h \]

\item $(\om'\tau)^4[(2^{5/3}/3)D_0 + 2 D_1/\al_1]^2 \ll 4(2^{2/3}-1)[(\om' J_0)^2 + 
D_1^2(\om'\tau)^4]/\al^2$, so that
\beqrs
& \mbox{} & \sqrt{4(2^{2/3}-1)[(\om' J_0)^2 + D_1^2(\om'\tau)^4]/\al_1^2 + 
(\om'\tau)^4[(2^{5/3}/3)D_0 + 2 D_1/\al_1]^2 } \\
& \simeq & 
2\sqrt{(2^{2/3}-1)[(\om' J_0)^2 + D_1^2(\om'\tau)^4]}/\al_1 + 
\frac{(\om'\tau)^4[(2^{5/3}/3)D_0 + 2 D_1/\al_1]^2}
{4\sqrt{(2^{2/3}-1)[(\om' J_0)^2 + D_1^2(\om'\tau)^4]}}\al_1
\eeqrs
In the special case that $D_1=0$, this implies 
\[ D_0 \ll \frac{\sqrt{(2^{2/3}-1)}}{2^{2/3}/3)}\frac{J_0}{\om' \tau^2} \]
while for the case $D_0=0$, this implies
\[ D_1 \ll \sqrt{\frac{2^{2/3}-1}{2 - 2^{2/3}}}\frac{J_0}{\om' \tau^2} \]

\item Terms of order $O(D_0^2), O(D_1^2), O(D_0 D_1)$ and higher can be dropped,
\eit

Under these assumptions, we find for the full width at half maximum
\beq
\Dl t_{FWHM} = 2\sqrt{2^{2/3}-1}(\frac{\al_1}{\om' J_0}) + 2(\frac{\al_1 \tau}{J_0})^2
\left[ \frac{2^{2/3}}{3}D_0 + \frac{D_1}{\al_1} \right]
\label{eq: FWHM_0_1}
\eeq

In terms of the coefficients $d_0, d_1$, we have 
\[ D_n = \frac{3}{2}(\frac{\eps}{\om_{rev}})^2 d_n, \;\;\; \al_1 = 1 + \mu^2 \tau^3 d_1 \]
and we can write
\beq
\Dl t_{FWHM} = 2\sqrt{2^{2/3}-1}(\frac{\al_1}{\om_{rev} \mu}) + 3(\frac{\al_1 \tau}{\om_{rev}})^2
\left[ \frac{2^{2/3}}{3}d_0 + \frac{d_1}{\al_1} \right]
\eeq
This can be used to solve for $d_0$ as
\beq
d_0 = \frac{3}{2^{2/3}}\left\{ \frac{1}{3}(\frac{\om_{rev}}{\al_1 \tau})^2
\left[ \Dl t_{FWHM} -2\sqrt{2^{2/3}-1}(\frac{\al_1}{\om_{rev} \mu})  \right] - \frac{d_1}{\al_1}
\right\}
\eeq

If $D_1=0$, then $\al_1=1$ and Eq.(\ref{eq: FWHM_0_1}) reduces to Eq.(\ref{eq: FWHM_0}) found in
the previous section.

If $D_0=0$, then
\beq
\Dl t_{FWHM} = 2\al_1\left[\sqrt{2^{2/3}-1}(\frac{1}{\om' J_0}) + D_1(\frac{\tau}{J_0})^2 \right]
 = \al_1\left[2\sqrt{2^{2/3}-1}(\frac{1}{\om_{rev} \mu}) + 3(\frac{\tau}{\om_{rev}})^2 d_1\right]
\eeq

\noi Vanishing echo pulse width 

The pulse width will vanish for delay times $\tau \ge \tau_{max}$ for a given $\mu$, when
\beq
\frac{2^{2/3}}{3}D_0 + D_1 = \sqrt{2^{2/3}-1} \frac{J_0}{\om' \tau_{max}^2}
\eeq
Here we dropped a term of $O(D_0 D_1)$, i.e. we approximated $\al_1 D_0 \simeq D_0$.
Hence, this relation can be used for another relation between $D_0, D_1$
given $\tau_{max}$. Alternatively, if the pulse width vanishes for $\mu\ge \mu_{max}$ given $\tau$,
the same relation above can be used. 

In terms of the scaled coefficients $d_0, d_1$ (they have dimension [$T^{-3}$], the above can be 
written as
\beq
\frac{2^{2/3}}{3}d_0 + d_1 = \frac{2}{3}\sqrt{2^{2/3}-1} \frac{\om_{rev}}{\mu \tau_{max}^2}
\eeq

However, given the many approximations needed to obtain the FWHM width in this case, it would
be preferred to use other constraints on obtaining the diffusion coefficients. 

\subsubsection{Escape time}

This time is given by
\beq
t_{esc} = \int_0^{J_a} dJ \frac{J}{D(J)} = \int_0^{J_a} dJ \frac{J}{D_0 + D_1 (J/J_0)}
\eeq
where $J_a$ is  the action at the aperture. 
Introducing a variable $y(J) = 1 + \frac{D_1}{D_0}\frac{J}{J_0}$, and $y_a = y(J_a)$, we have
\beqr
 \tau_{esc} & = & D_0 (\frac{J_0}{D_1})^2 \int_1^{y_a} (1 - \frac{1}{y}) dy \nonumber \\
& = & D_0 (\frac{J_0}{D_1})^2[ \frac{D_1}{D_0}\frac{J_a}{J_0} - \ln(1 + \frac{D_1}{D_0}\frac{J_a}{J_0}) ]
\eeqr
Ignoring the slowly growing log term, to leading order the escape time is
\[ \tau_{esc} \approx \frac{J_0 J_a}{D_1} \]
Note that this leading order term does not depend on the constant diffusion coefficient $D_0$.

\subsubsection{Emittance growth}

The diffusion coefficients determine the emittance growth as follows.
From the density distribution function $\psi(J)$ of an unperturbed beam with diffusion coefficient
$D(J)$, the average action as a function of time is
\beq
\lan J \ran(t) = \int_0^{J_a} J \psi(J,t) dJ
\eeq
where $J_a$ is the action at the absorbing boundary. 

Hence
\beqr
\frac{d}{dt}\lan J\ran & = & \int J \frac{\del \psi}{\del t} dJ = \int J 
\frac{\del }{\del J}[D(J)\frac{\del\psi}{\del J}] dJ \nonumber \\
 & = & JD(J)\frac{\del\psi}{\del J}|_0^{J_a} - \int D(J)\frac{\del\psi}{\del J} =
    - \int D(J)\frac{\del\psi}{\del J} \nonumber \\
 & = & -D(J)\psi|_0^{J_a} + \int D'(J) \psi dJ \nonumber \\
& = &   D(0) \psi(0) + \int D'(J) \psi dJ 
\eeqr
In the 2nd and 3rd lines, the integrated terms at $J=J_a$ vanish if we assume that the slope 
$\del\psi/\del J $ and $\psi$ vanish at the boundary faster than the combination $JD(J)$ and
$D(J)$ respectively. 

Now, considering the case where $D(J) = D_0 + D_1(J/J_0)$, and $\psi(J) = \exp(-J/J_0)/(2\pi J_0)$,
$\int \psi dJ =  1/(2\pi)$, we have for the emittance growth rate
\beq
\frac{d\eps}{dt} = \frac{d \lan J \ran}{dt} = \frac{1}{\pi \eps_0}(D_0 + D_1)
\eeq

{\tt The above analysis needs to be revised for the following reasons}

\bit
\item The distribution function with diffusion will in general also depend on the phase variable
$\phi$
\item Assuming that the action is still conserved, we would then have
\beqrs
\frac{d}{dt}\lan J\ran & = & \int J \frac{d \psi(J,\phi,t)}{d t} dJ \\
 & = & \int J (\frac{\del \psi(J,\phi,t)}{\del t} + \frac{\del \psi(J,\phi,t)}{\del \phi}
\frac{d\phi}{d t}) dJ
\eeqrs
This involves the equation of motion for $\phi$.

\item Another concern is that with only $D_0$, this predicts that the emittance grows linearly
with time. However, with a constant diffusion, one expects the underlying variable to grow
with time as $\sqrt{t}$. 

That is a shortcoming of this calculation. Does the inclusion of the $\dot{\phi}(\del\psi/\del\phi)$ term correct this problem?
\eit

\subsubsection{Multiple Echoes}

Multiple echoes may be observed at multiples of $2\tau$. We consider here the 2nd echo at time
$4\tau$. 

We write the echo amplitudes in terms of the scaled coefficients $d_0, d_1$. We have
\beq
\lan x \ran^{amp} (t) = \frac{\bt \theta q \om_{rev}\mu \tau}{(\al^2 + \xi^2)^{3/2}} 
 \exp[- \half d_0\mu^2 t_1^3] , \;\;\;\; \al = 1 + \half d_1\mu^2 \tau^3
\eeq
Hence the echo amplitude at $t=4\tau$ ($t_1^3 = 28\tau^3$) is
\beqr
\lan x \ran^{amp} (4\tau) & = & \frac{\bt \theta q \om_{rev}\mu \tau   }
{(\al_2^2 + \xi_2^2)^{3/2}}  \exp[-14 d_0\mu^2 \tau^3] \\
\al_2 &  = & 1 + 14 d_1\mu^2 \tau^3, \;\;\; \xi_2 = 2\om_{rev} \mu\tau 
\eeqr
while for the third echo at $t=6\tau$, $t_1^3 = 126\tau^3$, and
\beqr
\lan x \ran^{amp} (6\tau) & = & \frac{\bt \theta q \om_{rev}\mu \tau   }
{(\al_3^2 + \xi_3^2)^{3/2}}  \exp[-63 d_0\mu^2 \tau^3] \\
\al_2 &  = & 1 + 63 d_1\mu^2 \tau^3, \;\;\; \xi_2 = 2\om_{rev} \mu\tau 
\eeqr

Hence the ratio of the amplitudes at the 2nd echo and 1st echo is 
\beq
\frac{\lan x \ran^{amp}(4\tau)}{\lan x \ran^{amp}(2\tau)} = 
\frac{(1 + d_1\mu^2 \tau^3)^3}{[(1+14d_1 \mu^2\tau^3)^2 + (2\om_{rev}\mu\tau)^2]^{3/2}}
\exp[-13 d_0 \mu^2 \tau^3]
\eeq
and the ratio of amplitudes of the 3rd echo and 1st echo is
\beq
\frac{\lan x \ran^{amp}(6\tau)}{\lan x \ran^{amp}(2\tau)} = 
\frac{(1 + d_1\mu^2 \tau^3)^3}{[(1+ 63d_1 \mu^2\tau^3)^2 + (4\om_{rev}\mu\tau)^2]^{3/2}}
\exp[- 62 d_0 \mu^2 \tau^3]
\eeq

Since fewer approximations were used in obtaining this, it may preferable to use this constraint on
the coefficients $d_0, d_1$ instead of the pulse width.

\noi \underline{Case $D_1=0$}
\beq
\frac{\lan x \ran^{amp}(4\tau)}{\lan x \ran^{amp}(2\tau)} = 
\frac{1 }{[1 + (2\om_{rev}\mu\tau)^2]^{3/2}} \exp[-13 d_0 \mu^2 \tau^3]
\eeq

\noi \underline{Case $D_0=0$}
\beq
\frac{\lan x \ran^{amp}(4\tau)}{\lan x \ran^{amp}(2\tau)} = 
\frac{(1 + d_1\mu^2 \tau^3)^3}{[(1+14d_1 \mu^2\tau^3)^2 + (2\om_{rev}\mu\tau)^2]^{3/2}}
\eeq

\subsection{Constant, Linear and quadratic diffusion coefficients}

Here we have 
\beq
D(J) = D_0 + D_2 (\frac{J}{J_0})^2
\eeq
From Eq.(\ref{eq: dipmoment_n}) it follows that the dipole moment is 
\beq 
\lan x(t) \ran = - \pi \bt \theta q \om' \tau \exp[ - \frac{1}{3}D_0 (\om')^2 t_1^3] 
{\rm Im}[e^{[i \Phi_0]} \int J^2 \psi_0' \exp[-\frac{1}{3}(\om')^2 t_1^3 D_2 (J/J_0)^2 ]
e^{[i\Phi_1 J]} dJ 
\eeq
The amplitude of the moment is given by setting $|e^{i\Phi_0}| = 1 = |e^{i\Phi_1 J}|$. Using
\[
 \int_0^{\infty} x^2 \exp[-(ax + b x^2)] dx = \frac{1}{8 b^{5/2}}[\sqrt{\pi}(a^2 + 2b)e^{a^2/(4b)}
{\rm Erfc}(\frac{a}{2\sqrt{b}}) - 2 a\sqrt{b}]
\]

The general time dependent form of the echo at time $t=2\tau + \Dl t$ where $\Dl t $ can have either
sign is
\beqr
\lan x(t) \ran^{amp} & = & \half \bt_G \theta q \om_{rev}\mu \tau \exp[ - \frac{1}{2}d_0 \mu^2 t_1^3]
{\rm Im}[ e^{i\Phi_0}H_{02}(\Dl t)] \\
H_{02}(\Dl t) & \equiv & \int z^2 \exp[- a_0 z - \half b_2 z^2] dz \nonumber \\
 & = &  \frac{1}{\sqrt{2}}(\frac{1}{b_2} )^{5/2}\left\{
 \sqrt{\pi}\left[a_0^2 + b_2\right]\exp(\frac{a_0^2}{2b_2})
{\rm Erfc}(\frac{a_0}{\sqrt{2 b_2}}) - a_0 \sqrt{2b_2}\right\}
\nonumber \\
a_0 & = & (1 - i\mu\Dl t\om_{rev}), \;\;\; b_2  =  d_2 \mu^2 t_1^3 =  
d_2 \mu^2 [(\tau + \Dl t)^3 + \tau^3] \nonumber \\
\eeqr
Here Erfc is the complementary error function.

Hence at time $t = 2\tau$ where $t_1^3 = 2\tau^3$
\beqr
\lan x(t) \ran^{amp} & = & \half \bt_G \theta q \om_{rev}\mu \tau \exp[ - \frac{1}{2}d_0 \mu^2 t_1^3]
H_{02}(0)] \\
H_{02}(0) & \equiv & \int z^2 \exp[- a_0 z - \half b_2 z^2] dz \nonumber \\
 & = &  \frac{1}{\sqrt{2}}(\frac{1}{b_2} )^{5/2}\left\{
 \sqrt{\pi}\left[1 + b_2\right]\exp(\frac{1}{2b_2})
{\rm Erfc}(\frac{1}{\sqrt{2 b_2}}) - \sqrt{2b_2}\right\}
\eeqr

In the case
\[ D(J) = D_1(\frac{J}{J_0}) + D_2(\frac{J}{J_0})^2 \]
The time dependent centroid position is 
\beqr
\lan x(t) \ran  & = & - \pi \bt \theta q \om' \tau 
{\rm Im}[e^{[i \Phi_0]} \int J^2 \psi_0' \exp[-\frac{1}{3}(\om')^2 t_1^3 (D_1 (J/J_0)+ D_2 (J/J_0)^2 )]
e^{[i\Phi_1 J]} dJ \\
& = & \half \bt \theta q \om' \tau {\rm Im}[e^{[i \Phi_0]} \int \frac{J}{J_0}^2 \exp[-J/J_0] 
\exp[-\frac{1}{3}(\om')^2 t_1^3 (D_1 (J/J_0)+ D_2 (J/J_0)^2 )]
e^{[i\Phi_1 J]} dJ \\
& = & \pi \bt \theta q \mu N_{\tau}  {\rm Im}[e^{[i \Phi_0]} \int z^2 
\exp[- (1 - i\Phi_1 J_0)z ] - \frac{1}{3}(\om')^2 t_1^3 (D_1 z + D_2 z^2 )] dz \\
\eeqr
where we replaced
\[ \tau = N_{\tau} T_{rev} \]
This evaluates to
\beqr
\lan x(t) \ran^{amp} & = &  \half \bt_G \theta q \om_{rev} \mu \tau {\rm Im}[e^{i \Phi_0} H_{12}(\Dl t)] \\
H_{12}(\Dl t) & \equiv & \int z^2 \exp[- a_1 z - \half b_2 z^2] dz \nonumber \\
 & = &  \frac{1}{\sqrt{2}}(\frac{1}{b_2} )^{5/2}\left\{
 \sqrt{\pi}\left[a_1^2 + b_2\right]\exp(\frac{a_1^2}{2b_2})
{\rm Erfc}(\frac{a_1}{\sqrt{2 b_2}}) - a_1 \sqrt{2b_2}\right\}
\nonumber \\
a_1 & = & (1 + b_1) - i\mu\Dl t\om_{rev}, \;\;\; b_1  =  d_1 \mu^2 t_1^3 =  
d_1 \mu^2 [(\tau + \Dl t)^3 + \tau^3] \nonumber \\
\eeqr
At $t=2\tau$, both $\Phi_0, \Phi_1$ vanish while $t_1^3 = 2\tau^3$.
Hence 
\beqr
\lan x(t) \ran^{amp} & = & \half \bt_G \theta q \om_{rev} \mu \tau H_{12}(0) \\
a_1(0) & = & 1 + b_1(0) = 1 + 2d_1\mu^2\tau^3, \;\;\; b_2(0) = 2 d_2\mu^2 \tau^3 \nonumber 
\eeqr

 At 
nearby times $t = 2\tau + \Dl t$, 
\[ \Phi_0 = \om_0 \Dl t, \;\;\; \Phi_1 J_0 = \om'J_0(t-2\tau) = \mu\om_{rev}\Dl t, \; \;\;\;
t_1^3 = (\tau+\Dl t)^3 + \tau^3 
\]
If we consider times $\Dl t/\tau \ll 1$, then we can approximate
\[
t_3 \approx 2\tau^3 + 3\tau^2 \Dl t  \]

\section{Scaled diffusion coefficient}

From the expressions for the diffusion coefficients $D_0, D_1$ when one of them vanishes, it is clear
that the scale of the diffusion coefficients is set by the parameter 
\beq
D_{sc} = (\frac{\eps}{\om_{rev}})^2 \frac{1}{\mu^2 \tau^3}
\eeq
The decoherence time $\tau_D = 1/(\om' J_0) = 1/(\om_{rev}\mu)$, hence 
\[ \om_{rev}\mu\tau = \frac{\tau}{\tau_D} \]
and the diffusion scale is therefore determined by
\begin{framed}
\beq
D_{sc} = [\frac{\eps^2}{\tau}] (\frac{\tau_D}{\tau})^2 
\eeq
\end{framed}
In most cases, $D_0, D_1 \approx (0.1 - 1) D_{sc}$. Clearly, larger delay times $\tau$ and smaller
decoherence times $\tau_D$ lead to smaller $D_{sc}$ which allows measuring weaker diffusion coefficients. 

With RHIC parameters, $\tau_D = 114$ turns and typically $\tau/\tau_D \approx 4$. This ratio should
preferably be about 10, as seems to have been possible in the SPS experiment on transverse echoes.
For RHIC parameters, 
\beq
D_{sc} = 2.8 \times 10^{-13} \; {\rm m^2/s}
\eeq

Check of numerical assumptions made in deriving the pulse width 
\[ \mu = 0.0014 \Rarw \om' = \frac{2}{\eps_0}\om_{rev}\mu = 8.6\times 10^9 {\rm rad/(m-s)} \]

\bit
\item $D_0 (\om' \tau)^2 \Dl t_h \ll 1$ 
If we take $\Dl t_h = 10$ turns
\[ \Rarw D_0 \ll 3.2 \times 10^{-12} {\rm m^2/s} \]

\item When $D_1=0$ then
\beqrs D_0 & \ll &  \frac{\sqrt{(2^{2/3}-1)}}{2^{2/3}/3)}\frac{J_0}{\om' \tau^2} \\
& \ll & 4.1 \times 10^{-13}  {\rm m^2/s} 
\eeqrs

\item When $D_0=0$ then,
\beqrs
D_1 & \ll & \sqrt{\frac{2^{2/3}-1}{2 - 2^{2/3}}}\frac{J_0}{\om' \tau^2} \\
& \ll & 3.3  \times 10^{-13}  {\rm m^2/s} 
\eeqrs

\eit

Since the echo amplitude as a function of the delay time is not available, here we use the following
two observations
\bit
\item Echo amplitude as a function of detuning with maximum at $\mu_m=0.001$.

\item Maximum delay time $\tau_{max}$ before the echo disappears. For gold, the number is 
$\tau_{max} = 550$ turns. For copper, the number is $\tau_{max}=1000$ turns.
\eit

We solve these two equations for $d_0, d_1$
\beqr
d_1 & = & \frac{1 - 2 d_0 c_{\mu}}{c_{\mu}(5 + 2 d_0 c_{\mu})} \\
d_1 & = & \frac{2}{3}\sqrt{2^{2/3}-1}\frac{\om_{rev}}{\mu \tau_{max}^2} - \frac{2^{2/3}}{3} d_0 
\eeqr
where as before : $c_{\mu} = \mu_m^2 \tau^3$, $d_i =  (8/3)(\om_{rev}/\eps)^2 D_i$

\clearpage

\section{Multiple quadrupole kicks}

We now analyze multiple quad kicks. Some of the assumptions in the analysis with a single quad
kick have to be reconsidered.

Recap on the distribution functions

\noi $\psi_0(J)$ is the initial distribution
\beq
\psi_0(J) = \frac{1}{2\pi J_0} \exp[-\frac{J}{J_0}]
\eeq
\noi $\psi_1(J,\phi)$ is the distribution function (DF) after the dipole kick
\beq
\psi_1(J,phi) = \psi_0(x,p - \bt_K\theta) \simeq 
\psi_0(J) + \bt_K \theta \psi_0'(J) \sqrt{\frac{2J}{\bt}}\sin\phi
\eeq
\noi $\psi_2(J,\phi,t)$ is the DF at time $t$ after the dipole kick
\beq
\psi_2(J,\phi,t) = \psi_0(J) + \bt_K \theta \psi_0'(J) \sqrt{\frac{2J}{\bt}}\sin(\phi - \om(J)t)
\eeq
\noi $\psi_3(J,\phi,\tau)$ is the DF at time $\tau$ after the dipole kick
\beq
\psi_3(J,\phi,\tau) = \psi_0(J) + \bt_K \theta \psi_0'(J) \sqrt{\frac{2J}{\bt}}\sin(\phi - \om(J)\tau)
\eeq
\noi $\psi_4(J,\phi,\tau)$ is the DF at time $\tau$ right after the first quad kick
\beqr
\psi_4(J,\phi,\tau) & = & \psi_3(x,p + q_1 x) \approx \psi_3(J,\phi,\tau) + 
q_1 x \frac{\del\psi_3}{\del p} \nonumber \\
& = & \psi_3 + q_1 x [\frac{\del \psi_3}{\del J} \frac{\del J}{\del p} +
\frac{\del \psi_3}{\del \phi} \frac{\del \phi}{\del p}] \nonumber \\
& = & \psi_3 - q_1 \sqrt{2\bt J}\cos\phi[ \sqrt{\frac{2J}{\bt}}\sin\phi \frac{\del \psi_3}{\del J}
 + \frac{1}{\sqrt{2\bt J}}\cos\phi \frac{\del \psi_3}{\del \phi} ]
\eeqr
In evaluating the terms in [], Chao had kept only the single term with $\om'\tau$ arguing that for
long $\tau$, that term dominates. That argument perhaps cannot be made here, since the next kick
may be applied soon after $\tau$. Keeping all terms, and using
\beqrs
\frac{\del\psi_3}{\del J}&  = &  \psi_0' + \bt_K\theta\sqrt{\frac{2}{\bt}}
\left[(\psi_0'\sqrt{J})'\sin(\phi-\om(J)\tau) - \om'(J)\tau \psi_0'\sqrt{J}\cos(\phi-\om(J)\tau)
\right] \\
\frac{\del\psi_3}{\del \phi}&  = & \bt_K\theta \sqrt{\frac{2J}{\bt}}\cos(\phi - \om(J)\tau)
\eeqrs
Putting all terms together
\beqr
\psi_4(J,\phi,\tau) & = & 
\psi_0(J) + \bt_K \theta \psi_0'(J) \sqrt{\frac{2J}{\bt}}\sin(\phi - \om(J)\tau) \nonumber \\
&  & - q_1 \sqrt{2\bt J}\cos\phi 
\left\{ \sqrt{\frac{2J}{\bt}}\sin\phi \left[ \psi_0' + \bt_K\theta\sqrt{\frac{2}{\bt}}
\right. \right. \nonumber \\
&  & \left. [(\psi_0'\sqrt{J})'\sin(\phi-\om(J)\tau) - \om'(J)\tau \psi_0'\sqrt{J}\cos(\phi-\om(J)
\tau)] \right]  \nonumber \\
&  & \left. + \frac{1}{\sqrt{2\bt J}}\cos\phi \bt_K\theta \sqrt{\frac{2J}{\bt}}\psi_0'
\cos(\phi - \om(J)\tau)
\right\}
\eeqr
We can drop the $\psi_0$ term as it does not contribute to the dipole moment. 
Hence the DF at time $m T_{rev}$ after time $\tau$ is
\beq
\psi_5(J,\phi,\tau+ m T_{rev}) = \psi_4(J,\phi - m\om(J)T_{rev}) 
\eeq
Introduce some shorthand notation
\beqrs
c_5 & = & \cos(\phi - m\om(J)T_{rev}) \;\;\; s_5 = \sin(\phi - m\om(J)T_{rev}) \\
c_{5\tau} & = & \cos(\phi - \om(J)\tau- m\om(J)T_{rev}) \;\;\; 
s_{5\tau} = \sin(\phi - \om(J)\tau- m\om(J)T_{rev}) 
\eeqrs
Then we can write
\beqr
\psi_5(J, \phi, \tau+ m T_{rev}) & = & 
\bt_K \theta \sqrt{\frac{2}{\bt}} \psi_0'(J)\sqrt{J}  s_{5\tau} \nonumber \\
&  & -q_1 \sqrt{2\bt J}c_5\left\{ \sqrt{\frac{2J}{\bt}}s_5 \left[ \psi_0' + \bt_K\theta\sqrt{\frac{2}{\bt}}
 [(\psi_0'\sqrt{J})'s_{5\tau} - \om'(J)\tau \psi_0'\sqrt{J}c_{5\tau}] \right] \right. \nonumber \\
&  & \left. + \frac{\bt_K}{\bt}\theta \psi_0' c_5 c_{5\tau} \right\} \nonumber \\
&  = &  \bt_K \theta \sqrt{\frac{2}{\bt}} \psi_0'(J)\sqrt{J}  s_{5\tau} \nonumber \\
& & - q_1 \left\{ J s_{25}\left[ \psi_0' + \bt_K\theta\sqrt{\frac{2}{\bt}}
 [(\psi_0'\sqrt{J})'s_{5\tau} - \om'(J)\tau \psi_0'\sqrt{J}c_{5\tau}] \right] \right. 
\nonumber \\
&  & \left. + \sqrt{\frac{2}{\bt}}\bt_K\theta \psi_0'\sqrt{J} c_5^2 c_{5\tau} \right\} 
\nonumber \\
&  = &  \bt_K \theta \sqrt{\frac{2}{\bt}} \psi_0'(J)\sqrt{J}  s_{5\tau} - q_1  J s_{25} \psi_0'
\nonumber \\
& & - q_1 \bt_K\theta\sqrt{\frac{2}{\bt}} \left\{  
J s_{25}\left[(\psi_0'\sqrt{J})'s_{5\tau} - \om'(J)\tau \psi_0'\sqrt{J}c_{5\tau} \right] 
 + \psi_0'\sqrt{J} c_5^2 c_{5\tau} \right\}
\eeqr
where we defined
\[ s_{25} = 2 c_5 s_5 = \sin 2(\phi - m\om(J)T_{rev}) \]
Anticipating that the echo must involve the dipole kick, we drop the second term so that
\beqr
\psi_5(J, \phi, \tau+ m T_{rev}) & \approx  & 
\bt_K \theta \sqrt{\frac{2}{\bt}}\left\{ \psi_0'(J)\sqrt{J}  s_{5\tau} \right. \nonumber \\
& & \left. - q_1 J s_{25}\left[(\psi_0'\sqrt{J})'s_{5\tau} - \om'(J)\tau \psi_0'\sqrt{J}c_{5\tau} \right] 
 - q_1 \psi_0'\sqrt{J} c_5^2 c_{5\tau} \right\}
\eeqr

At time $\tau + m T_{rev}$, there is a second quad kick $q_2$ which changes the DF to
\[
\psi_6(J,\phi,\tau + m T_{rev}) = \psi_5(x,p + q_2 x) \approx
\psi_5(J,\phi,\tau + m T_{rev}) + q_2 \sqrt{2\bt J}\cos\phi
[\frac{\del \psi_5}{\del J} \frac{\del J}{\del p} +
\frac{\del \psi_5}{\del \phi} \frac{\del \phi}{\del p}]
\]
Using
\beqrs
\frac{\del }{\del J}c_5 & = & \om'(J) m T_{rev}s_5 , \;\;\; 
\frac{\del }{\del J}s_5 = - \om'(J)m T_{rev}c_5  \\
\frac{\del }{\del J}c_{5\tau} & = & \om'(J)(\tau + m T_{rev}) s_{5\tau} , \;\;\; 
\frac{\del }{\del J}s_{5\tau} = - \om'(J)(\tau + m T_{rev}) c_{5\tau}
\eeqrs
Hence
\beqrs
\frac{\del\psi_5}{\del J} & = & 
\bt_K \theta\sqrt{\frac{2}{\bt}} \left\{ (\psi_0'\sqrt{J})' s_{5\tau} - 
\om'(J)(\tau + m T_{rev}) \psi_0'\sqrt{J} c_{5\tau}  \right. \\
& & - q_1 Js_{25}\left[
(\psi_0'\sqrt{J})''s_{5\tau} - \om'(J)(\tau + m T_{rev}) (\psi_0'\sqrt{J})' c_{5\tau} \right. \\
& &  - \om''(J)\tau  (\psi_0'\sqrt{J}) c_{5\tau} - \om'(J)\tau (\psi_0'\sqrt{J})' c_{5\tau}  \\
& &   - \left. (\om'(J))^2 \tau (\tau + m T_{rev}) (\psi_0'\sqrt{J}) s_{5\tau}]  \right] \\
&  & - q_1 ( s_{25} - 2 J \om'(J)m T_{rev}c_{25}) 
\left[(\psi_0'\sqrt{J})'s_{5\tau} - \om'(J)\tau \psi_0'\sqrt{J}c_{5\tau} \right] 
\\
&  &  \left. - q_1 \left[ \psi_0'\sqrt{J})'  c_5^2 c_{5\tau} + 
\psi_0'\sqrt{J} \om'(J)(2 m T_{rev}s_5 c_5 c_{5\tau} + (\tau + m T_{rev})c_5^2 s_{5\tau}) 
\right] \right\}
\eeqrs

Now I drop terms that are independent of $\om'(J) \tau$. I also now assume the following model
for the action dependence of $\om(J)$
\beq
\om(J) = \om_{\bt} + \om' J 
\eeq
where $\om'$ is a constant. Now we have
\beqrs
\frac{\del\psi_5}{\del J} & \approx & 
- \bt_K \theta\sqrt{\frac{2}{\bt}} \om'  \\
& & \left\{ \psi_0'\sqrt{J} \left[ (\tau + m T_{rev})c_{5\tau}  
+ q_1 \left( 2 m T_{rev}s_5 c_5 c_{5\tau} + (\tau + m T_{rev})c_5^2 
s_{5\tau} -  \tau s_{25} c_{5\tau} \right) \right] \right.  \\
& & - q_1 J s_{25}\left[
(\psi_0'\sqrt{J})' \left( (\tau + m T_{rev}) c_{5\tau} + \tau  c_{5\tau}\right) 
+ \om' \tau (\tau + m T_{rev}) (\psi_0'\sqrt{J}) s_{5\tau} \right] \\
&  &  \left.  -  2 q_1 J m T_{rev}c_{25} 
\left[(\psi_0'\sqrt{J})'s_{5\tau} - \om'\tau \psi_0'\sqrt{J}c_{5\tau} \right] \right\}
\eeqrs

The derivative with respect to the phase $\phi$ under the same approximation of dropping
terms independent of $\om'$ is given by
\beqrs
\frac{\del\psi_5}{\del \phi} & \approx & 
 \bt_K \theta\sqrt{\frac{2}{\bt}} q_1 \om' J \tau (\psi_0'\sqrt{J})\frac{\del}{\del \phi}
\left[ s_{25}c_{5\tau} \right] \\
& = &  \bt_K \theta\sqrt{\frac{2}{\bt}} q_1\om' J \tau (\psi_0'\sqrt{J})
\left[ 2 c_{25}c_{5\tau} - s_{25}s_{5\tau} \right]
\eeqrs

Hence the DF right after the 2nd quadrupole kick is
\beq
\psi_6(J,\phi,\tau + m T_{rev}) = \psi_5 - q_2( J \sin 2\phi \frac{\del\psi_5}{\del J}
 + \cos^2\phi \frac{\del\psi_5}{\del\phi})
\eeq
while the DF at time $t$ after the dipole kick and after the 2nd quadrupole kick is
\beq
\psi_7(J,\phi,t > \tau + m T_{rev}) = \psi_6(J,\phi - \om(J)(t - (\tau+mT_{rev})))
\eeq

Under the transform $\phi \rarw \phi - \om(J)(t - (\tau+mT_{rev}))$, the different
trigonometric terms transform as 
\beqrs
\cos\phi & \rarw & \cos(\phi - \om(J)(t - \tau - m T_{rev})) \equiv c_{\tau,m} \\
c_5  = \cos( \phi - \om(J) mT_{rev}) & \rarw & \cos(\phi - \om(J)(t - \tau)) \equiv c_{\tau} \\
c_{5\tau} = \cos(\phi - \om(J)(\tau+ mT_{rev}) & \rarw & \cos(\phi - \om(J)t) \equiv c_t\\
c_{25} = \cos(2(\phi- \om(J) mT_{rev})) & \rarw & \cos(2(\phi - \om(J)(t-\tau))) \equiv 
c_{2\tau}
\eeqrs
and similarly the sine terms. Also, define
\[ \sin 2 \phi \rarw 2 c_{\tau,m}s_{\tau_m} \equiv s_{2\tau,m} \]

The DF $\psi_5$ transforms to (after keeping only the $\om'$ dependent terms)
\beq
\psi_5 = 
q_1 \bt_K \theta \sqrt{\frac{2}{\bt}} J s_{2\tau} \om'\tau (\psi_0'\sqrt{J}) c_{t}
\eeq

while the derivative terms transform as
\beqrs
\frac{\del\psi_5}{\del J} & = & 
- \bt_K \theta\sqrt{\frac{2}{\bt}} \om'  \\
& & \left\{ \psi_0'\sqrt{J} \left[ (\tau + m T_{rev})c_t 
+ q_1 \left( 2 m T_{rev}s_{\tau} c_{\tau} c_t + (\tau + m T_{rev})c_{\tau}^2 
s_t - \tau s_{2\tau} c_t \right) \right] \right.  \\
& & - q_1 J(\psi_0'\sqrt{J})'\left[
 \left( (\tau + m T_{rev}) c_t + \tau  c_t\right) s_{2\tau} +  2 m T_{rev}c_{2\tau}s_t 
\right] \\
& & \left. - q_1 \om' \tau J (\psi_0'\sqrt{J}) \left[ (\tau + m T_{rev}) s_{2\tau} s_t 
-  2  m T_{rev}c_{2\tau} c_t \right] \right\} \\
\frac{\del\psi_5}{\del \phi} & = & 
 \bt_K \theta\sqrt{\frac{2}{\bt}} q_1 \om' \tau J (\psi_0'\sqrt{J})
\left[ 2 c_{2\tau}c_{t} - s_{2\tau}s_{t} \right]
\eeqrs
and the DF is
\beq
\psi_7(J,\phi,t) = 
\psi_5 - q_2( J s_{2\tau,m}\frac{\del\psi_5}{\del J} + c_{\tau,m}^2 \frac{\del\psi_5}{\del\phi})
\eeq

The dipole moment at time $t$ is 
\beqr
\lan x\ran(t)&  =  & \sqrt{2\bt} \int \sqrt{J}\cos\phi \psi_7(J,\phi,t) d\phi dJ \nonumber \\
& = & \sqrt{2\bt} \int \sqrt{J}\cos\phi \left[ 
\psi_5 - q_2( J s_{2\tau,m}\frac{\del\psi_5}{\del J} + c_{\tau,m}^2 \frac{\del\psi_5}{\del\phi})
\right] d\phi dJ
\eeqr
The term that is independent of $q_1, q_2$ vanishes after integrating over $\phi$. The
remaining terms can be categorized into three groups as $O(q_1), O(q_2), O(q_1q_2)$. Since
the dimensionless quadrupole kicks $q_1, q_2 \ll 1$, we may expect the terms in the last
group to be negligible. We label these terms as $T(q_1), T(q_2)$ and $T(q_1,q_2)$ where
\[
 \lan x\ran(t) \equiv T(q_1) + T(q_2) + T(q_1 q_2)
\]
Writing
\beqrs
\frac{\del\psi_5}{\del J} & \equiv & - \bt_K \theta\sqrt{\frac{2}{\bt}} \om' 
\left\{ \psi_0'\sqrt{J} (\tau + m T_{rev})c_t 
+ q_1(\frac{\del\psi_5}{\del J})_{q1}  \right\} \\
&  \equiv & (\frac{\del\psi_5}{\del J})_0 - [\bt_K \theta\sqrt{\frac{2}{\bt}} \om' ]
q_1(\frac{\del\psi_5}{\del J})_{q1}  \\
 \frac{\del\psi_5}{\del \phi} & \equiv & [\bt_K \theta\sqrt{\frac{2}{\bt}} \om'] q_1 
(\frac{\del\psi_5}{\del \phi})_{q1}
\eeqrs
We have
\beqr
T(q_1) & = & \sqrt{2\bt} \int \sqrt{J}\cos\phi \psi_5(J,\phi,t) d\phi dJ \\
& = & q_1 \sqrt{2\bt} \bt_K \theta\sqrt{\frac{2}{\bt}}\om'\tau \int dJ d\phi 
\sqrt{J}\cos\phi J s_{2\tau} \psi_0'\sqrt{J}c_{t}  \nonumber  \\
& = & 2 q_1 \bt_K \theta \om'\tau \int dJ J^2 \psi_0' \int d\phi \cos\phi  s_{2\tau} c_{t}  \\
T(q_2) & = & -q_2\sqrt{2\bt} \int \sqrt{J}\cos\phi J s_{2\tau,m}
(\frac{\del\psi_5}{\del J})_0 d\phi dJ \\
& = & -q_2 \sqrt{2\bt}  (-\bt_K \theta\sqrt{\frac{2}{\bt}} \om')
\int dJ d\phi \sqrt{J}\cos\phi 
J s_{2\tau,m} \left\{ \psi_0'\sqrt{J} (\tau + m T_{rev})c_t \right\} \nonumber \\
& = & 2 q_2 \bt_K \theta \om' \int dJ J^2 \psi_0' \int d\phi
\cos\phi s_{2\tau,m} \left[ (\tau + m T_{rev})c_t \right]
\eeqr
and
\beqr
T(q_1, q_2)  & = & -q_2\sqrt{2\bt} \int \sqrt{J}\cos\phi (-\bt_K \theta\sqrt{\frac{2}{\bt}} \om')
\left\{ J s_{2\tau,m}q_1(\frac{\del\psi_5}{\del J})_{q1} - 
q_1 c_{\tau,m}^2(\frac{\del\psi_5}{\del \phi})_{q1}  \right\}  d\phi dJ \nonumber \\
& = & 2 q_1 q_2 \bt_K \theta \om' \int dJ \sqrt{J} \int d\phi \cos\phi
\left\{ Js_{2\tau,m}(\frac{\del\psi_5}{\del J})_{q1} - c_{\tau,m}^2(\frac{\del\psi_5}{\del \phi})_{q1} \right\} \\
(\frac{\del\psi_5}{\del J})_{q1} & = & 
\psi_0'\sqrt{J}\left( 2 m T_{rev}s_{\tau} c_{\tau} c_t + (\tau + m T_{rev})c_{\tau}^2 
s_t - \tau s_{2\tau} c_t \right)  \nonumber \\
& & - J(\psi_0'\sqrt{J})'\left[
 \left( (\tau + m T_{rev}) c_t + \tau  c_t\right) s_{2\tau} +  2 m T_{rev}c_{2\tau}s_t 
\right]  \\
& &  - \om' \tau J (\psi_0'\sqrt{J}) \left[ (\tau + m T_{rev}) s_{2\tau} s_t 
-  2  m T_{rev}c_{2\tau} c_t \right]  \label{eq: psi5_J} \\
(\frac{\del\psi_5}{\del \phi})_{q1} & = & 
- \tau J (\psi_0'\sqrt{J})\left[ 2 c_{2\tau}c_{t} - s_{2\tau}s_{t} \right] \label{eq: psi5_phi}
\eeqr
Writing $\om' = (\om_{rev}/J_0)\mu$ where $\mu$ is the detuning parameter,
 $\om'\tau J_0 = 2\pi N_d \mu$ where $N_d$ is the delay in turns. With $N_d \simeq 400, 
\mu \simeq 0.001$, $\om'\tau J_0 \simeq O(1)$, hence it is of the same order as the other
terms. It appears that all terms in $T(q_1,q2)$ have to be kept. 

The $\phi$ integration in $T(q_1)$ yields
\[
 \int d\phi \cos\phi  \sin(2(\phi - \om(J)(t-\tau)))\cos(\phi-\om(J)t) = 
-\frac{\pi}{2}\sin(\om(J)(t - 2\tau))
\]
with a similar $\phi$ integration in $T(q_2)$ 
\[
\int  d\phi  \cos\phi \sin(2(\phi - \om(J)(t-\tau- m T_{rev}))) \cos(\phi-\om(J)t) \\
 =  -\frac{\pi}{2}\sin(\om(J)(t - 2(\tau + m T_{rev}))) \\
\]
Hence
\beqrs
T(q_1) & = & -\pi q_1 \bt_K \theta \om'\tau \int dJ \; J^2 \psi_0' \sin(\om(J)(t - 2\tau)) \\
T(q_2) & = & -\pi q_2 \bt_K \theta \om' (\tau + m T_{rev}) \int dJ \; J^2 \psi_0' 
\sin[\om(J)(t - 2(\tau + m T_{rev}))]
\eeqrs

To 1st order in the quad strength, each quad has a similar contribution with a time dependent
factor depending on the time the kick is applied.

If there are $N_q$ kicks applied at times $\tau + m T_{rev}$, $m=0,..., N_q$, then again to
1st order in the kicks, the dipole moment is
\begin{framed}
\beq
\lan x\ran(t) = -\pi \bt_K \theta \om' \int dJ \; J^2 \psi_0'
\sum_{m=0}^{N_q} q_m (\tau + m T_{rev}) \sin[\om(J)(t - 2(\tau + m T_{rev}))] 
\eeq
\end{framed}

\noi We consider two cases of the above \newline
\noi Case 1:  All kicks have the same strength: $q_m = q$, then
\beq
\lan x\ran(t) = -\pi q \bt_K \theta \om' \int dJ \; J^2 \psi_0'
\sum_{m=0}^{N_q} (\tau + m T_{rev}) \sin[\om(J)(t - 2(\tau + m T_{rev}))] 
\eeq
Using
\beqr
\sum_{m=0}^{N_q} \sin[\om(J)(t - 2(\tau + m T_{rev}))] & = & 
\frac{ \sin[\om(J)(N_q + 1)T_{rev}]\sin[\om(J)(t - 2\tau - N_q T_{rev})]}{\sin(\om(J)T_{rev})}  
\nonumber \\
\sum_{m=0}^{N_q} m \sin[\om(J)(t - 2(\tau + m T_{rev}))] & = & (\frac{1}{2\sin(\om(J)T_{rev})})^2
\left\{ -\sin[\om(J)(t - 2\tau)]   \right.  \nonumber \\
& &  + (N_q+1)\sin[\om(J)(t - 2(\tau + N_q T_{rev}))] \nonumber \\
& & \left. - N_q \sin[\om(J)(t - 2(\tau + (N_q +1)T_{rev}))] \right\} 
\eeqr
$N_q = 0$ corresponds to the case with the single quad kick at $\tau$. In this case, the second
sum vanishes and the first sum contributes $\sin[\om(J)(t - 2\tau)]$, the same as before. 

Case 2: Alternating sign quad kicks: $q_m = (-1)^m q$
\beq
\lan x\ran(t) = -\pi q \bt_K \theta \om' \int dJ \; J^2 \psi_0'
\sum_{m=0}^{N_q} (-1)^m (\tau + m T_{rev}) \sin[\om(J)(t - 2(\tau + m T_{rev}))] 
\eeq
Using
\beqr
\sum_{m=0}^{N_q} (-1)^m \sin[\om(J)(t - 2(\tau + m T_{rev}))] & = & 
\frac{\cos[N_q\pi/2 - (N_q+1)\om(J)T_{rev}]}{\cos(\om(J)T_{rev})} \nonumber \\
 & &  \times \sin[N_q\pi/2 + \om(J)(t-2\tau-N_q T_{rev})] \nonumber \\
\sum_{m=0}^{N_q} (-1)^m m \sin[\om(J)(t - 2(\tau + m T_{rev}))] & = & 
(\frac{1}{2\cos(\om(J)T_{rev})})^2 
\left\{ -\sin[\om(J)(t-2\tau)] \right. \nonumber \\
& & + (N_q+1)\sin[N_q\pi + \om(J)(t-2(\tau + N_q T_{rev}))] \nonumber \\
& & \!\!\! \!\!\! \!\!\! \!\!\! \!\!\!  \left. + N_q \sin[N_q\pi + \om(J)(t - 2(\tau + (N_q+1)T_{rev}))] \right\}
\eeqr

The integration over the action however cannot be done analytically. 
So instead of summing over the trigonometric term, it may be preferable to do the integration 
first and then do a numerical sum over the integrated terms. Since
\[ 
\psi_0(J) = \frac{1}{2\pi J_0}\exp[-\frac{J}{J_0}], \;\;\; 
\psi_0'(J) = -\frac{1}{2\pi J_0^2}\exp[-\frac{J}{J_0}]
\]
we have
\beqr
\lan x\ran(t) & = & 
\half \bt_K \theta \om' J_0 \sum_{m=0}^{N_q} q_m (\tau + m T_{rev}) \times \nonumber \\
& & {\rm Im}\left[ e^{[i \om_{\bt}(t - 2(\tau + m T_{rev}))]} \int dz \; z^2 \exp[-z]
\exp[i\om' J_0 z(t - 2(\tau + m T_{rev}))] \right] \nonumber \\
& = &  \bt_K \theta \om' J_0 \sum_{m=0}^{N_q} q_m (\tau + m T_{rev}) 
{\rm Im}\left[ \frac{\exp[i\Phi_m]}{(1 - i\xi_m)^3} \right]  \nonumber \\
& = & \bt_K \theta \om' J_0 \sum_{m=0}^{N_q} q_m (\tau + m T_{rev}) 
\frac{\left[\xi_m(3 - \xi_m^2)\cos\Phi_m + (1 - 3\xi_m^2)\sin\Phi_m \right] }{(1 + \xi_m^2)^3}
 \nonumber \\
& = & \bt_K \theta \om' J_0 \sum_{m=0}^{N_q} q_m (\tau + m T_{rev}) 
\frac{\sin(\Phi_m + 3\tan^{-1}[\xi_m]) }{(1 + \xi_m^2)^{3/2}}
\label{eq: amp_multquads_linear_1}  \\
\Phi_m & = & \om_{\bt}(t - 2(\tau + m T_{rev}))  \nonumber \\
\xi_m & = & \om' J_0(t - 2(\tau + m T_{rev}))  \nonumber 
\eeqr
In doing the integration over $z$, we used
\[ \int_0^{\infty} dz \; z^2 \exp[-a z] = \frac{2}{a^3} 
\]
In the above we assumed that the quad kicks are applied every turn from $2\tau$ to 
$2\tau + N_q T_{rev}$. If instead the kicks are applied with the same gap $n_{gap}$ between kicks,
then the above formula is easily generalized by replacing $T_{rev}$ by $ n_{gap} T_{rev}$ with 
$n_{gap} \ge 1$.
Thus if $n_{gap} = 2$, there is a gap of a single turn between kicks. If the gaps between 
successive kicks are different, then $n_{gap}$ will depend on the $m$th kick. 

As a special case of the above, consider a {\bf stimulated echo} at later times with 2 quad 
kicks, i.e. $N_q=1$ and
the 2nd kick is applied at time $p\tau$ after the 1st kick. In this case, we have
$n_{gap} T_{rev} = p\tau$ and
\beqr
\lan x(t) \ran & = & \bt_K \theta \om' J_0 [q_0 (\tau) \frac{\sin(\Phi_0 + 3\tan^{-1}[\xi_0])}{(1 + \xi_0^2)^{3/2}}
+ q_1 ((p+1)\tau) \frac{\sin(\Phi_1 + 3\tan^{-1}[\xi_1]) }{(1 + \xi_1^2)^{3/2}} ]
\nonumber \\
\Phi_0 & = & \om_{\bt}(t - 2\tau), \;\;\; \Phi_1  =  \om_{\bt}(t - 2(p+1)\tau)  \nonumber \\
\xi_0 & = & \om' J_0(t - 2\tau ), \;\;\; \xi_1  =  \om' J_0(t - 2(p+1)\tau )  \nonumber 
\eeqr
Note that the second term is only applicable for times $t \ge (p+1)\tau$. If we assume that the
two kicks have the same sign, $q_0 = q_1 = q$ and introducing the dimensionless quad strength 
parameter
\[  Q = q \om' J_0 \tau \]
We have therefore 
\beq
\lan x(t) \ran  =  \bt_K \theta Q [ \frac{\sin(\Phi_0 + 3\tan^{-1}[\xi_0])}{(1 + \xi_0^2)^{3/2}}
+  (p+1) \frac{\sin(\Phi_1 + 3\tan^{-1}[\xi_1]) }{(1 + \xi_1^2)^{3/2}} ]
\eeq
The stimulated echo at time $t = 4\tau$ is given by this expression with
\beqrs
\Phi_0 & = & 2\om_{\bt}\tau3, \;\;\; \Phi_1  =  2\om_{\bt}\tau(2 - (p+1))   \\
\xi_0 & = & 2\om' J_0\tau , \;\;\; \xi_1  =  2\om' J_0\tau(2 - (p+1))  
\eeqrs

\subsection{2nd order in quad kicks}

Now we consider the terms in $T(q_1, q_2)$. We had
\beq
T(q_1, q_2) = 2 q_1 q_2 \bt_K \theta \om' \int dJ \sqrt{J}\int d\phi \cos\phi
\left\{ Js_{2\tau,m}(\frac{\del\psi_5}{\del J})_{q1} - c_{\tau,m}^2
(\frac{\del\psi_5}{\del \phi})_{q1} \right\} \equiv T(q_1, q_2)|_{J} + T(q_1, q_2)|_{\phi} 
\label{eq: T_q1q2}
\eeq
We simplify the earlier expression Eq.(\ref{eq: psi5_J}) for the partial derivative
\beqrs
(\frac{\del\psi_5}{\del J})_{q1} & = & 
\psi_0'\sqrt{J}\left( (m T_{rev} - \tau) s_{2\tau} c_t + (\tau + m T_{rev})c_{\tau}^2 s_t  \right) \\
& & - J(\psi_0'\sqrt{J})'\left[
  (2\tau + m T_{rev}) s_{2\tau} c_t  +  2 m T_{rev}c_{2\tau}s_t \right]  \\
& &  - \om' \tau J (\psi_0'\sqrt{J}) \left[ (\tau + m T_{rev}) s_{2\tau} s_t 
-  2  m T_{rev}c_{2\tau} c_t \right]
\eeqrs
There are six distinct terms in the derivative w.r.t the action. Doing the integrations 
over the phase, we have the five distinct integrals
\beqrs
I_1 & = & \int d\phi \cos\phi s_{2\tau,m} s_{2\tau} c_t \\
 & =  & \int d\phi \cos\phi \sin 2[\phi - \om(J)(t - \tau - m T_{rev})]
 \sin 2[\phi - \om(J)(t - \tau)] \cos[\phi - \om(J)t] \\
& = & \frac{\pi}{2}\cos [\om(J)t] \cos [2\om(J) m T_{rev}] \\
& = & \frac{\pi}{4}\left(\cos[\om(J)(t - 2 m T_{rev})] + \cos[\om(J)(t + 2 m T_{rev})] \right) \\
I_2 & = & \int d\phi \cos\phi s_{2\tau,m} c_{\tau}^2 s_t \\
 & =  & \int d\phi \cos\phi \sin 2[\phi - \om(J)(t - \tau - m T_{rev})]
 \cos^2[\phi - \om(J)(t - \tau)] \sin[\phi - \om(J)t] \\
& = & \frac{\pi}{4}[\cos[\om(J)(t - 2(\tau+ m T_{rev}))] - \sin[\om(J)t] \sin [2\om(J) m T_{rev}]]
 \\
& = & \frac{\pi}{4}\left[ \cos[\om(J)(t - 2(\tau+ m T_{rev}))] - 
\half\left\{\cos[\om(J)(t - 2 m T_{rev})] - \cos[\om(J)(t + 2 m T_{rev})] \right\} \right] \\
I_3 & = & \int d\phi \cos\phi s_{2\tau,m} c_{2\tau} s_t \\
 & =  & \int d\phi \cos\phi \sin 2[\phi - \om(J)(t - \tau - m T_{rev})]
 \cos 2[\phi - \om(J)(t - \tau)] \sin[\phi - \om(J)t] \\
& = & -\frac{\pi}{2}\sin [\om(J)t] \sin [2\om(J) m T_{rev}] \\
& = & -\frac{\pi}{4}\left[\cos[\om(J)(t - 2 m T_{rev})] - \cos[\om(J)(t + 2 m T_{rev})] \right] \\
I_4 & = & \int d\phi \cos\phi s_{2\tau,m} s_{2\tau} s_t \\
 & =  & \int d\phi \cos\phi \sin 2[\phi - \om(J)(t - \tau - m T_{rev})]
 \sin 2[\phi - \om(J)(t - \tau)] \sin[\phi - \om(J)t] \\
& = & -\frac{\pi}{2}\sin [\om(J)t] \cos [2\om(J) m T_{rev}] \\
& = & -\frac{\pi}{4}\left[ \sin[\om(J)(t - 2 m T_{rev})] + \sin[\om(J)(t + 2 m T_{rev})] \right] \\
I_5 & = & \int d\phi \cos\phi s_{2\tau,m} c_{2\tau} c_t \\
 & =  & \int d\phi \cos\phi \sin 2[\phi - \om(J)(t - \tau - m T_{rev})]
 \cos 2[\phi - \om(J)(t - \tau)] \cos[\phi - \om(J)t] \\
& = & \frac{\pi}{2}\cos [\om(J)t] \sin [2\om(J) m T_{rev}] \\
& = & -\frac{\pi}{4}\left[ \sin[\om(J)(t - 2 m T_{rev})] - \sin[\om(J)(t + 2 m T_{rev})] \right] \\
\eeqrs
Combining all the terms in $(\del\psi_5/\del J)_{q1}$, 
\beqrs
\frac{T(q1,q2)|_J}{2 q_1 q_2 \bt_K \theta \om'}
& =  & \int dJ \sqrt{J} \int d\phi \cos\phi Js_{2\tau,m}(\frac{\del\psi_5}{\del J})_{q1}\\
& = & \int J^{3/2}  \left[
\psi_0'\sqrt{J}\left\{[ (m T_{rev} - \tau)I_1 + (\tau + m T_{rev})I_2]  
- \om' \tau J  [ (\tau + m T_{rev}) I_4 - 2  m T_{rev}I_5 ] \right\}\right. \\
& & - \left. J(\psi_0'\sqrt{J})' [ (2\tau + m T_{rev}) I_1  +  2 m T_{rev}I_3 ] \right] \; dJ
\eeqrs

\vspace{2em}

Now the term from $(\del\psi_5/\del \phi)_{q1}$, 
Substituting from Eq.(\ref{eq: T_q1q2}) and Eq.(\ref{eq: psi5_phi}),
\beqrs
T(q_1, q_2)|_{\phi} & = & -
2 q_1 q_2 \bt_K \theta \om' \int dJ \sqrt{J}\int d\phi \cos\phi
c_{\tau,m}^2 (\frac{\del\psi_5}{\del \phi})_{q1} \\ 
& = & 
2 q_1 q_2 \bt_K \theta \om' \tau \int dJ J^{3/2} \int d\phi \cos\phi
c_{\tau,m}^2 (\psi_0'\sqrt{J})\left[ 2 c_{2\tau}c_{t} - s_{2\tau}s_{t} \right]
\eeqrs

The $\phi$ integrations involve
\beqrs
I_6 & = & \int d\phi \; \cos\phi c_{\tau,m}^2 c_{2\tau}c_{t} \\
  & = & \int d\phi \; \cos\phi \cos^2[\phi - \om(J)(t - \tau - m T_{rev})]
 \cos 2[\phi - \om(J)(t - \tau)] \cos[\phi - \om(J)t] \\
 & = & \frac{\pi}{2}\cos[2\om(J)m T_{rev}]\cos[\om(J)t] \\
 & = & \frac{\pi}{4}[\cos[\om(J)(t - 2 m T_{rev})] + \cos[\om(J)(t + 2 m T_{rev})] ]
\\
I_7 & = & \int d\phi \; \cos\phi c_{\tau,m}^2 s_{2\tau} s_{t} \\
  & = & \int d\phi \; \cos\phi \cos^2[\phi - \om(J)(t - \tau - m T_{rev})]
 \sin 2[\phi - \om(J)(t - \tau)] \sin[\phi - \om(J)t] \\
 & = & \frac{\pi}{2}\sin[\om(J)t] \sin[2\om(J) m T_{rev}] \\
 & = & \frac{\pi}{4}[\cos[\om(J)(t - 2 m T_{rev})] - \cos[\om(J)(t + 2 m T_{rev})] ]
\eeqrs

Combining the two terms, we have
\beqrs
T(q_1, q_2)|_{\phi} & = &
2 q_1 q_2 \bt_K \theta \om' \tau \int dJ J^{2}\psi_0' [ 2 I_6 - I_7] \\
& = & \frac{\pi}{2}q_1 q_2 \bt_K \theta (\om')^2 \tau \int dJ J^{2}(\psi_0'
\{ \cos[\om(J)(t - 2 m T_{rev})] + 3\cos[\om(J)(t + 2 m T_{rev})] \}
\eeqrs

In the expressions for the integrated terms $I_j$, $j=1,...7$, there are only three different arguments.
Define
\[
\Phi_{\tau,m} = \om(J)(t - 2(\tau+ m T_{rev})), \;\;\; \Phi_{m,\pm}= \om(J)(t \pm 2 m T_{rev})
\]
Then we have
\beqrs
I_1 & = & \frac{\pi}{4}\left\{ {\rm Re}[\exp(i\Phi_{m,-})] +  {\rm Re}[\exp(i\Phi_{m,+})]\right\} \\
I_2 & = & \frac{\pi}{4}\left\{ {\rm Re}[\exp(i\Phi_{\tau,m})] -\half {\rm Re}[\exp(i\Phi_{m,-})] 
+  \half{\rm Re}[\exp(i\Phi_{m,+})]\right\} \\
I_3 & = & \frac{\pi}{4}\left\{ {\rm Re}[\exp(i\Phi_{m,+})] -  {\rm Re}[\exp(i\Phi_{m,-})]\right\} \\
I_4 & = & -\frac{\pi}{4}\left\{ {\rm Im}[\exp(i\Phi_{m,-})] +  {\rm Im}[\exp(i\Phi_{m,+})]\right\} \\
I_5 & = & -\frac{\pi}{4}\left\{ {\rm Im}[\exp(i\Phi_{m,-})] -  {\rm Im}[\exp(i\Phi_{m,+})]\right\} \\
I_6 & = & I_1 = \frac{\pi}{4}\left\{ {\rm Re}[\exp(i\Phi_{m,-})] +  {\rm Re}[\exp(i\Phi_{m,+})]\right\} \\
I_7 & = &  - I_3 = \frac{\pi}{4}\left\{ {\rm Re}[\exp(i\Phi_{m,-})] -  {\rm Re}[\exp(i\Phi_{m,+})]\right\} 
\eeqrs

Defining
\[ \tau_{\pm} = \tau \pm m T_{rev} \]
and combining all terms, we have
\beqr
T(q_1, q_2) & = & T(q_1, q_2)|_J + T(q_1, q_2)|_{\phi} \nonumber \\
& = & 2 q_1 q_2 \bt_K \theta \om'
\int  \left[ J^2 \psi_0'\left\{[ \tau_{+} I_2 -\tau_{-}I_1 ]  
- \om' \tau J  [  \tau_{+} I_4 - 2mT_{rev}I_5 ] \right\}\right. \nonumber \\
& & - \left. J^{5/2}(\psi_0'\sqrt{J})' [ (\tau + \tau_+) I_1  +  2 m T_{rev}I_3 ] \right] \; dJ \nonumber \\
& & + 2 q_1 q_2 \bt_K \theta \om' \tau \int  J^{2}\psi_0' [ 2 I_1 + I_3] dJ \nonumber \\
& = & 2 q_1 q_2 \bt_K \theta \om' \times \nonumber \\
& & \int  \left[ J^2 \psi_0'\left\{[ \tau_{+} (I_2 + I_1) + \tau I_3]  
- \om' \tau J  [  \tau_{+} I_4 - 2mT_{rev}I_5 ] \right\} \right. \nonumber \\
& & - \left. J^{5/2}(\psi_0'\sqrt{J})' [ (\tau + \tau_+) I_1  +  2 m T_{rev}I_3 ] \right] \; dJ \\
 & = & 2 q_1 q_2 \bt_K \theta \om' (T_1 + T_2 + T_3) \nonumber
\eeqr
There are three types of terms in the $J$ integration:
\[ \int dJ J^2 \psi_0' \exp[i a J], \;\;\; \int dJ J^3 \psi_0' \exp[i a J] \;\;\;
\int dJ J^{5/2} (\psi_0'\sqrt{J})' \exp[i a J]
\]
With
\beqrs
\psi_0 & = & \frac{1}{2\pi J_0}\exp[-J/J_0], \;\;\; \psi_0' = -\frac{1}{2\pi J_0^2}\exp[-J/J_0] \\
J^{5/2}(\psi_0' \sqrt{J})' & = & -\frac{J^{5/2}}{2\pi J_0^2}[-\sqrt{\frac{J}{J_0}} + 
\frac{1}{2\sqrt{J}}]\exp[-J/J_0] \\
 & = & -\frac{1}{2\pi J_0^2}[-\frac{J^3}{J_0} + \half J^2]\exp[-J/J_0]
\eeqrs
we have
\beqrs
\int dJ J^2 \psi_0' \exp[i a J] & = & -\frac{1}{2\pi J_0^2}\int dJ \; J^2 \exp[-J/J_0] \exp[iaJ]\\
 & = & -\frac{J_0}{2\pi}\int dz \; z^2 \exp[-(1 - i a J_0)z] \\
 & = & -\frac{J_0}{\pi}\frac{1}{(1 - i a J_0)^3}  \\
\int dJ J^3 \psi_0' \exp[i a J] & = & -\frac{3 J_0}{\pi} \frac{1}{(1 - i a J_0)^4} \\
\eeqrs

Hence
\beqrs
T_1 & \equiv & \int dJ\; J^2 \psi_0'\left\{[ \tau_{+} (I_2 + I_1) + \tau I_3] \right\}   \\
& = & \frac{\pi}{4}\int dJ\; J^2 \psi_0'\left\{[ \tau_{+}\left( {\rm Re}[\exp(i\Phi_{\tau,m})] + 
\half {\rm Re}[\exp(i\Phi_{m,-})] +  \frac{3}{2}{\rm Re}[\exp(i\Phi_{m,+})]\right) \right. \\
 & & \left. + \tau \left({\rm Re}[\exp(i\Phi_{m,+})] -  {\rm Re}[\exp(i\Phi_{m,-})]\right) \right\} \\
 & = & \frac{\pi}{4}\int dJ\; J^2 \psi_0'\left\{ \tau_{+}{\rm Re}[\exp(i\Phi_{\tau,m})] + 
+ (\half\tau_+ - \tau) {\rm Re}[\exp(i\Phi_{m,-})] \right. \\
& & \left. + (\frac{3}{2}\tau_+ + \tau){\rm Re}[\exp(i\Phi_{m,-})] \right\} \\
& = & -\frac{J_0}{4}\left\{ \tau_+ {\rm Re}[\frac{e^{i\om_{\bt}\Dl t_m}}{(1 - i\xi_{\tau,m})^3}]
 + (\half\tau_+ - \tau){\rm Re}[\frac{e^{i\om_{\bt}t_{m,-}}}{(1 - i\xi_{m,-})^3}] 
 + (\frac{3}{2}\tau_+ + \tau){\rm Re}[\frac{e^{i\om_{\bt}t_{m,+}}}{(1 - i\xi_{m,+})^3}] \right\}
\eeqrs
where $\Dl t_m, t_{m,\pm}, \xi_{\tau,m}, \xi_{m,\pm}$ are defined by 
\beqrs
\Phi_{\tau,m} & = & \om(J) (t - 2(\tau+ m T_{rev})) \equiv (\om_{\bt}+ \om' J)\Dl t_m \\
\Phi_{m,\pm} & = & \om(J)(t \pm 2 m T_{rev}) \equiv (\om_{\bt}+ \om' J)t_{m,\pm}  \\
\xi_{\tau,m} & = & \om' J_0 \Dl t, \;\;\; \xi_{m,\pm}  =  \om' J_0 t_{m,\pm}
\eeqrs
Next
\beqrs
T_2 & = & 
- \om' \tau \int dJ\; J^3 \psi_0'  [  \tau_{+} I_4 - 2mT_{rev}I_5 ] \\
 & = & - \om' \tau(-\frac{\pi}{4})\int dJ\; J^3 \psi_0' \left\{
\tau_+({\rm Im}[\exp(i\Phi_{m,-})] +  {\rm Im}[\exp(i\Phi_{m,+})]) \right. \\
&  & \left. - 2mT_{rev}({\rm Im}[\exp(i\Phi_{m,-})] -  {\rm Im}[\exp(i\Phi_{m,+})]) \right\} \\
& = &  \frac{\pi}{4}\om' \tau\int dJ\; J^3 \psi_0' \left\{
(\tau_+ - 2mT_{rev}){\rm Im}[\exp(i\Phi_{m,-})] + (\tau_+ + 2mT_{rev}){\rm Im}[\exp(i\Phi_{m,+})]
\right\}
\eeqrs 

\clearpage

\section{Nonlinear quad kicks} \label{sec: NL_quad}

Here a theory to find the echo amplitude with a nonlinear dependence on the quad strength is developed.
A Lagrangian theory was developed in \cite{Stup_Kauf}. Here instead we develop an Eulerian theory
by following the flow of the density distribution, similar to that in \cite{Chao}.

We start with the usual definitions  of the phase space variables
\[ x = \sqrt{2 \bt J} \cos\phi, \;\;\;\;  p = \bt x' + \al x = - \sqrt{2 \bt  J} \sin\phi  \]
and the inverse relations
\[ J = \frac{1}{2\bt}[x^2 + p^2], \;\;\;\; \phi = \tan^{-1}(\frac{-p}{x}) \]

\noi $\psi_0(J)$ is the initial distribution with initial emittance $\eps_0 = J_0$
\beq
\psi_0(J) = \frac{1}{2\pi J_0} \exp[-\frac{J}{J_0}]
\eeq
\noi $\psi_1(J,\phi)$ is the distribution function (DF) after the dipole kick $\Dl p = \bt \Dl x' = \bt \theta$
\beq
\psi_1(J, \phi) = \psi_0(x,p - \bt_K\theta) \simeq 
\psi_0(J) + \bt_K \theta \psi_0'(J) \sqrt{\frac{2J}{\bt}}\sin\phi
\eeq
In the second equality, the DF was expanded to first order in $\theta$. 

\noi $\psi_2(J,\phi,t)$ is the DF at time $t$ after the dipole kick
\beq
\psi_2(J,\phi,t) = \psi_0(J) + \bt_K \theta \psi_0'(J) \sqrt{\frac{2J}{\bt}}\sin(\phi - \om(J)t)
\eeq
\noi $\psi_3(J,\phi,\tau)$ is the DF at time $\tau$ after the dipole kick
\beq
\psi_3(J,\phi,\tau) = \psi_0(J) + \bt_K \theta \psi_0'(J) \sqrt{\frac{2J}{\bt}}\sin(\phi - \om(J)\tau)
\eeq
Since $\psi_0(J)$ will not contribute to the dipole moment, it will be dropped. 
The quad kick $\Dl p = - qx$ changes the distribution to
\beq
\psi_4(x,p,\tau) = \psi_3(x, p + q x,\tau) 
\eeq
Under this change, we have in the argument of the density distribution,
\beqrs
J & \rarw & \frac{1}{2\bt}[x^2 + (p+qx)^2] = J + \frac{1}{2\bt}(2q px + q^2 x^2) \\
  & \rarw & J + J(-q \sin 2\phi + q^2 \cos^2\phi) \equiv J + A(q,\phi)J \\
\phi & \rarw &  \tan^{-1}(-\frac{p + qx}{x}) = \tan^{-1}(\tan\phi - q)
\eeqrs
where
\[ A(q,\phi) = (-q \sin 2\phi + q^2 \cos^2\phi) \]
Hence
\beqr
\psi_4(J, \phi,\tau) =  \psi_0(J+A(q,\phi)J) +  \bt_K\theta \psi_0'(J+A(q,\phi)J) 
\sqrt{\frac{2(J+A(q,\phi)J)}{\bt}}\sin\left[\tan^{-1}(\tan\phi - q) - \om(J+A(q,\phi)J)\tau\right] 
\nonumber \\
& &  \mbox{}
\eeqr
At any time after the quad kick, the distribution function at time $t$ (measured from the start of the
dipole kick) is simply a rotation at the betatron frequency at the action $J$ (and not at 
$J+ A(q,\phi)J$). In the Eulerian
description, we stay at a fixed phase space location and follow the change of density at that 
location. 
\beq
\psi_5(J, \phi,t) = \psi_4(J, \phi - \om(J)(t - \tau))
\eeq
and the dipole moment is
\beqr
\lan x \ran(t) & = & \bt_K \theta \sqrt{2\bt}\int dJ \int d\phi \sqrt{J}\cos\phi \phi_5(J,\phi,t) \nonumber \\
& = & 2 \bt_K \theta \int dJ \int d\phi \sqrt{J}\cos\phi  \psi_0'(J+A(q,\phi- \om(J)(t - \tau))J) 
 \nonumber \\
& & \times \sqrt{(J+A(q,\phi- \om(J)(t - \tau))J)} \nonumber \\
& & \times \sin\left[\tan^{-1}\left(\tan(\phi- \om(J)(t - \tau)) - q\right) \right. 
\nonumber \\ 
& & \left. - \om(J+A(q,(\phi- \om(J)(t - \tau)))J)\tau\right]
\label{eq: dipmom_theta1}
\eeqr

Now we start making approximations. 

We assume that $q \ll 1$, this is almost always satisfied in experiments. Hence we can approximate
\[ A(q,\phi) = -q\sin 2\phi + q^2 \cos^2\phi \approx -q\sin 2\phi \]
And
\[ A(q,(\phi- \om(J)(t - \tau))) \approx - q\sin 2[(\phi- \om(J)(t - \tau))] \]

Previously I made these approximations
\bit
\item In the pure action term
\beqrs
 \sqrt{(J+A(q,\phi- \om(J)(t - \tau))J)} & = & \sqrt{J[1 + A(q,\phi- \om(J)(t - \tau))]} \\
& \approx & \sqrt{J\{ 1 - q\sin 2[(\phi- \om(J)(t - \tau))]\}} < \sqrt{J\{ 1 + |q| \}} \approx \sqrt{J}
\eeqrs

\item In the same spirit
\[  \psi_0'(J+A(q,\phi- \om(J)(t - \tau))J) \approx \psi_0'(J) \]
\eit
These will be improved here
Recall that
\[ A(q,\phi) = -B(q)\sin 2\phi + \half q^2 , \;\;\;\;  B(q) = \sqrt{q^2 + q^4/4} \]
Now I keep terms to the next leading order in $A(q,\phi)$ assuming that $A(q,\phi) \ll 1$
First we introduce some shorthand notation
\beq
 \Dl\phi = \om(J)(t - \tau) , \;\;\; \phi_- = \phi -  \Dl\phi  
\eeq
Expand the square root to first order in $A(q)$ as
\beqr
\sqrt{J[1 + A(q,\phi- \om(J)(t - \tau))]} & \approx & \sqrt{J}\left[ 1 + \half  A(q,\phi_- \right]
\nonumber \\
 & \approx & \sqrt{J}\left[ 1 + \frac{q^2}{4} - \half  B(q)\sin(2(\phi_- )) \right] \\
\psi_0(J+A(q,\phi_- )J) & = & \frac{1}{2\pi J_0}\exp[-\frac{J}{J_0}(1 + A(q,\phi_- ))] 
\eeqr
$A(q,\phi_-)$ depends on $J$ via $\Dl\phi$. Thus
\[ \fr{\del}{\del J}A(q,\phi_-) = - 2B(q)\cos 2\phi_- (-\fr{\del}{\del J}\Dl\phi) = 2B(q)\cos 2\phi_- \om'(t-\tau) 
\equiv  2B(q)\cos 2\phi_- \Dl \phi' \]
Hence the slope of the distribution function is
\beqr
\psi_0'(J+A(q, \phi_-)J) & = & -\frac{1}{2\pi J_0^2}\exp[-\frac{J}{J_0}(1 + A(q,\Dl\phi))]
\left[ 1 + A(q,\phi_-) + J (\frac{\del }{\del J}A(q,\phi_-)) \right] \nonumber \\
& = & -\frac{1}{2\pi J_0^2}\exp[-\frac{J}{J_0}(1 + A(q,\Dl\phi))]  \nonumber \\
&  &  \times \left[ 1 + A(q,\phi_-) + 2 B(q)\cos 2\phi_- \Dl\phi' J  \right]
\eeqr
where 
\[ \Dl\phi = \om(J)(t-\tau), \;\;\;  \Dl\phi' = \om'(t - \tau) \]
and assumed that $\om(J) = \om_{\bt} + \om' J$. 

The dipole moment is now from Eq. \ref{eq: dipmom_theta1}
\beqrs
\lan x(t)\ran & = & 2 \bt_K \theta \int dJ \int d\phi \sqrt{J}\cos\phi  \psi_0'(J+A(q,\phi_-)J)  \\
& & \times \sqrt{(J+A(q,\phi_-)J)} \sin\left[\phi_-  - \om(J+A(q,\phi_-)J)\tau\right]
\eeqrs
and  introduce $\om_+(J)$ as
\beqrs
 \om(J+A(q,\phi_-)J) & = & \om_{\bt} + \om'J(1 + A(q,\phi_-)) = \om_{\bt} + \om'J(1 + \half q^2 - B(q)\sin(2\phi_-)) \\
& = & \om_+ - \om' J B(q) \sin 2\phi_- , \;\;\;  \om_+ = \om_{\bt} + \om'J(1 + \half q^2)
\eeqrs

With the above approximations, we have
\beqr
\lan x(t)\ran & = & 2 \bt_K \theta (-\frac{1}{2\pi J_0^2}) \int dJ \; J \exp[-\frac{J}{J_0}(1 +\half q^2)]\int d\phi 
\exp[B(q)\frac{J}{J_0}\sin 2\phi_-] \cos\phi  \nonumber \\
&  & \times [ 1 + \half A(q,\phi_-)] \left( 1 + A(q,\phi_-) + 2 B(q)\cos 2\phi_- \Dl\phi' J  \right)
 \sin\left[\phi_-  - \tau\om(J +  J A(q, \phi_-)) \right]  
\eeqr
Using
\[ \sin A \cos B = \half[ \sin(A+B) + \sin(A - B)]  \]
we have
\beqrs
\cos\phi  \sin\left[\phi_-  - \tau\om(J+J A(q,\phi_-))\right] & = & \half \left\{
\sin[2\phi - \Dl\phi  - \tau\om(J+J A(q,\phi_-))] - \sin[\Dl\phi + \tau\om(J+J A(q,\phi_-))] \right\}
\eeqrs
Multiplying out the other terms
\beqrs
& \mbox{} &  [ 1 + \half A(q,\phi_-)]\left( 1 + A(q,\phi_-) + 2 B(q)\cos 2\phi_- \Dl\phi' J  \right) \\
& = & [ 1 + \qrtr q^2 - B(q)\sin 2\phi_-]\left( 1 + \qrtr q^2  - B(q)\sin 2\phi_- + 2 B(q)\cos 2\phi_- \Dl\phi' J  \right) \\
& = & (1 + \qrtr q^2)(1 + \half q^2)  - \half B(q)[3 + q^2 ]\sin 2\phi_-  \\
& & + 2(1 + \qrtr q^2)B(q) \Dl\phi' J \cos 2\phi_- + \half B^2(q) \sin^2 2 \phi_- \\
& & - B^2(q) \Dl\phi' J \sin 2\phi_- \cos 2\phi_-  \\
& \equiv & C_1 + C_3 \sin 2\phi_- + C_5 \fr{J}{J_0} \cos 2\phi_- + C_7 \sin^2 2\phi_- + C_9 \fr{J}{J_0} \sin 2\phi_- \cos 2\phi_- 
\eeqrs
where the dimensionless constants $C_i$ independent of $\phi, J$ are
\beqr
C_1 & = & (1 + \qrtr q^2)(1 + \half q^2)   \sim O(1)   \nonumber \\
C_3 & = & - \half B(q)[3 + q^2 ]   \sim O(q)   \nonumber  \\
C_5 & = & 2(1 + \qrtr q^2)B(q) \Dl\phi' J_0 \sim O(q)   \nonumber   \\
C_7 & = & \half B^2(q)  \nonumber \sim O(q^2)  \\
C_9 & = & - B^2(q) \Dl\phi' J_0  \sim O(q^2)
\eeqr
 where $\Dl \phi' = \om'(t - \tau)$ is time dependent. 

Hence we can write
\beq
\lan x(t)\ran = -\frac{\bt_K \theta}{2\pi J_0^2} \int dJ J \exp[-\frac{J}{J_0}(1 +\half q^2)]
\ \left\{ \Phi_1 - \Phi_2 + \Phi_3 - \Phi_4 + \fr{J}{J_0}( \Phi_5 - \Phi_6) +  \Phi_7  - \Phi_8 + \fr{J}{J_0}(\Phi_9 - \Phi_{10}) \right\}
\eeq
where 
\beqr
\Phi_1 & = & C_1 \int d\phi \sin[2\phi - \Dl\phi  - \tau\om(J+J A(q,\phi_-))]  \exp[B(q)\frac{J}{J_0}\sin 2\phi_-] \\
\Phi_2 & = & C_1 \int d\phi \sin[\Dl\phi + \tau\om(J+J A(q,\phi_-))]  \exp[B(q)\frac{J}{J_0}\sin 2\phi_-] \\
\Phi_3 & = & C_3 \int d\phi \sin 2\phi_- \sin[2\phi - \Dl\phi  - \tau\om(J+J A(q,\phi_-))]  \exp[B(q)\frac{J}{J_0}\sin 2\phi_-] \\
\Phi_4 & = & C_3 \int d\phi \sin 2\phi_- \sin[\Dl\phi + \tau\om(J+J A(q,\phi_-))]  \exp[B(q)\frac{J}{J_0}\sin 2\phi_-] \\
\Phi_5 & = & C_5 \int d\phi \cos 2\phi_- \sin[2\phi - \Dl\phi  - \tau\om(J+J A(q,\phi_-))]  \exp[B(q)\frac{J}{J_0}\sin 2\phi_-] \\
\Phi_6 & = & C_5 \int d\phi \cos 2\phi_- \sin[\Dl\phi + \tau\om(J+J A(q,\phi_-))]  \exp[B(q)\frac{J}{J_0}\sin 2\phi_-] \\
\Phi_7 & = & C_7 \int d\phi \sin^2 2\phi_- \sin[2\phi - \Dl\phi  - \tau\om(J+J A(q,\phi_-))]  \exp[B(q)\frac{J}{J_0}\sin 2\phi_-] \\
\Phi_8 & = & C_7 \int d\phi \sin^2 2\phi_- \sin[\Dl\phi + \tau\om(J+J A(q,\phi_-))]  \exp[B(q)\frac{J}{J_0}\sin 2\phi_-] \\
\Phi_9 & = & C_9 \int d\phi \sin 2\phi_-\cos 2\phi_- \sin[2\phi - \Dl\phi  - \tau\om(J+J A(q,\phi_-))]  \exp[B(q)\frac{J}{J_0}\sin 2\phi_-] \\
\Phi_{10} & = & C_9 \int d\phi \sin 2\phi_- \cos 2\phi_- \sin[\Dl\phi + \tau\om(J+J A(q,\phi_-))]  \exp[B(q)\frac{J}{J_0}\sin 2\phi_-] 
\eeqr
From the definition of $\om_+$, we have $\om(J + J A(q,\phi_-)) = \om_+ - B(q)\om' J \sin 2\phi_-$. All the
above integrals are of the form
\[ \int d\phi \exp[i(m \phi + a \sin (2 \phi - 2\Dl\phi))] \]
for different integer values $m$ and complex constants $a$. We first expand into Bessel functions
\beqr
\int d\phi \exp[i m\phi] \exp[i a \sin (2\phi - 2\Dl\phi)] & = & \int d\phi \exp[i m\phi]  \sum_k  J_k(a) \exp[i k (2\phi - 2\Dl\phi)]  \nonumber \\
& = &  \sum_k  J_k(a)  \exp[-2i k \Dl\phi] 2\pi \dl(m + 2k,0) \nonumber \\
& = & 2\pi J_{-m/2}(a) \exp[i m \Dl\phi] 
\eeqr

Working through the integrals
\beqrs
\Phi_1 & = & C_1 {\rm Im}\left\{ \int d\phi \exp[i (2\phi - \Dl\phi  - \tau \om_+ + \tau B(q)\om' J\sin 2\phi_-)+
 B(q)\frac{J}{J_0}\sin 2\phi_-] \right\} \\
& = & C_1 {\rm Im}\left\{\exp[-i( \Dl\phi + \tau \om_+ )]\int d\phi \exp[i(2\phi + z_1 J \sin(2\phi-2\Dl\phi))] \right\} \\
& = & 2\pi C_1 {\rm Im}\left\{\exp[i( \Dl\phi - \tau \om_+ )] J_{-1}(z_1 J) \right\} \\
& = & - 2\pi C_1 {\rm Im}\left\{\exp[i( \Dl\phi - \tau \om_+ )] J_{1}(z_1 J) \right\}
\eeqrs
where we defined the complex parameter
\[ z_1 = \frac{B(q)}{J_0}[\tau \om' J_0 - i]  \]
and used $J_{-1}(z) = - J_1(z)$. Noting that $\om' J_0 = 1/\tau_D$, we can write
\beq 
z_1 = \frac{B(q)}{J_0}[\fr{\tau}{\tau_D} - i], \;\;\;\; \Rarw |\fr{{\rm Im}[z_1]}{{\rm Re}[z_1]}| \ll 1
\eeq

Next
\beqrs
\Phi_2 & = & C_1 {\rm Im}\left\{ \int d\phi \exp[i (\Dl\phi + \tau\om_+ - \tau B(q)\om' J\sin 2\phi_- )+B(q)\frac{J}{J_0}\sin 2\phi_- ]\right\} \\
 & = & C_1 {\rm Im}\left\{ \exp[i (\Dl\phi + \tau\om_+)] \int d\phi \exp[-i z_1^* J \sin (2\phi - 2\Dl\phi)] \right\} \\
& = & 2\pi C_1 {\rm Im}\left\{ \exp[i (\Dl\phi + \tau\om_+)] J_0(z_1^* J) \right\}
\eeqrs
where we used $J_0(-z) = J_0(z)$. 

Next we use
\[ \sin A \sin B = \half[ \cos(A-B) - \cos(A+B) ]  \]
to decompose $\Phi_3, \Phi_4$ as
\[ \Phi_3 = \Phi_{3.1} -  \Phi_{3.2} \]
where
\beqrs
\Phi_{3.1} & = & \half C_3 \int d\phi \cos[-\Dl\phi + \tau\om_+ - \tau\om' J B(q)\sin 2\phi_-]
\exp[B(q)\frac{J}{J_0}\sin 2\phi_-] \\
& = & \half C_3 {\rm Re}\left\{ \exp[i(-\Dl\phi + \tau\om_+)] \int d\phi \exp[-z_1^* J \sin 2\phi_-] \right\} \\
& = & \pi C_3 {\rm Re}\left\{ \exp[i(-\Dl\phi + \tau\om_+)] J_0(z_1^* J) \right\}
\eeqrs
and
\beqrs
\Phi_{3.2} & = & \half C_3 \int d\phi \cos[4\phi - 3\Dl\phi - \tau\om_+ + \tau\om' J B(q)\sin 2\phi_-]
\exp[B(q)\frac{J}{J_0}\sin 2\phi_-] \\
& = & \half C_3 {\rm Re}\left\{ \exp[-i(3\Dl\phi + \tau\om_+)]\int d\phi \exp[i(4\phi + z_1 J \sin 2\phi_-)] \right\} \\
& = & \pi C_3 {\rm Re}\left\{ \exp[-i(3\Dl\phi + \tau\om_+)] J_{-2}(z_1 J) \exp[i 4\Dl\phi] \right\} \\
& = & \pi C_3 {\rm Re}\left\{ \exp[i(\Dl\phi - \tau\om_+)] J_{2}(z_1 J) \right\}
\eeqrs
and
\[
\Phi_3 = \pi C_3{\rm Re}\left\{ \exp[i(-\Dl\phi + \tau\om_+)] J_0(z_1^* J)  - \exp[i(\Dl\phi - \tau\om_+)] J_{2}(z_1 J) \right\}
\]
Similarly
\[ \Phi_4 =  \Phi_{4.1} - \Phi_{4.2} \]
where
\beqrs
\Phi_{4.1} & = & \half C_3 \int d\phi  \cos[2\phi - 3\Dl\phi - \tau\om_+ + \tau\om' J B(q)\sin 2\phi_-]  \exp[B(q)\frac{J}{J_0}\sin 2\phi_-] \\
& = & \half  C_3 \; {\rm Re}\left\{ \exp[i(- 3\Dl\phi - \tau\om_+)]\int d\phi  \exp[i(2\phi + z_1 J\sin 2\phi_-)] \right\} \\
& = & \pi C_3 \; {\rm Re} \left\{ \exp[i(- 3\Dl\phi - \tau\om_+)] J_{-1}(z_1 J) \exp[2i\Dl\phi] \right\} \\
& = & - \pi  C_3 \; {\rm Re}\left\{ \exp[-i(\Dl\phi + \tau\om_+)] J_{1}(z_1 J) \right\} 
\eeqrs
using $J_{-1}(z) = - J_1(z)$.

Next
\beqrs
\Phi_{4.2} & = & \half C_3 \int d\phi  \cos[2\phi - \Dl\phi + \tau\om_+ - \tau\om' J B(q)\sin 2\phi_-]  \exp[B(q)\frac{J}{J_0}\sin 2\phi_-] \\
& = & \half  C_3 \; {\rm Re} \left\{ \exp[i(- \Dl\phi + \tau\om_+)]\int d\phi  \exp[i(2\phi - z_1^* J\sin 2\phi_-)] \right\} \\
& = & \pi  C_3 \; {\rm Re} \left\{ \exp[i(- \Dl\phi + \tau\om_+)] J_{-1}(-z_1^* J)\exp[2i\Dl\phi] \right\} \\
& = & \pi  C_3\;  {\rm Re} \left\{ \exp[i(\Dl\phi + \tau\om_+)] J_{1}(z_1^* J) \right\} 
\eeqrs
using $J_{-1}(-z) = J_1(z)$. Hence
\[ \Phi_4 =  -\pi C_3 \; {\rm Re} \left\{ \exp[-i(\Dl\phi + \tau\om_+)] J_{-1}(z_1 J) + \exp[i(\Dl\phi + \tau\om_+)] J_{1}(z_1^* J) \right\} 
\]

Next
\[ \Phi_5 = \Phi_{5.1} + \Phi_{5.2}  \]
where
\beqrs
\Phi_{5.1} & = & \half C_5 \int d\phi \sin[4\phi - 3\Dl\phi  - \tau\om(J+J A(q,\phi_-))]  \exp[B(q)\frac{J}{J_0}\sin 2\phi_-] \\
& = & \half C_5 \; {\rm Im}\left\{\exp[-i(3\Dl\phi + \tau\om_+)]\int d\phi \exp[i(4\phi + z_1 J \sin 2\phi_-)] \right\} \\
& = & \pi C_5 \; {\rm Im}\left\{\exp[-i(3\Dl\phi + \tau\om_+)] J_{-2}(z_1 J)\exp[i 4\Dl\phi] \right\} \\
& = & \pi C_5 \; {\rm Im}\left\{\exp[i(\Dl\phi - \tau\om_+)] J_{2}(z_1 J) \right\} 
\eeqrs
\beqrs
\Phi_{5.2} & = & \half C_5 \int d\phi \sin[\Dl\phi  - \tau\om(J+J A(q,\phi_-))]  \exp[B(q)\frac{J}{J_0}\sin 2\phi_-] \\
& = & \half C_5 \; {\rm Im}\left\{\exp[i(\Dl\phi - \tau\om_+)]\int d\phi \exp[i(z_1 J \sin 2\phi_-)] \right\} \\
& = & \pi C_5 \; {\rm Im}\left\{\exp[i(\Dl\phi - \tau\om_+)] J_{0}(z_1 J) \right\} 
\eeqrs
Hence
\[ \Phi_5 = \pi C_5 \; {\rm Im}\left\{\exp[i(\Dl\phi - \tau\om_+)] J_{2}(z_1 J) +
\exp[i(\Dl\phi - \tau\om_+)] J_{0}(z_1 J) \right\} 
\]

Next
\[ \Phi_6  = \Phi_{6.1} + \Phi_{6 .2} \]
where
\beqrs
\Phi_{6.1} & = & \half C_5 \int d\phi \sin[2\phi - \Dl\phi + \tau\om(J+J A(q,\phi_-))]  \exp[B(q)\frac{J}{J_0}\sin 2\phi_-] \\
& = & \half C_5 \; {\rm Im}\left\{\exp[i(-\Dl\phi + \tau\om_+)]\int d\phi \exp[i(2\phi - z_1^* J \sin 2\phi_-)] \right\} \\
& = & \pi C_5 \; {\rm Im}\left\{\exp[i(-\Dl\phi + \tau\om_+)]J_{-1}(- z_1^* J) \exp[2i\Dl\phi] \right\} \\
& = & \pi C_5 \; {\rm Im}\left\{\exp[i(\Dl\phi + \tau\om_+)]J_{1}(z_1^* J) \right\} 
\eeqrs
\beqrs
\Phi_{6.2} & = & \half C_5 \int d\phi \sin[-2\phi + \Dl\phi + \tau\om(J+J A(q,\phi_-))]  \exp[B(q)\frac{J}{J_0}\sin 2\phi_-] \\
 & = & -\half C_5\; {\rm Im}\left\{ \exp[i(- \Dl\phi - \tau\om_+)]\int d\phi \exp[i (2\phi + z_1 J\sin2\phi_-)] \right\} \\
 & = & -\pi C_5\; {\rm Im}\left\{ \exp[i(- \Dl\phi - \tau\om_+)]J_1(z_1 J) \exp[2i\Dl\phi] \right\} \\
 & = & -\pi C_5\; {\rm Im}\left\{ \exp[i( \Dl\phi - \tau\om_+)]J_1(z_1 J)  \right\} 
\eeqrs
Hence
\[ \Phi_6 = \pi C_5 \; {\rm Im}\left\{\exp[i(\Dl\phi + \tau\om_+)]J_{1}(z_1^* J)  - 
\exp[i( \Dl\phi - \tau\om_+)]J_1(z_1 J)  \right\} 
\]

For $\Phi_7, \Phi_8$, we decompose
\[ \sin^2 2\phi_- = \half ( 1 - \cos 4\phi_-) \]
and then we have
\beqrs
\Phi_7 & = & \half C_7\left[ \frac{1}{C_1}\Phi_1 - \int d\phi \cos 4\phi_-
 \sin[2\phi - \Dl\phi  - \tau\om(J+J A(q,\phi_-))]  \exp[B(q)\frac{J}{J_0}\sin 2\phi_-] \right] \\
& \equiv & \half \frac{C_7}{C_1}\Phi_1 -  \half  (\Phi_{7.1} + \Phi_{7.2}) \\
\Phi_8 & = &  \half C_7\left[ \frac{1}{C_1}\Phi_2 - \int d\phi \cos 4\phi_-
 \sin[\Dl\phi  + \tau\om(J+J A(q,\phi_-))]  \exp[B(q)\frac{J}{J_0}\sin 2\phi_-]  \right] \\
& \equiv &  \half \frac{C_7}{C_1}\Phi_2 -  \half (\Phi_{8.1} + \Phi_{8.2}) 
\eeqrs
\beqrs
\Phi_{7.1} & = & \half C_7 \int d\phi \sin[6\phi - 5\Dl\phi - \tau\om_+ + \tau\om' J B(q)\sin 2\phi_-]
\exp[B(q)\frac{J}{J_0}\sin 2\phi_-]  \\
& = & \half C_7\; {\rm Im}\left\{\exp[i(- 5\Dl\phi - \tau\om_+ )] \int d\phi \exp[i(6\phi + z_1 J \sin 2\phi_-)]\right\} \\
& = & \pi C_7\; {\rm Im}\left\{\exp[i(- 5\Dl\phi - \tau\om_+ )]J_{-3}(z_1 J) \exp[6i\Dl\phi] \right\} \\
& = & -\pi C_7\; {\rm Im}\left\{\exp[i(\Dl\phi - \tau\om_+ )]J_{3}(z_1 J) \right\} 
\eeqrs
and
\beqrs
\Phi_{7.2} & = & \half C_7 \int d\phi \sin[-2\phi + 3\Dl\phi - \tau\om_+ + \tau\om' J B(q)\sin 2\phi_-]
\exp[B(q)\frac{J}{J_0}\sin 2\phi_-]  \\
& = & -\half C_7\; {\rm Im}\left\{\exp[i(-3\Dl\phi + \tau\om_+ )] \int d\phi \exp[i(2\phi - z_1^* J \sin 2\phi_-)]\right\} \\
& = & -\pi C_7\; {\rm Im}\left\{\exp[i(-3\Dl\phi + \tau\om_+ )] J_{-1}(-z_1^* J) \exp[2i\Dl\phi]\right\} \\
& = & -\pi C_7\; {\rm Im}\left\{\exp[i(-\Dl\phi + \tau\om_+ )] J_{1}(z_1^* J) \right\}
\eeqrs
which implies
\beqrs
\Phi_7 & = &  \pi C_7 {\rm Im}\left\{- \exp[i( \Dl\phi - \tau \om_+ )] J_{1}(z_1 J) \right. \\
& &  \left. +  \half  \exp[i(\Dl\phi - \tau\om_+ )]J_{3}(z_1 J)  + 
 \half  \exp[-i(\Dl\phi - \tau\om_+ )] J_{1}(z_1^* J) \right\} 
\eeqrs

Next
\beqrs
\Phi_{8.1} & = & \half C_7 \int d\phi \sin[4\phi - 3\Dl\phi + \tau\om_+ + \tau\om' J B(q)\sin 2\phi_-]
\exp[B(q)\frac{J}{J_0}\sin 2\phi_-]  \\
& = &  \half C_7\; {\rm Im}\left\{\exp[i(-3\Dl\phi + \tau\om_+)] \int d\phi \exp[i(4\phi -z_1^* J \sin 2\phi_-)] \right\} \\
& = &  \pi C_7\; {\rm Im}\left\{\exp[i(-3\Dl\phi + \tau\om_+)] J_{-2}(-z_1^* J) \exp[4i\Dl\phi] \right\} \\
& = &  \pi C_7\; {\rm Im}\left\{\exp[i(\Dl\phi + \tau\om_+)] J_2(z_1^* J)  \right\}
\eeqrs
\beqrs
\Phi_{8.2} & = & \half C_7 \int d\phi \sin[-4\phi + 3\Dl\phi + \tau\om_+ + \tau\om' J B(q)\sin 2\phi_-]
\exp[B(q)\frac{J}{J_0}\sin 2\phi_-]  \\
& = & - \half C_7\; {\rm Im}\left\{\exp[i(-3\Dl\phi - \tau\om_+)] \int d\phi \exp[i(4\phi + z_1 J \sin 2\phi_-)] \right\} \\
& = & - \pi C_7\; {\rm Im}\left\{\exp[i(-3\Dl\phi - \tau\om_+)] J_{-2}(z_1 J) \exp[4i\Dl\phi] \right\} \\
& = &  -\pi C_7\; {\rm Im}\left\{\exp[i(\Dl\phi - \tau\om_+)] J_2(z_1 J)  \right\}
\eeqrs
Hence
\beqrs
\Phi_8 & = & \pi C_7\; {\rm Im}\left\{  \exp[i (\Dl\phi + \tau\om_+)] J_0(z_1^* J)  \right. \\
& & \left. - \half \exp[i(\Dl\phi + \tau\om_+)] J_2(z_1^* J) + \half \exp[i(\Dl\phi - \tau\om_+)] J_2(z_1 J)  \right\}
\eeqrs

CHECK
\bit
\item If the exponent has $\exp[+i\tau\om_+]$, then arg. of Bessel function should be $z_1^*J$

\eit

Next
\beqrs
\Phi_9 & = & \half C_9 \int d\phi \sin 4\phi_- \sin[2\phi - \Dl\phi  - \tau\om(J+J A(q,\phi_-))]  \exp[B(q)\frac{J}{J_0}\sin 2\phi_-] \\
& \equiv & \Phi_{9.1} - \Phi_{9.2} 
\eeqrs
where
\beqrs
\Phi_{9.1} & = & \qrtr C_9 \int d\phi \cos[2\phi - 3\Dl\phi + \tau\om_+ -
\tau\om'J B(q)\sin 2\phi_-]  \exp[B(q)\frac{J}{J_0}\sin 2\phi_-] \\
& = & \qrtr C_9 \; {\rm Re}\left\{\exp[i(-3\Dl\phi + \tau\om_+)]\int d\phi 
\exp[i(2\phi - z_1^*J \sin 2\phi_-)] \right\} \\
& = & \fr{\pi}{2} C_9 \; {\rm Re}\left\{\exp[i(-3\Dl\phi + \tau\om_+)]J_{-1}(-z_1^* J)\exp[2i\Dl\phi] 
\right\} \\
& = & \fr{\pi}{2} C_9 \; {\rm Re}\left\{\exp[i(-\Dl\phi + \tau\om_+)]J_{-1}(-z_1^* J) \right\} \\
& = & \fr{\pi}{2} C_9 \; {\rm Re}\left\{\exp[-i(\Dl\phi - \tau\om_+)]J_{1}(z_1^* J) \right\} \\
\Phi_{9.2} & = & \qrtr C_9 \int d\phi \cos[6\phi - 5\Dl\phi - \tau\om_+ +
\tau\om'J B(q)\sin 2\phi_-]  \exp[B(q)\frac{J}{J_0}\sin 2\phi_-] \\
& = & \qrtr C_9 {\rm Re}\; \left\{\exp[i(-5\Dl\phi - \tau\om_+)]\int d\phi 
\exp[i(6\phi + z_1J \sin 2\phi_-)] \right\} \\
& = & \fr{\pi}{2} C_9 \; {\rm Re}\left\{\exp[i(-5\Dl\phi - \tau\om_+)]J_{-3}(z_1 J)\exp[6i\Dl\phi] 
\right\} \\
& = & - \fr{\pi}{2} C_9 \; {\rm Re}\left\{\exp[i(\Dl\phi - \tau\om_+)]J_{3}(z_1 J)\right\}
\eeqrs
Hence
\[
\Phi_9 = \fr{\pi}{2} C_9 {\rm Re}\left\{\exp[-i(\Dl\phi - \tau\om_+)]J_{1}(z_1^* J)
 + \exp[i(\Dl\phi - \tau\om_+)]J_{3}(z_1 J)\right\}
\]
Next
\beqrs
\Phi_{10} & = & \half C_9 \int d\phi \sin 4\phi_- \sin[\Dl\phi + \tau\om(J+J A(q,\phi_-))]  \exp[B(q)\frac{J}{J_0}\sin 2\phi_-] \\
& \equiv & \Phi_{10.1} - \Phi_{10.2}
\eeqrs
where
\beqrs
\Phi_{10.1} & = & \qrtr C_9\int d\phi \cos[4\phi - 5\Dl\phi - \tau\om_+ + 
\tau\om' J B(q)\sin 2\phi_-]  \exp[B(q)\frac{J}{J_0}\sin 2\phi_-] \\
& = & \qrtr C_9 \; {\rm Re}\left\{ \exp[i(- 5\Dl\phi - \tau\om_+)]\int d\phi 
\exp[i(4\phi  + z_1 J \sin 2\phi_-)] \right\} \\
& = & \qrtr C_9 \; {\rm Re}\left\{ \exp[i(- 5\Dl\phi - \tau\om_+)]J_{-2}(z_1 J)\exp[4i\Dl\phi]
\right\} \\
& = & \qrtr C_9 \; {\rm Re}\left\{ \exp[-i(\Dl\phi +\tau\om_+)]J_2(z_1 J)\right\} \\
\Phi_{10.2} & = & \qrtr C_9\int d\phi \cos[4\phi - 3\Dl\phi + \tau\om_+ -
\tau\om' J B(q)\sin 2\phi_-]  \exp[B(q)\frac{J}{J_0}\sin 2\phi_-] \\
& = & \qrtr C_9 \; {\rm Re}\left\{ \exp[i(- 3\Dl\phi + \tau\om_+)]J_2(-z_1^* J)\exp[4i\Dl\phi]
\right\} \\
& = & \qrtr C_9 \; {\rm Re}\left\{ \exp[i(\Dl\phi + \tau\om_+)]J_2(z_1^* J)\right\} 
\eeqrs
Hence
\[ \Phi_{10} = \qrtr C_9 \; {\rm Re}\left\{ \exp[-i(\Dl\phi +\tau\om_+)]J_2(z_1 J)
 - \exp[i(\Dl\phi + \tau\om_+)]J_2(z_1^* J)\right\} 
\]

Gathering the results for the terms in $\lan x(t)\ran$,
\beqr
\lan x(t)\ran_{1-4} &  = &  -\frac{\bt_K \theta}{2\pi J_0^2} \int dJ J \exp[-(1 +\half q^2)\frac{J}{J_0}]
\left\{ \Phi_1 - \Phi_2 + \Phi_3 - \Phi_4 + J( \Phi_5 - \Phi_6) +  \Phi_7  - \Phi_8 + J(\Phi_9 - \Phi_{10}) \right\} \\
 & \equiv & I_1 - I_2 + I_3 - I_4 + I_5 - I_6 + I_7 - I_8 + I_9 - I_{10}
\eeqr
where
\beqrs
\Phi_1 & = & - 2\pi C_1 \; {\rm Im}\left\{\exp[i( \Dl\phi - \tau \om_+ )] J_{1}(z_1 J) \right\} \\
\Phi_2 & = & 2\pi C_1 {\rm Im}\left\{ \exp[i (\Dl\phi + \tau\om_+)] J_0(z_1^* J) \right\} \\
\Phi_3 & = & \pi C_3{\rm Re}\left\{ \exp[i(-\Dl\phi + \tau\om_+)] J_0(z_1^* J)  - \exp[i(\Dl\phi - \tau\om_+)] J_{2}(z_1 J) \right\} \\
\Phi_4 & = &  -\pi C_3 \; {\rm Re} \left\{ \exp[-i(\Dl\phi + \tau\om_+)] J_{-1}(z_1 J) + 
\exp[i(\Dl\phi + \tau\om_+)] J_{1}(z_1^* J) \right\}  \\
\Phi_5 & = & \pi C_5 \; {\rm Im}\left\{\exp[i(\Dl\phi - \tau\om_+)] J_{2}(z_1 J) +
\exp[i(\Dl\phi - \tau\om_+)] J_{0}(z_1 J) \right\}  \\
\Phi_6 & = & \pi C_5 \; {\rm Im}\left\{\exp[i(\Dl\phi + \tau\om_+)]J_{1}(z_1^* J)  - 
\exp[i( \Dl\phi - \tau\om_+)]J_1(z_1 J)  \right\} \\
\Phi_7 & = &  \pi C_7 {\rm Im}\left\{- \exp[i( \Dl\phi - \tau \om_+ )] J_{1}(z_1 J) \right. \\
& &  \left. +  \half  \exp[i(\Dl\phi - \tau\om_+ )]J_{3}(z_1 J)  + 
 \half  \exp[-i(\Dl\phi - \tau\om_+ )] J_{1}(z_1^* J) \right\}  \\
\Phi_8 & = & \pi C_7\; {\rm Im}\left\{  \exp[i (\Dl\phi + \tau\om_+)] J_0(z_1^* J)  \right. \\
& & \left. - \half \exp[i(\Dl\phi + \tau\om_+)] J_2(z_1^* J) + \half \exp[i(\Dl\phi - \tau\om_+)] J_2(z_1 J)  \right\} \\
\Phi_9 & = & \fr{\pi}{2} C_9 {\rm Re}\left\{\exp[-i(\Dl\phi - \tau\om_+)]J_{1}(z_1^* J)
 + \exp[i(\Dl\phi - \tau\om_+)]J_{3}(z_1 J)\right\}  \\
\Phi_{10} &  = & \qrtr C_9 \; {\rm Re}\left\{ \exp[-i(\Dl\phi +\tau\om_+)]J_2(z_1 J)
 - \exp[i(\Dl\phi + \tau\om_+)]J_2(z_1^* J)\right\} 
\eeqrs
Consider the exponent terms in the $\Phi_i$
\beqrs
\Dl \phi + \tau\om_+ & = & (\om_{\bt} + \om' J)(t-\tau) + \tau (\om_{\bt} + \om'J(1 + \half q^2))
 = \om_{\bt} t + (t + \half q^2 \tau)\om' J \\
\Dl \phi - \tau\om_+ & = & (\om_{\bt} + \om' J)(t-\tau) - \tau (\om_{\bt} + \om'J(1 + \half q^2))
 = \om_{\bt} (t - 2\tau) + (t - 2\tau - \half q^2\tau) \om' J 
\eeqrs
Then
\beqrs
I_1 & = & \frac{\bt_K \theta}{J_0^2}C_1 \; {\rm Im}\left\{\int dJ \; J \exp[-\frac{J}{J_0}(1 +\half q^2)]
\exp[i(\om_{\bt} (t - 2\tau) + (t - 2\tau - \half q^2\tau) \om' J)] J_{1}(z_1 J) \right\}  \\
I_2 & = & -\frac{\bt_K \theta}{J_0^2}C_1 \; {\rm Im}\left\{\int dJ \; J \exp[-\frac{J}{J_0}(1 +\half q^2)]
\exp[i(\om_{\bt} t  + (t + \half q^2\tau) \om' J)] J_{0}(z_1 J) \right\} \\
\eeqrs
Introduce the dimensionless integration variable $u = J/J_0$ and define
\beqr
 \Phi & = & \om_{\bt} (t - 2\tau) \\
 a_1 & = & (1 +\half q^2) - i (t - 2\tau - \half q^2\tau) \om' J_0 \\
a_2 & = & (1 +\half q^2) - i (t + \half q^2\tau) \om' J_0 \\
b & = & z_1 J_0 = B(q)\left( \tau \om' J_0 - i \right) 
\eeqr
where $a_1, a_2, b$ are complex dimensionless parameters, independent of $J$. 
It follows that
\beqrs
I_1 & = & \bt_K \theta C_1 \; {\rm Im}\left\{ \exp[i\Phi] \int du \; u \exp[- a_1 u] J_{1}(b u) \right\} \\
I_2 & = & - \bt_K \theta C_1 \; {\rm Im}\left\{ \exp[i\om_{\bt} t] \int du  \; u \exp[- a_2 u] J_{0}(b u) \right\} \\
I_3 & = & \half \bt_K \theta C_3 \; {\rm Re}\left\{ \exp[-i\Phi]
\int du \; u \exp[- a_1^* u] J_{0}(b^* u) - \exp[i\Phi] \int du \; u \exp[- a_1 u] J_{2}(b u)\right\} \\
I_4 & = & -\half \bt_K \theta J_0^2 C_3 \; {\rm Re}\left\{ \exp[-i\om_{\bt} t]
\int du \; u \exp[- a_2^* u] J_{1}(b u) + \exp[i\om_{\bt} t] \int du \; u \exp[- a_2 u] J_{1}(b^* u)\right\} \\
\eeqrs

Using the integration results in the appendix
\beqrs
I_1 & = &  \bt_K \theta C_1 \; {\rm Im}\left\{ \exp[i\Phi] \frac{b}{(a_1^2 + b^2)^{3/2}} \right\} \\
I_2 & = & -\bt_K \theta C_1 \; {\rm Im}\left\{ \exp[i\om_{\bt} t]\frac{a_2}{(a_2^2 + b^2)^{3/2}} \right\}
\eeqrs

Consider only the terms that depend on $t - 2\tau$ rather than on $t$ alone. These are likely to be the
dominant terms at long times.  Besides $I_1, I_3$, these are
\beqrs
I_5 & = & -\half \bt_K J_0 \theta C_5 \; {\rm Im}\left\{ \exp[i\Phi]
\int du \; u^2 \exp[- a_1 u] [ J_0(b u) + J_2(b u J)] \right\} \\
I_{6.2} & = & \half \bt_K \theta J_0 C_5 \; {\rm Im}\left\{ \exp[i\Phi]\int du \; u^2 \exp[- a_1 u]  J_1(b u)  \right\} \\
I_7 & = & -\half \bt_K \theta C_7 \; {\rm Im}\left\{ \exp[i\Phi]
\int du \; u  \exp[- a_1 u] [ J_1(b u) + \half J_3(b u)]  \right. \\
&  &  \left.  + \half \exp[- i\Phi] \int du \; u  \exp[- a_1 u]  J_1(b^* u)  \right\} \\
I_{8.3} & = & -\qrtr \bt_K \theta C_7 \; {\rm Im}\left\{ \exp[i\om_{\bt} (t - 2\tau)]
\int du \; u  \exp[- a_1 u]  J_2 (b u) \right\} \\
I_9 & = & -\qrtr \bt_K \theta J_0 C_9 {\rm Re}\left\{\exp[-i\Phi]  \int du \; u^2  \exp[- a_1 u] J_{1}(b^* u) 
\right. \\
& & \left.   + \exp[i\Phi]  \int du \; u^2  \exp[- a_1 u]  J_{3}(b u) \right\} \\
\eeqrs
The different integrals can be represented as the complex function
\beq
H_{m, n}(a, b) = \int du \; u^m \exp[- a u] J_n(b u) 
\eeq
where the integers $m, n$ take one of the values $m=1, 2$ and $n=0, 1, 2, 3$ and $(a, z)$ are complex. 
In  terms of this function $H_{m, n}$ and phase, we can express the different terms as
\beqrs
I_1 & = & \bt_K \theta C_1 \; {\rm Im}\left\{ \exp[i\Phi] H_{1, 1}(a_1, b)\right\} \\
I_3 & = & \half \bt_K \theta C_3 \; {\rm Re}\left\{ \exp[-i\Phi] H_{1,0}(a_1^*, b^*)  
 - \exp[i\Phi] H_{1, 2}(a_1, b) \right\} \\
I_5 & = & -\half \bt_K \theta J_0 C_5 \; {\rm Im}\left\{ \exp[i\Phi](H_{2, 0}(a_1, b) +H_{2, 2}(a_1, b)) \right\} \\
I_{6.2} & = & \half \bt_K \theta J_0 C_5 \; {\rm Im}\left\{ \exp[i\Phi]H_{2, 1}(a_1, b)\right\} \\
I_7 & = & -\half \bt_K \theta C_7 \; {\rm Im}\left\{ \exp[i\Phi] [H_{1, 1}(a_1,  b)
 + \half H_{1, 3}(a_1, b)] + \half \exp[-i\Phi] H_{1, 1}(a_1,  b^*) \right\} \\
I_{8.3} & = & -\qrtr \bt_K \theta C_7 \; {\rm Im}\left\{ \exp[i\Phi] H_{1, 2}(a_1, b)\right\} \\
I_9 & = & -\qrtr \bt_K \theta J_0 C_9 {\rm Re}\left\{\exp[-i\Phi]  H_{2, 1}(a_1, b^*)
 + \exp[i\Phi]  H_{2,  3}(a_1,  b) \right\}
\eeqrs

Using the integrations from the appendix, we have
\beqrs
I_1 & = &  \bt_K \theta C_1 \; {\rm Im}\left\{ \exp[i\Phi] \frac{b}{(a_1^2 + b^2)^{3/2}} \right\} \\
I_3 & = & \half \bt_K \theta C_3 \; {\rm Re}\left\{ \exp[-i\Phi] \fr{a_1^*}{((a_1^*)^2 + (b^*)^2)^{3/2}} 
- \exp[i\Phi]   \fr{2 (a_1^2 + b^2)^{3/2} - a_1(2a_1^2 +3 b^2)}{b^2 (a_1^2 + b^2)^{3/2}} \right\} \\
I_5 & = & -\half \bt_K \theta J_0 C_5 \; {\rm Im}\left\{ \exp[i\Phi](\frac{(2 a_1^2 - b^2)}{(a_1^2 + b^2)^{5/2}}  +  \fr{3  b^2 }{(a_1^2 + b^2)^{5/2}} ) \right\} \\
I_{6.2} & = & \half \bt_K \theta J_0 C_5 \; {\rm Im}\left\{ \exp[i\Phi] \fr{3 a_1 b}{(a_1^2 + b^2)^{5/2}}
\right\} \\
I_7 & = & -\half \bt_K \theta C_7 \; {\rm Im}\left\{ \exp[i\Phi] \left( \frac{b}{(a_1^2 + b^2)^{3/2}} 
+ \half \fr{8a_1^4+12a_1^2 b^2 + 3b^4 - 8 a_1(a_1^2+b^2)^{3/2}}{b^3 (a_1^2 + b^2)^{3/2}} \right) \right.
\\
&  & \left. + \half \exp[-i\Phi] \frac{b^*}{(a_1^2 + (b^*)^2)^{3/2}} \right\} \\
I_{8.3} & = & -\qrtr \bt_K \theta C_7 \; {\rm Im}\left\{ \exp[i\Phi]  \fr{2 (a_1^2 + b^2)^{3/2} -a_1(2a_1^2 +3 b^2)}{b^2 (a_1^2 + b^2)^{3/2}}   \right\} \\
I_9 & = & -\qrtr \bt_K \theta J_0 C_9 {\rm Re}\left\{\exp[-i\Phi]  \fr{3 a_1 b^*}{(a_1^2 + (^*)^2)^{5/2}} 
 + \exp[i\Phi]  \fr{8 (a_1^2 + b^2)^{5/2} - a_1(8a_1^4 + 20a_1^2 b^2 + 15 b^4)}{b^3(a_1^2  + b^2)^{5/2}} \right\}
\eeqrs

\subsection{FWHM of the 1D pulse in the nonlinear quad theory}

\bit
\item conservation of pulse area
\item  minimum of fwhm as a function of parameters
  \eit

  Definitions of notation
  \beqrs
  \tau_D & = & \frac{1}{\om' \eps_0}, \;\;\; \eta = \om' \eps_0 \tau \\
  \Rarw Q & \equiv & q\om' \eps_0 \tau  =  q \eta \\
  \xi(t) & = & (t - 2\tau)\om' \eps_0 
  \eeqrs

\noi  {\bf Recap of the linear quad kick theory} \\

The echo amplitude un the linear quad  kick approximation is 
\beq
\lan x(t) \ran_{linear} = 
\bt_K \theta \frac{Q}{[(1 + \xi^2(t) ]^{3/2}} \sin (\Phi(t) + 3{\rm Arctan}[\xi(t)] - \half q )
\label{eq: amp_linear}
\eeq
At $t \simeq 2\tau$, the amplitude takes the maximum value
\beq
{\rm Ampl}= \bt_K \theta \frac{Q}{[(1 + \xi^2(t) ]^{3/2}}
  \eeq
  The phase term in the echo shape leads only to fast oscillations at the betatron frequency, The shape of the
  echo is determined by the envelope
  The max echo amplitude is therefore
  \beq
      {\rm Ampl}^{max} = \bt_K \theta Q \;\;\;\; \Rarw \;\; \fr{ {\rm Ampl}^{max}}{\bt_K \theta } = Q
      \eeq 
The second equation above defines the echo amplitude scaled  by the dipole kick. 
 The echo amplitude falls to half the max value at times $\Dl t_{HFHM}$ (the half width at half max) before and after the time $2\tau$.
      Hence this time is given by
\beqr
 \frac{1}{[1 + \xi^2(2\tau  \pm \Dl t_{HFHM}^{3/2}]} & = & \half \\
 \Rarw \fr{1}{1 + \Dl t_{HFHM}\om' \eps_0}     & = & (\half)^{2/3} \\
 \Rarw  \Dl t_{HFHM} & = & \frac{\sqrt{2^{2/3} -1 }}{\om'\eps_0} 
 \eeqr 
 Hence $\Dl t_{FWHM} =  2\Dl t_{HFHM} = \fr{1.532}{\om'\eps_0}$; hence the pulse width is independent of
 the quad strength. 
\beq
{\rm Scaled \; Echo \; amplitude} \times \Dl t_{FWHM}   =  2\sqrt{2^{2/3} -1 }q \tau 
\eeq
The term in the LHS is roughly the pulse area , In this linear quad kick theory, the pulse area depends  only on the
dipole and quad kicks and the delay time, i.e only on the external parameters. The area is independent of the
internal variables, the detuning and the emittance. 

\noi  {\bf Echo amplitude and FWHM in the nonlinear theory} 

The time dependent echo pulse in the nonlinear quad theory is given by Eq.(2.39) and (2.40)  in the 2018
PRAB paper
\beqr
\lan x(t)\ran & \approx & \bt_K \theta 
\sin[\Phi(t) + 3\Theta(t) - \half q ]  \label{eq: pulse_nonlinq}
\eeqr
The echo amplitude at $t=2\tau$ is approximated by 
\beq
 \lan x(t=2\tau) \ran^{amp} \approx \bt_k \theta  \frac{Q}{(1 + Q^2 )^{3/2}} 
\label{eq: amp_nonlinq}
\eeq

The FWHM time can therefore be found from
\beq \frac{ Q}{[(1 -  \xi^2(\Dl t_{HFHM}) + Q^2)^2 + 4 \xi^2(\Dl t_{HFHM})]^{3/4}} =
\half \frac{Q}{(1 + Q^2 )^{3/2}}
\eeq
which leads to (writing $r_{HF} = \xi^2(\Dl t_{HFHM})$ and $ y = (1 + Q^2)$
\beqrs
    [  (y - r_{HF})^2 + 4 r_{HF} ]^{3/4} & = & 2 y^{3/2}
    \eeqrs
    which (using Mathematica) has the solutions  (the positive root is $\Dl t^+_{HWFM}$)
    \beq
    \xi^+_{HFHM} = \left\{ Q^2 - 1 + \sqrt{2}\left[ 2^{1/3}(1 + Q^2)^2 -  2Q^2\right]^{1/2}\right\}^{1/2} \\
\eeq    
Hence
which results in the expression for the FWHM
\beqr
\Dl t^+_{HFHM} =  \xi^+_{HFHM}\tau_{Dec} 
 =   \tau_{Dec}\left\{ Q^2 - 1 + \sqrt{2}\left[ 2^{1/3}(1 + Q^2)^2 -  2Q^2\right]^{1/2}\right\}^{1/2} \\
\eeqr
Expanding this expression to $O(Q^2)$ for small $Q$, we have
\beqr
\fr{\Dl t^+_{HFHM}}{\tau_{Dec}} &  = &
\sqrt{2^{2/3} - 1}  +  \fr{ 2^{2/3} - 2^{1/3} + 1 } {2 \sqrt{2^{2/3} -1 } } Q^2 + O(Q^4)  \\
  & = & 0.766 + 0.866 Q^2 + O(Q^4)
  \eeqr

\clearpage

\subsection{Superposition of Nonlinear Quadrupole Kicks}

We consider the case of two quadrupole kicks $q_0, q_1$ applied at times $\tau, \tau + \Dl t_1$ 
respectively. In most cases we expect $\Dl t_1$ to be within a few turns, so $\Dl t_1 = m T_{rev}, m= 1, 2, ...$.  After the 1st quad kick, the action and phase in the arguments of the distribution function had
transformed to 
\[
J \rarw J + A(q_0,\phi), \;\;\;\; \phi \rarw \tan^{-1}(\tan \phi - q_0) 
\]
and the distribution function at any time $t$ after $\tau$ was written as
\beqr
\psi_5(J,\phi,t) 
& = & \sqrt{\frac{2}{\bt}} \bt_K \theta  \psi_0'(J+A(q,\phi- \om(J)(t - \tau))J) 
\sqrt{(J+A(q,\phi- \om(J)(t - \tau))J)} \nonumber \\
& & \times \sin\left[\tan^{-1}\left(\tan(\phi- \om(J)(t - \tau)) - q\right) \right. 
\nonumber \\ 
& & \left. - \om(J+A(q,(\phi- \om(J)(t - \tau)))J)\tau\right]
\eeqr
We define
\beq
\dl\varphi_0(J,t) = \om(J)(t - \tau) 
\eeq
which we note is independent of $\phi$. 
Making the same approximations as before, we simplify $\psi_5$ to
\beqr
\psi_5(J,\phi,t) 
& = & \sqrt{\frac{2}{\bt}} \bt_K \theta  \psi_0'(J) \sqrt{J} \nonumber \\
& & \times \sin\left[(\phi- \dl\varphi_0(t)) 
 - \tau\om(J+A(q_0,(\phi- \dl\varphi_0(t))J)\right]
\eeqr
Hence at time $t= t_1= \tau + \Dl t_1$, we have
\beqr
\psi_5(J,\phi,\Dl t_1) 
& = & \sqrt{\frac{2}{\bt}} \bt_K \theta  \psi_0'(J) \sqrt{J} \nonumber \\
& & \times \sin\left[(\phi- \dl\varphi_0)  - \tau\om(J+A(q_0,(\phi- \dl\varphi_0))J)\right] \nonumber \\
& = & \sqrt{\frac{2}{\bt}} \bt_K \theta  \psi_0'(J) \sqrt{J} \nonumber \\
& & \times \sin\left[(\phi- \dl\varphi_0)  - \tau\left\{\om_{\bt} + \om'J[1+A(q_0,(\phi- \dl\varphi_0))]
\right\} \right] 
\eeqr

At this time, the second quadrupole kick with strength $q_1$ is applied changing the distribution
function to 
\beq
\psi_6(J,\phi, t_{1,+}) = \psi_5(x,p+q_1 x, t_1) \simeq \psi_5(J + A(q_1,\phi)J,\phi,t_1) 
\eeq
where as before we dropped the small change in the phase $\phi$. 
Under this transformation, $\dl\varphi_0$ transforms as
\beq
\dl\varphi_0(J,\tau + \Dl t_1) \rarw \Dl\phi_0(J, \phi,\Dl t_1) = \om(J+ A(q_1,\phi)J)\Dl t_1 = [\om_{\bt} + \om'J(1+A(q_1,\phi))]\Dl t_1
\eeq
Here, unlike $\dl\varphi_0$,  $\Dl\phi_0$ is a function of $\phi$ but evaluated at the fixed time 
$t=t_1 = \tau+\Dl t_1$. 
Hence,
\beqrs
\psi_6(J,\phi,t_{1,+}) 
& = & \sqrt{\frac{2}{\bt}} \bt_K \theta  \psi_0'(J) \sqrt{J} \nonumber \\
& & \times \sin\left[\phi- \Dl\phi_0(\phi)  - \tau\om\left(J+A(q_1,\phi)J + 
A(q_0,\phi- \Dl\phi_0(\phi))(J+A(q_1,\phi)J)\right)\right]
\eeqrs
Expanding the argument of $\om$
\beqrs
J & +  & A(q_1,\phi)J + A(q_0,(\phi- \Dl\phi_0))(J+ A(q_1,\phi)J )     \\
& = & J+A(q_1,\phi)J + A(q_0,(\phi- \Dl\phi_0)J + A(q_0,\phi- \Dl\phi_0)A(q_1,\phi)J
\eeqrs
Since $A(q,\phi) = -q\sin 2\phi + q^2\cos^2\phi$ and we drop terms of $O(q^2)$ from the argument of
the sine function, we can drop the product terms $A(q_0,\phi- \Dl\phi_0)A(q_1,\phi)$, so we can
write
\beqrs
\psi_6(J,\phi) 
& = & \sqrt{\frac{2}{\bt}} \bt_K \theta  \psi_0'(J) \sqrt{J} \nonumber \\
& & \times \sin\left[\phi- \Dl\phi_0  - \tau\left\{ \om_{\bt} + 
\om' J\left(1 + +A(q_1,\phi)J + A(q_0,(\phi- \Dl\phi_0))\right) \right\} \right]
\eeqrs
At any time $t > t_1 = \tau + \Dl t_1$, the distribution evolves by rotation to
\beq
\psi_7(J,\phi,t)  =  \psi_6(J,\phi - \om(J)(t - (\tau + \Dl t_1)))
\eeq
Define 
\beq
\dl \varphi_1(J,t) = \om(J)(t - (\tau + \Dl t_1))
\eeq
which like $\dl\varphi_0$ is also independent of $\phi$. 
Then at times $t \ge t_1 = \tau + \Dl t_1$
\beqr
\psi_7(J,\phi,t &  \ge &  t_1) 
 =  \sqrt{\frac{2}{\bt}} \bt_K \theta  \psi_0'(J) \sqrt{J} \nonumber \\
& & \times \sin\left[\phi- \dl \varphi_1(J,t) - \Dl\phi_0(\phi - \dl \varphi_1(J,t)) 
\right.   \nonumber \\
& & - \tau\left\{ \om_{\bt} +  \om' J\left[1 + A(q_1,\phi-\dl \varphi_1(J,t)) \right.  \right. 
\nonumber \\
& & \left. \left. \left. + A(q_0,\phi - \dl \varphi_1(J,t) - \Dl\phi_0(\phi-\dl \varphi_1(J,t))) 
\right] \right\} \right]
\eeqr

I need to consider a third kick at time $t = \tau + \Dl t_1 + \Dl t_2$ to get the correct pattern. 
From the above, it is clear that $\phi - \dl\varphi_1(J,t)$ will be replaced by 
$\phi - \dl\varphi_2(J,t)$ where $\dl\varphi_2(J,t) = \om(J)(t - [\tau + \Dl t_1 + \Dl t_2])$. 
It is not yet clear how $\Dl\phi_0$ will be replaced. The next variable in this series will be
\[ \Dl\phi_1(J,\phi, \Dl t_2) = \om(J+ A(q_2,\phi) J)\Dl t_2 = [\om_{\bt} + \om'J(1+A(q_2,\phi))]\Dl t_2 \]

At time $t_2 = \tau + \Dl t_1 + \Dl t_2$, a quadrupole kick with strength $q_2$ is applied. 
The distribution changes to 
\beq
\psi_8(J,\phi, t_2) = \psi_7(x,p+q_2 x,t) = \psi_7(J+A(q_2,\phi)J,\phi, t_2)
\eeq
Now $\dl\varphi_1$ transforms to
\beq
\dl\varphi_1(J,t-(\tau + \Dl t_1 + \Dl t_2)) \rarw \Dl\phi_1(J,\phi,\Dl t_2) = 
\om(J+ A(q_2,\phi)J)\Dl t_2 = [\om_{\bt} + \om'J(1+A(q_2,\phi))]\Dl t_2
\eeq
Hence
\beqr
\psi_8(J,\phi,t & = & t_2) 
 =  \sqrt{\frac{2}{\bt}} \bt_K \theta  \psi_0'(J) \sqrt{J} \nonumber \\
& & \times \sin\left[\phi- \Dl \phi_1(J,\phi,\Dl t_2) - \Dl\phi_0(\phi - \Dl \phi_1(J,\phi,\Dl t_2)) 
\right.   \nonumber \\
& & - \tau\left\{ \om_{\bt} +  \om' J[1+A(q_2,\phi)]\left[1 + A(q_1,\phi-\Dl \phi_1(J,\phi,\Dl t_2)) \right.  \right. 
\nonumber \\
& & \left. \left. \left. + A(q_0,\phi - \Dl \phi_1(J,\phi,\Dl t_2) - \Dl\phi_0(\phi-\Dl \phi_1(J,\phi,\Dl t_2))) 
\right] \right\} \right]
\eeqr
Expanding the terms in [] and keeping only terms linear in $A$, so dropping terms $A(q_2,...)A(q_1,..)$
and  $A(q_2,...)A(q_0,..)$ 
\beqrs 
[1+A(q_2,\phi)] &  &  \left[1 + A(q_1,\phi-\Dl \phi_1(J,\phi,\Dl t_2)) \right.   \\
& & \left. + A(q_0,\phi - \Dl \phi_1(J,\phi,\Dl t_2) - \Dl\phi_0(\phi-\Dl \phi_1(J,\phi,\Dl t_2))) \right] \\
& = & 1 + A(q_2,\phi) + A(q_1,\phi-\Dl \phi_1(J,\phi,\Dl t_2)) \\
& & A(q_0,\phi - \Dl \phi_1(J,\phi,\Dl t_2) - \Dl\phi_0(\phi-\Dl \phi_1(J,\phi,\Dl t_2))) 
\eeqrs
Then
\beqrs
\psi_8(J,\phi,t & = & t_2) 
 =  \sqrt{\frac{2}{\bt}} \bt_K \theta  \psi_0'(J) \sqrt{J} \nonumber \\
& & \times \sin\left[\phi- \Dl \phi_1(J,\phi,\Dl t_2) - \Dl\phi_0(\phi - \Dl \phi_1(J,\phi,\Dl t_2)) 
\right.   \nonumber \\
& & - \tau\left\{ \om_{\bt} +  \om' J\left[1 + A(q_2,\phi) + A(q_1,\phi-\Dl \phi_1(J,\phi,\Dl t_2)) \right.  \right. 
\nonumber \\
& & \left. \left. \left. + A(q_0,\phi - \Dl \phi_1(J,\phi,\Dl t_2) - \Dl\phi_0(\phi-\Dl \phi_1(J,\phi,\Dl t_2))) 
\right] \right\} \right]
\eeqrs
At times $ t \ge \tau+ \Dl t_1 + \Dl t_2$, we define
\beq
\dl\varphi_2(J,t) = \om(J)(t - [\tau + \Dl t_1 + \Dl t_2])
\eeq
The distribution function at these times is therefore
\beq
\psi_9(J,\phi,t\ge  t_2) = \psi_8(J,\phi - \dl\varphi_2(J,t))
\eeq
Hence
\beqrs
\psi_9( & J & ,\phi,t \ge t_2)  
 =  \sqrt{\frac{2}{\bt}} \bt_K \theta  \psi_0'(J) \sqrt{J} \nonumber \\
& & \times \sin\left[\phi  - \dl\varphi_2(J,t) - \Dl \phi_1(J,\phi - \dl\varphi_2(J,t)) - 
\Dl\phi_0(\phi - \dl\varphi_2(J,t) - \Dl \phi_1(J,\phi - \dl\varphi_2(J,t))) \right.   \nonumber \\
& & - \tau\left\{ \om_{\bt} +  \om' J\left[1 + A(q_2,\phi - \dl\varphi_2(J,t)) + 
A(q_1,\phi  - \dl\varphi_2(J,t) - \Dl \phi_1(J,\phi - \dl\varphi_2(J,t))) \right.  \right. 
\nonumber \\
& & \left. \left. \left. + A(q_0,\phi - \dl\varphi_2(J,t) - \Dl \phi_1(J,\phi - \dl\varphi_2(J,t)) 
- \Dl\phi_0(\phi - \dl\varphi_2(J,t)-\Dl \phi_1(J,\phi - \dl\varphi_2(J,t)))) 
\right] \right\} \right] 
\eeqrs

Let
\beq
\Phi(J,\phi, t) = \phi - \dl\varphi_2(J,t)
\eeq
We can rewrite
\beqr
\psi_9( & J & ,\phi,t \ge  t_2) 
 =  \sqrt{\frac{2}{\bt}} \bt_K \theta  \psi_0'(J) \sqrt{J} \nonumber \\
& & \times \sin\left[\Phi- \Dl \phi_1(J,\Phi) - 
\Dl\phi_0(\Phi - \Dl \phi_1(J,\Phi)) \right.   \nonumber \\
& & - \tau\left\{ \om_{\bt} +  \om' J\left[1 + A(q_2,\Phi) + 
A(q_1,\Phi  - \Dl \phi_1(J,\Phi)) \right.  \right.  \nonumber \\
& & \left. \left. \left. + A(q_0,\Phi - \Dl \phi_1(J,\Phi) 
- \Dl\phi_0(\Phi - \Dl \phi_1(J,\Phi))) \right] \right\} \right] \\
\eeqr

At time $t= \tau + \Dl t_1 + \Dl t_2 + \Dl t_3$, a kick $q_3$ is applied. Due to the replacement of
$ J \rarw J[1 + A(q_3,\phi)]$, we have
\beq
\dl \varphi_2(J,t) \rarw \Dl\phi_2(J,\phi,\Dl t_3) = [\om_{\bt} + \om' J(1+A(q_3,\phi))]\Dl t_3
\eeq
We had
\beqrs
\psi_7(J,\phi,t &  \ge &  t_1) 
 =  \sqrt{\frac{2}{\bt}} \bt_K \theta  \psi_0'(J) \sqrt{J} \nonumber \\
& & \times \sin\left[\phi- \dl \varphi_1(J,t) - \Dl\phi_0(\phi - \dl \varphi_1(J,t)) 
\right.   \nonumber \\
& & - \tau\left\{ \om_{\bt} +  \om' J\left[1 + A(q_1,\phi-\dl \varphi_1(J,t)) \right.  \right. 
\nonumber \\
& & \left. \left. \left. + A(q_0,\phi - \dl \varphi_1(J,t) - \Dl\phi_0(\phi-\dl \varphi_1(J,t))) 
\right] \right\} \right] \\
\psi_8(J,\phi,t & = &  t_2) 
 =  \sqrt{\frac{2}{\bt}} \bt_K \theta  \psi_0'(J) \sqrt{J} \nonumber \\
& & \times \sin\left[\phi- \Dl \phi_1(J,\phi,\Dl t_2) - \Dl\phi_0(\phi - \Dl \phi_1(J,\phi,\Dl t_2)) 
\right.   \nonumber \\
& & - \tau\left\{ \om_{\bt} +  \om' J\left[1 + A(q_2,\phi) + A(q_1,\phi-\Dl \phi_1(J,\phi,\Dl t_2)) \right.  \right. 
\nonumber \\
& & \left. \left. \left. + A(q_0,\phi - \Dl \phi_1(J,\phi,\Dl t_2) - \Dl\phi_0(\phi-\Dl \phi_1(J,\phi,\Dl t_2))) 
\right] \right\} \right]
\eeqrs
Hence, by direct inspection, we can write $\psi_{10}$ by replacing $\dl\varphi_2$ in $\psi_9$ above by 
$\Dl\phi_2$  and add the additional term $A(q_3,\phi)$
\beqrs
\psi_9( & J & ,\phi,t \ge  t_2) 
 =  \sqrt{\frac{2}{\bt}} \bt_K \theta  \psi_0'(J) \sqrt{J} \nonumber \\
& & \times \sin\left[\Phi- \Dl \phi_1(J,\Phi) - 
\Dl\phi_0(\Phi - \Dl \phi_1(J,\Phi)) \right.   \nonumber \\
& & - \tau\left\{ \om_{\bt} +  \om' J\left[1 + A(q_2,\Phi) + 
A(q_1,\Phi  - \Dl \phi_1(J,\Phi)) \right.  \right.  \nonumber \\
& & \left. \left. \left. + A(q_0,\Phi - \Dl \phi_1(J,\Phi) 
- \Dl\phi_0(\Phi - \Dl \phi_1(J,\Phi))) \right] \right\} \right] \\
\psi_{10}( & J & ,\phi,t= t_3) 
 =  \sqrt{\frac{2}{\bt}} \bt_K \theta  \psi_0'(J) \sqrt{J} \nonumber \\
& & \times \sin\left[\Phi' - \Dl \phi_1(J,\Phi') - 
\Dl\phi_0(\Phi' - \Dl \phi_1(J,\Phi')) \right.   \nonumber \\
& & - \tau\left\{ \om_{\bt} +  \om' J\left[1 + A(q_3,\phi) + A(q_2,\Phi') + 
A(q_1,\Phi'  - \Dl \phi_1(J,\Phi')) \right.  \right.  \nonumber \\
& & \left. \left. \left. + A(q_0,\Phi' - \Dl \phi_1(J,\Phi') 
- \Dl\phi_0(\Phi' - \Dl \phi_1(J,\Phi'))) \right] \right\} \right] \\
\Phi' & \equiv & \phi - \Dl\phi_2(J,\phi, \Dl t_3)
\eeqrs
Now define $\dl\varphi_3$ and redefine $\Phi$
\beq
\dl\varphi_3(J,t) = \om(J)(t - (\tau + \Dl t_1+\Dl t_2 + \Dl t_3)), \;\;\; \Phi = \phi - \dl\varphi_3(J,t)
\eeq
We observe that $\psi_9$ is obtained from $\psi_8$ by the replacement $\phi \rarw \Phi$. 
We can obtain the DF $\psi_{11}$ at any time $t \ge  t_3 = \tau + \Dl t_1 + \Dl t_2 + \Dl t_3$ by replacing 
$\Phi' \rarw \Phi - \Dl\phi_2$ in $\psi_{10}$. Hence
\beqr
\psi_{11}( & J & ,\phi,t \ge  t_3) 
 =  \sqrt{\frac{2}{\bt}} \bt_K \theta  \psi_0'(J) \sqrt{J} \nonumber \\
& & \times \sin\left[\Phi - \Dl\phi_2(J,\Phi) - \Dl \phi_1(J,\Phi - \Dl\phi_2(J,\Phi)) \right.
 \nonumber \\
& & -\Dl\phi_0(\Phi - \Dl\phi_2(J,\Phi) - \Dl \phi_1(J,\Phi - \Dl\phi_2(J,\Phi)))    \nonumber \\
& & - \tau\left\{ \om_{\bt} +  \om' J\left[1 + A(q_3,\Phi) + A(q_2,\Phi - \Dl\phi_2(J,\Phi)) 
\right. \right. \nonumber\\
& & + A(q_1,\Phi - \Dl\phi_2(J,\Phi)  - \Dl \phi_1(J,\Phi - \Dl\phi_2(J,\Phi)))   \nonumber \\
& &  + A(q_0,\Phi - \Dl\phi_2(J,\Phi) - \Dl \phi_1(J,\Phi - \Dl\phi_2(J,\Phi)) 
\nonumber \\
& & \left. \left. \left. - \Dl\phi_0(\Phi - \Dl\phi_2(J,\Phi) - \Dl \phi_1(J,\Phi - \Dl\phi_2(J,\Phi)))) \right] \right\} \right]
\eeqr

Let us consider the DF at times $t$ after each quadrupole kick
\beqr
\psi_5(J,\phi,t \ge \tau ) 
& = & \sqrt{\frac{2}{\bt}} \bt_K \theta  \psi_0'(J) \sqrt{J} 
\sin\left[\Phi  - \tau\left\{\om_{\bt} + \om'J[1 + A(q_0,\Phi)] \right\} \right] 
\label{eq: psi5} \\
\psi_7(J,\phi,t \ge   t_1) 
& = & \sqrt{\frac{2}{\bt}} \bt_K \theta  \psi_0'(J) \sqrt{J} \nonumber \\
& & \times \sin\left[\Phi - \Dl\phi_0(\Phi )  - \tau\left\{ \om_{\bt} +  \om' J\left[1 + A(q_1,\Phi) 
\right.  \right. \right. \nonumber \\
& & \left. \left. \left. + A(q_0,\Phi  - \Dl\phi_0(\Phi)) \right] \right\} \right] 
\label{eq: psi7}\\
\psi_9( J ,\phi,t \ge  t_2) 
& = & \sqrt{\frac{2}{\bt}} \bt_K \theta  \psi_0'(J) \sqrt{J} \nonumber \\
& & \times \sin\left[\Phi- \Dl \phi_1(J,\Phi) - 
\Dl\phi_0(\Phi - \Dl \phi_1(J,\Phi)) \right.   \nonumber \\
& & - \tau\left\{ \om_{\bt} +  \om' J\left[1 + A(q_2,\Phi) + 
A(q_1,\Phi  - \Dl \phi_1(J,\Phi)) \right.  \right.  \nonumber \\
& & \left. \left. \left. + A(q_0,\Phi - \Dl \phi_1(J,\Phi) 
- \Dl\phi_0(\Phi - \Dl \phi_1(J,\Phi))) \right] \right\} \right] \label{eq: psi9} \\
\psi_{11}( J,\phi,t \ge  t_3) 
& = & \sqrt{\frac{2}{\bt}} \bt_K \theta  \psi_0'(J) \sqrt{J} \nonumber \\
& & \times \sin\left[\Phi - \Dl\phi_2(J,\Phi) - \Dl \phi_1(J,\Phi - \Dl\phi_2(J,\Phi))\right. \nonumber 
\\
& &  - \Dl\phi_0\left(\Phi - \Dl\phi_2(J,\Phi) - \Dl \phi_1(J,\Phi - \Dl\phi_2(J,\Phi))\right)
 \nonumber \\
& & - \tau\left\{ \om_{\bt} +  \om' J\left[1 + A(q_3,\Phi) + A(q_2,\Phi - \Dl\phi_2(J,\Phi)) 
\right.  \right.\nonumber\\
& & + A(q_1,\Phi - \Dl\phi_2(J,\Phi) - \Dl \phi_1(J,\Phi - \Dl\phi_2(J,\Phi)))   \nonumber \\
& & + A\left(q_0,\Phi - \Dl\phi_2(J,\Phi) - \Dl \phi_1(J,\Phi - \Dl\phi_2(J,\Phi))  \right. 
\nonumber \\
& & \left. \left. \left. \left.- \Dl\phi_0(\Phi - \Dl\phi_2(J,\Phi) - \Dl \phi_1(J,\Phi - 
\Dl\phi_2(J,\Phi)))\right) \right] \right\} \right] \label{eq: psi11}
\eeqr
In each case, the definition of $\Phi$ is determined by the number of kicks
\beq
\Phi(J,\phi,t) = \phi - \om(J)[t - (\tau + \sum_{n=0} \Dl t_n)]
\eeq
We can write the last DF in the form
\beqr
\psi_{11}( J,\phi,t \ge  t_3) 
& = & \sqrt{\frac{2}{\bt}} \bt_K \theta  \psi_0'(J) \sqrt{J} 
\sin\left[\Phi - \Dl\Phi_2 - \Dl \Phi_1 - \Dl\Phi_0 \right.  \nonumber  \\
& & - \tau\left\{ \om_{\bt} +  \om' J\left[1 + A(q_3,\Phi) + A(q_2,\Phi - \Dl\Phi_2) 
+ A(q_1,\Phi - \Dl\Phi_2 - \Dl \Phi_1)  \right. \right. \nonumber \\
& & + \left. \left. \left. A(q_0,\Phi - \Dl\Phi_2 - \Dl \Phi_1 - \Dl\Phi_0) \right] \right\} \right]
\label{eq: psi11_2}
\eeqr
where the $\Dl\Phi_{N-j}$ is obtained recursively from $\Dl\Phi_{N-j+1}$. 
Now the pattern is clear. 

Define the functions $\Dl\Phi$ as follows
\beq
\Dl \Phi_{N-j} = \Dl\phi_{N-j}(J,\Phi - \sum_{k=0}^{j-1}\Dl\Phi_{N-k}), \;\;\; j=0,1,...N
\eeq
The first function in the chain $\Dl\phi_N \propto \Dl t_{N+1} \equiv 0$, hence
\beqrs
\Dl \Phi_N & = & 0 \\
\Dl\Phi_{N-1} & = & \Dl\phi_{N-1}(J,\Phi) \\
\Dl\Phi_{N-2} & = & \Dl\phi_{N-2}(J,\Phi - \Dl\Phi_{N-1}) = \Dl\phi_{N-2}(J,\Phi - \Dl\phi_{N-1}(J,\Phi)) \\
\Dl\Phi_{N-3} & = & \Dl\phi_{N-2}(J,\Phi - \Dl\Phi_{N-1} - \Dl\Phi_{N-2}) = ...
\eeqrs

After $(N+1)$ kicks applied at times $t_0 \equiv \tau, t_1= t_0 + \Dl t_1, ..., t_N = t_{N-1}+\Dl t_{N}$, the distribution function is
\beqr
\psi(& J &,\phi,t \ge t_N) 
 =  \sqrt{\frac{2}{\bt}} \bt_K \theta  \psi_0'(J) \sqrt{J}
 \sin\left[\Phi - \sum_{j=0}^{N} \Dl\Phi_{N-j} \right. \nonumber \\
& & \left. -\tau\left\{ \om_{\bt} + \om' J \left[ 1 + 
\sum_{j=0}^{N}A \left(q_j,\Phi -  \sum_{k=j}^{N}\Dl\Phi_{k} \right) \right] \right\}\right] \\
\Phi(J,\phi,t) & = & \phi - \om(J)[t - (\tau + \sum_{n=0^{N-1}} \Dl t_n)] \nonumber \\
\Dl \Phi_{N-j} & = & \Dl\phi_{N-j}(J,\Phi - \sum_{k=0}^{j-1}\Dl\Phi_{N-k}) \\
\Dl\phi_j(J,\phi) & = & [\om_{\bt} + \om' J(1+A(q_{j+1},\phi))]\Dl t_{j+1} \\
t_j & = & t_{j-1} + \Dl t_j = \tau + \sum_{k = 1}^j \Dl t_k 
\eeqr

Check the expansion for the different cases. 

\noi $N=0$. Then after the kick at $t_0 = \tau$, Here by definition $\Dl\Phi_0=0$ and there are no
lower members in the recursive chain. 
\beqrs
\psi(J,\phi,t \ge t_0) 
& = & \sqrt{\frac{2}{\bt}} \bt_K \theta  \psi_0'(J) \sqrt{J} 
\sin\left[\Phi -  \tau\{\om_{\bt} + \om' J [ 1 + A(q_0,\Phi) ] \}\right] \\
\Phi(J,\phi,t) & = & \phi - \om(J)[t - \tau]
\eeqrs
This agrees with Eq.(\ref{eq: psi5}). 

\noi $N = 1$. Here we have  
\beqrs
\Dl\Phi_{0} & = & \Dl\phi_{0}(J,\Phi), \;\;\;\; \Dl\Phi_1 = 0 \\
\sum_{j=0}^{1}A( & q_j, & \Phi - \sum_{k=j}^{1}\Dl\Phi_{k}) = A(q_0,\Phi - \Dl\Phi_0) +
A(q_1, \Phi) \\
\psi( & J & ,\phi,t \ge t_1) 
= \sqrt{\frac{2}{\bt}} \bt_K \theta  \psi_0'(J) \sqrt{J}\sin\left[\Phi - \Dl\phi_0(J,\Phi)
 \right. \nonumber \\
& & \left. -\tau\left\{ \om_{\bt} + \om' J \left[ 1 + 
A(q_0,\Phi - \Dl\phi_0(J,\Phi))+ A(q_1,\Phi) \right]
 \right\}\right]
\eeqrs
This agrees with Eq.(\ref{eq: psi7}). 

\noi $N-2$. Now $\Dl\Phi_2 = 0$, 
\beqrs
\Dl\Phi_1` & = & \Dl \phi_1(J,\Phi - \sum_{k=0}^0 \Dl\Om_{2-k}) = \Dl \phi_1(J,\Phi) \\
\Dl\Phi_0 & = & \Dl \phi_0(J,\Phi - \sum_{k=0}^1 \Dl\Phi_{2-k}) = \Dl \phi_0(J,\Phi - \Dl\Phi_{1})\\
& = & \Dl \phi_0(J,\Phi - \Dl\phi_1(J,\Phi))  \\
\psi( J ,\phi,t \ge t_2) 
& = & \sqrt{\frac{2}{\bt}} \bt_K \theta  \psi_0'(J) \sqrt{J}\sin\left[\Phi - \Dl\Phi_1 - \Dl\Phi_0 
\right. \\
& & \left. -\tau\left\{ \om_{\bt} + \om' J \left[ 1 +  A(q_2,\Phi) + 
 A(q_1,\Phi - \sum_{k=1}^2\Dl\Phi_j ) + A(q_0,\Phi -  \sum_{k=0}^2\Dl\Phi_j)  \right] 
\right\}\right] \\
& = & \sqrt{\frac{2}{\bt}} \bt_K \theta  \psi_0'(J) \sqrt{J}\sin\left[\Phi - \Dl\Phi_1 - \Dl\Phi_0
\right. \\
& & \left. -\tau\left\{ \om_{\bt} + \om' J \left[ 1  + A(q_2,\Phi) 
+ A(q_1,\Phi - \Dl\Phi_1) + A(q_0,\Phi - \Dl\Phi_0 - \Dl\Phi_1)  \right] \right\}\right] 
\eeqrs
This agrees with Eq.(\ref{eq: psi9}). 

\noi $N=3$. Here $\Dl\Phi_3 = 0$, and
\beqrs
\Dl\Phi_2 & = & \Dl\phi_2(J,\Phi) \\
\Dl\Phi_1 & = & \Dl \phi_1(J,\Phi - \sum_{k=0}^1 \Dl\Phi_{3-k}) = \Dl \phi_1(J,\Phi - \Dl\Phi_2) \\
 & = &  \Dl \phi_1(J,\Phi - \Dl\phi_2(J,\Phi)) \\
\Dl\Phi_0 & = & \Dl \phi_0(J,\Phi - \sum_{k=0}^2 \Dl\Phi_{3-k}) = \Dl \phi_0(J,\Phi - \Dl\Phi_2  - 
\Dl\Phi_1) \\
& = &  \Dl \phi_0(J,\Phi - \Dl\phi_2(J,\Phi) - \Dl \phi_1(J,\Phi - \Dl\phi_2(J,\Phi))) \\
\psi( J ,\phi,t \ge t_3) 
& = & \sqrt{\frac{2}{\bt}} \bt_K \theta  \psi_0'(J) \sqrt{J}\sin\left[\Phi - \Dl\Phi_2 - \Dl\Phi_1 
-  \Dl\Phi_0 \right. \\
& &  -\tau\left\{ \om_{\bt} + \om' J \left[ 1 + A(q_3,\Phi) + A(q_2,\Phi - \Dl\Phi_2)
 \right. \right. \\
& & \left.\left.\left. + A(q_1,\Phi - \Dl\Phi_2 - \Dl\Phi_1) + A(q_0,\Phi -  \Dl\Phi_2 - \Dl\Phi_1 - \Dl\Phi_0)
\right] \right\}\right] 
\eeqrs
This agrees with Eq.(\ref{eq: psi11_2}) which is the symbolic form of the expanded form 
Eq.(\ref{eq: psi11}).

The dipole moment at a time $t \ge t_N$ is given by
\beqr
\lan x(t \ge t_N) \ran & = & \int dJ d\phi \; \sqrt{2\bt J}\cos\phi \psi(J ,\phi,t \ge t_N) \nonumber \\
& = & 2\bt_K\theta \int dJ d\phi \; J \psi_0'(J) \cos\phi \sin\left[\Phi - B(J,\phi)
 -\tau\left\{ \om_{\bt} + \om' J [ 1 + C(J,\phi)]\right\} \right] \nonumber \\
\mbox{} \\
B(J,\phi) & = & \sum_{j=0}^{N} \Dl\Phi_{N-j} \\
C(J,\phi) & = & \sum_{j=0}^{N}A \left(q_j,\Phi -  \sum_{k=j}^{N}\Dl\Phi_{k} \right)  
\eeqr

Caveats
\bit
\item The approximations made at each kick of replacing 
$\sqrt{(J+A(q,\phi- \om(J)(t - \tau))J)} \approx \sqrt{J}$ and similarly in the argument of 
$\psi_0(J)'$ will start to accumulate larger errors with increasing kicks.
\eit

We will also approximate
\beqrs
\cos\phi \sin[\Phi + f(J,\phi)] & = & \half[\sin(\Phi + f(J,\phi) + \phi) + 
\sin(\Phi + f(J,\phi) - \phi)] \\
 & \approx & \half \sin(\Phi + f(J,\phi) + \phi)
\eeqrs
since the dropped second term decreases rapidly with time, compared to the first term.
Hence
\beqr
\lan x(t \ge t_N) \ran 
& \approx &  \bt_K\theta \int dJ d\phi \; J \psi_0'(J) \sin\left[\Phi + \phi - B(J,\phi) \right. 
\nonumber \\
& & \left. -\tau\left\{ \om_{\bt} + \om' J [ 1 + C(J,\phi)]\right\} \right] \\
\eeqr

Now, I will need to consider specific cases of $N$ in order to do the integration. 
First consider $N=1$ with the distribution function
\beqrs
\psi(J,\phi,t \ge   t_1) 
& = & \sqrt{\frac{2}{\bt}} \bt_K \theta  \psi_0'(J) \sqrt{J} 
\sin \zeta \\ 
\zeta & = & \left[\Phi - \Dl\phi_0(\Phi )  - \tau\left\{ \om_{\bt} +  \om' J\left[1 + A(q_1,\Phi) 
+ A(q_0,\Phi  - \Dl\phi_0(\Phi)) \right] \right\} \right]  \\ 
\Phi & = & \phi - \om(J)(t - t_1) \\
\Dl\phi_0(\Phi) & = & [\om_{\bt} + \om' J(1+A(q_{1},\Phi))]\Dl t_{1} = [\om(J)+ \om' J A(q_{1},\Phi)]
\Dl t_{1}\\ 
A(q_1,\Phi) & = & -q_1\sin 2\Phi \\
A(q_0,\Phi  - \Dl\phi_0(\Phi)) & = & -q_0\sin 2(\Phi - \Dl\phi_0)\Phi))
\eeqrs
Using $t_1 = \tau + \Dl t_1$, we have
\[ \Phi - \Dl\phi_0(\Phi) = \phi - \om(J)(t- \tau) - \om' J A(q_{1},\Phi)\Dl t_1 \]
Now approximate $A(q_0, \Phi  - \Dl\phi_0(\Phi))$ as
\beqrs
A(q_0,\Phi  - \Dl\phi_0(\Phi)) & = & -q_0\sin 2(\phi - \om(J)(t- \tau) - \om' J A(q_{1},\Phi)\Dl t_1)\\
 & \approx & -q_0 \sin 2(\phi - \om(J)(t- \tau))
\eeqrs
by dropping the last term. We can estimate this term as follows by replacing $J \rarw \eps$ and
\[ \om' \eps A(q_1,\Phi) \Dl t_1 = -\frac{\Dl t_1}{\tau_D} q_1\sin 2\Phi \ll \sin 2\Phi \]
Both factors  $q_1 \ll 1$ and $\Dl t_1/\tau_D \ll 1$ 
for time intervals $\Dl t_1 \ll \tau_D$, the decoherence time justifying dropping this term. 
Hence we have
\beqrs
\zeta & = & \phi - \om(J)(t- \tau) - \om' J A(q_{1},\Phi)\Dl t_1 - \om(J)\tau - 
\tau[A(q_1,\Phi) + A(q_0,\Phi  - \Dl\phi_0(\Phi))] \\
& = &  \phi - \om(J)t + \om' J[q_0\tau\sin 2(\phi - \om(J)(t- \tau)) + 
q_1 t_1\sin 2(\phi - \om(J)(t - t_1))]
\eeqrs
The last form shows how it can be generalized for any $N$. 

The dipole moment is (after dropping the term that decreases rapidly with time)
\beqr
\lan x(t \ge t_1)\ran & = & \bt_K\theta \int dJ \; d\phi \; J \psi_0'(J)\sin[\phi + \zeta] \nonumber\\
& = & \half \bt_K\theta \int dJ \; d\phi \; J \psi_0'(J) \sin\left[2\phi - \om(J)t  \right. \nonumber \\
& & \left. + \om' J\left\{q_0\tau\sin 2(\phi - \om(J)(t- \tau)) + 
q_1 t_1\sin 2(\phi - \om(J)(t - t_1))\right\} \right]
\eeqr
This form suggests that after $N+1$ kicks at times $t_0=\tau, t_1,... , t_N$, the dipole moment will be
\beqr
\lan x(t \ge t_N)\ran & = & \bt_K\theta \int dJ \; d\phi \; J \psi_0'(J) \nonumber \\ 
& & \times\sin\left[2\phi - \om(J)t + \om' J\sum_{j=0}^N q_j t_j \sin 2(\phi - \om(J)(t - t_j)) \right]
\eeqr
In order to do the integration over $\phi$, first I need to combine the sum over the sine functions into
a single sine function. Writing
\beqrs
\sum_j B_j \sin (2\phi + C_j) & = & \sum_j (B_j\cos C_j)\sin 2\phi + \sum_j (B_j \sin C_j)\cos 2\phi \\
& = & Q \sin (2\phi + \xi) \\
Q & = & \left[ (\sum_j B_j\cos C_j)^2 + (\sum_j B_j\sin C_j)^2 \right]^{1/2} \\
\xi & = & (\tan)^{-1} \left[ \frac{\sum_j B_j\sin C_j}{\sum_j B_j\cos C_j} \right]
\eeqrs
The expressions for $Q,\xi$ show the nonlinear superposition of sinusoidal harmonic forces. 

In our case, we have
\[ B_j \equiv \om' J q_j t_j, \;\;\;\; C_j = -2\om(J)(t - t_j) \]
Hence here we have
\beqr
Q & = & \om' J \left[ \sum_{j=0}^N (q_j t_j)^2 \cos^2[2\om(J)(t - t_j)] +
 \sum_{j=0}^N (q_j t_j)^2 \sin^2[2\om(J)(t - t_j)] \right]^{1/2} \equiv \om' J QT_N \nonumber \\
\mbox{} \\
\xi(J) & = & -\tan^{-1}\left[\frac{ \sum_{j=0}^N q_j t_j \sin 2\om(J)(t - t_j)}{ \sum_{j=0}^N q_j t_j \cos 2\om(J)(t - t_j)}\right]
\eeqr
where $QT_N$ has the dimension of time but is independent of $J$. 
Hence
\beqr
\lan x(t \ge t_N)\ran & = & \bt_K\theta \int dJ \;  J \psi_0'(J) 
{\rm Im}\left[e^{-i\om(J) t} \int d\phi \; \exp[i\{2\phi + Q \sin(2\phi + \xi)\}]\right] \nonumber \\
& = & -2\pi \bt_K\theta {\rm Im}\left[\int dJ \;  J \psi_0'(J) e^{-i\om(J) t} e^{-i\xi(J)}J_1(Q)\right]
\nonumber \\
& = & \bt_K\theta {\rm Im}\left[\int dz \;  z e^{-z} e^{-i\om(\eps z) t} e^{-i\xi(\eps z)}J_1(\om'\eps QT_N z)
\right]
\eeqr
where as before, $J_1$ is the Bessel function and we substituted the form of $\psi_0$. $QT_N$ is 
independent of $z$. 

Evaluating this analytically may not be possible because of the complicated dependence of $\xi(\eps z)$.

The maximum amplitude of this can be estimated by setting the phase factors to unity. In that case
\beqr
\lan x(t \ge t_N)\ran^{max} & = &  \bt_K\theta \int dz \;  z e^{-z} J_1(\om'\eps QT_N z) \nonumber \\
 & = &  \bt_K \theta \frac{\om' \eps QT_N}{[1 + (\om' \eps QT_N)^2]^{3/2}}
\eeqr

With only the single kick at time $t_0= \tau$, we have $QT_0 = q_0\tau$ and the above maximum amplitude
agrees with the result obtained earlier for the amplitude at the time $t=2\tau$, see 
Eq.(\ref{eq: amp_nonlinq}). 

Now the depressing part: As a function of $Q = \om' \eps QT_N$, this is exactly the same
functional form of $Q$ as with a single quadrupole kick. Hence the optimum value of $Q$ is the same as
before, i.e. $Q^2 = 1/2$ and the maximum possible amplitude is the same as before
\[ \lan x(t \ge t_N)\ran^{max, amp}/(\bt_K\theta) = \frac{2}{3\sqrt{3}} = 0.38 \]

\subsection{Stimulated Echoes}

In the field of magnetic resonance imaging, the standard spin echo is generated by using two rf pulses, the first to 
excite the spins and the second to refocus the dephasing spins\cite{Hahn}. The addition of a third rf pulse again 
refocuses the spins again after the first echo and leads to additional echoes. In 
Section \ref{sec: NL_dip}, we saw that a single quadrupole kick
can lead to multiple echoes at times $4\tau, 6\tau$ etc. In this section, we briefly 
consider  the possibility of  amplifying the echoes at $2\tau, 4\tau, ...$ with additional quadrupole  kicks. 

We consider first the linear theory, where the echo amplitude is given by Eq.(\ref{eq: amp_linear}) in 
Section \ref{sec: NL_quad}
By linear superposition, it follows that if there are $N_q$ kicks applied at times $\tau_m$, $m=1,..., N_q$

\beqr
\lan x\ran(t) 
& = & \bt_K \theta \om' \eps_0 \sum_{m=1}^{N_q} q_m \tau_m 
\frac{\sin(\Phi_m + 3\tan^{-1}[\xi_m]) }{(1 + \xi_m^2)^{3/2}}
\label{eq: amp_multquads_linear}  \\
\Phi_m & = & \om_{\bt}(t - 2\tau_m)  \; \; \; \xi_m  =  \om' \eps(t - 2\tau_m )  \nonumber 
\eeqr
We know that this is valid only in the regime of small $q < q_{opt}$. We consider first the case of amplifying the
echo at 2$tau$ by using several small quadrupole kicks. 
We consider two cases: 1) all kicks have the same strength: $q_m = q$; 2) 
alternating sign quadrupole kicks: $q_m = (-1)^m q$. 

As a
 special case of the above, consider a {\bf stimulated echo} at later times with 2 quadrupole 
kicks, i.e. $N_q=1$ and
the 2nd kick is applied at time $p\tau$ after the 1st kick. In this case, we have
$n_{gap} T_{rev} = p\tau$ and
\beqr
\lan x(t) \ran & = & \bt_K \theta \om' \eps_0 [q_0 (\tau) \frac{\sin(\Phi_0 + 3\tan^{-1}[\xi_0])}{(1 + \xi_0^2)^{3/2}}
+ q_1 ((p+1)\tau) \frac{\sin(\Phi_1 + 3\tan^{-1}[\xi_1]) }{(1 + \xi_1^2)^{3/2}} ]
\nonumber \\
\Phi_0 & = & \om_{\bt}(t - 2\tau), \;\;\; \Phi_1  =  \om_{\bt}(t - 2(p+1)\tau)  \nonumber \\
\xi_0 & = & \om' \eps_0(t - 2\tau ), \;\;\; \xi_1  =  \om' \eps_0(t - 2(p+1)\tau )  \nonumber 
\eeqr
Note that the second term is only applicable for times $t \ge (p+1)\tau$. If we assume that the
two kicks have the same sign, $q_0 = q_1 = q$ and introducing the dimensionless quadrupole strength 
parameter
\[  Q = q \om' \eps_0 \tau \]
We have therefore 
\beq
\lan x(t) \ran  =  \bt_K \theta Q [ \frac{\sin(\Phi_0 + 3\tan^{-1}[\xi_0])}{(1 + \xi_0^2)^{3/2}}
+  (p+1) \frac{\sin(\Phi_1 + 3\tan^{-1}[\xi_1]) }{(1 + \xi_1^2)^{3/2}} ]
\eeq
The stimulated echo at time $t = 4\tau$ is given by this expression with
\beqrs
\Phi_0 & = & 2\om_{\bt}\tau3, \;\;\; \Phi_1  =  2\om_{\bt}\tau(2 - (p+1))   \\
\xi_0 & = & 2\om' \eps_0\tau , \;\;\; \xi_1  =  2\om' \eps_0\tau(2 - (p+1))  
\eeqrs

\section{Nonlinear dipole and quad kicks theory}\label{sec: NL_dip}

Here I consider the complete distribution function following the dipole kick, instead of the 1st order Taylor expansion. At time $\tau$ after the dipole kick, the DF is
\beq
\psi_3(J,\phi \tau) = \psi_0(J + \bt_k\theta \sqrt{2 J/\bt}\sin \phi_{-\tau} + (1/2)\bt_k\theta^2) , \;\;\;
\phi_{-\tau} \equiv \phi - \om(J) \tau
\eeq
After the quad kick,
\[ J \rarw J( 1 + A(q,\phi)), \;\;\; \phi \rarw \phi - q \]
Hence the DF right after the quad kick is 
\beq
\psi_4(J,\phi, \tau) = \psi_0(J( 1 + A(q,\phi)) + \bt_k\theta \sqrt{2 /\bt}\sqrt{J( 1 + A(q,\phi))}\sin(\phi_{-\tau} - q) + (1/2)\bt_k\theta^2)
\eeq
where now 
\beq
\phi_{-\tau} = \phi - \tau [\om_{\bt} + \om' J(1 + A(q,\phi))]
\eeq
At a time $t > \tau$, the DF is
\beq
\psi_5(J,\phi, t > \tau) = \psi_4(J, \phi_{-\Dl \phi}), \;\;\; \phi_{-\Dl \phi} = \phi - \om(J)(t - \tau) \equiv \phi - \Dl \phi
\eeq
The dipole moment is 
\beq
\lan x(t) \ran = \sqrt{2\bt} \int dJ \; \sqrt{J} \int d\phi \cos\phi \psi_5(J, \phi,t) 
\eeq
Under the change $\phi \rarw \phi_{-\Dl \phi}$, 
\beqrs
\phi_{-\tau} & \rarw & \phi - \om(J)(t - \tau) - \om(J + A(q,\phi)J) \tau = \phi_{-\Dl \phi} -  \tau [\om_{\bt} + J(1+\half q^2) - \om' J B(q) 
\sin 2\phi_{-\Dl \phi}]   \\
& = &  \phi - \Dl\phi -  \tau \om_+ + Q z\sin 2\phi_{-\Dl \phi} , \;\;\; \; Q = \tau \om' B(q) \eps_0, \;\;\; 
z= J/\eps_0, \;\;\; \om_+ = \om_{\bt} + (1 + \half q^2)\om' J
\eeqrs
Note that $\Dl\phi$ is independent of the phase $\phi$ but depends on $J$. Let
\beq
D(z) = \Dl\phi(z) + \tau \om_+(z) + q, \;\;\; \Rarw \phi_{-\tau} - q \rarw \phi_{-\Dl\phi} - D(z) + Qz\sin 2\phi_{-\Dl\phi}
\eeq
and  we have
\beq
\psi_5(J,\phi, t) = \psi_0(J( 1 
+ A(q,\phi_{-\Dl \phi})) + \bt_k\theta \sqrt{2 /\bt}\sqrt{J( 1 + A(q,\phi_{-\Dl \phi}))}\sin (\phi - D(z) + Q z\sin 2\phi_{-\Dl \phi} ) 
+ (1/2)\bt_k\theta^2)
\eeq
Since
\[ \psi_0(J)  = \frac{1}{2\pi J_0} \exp[ - \frac{J}{J_0}] \]
we have
\beqr
\psi_5(J,\phi,t) & = & \frac{1}{2\pi J_0} \exp[-\frac{\bt_k\theta^2}{2 J_0}]\exp\left\{-\frac{1}{J_0}
[J( 1 + A(q,\phi_{-\Dl \phi})) + \bt_k\theta \sqrt{2 /\bt}\sqrt{J( 1 + A(q,\phi_{-\Dl \phi}))} \right. \nonumber \\
& &  \left.  \times \sin (\phi - D(z) + Q z\sin 2\phi_{-\Dl \phi}) ] \right\}
\eeqr
Making the approximation
\[ \sqrt{1 + A(q, \phi)} \approx 1 + \half A(q, \phi)  = 1 + \qrtr q^2 - \half B(q) \sin (2 \phi - q/2)  \equiv C_1 + C_3\sin (2 \phi - q/2)
\]
Caution: Does this approximation hold for $A(q, \phi_{-\Dl \phi}) = -B(q)\sin [2(\phi - \om(t - \tau)) - q/2]+ \half q^2 $ when $t \gg \tau$? Yes, since that occurs in the argument of the sine.

The argument of the exponential inside the $\phi$ integration is
\beqrs
\mbox{} & = & -\frac{1}{J_0}\left\{ J(-B(q)\sin (2\phi_{-\Dl \phi}-q/2)+\half q^2) + \bt_k\theta\sqrt{\frac{2J}{\bt}}\left(1 + \qrtr q^2 - 
\half B(q) \sin (2\phi_{-\Dl \phi} - q/2) \right) \right. \\
&  &  \left.  \times \sin( \phi - D(z) + Q z \sin 2\phi_{-\Dl \phi} ) \right\}  \\
& = &  -\half q^2 \frac{J}{J_0}-\frac{1}{J_0}\left\{ - B(q) J \sin (2\phi_{-\Dl \phi}-q/2) + \bt_k\theta\sqrt{\frac{2J}{\bt}}(1 + \qrtr q^2)
\sin( \phi - D(z) + Q z\sin 2\phi_{-\Dl \phi}  )  \right. \\
& &  \left. - \half \bt_k\theta\sqrt{\frac{2J}{\bt}}B(q ) \sin (2\phi_{-\Dl \phi}-q/2)\sin( \phi - D(z) + Q z\sin 2\phi_{-\Dl \phi} ) \right\} \\
& = & -\half q^2 z + b_1z \sin (2\phi_{-\Dl \phi}-q/2) - b_2\sqrt{z} \sin( \phi - D(z) + Q z \sin 2\phi_{-\Dl \phi} ) \\ 
&  &  +  2 b_3\sqrt{z} \sin (2\phi_{-\Dl \phi} - q/2)  \sin( \phi - D(z) + Q z \sin 2\phi_{-\Dl \phi} ) 
\eeqrs
where we replaced $z = J/J_0$ and defined the positive definite dimensionless parameters
\beqrs
a_{\theta} & = &  \frac{\bt_k\theta}{\sg_0}, \;\;\;\;\; C_1 = 1 + \qrtr q^2   \\
b_1 & = &  B(q)  \\
b_2 &  = & \bt_k\theta\sqrt{\frac{2}{\bt J_0}}(1 + \qrtr q^2)= \sqrt{2}\frac{\bt_k\theta}{\sg_0}(1+ \qrtr q^2)= 
\sqrt{2}C_1 a_{\theta} \\
b_3 & = &  \qrtr \bt_k\theta\sqrt{\frac{2}{\bt J_0}}B(q ) = \frac{\sqrt{2}}{4}\frac{\bt_k\theta}{\sg_0} B(q) = \frac{\sqrt{2}}{4}
B(q) a_{\theta} , \;\;\;\;\; b_i \ge 0 \\
\eeqrs
The dimensionless parameter $a_{\theta}$ is the rms dipole kick in units of the rms beam size. 
Clearly for $q \ll 1$, we have $b_1 \ll 1$. Now using
\[ \sin A \sin B = \half(\cos(A-B) - \cos(A+B) ) \]
and
\beqrs
2\phi_{-\Dl \phi} - q/2 - (\phi - D(z) + Q z \sin 2\phi_{-\Dl \phi} ) & = & 2(\phi - \om(t - \tau)) - q/2 - (\phi - \om(t - \tau) - \tau \om_+
 - q + Q z \sin 2\phi_{-\Dl \phi})  \\
&  = &  \phi - \om(t - \tau) + \tau\om_+ - Q z \sin 2\phi_{-\Dl \phi} + q/2 \\
 2\phi_{-\Dl \phi} - q/2 + (\phi - D(z) + Q z\sin 2\phi_{-\Dl \phi} ) & = & 3\phi - 3\om(t - \tau) - \tau\om_+ + Q z \sin 2\phi_{-\Dl \phi} -3q/2
\eeqrs
Transforming from the variable $J$ to $z = J/J_0$, we can write the exponent as 
\beqrs
arg & = & -\half q^2 z + b_1 z \sin 2\phi_{-\Dl \phi} - b_2\sqrt{z} \sin( \phi - \Dl\phi - \tau \om_+ - q + Q z \sin 2\phi_{-\Dl \phi} ) \\
&  &  + b_3\sqrt{z} \cos[\phi - \om(t - \tau) + \tau\om_+ - Q z \sin 2\phi_{-\Dl \phi}+ q/2]  \\
&  & - b_3 \sqrt{z} \cos [3\phi - 3\om(t - \tau) - \tau\om_+ + Q z \sin 2\phi_{-\Dl \phi} - 3q/2]
\eeqrs
Thus we have for the dipole moment
\beqr
\lan x(t) \ran & = & \frac{\sqrt{2\bt J_0}}{2\pi} \exp[-\frac{\bt_k\theta^2}{2 J_0}]\int dz \; \sqrt{z} \exp[-(1 + \half q^2)z] 
T_{\phi}(z) \\
T_{\phi}(z)  & = & \int d\phi \cos\phi \exp\left[ b_1 z \sin (2\phi_{-\Dl \phi} - q/2) - b_2\sqrt{z} \sin( \phi - \Dl\phi - \tau \om_+ - q + Q z\sin 2\phi_{-\Dl \phi})  \right.   \nonumber \\
& &  +  b_3\sqrt{z} \cos\left(\phi - \om(t - \tau) + \tau\om_+ - Q z \sin 2\phi_{-\Dl \phi}+ q/2\right)   \nonumber \\
& &  \left.   - b_3\cos \left(3\phi -  3\om(t - \tau) - \tau\om_+ + Q z \sin 2\phi_{-\Dl \phi} - 3q/2\right) \right]  \nonumber \\
& = &  {\rm Re} \left\{  \int d\phi e^{i\phi} \exp\left[
b_1z\sin (2\phi_{-\Dl \phi} - q/2) - b_2\sqrt{z} \sin( \phi - \Dl\phi - \tau \om_+ - q + Q z \sin 2\phi_{-\Dl \phi} ) \right. \right. \\ 
&  &  + b_3\sqrt{z} \cos\left(\phi - \om(t - \tau) + \tau\om_+ - Q z \sin 2\phi_{-\Dl \phi}+ q/2\right)   \nonumber  \\
& & \left.  \left.   - b_3\sqrt{z} \cos \left(3\phi -  3\om(t - \tau) - \tau\om_+ + Q z \sin 2\phi_{-\Dl \phi} - 3q/2\right) \right] \right\}
\eeqr

One way to do the integration over $\phi$ is to use the generating function for the modified Bessel functions
\[ \exp[\half(t + \frac{1}{t})z] = \sum_{n=-\infty}^{\infty} I_n(z) t^n \]
Firs setting $t = \pm e^{i \theta}$ and then $t = i e^{\mp i \theta}$, we obtain
\[ e^{ \pm z \cos\theta} = \sum_{n=-\infty}^{\infty} (\pm 1)^n  I_n(z) e^{i n \theta}, \;\;\;
e^{ \pm z \sin\theta} = \sum_{n=-\infty}^{\infty}  i^n I_n(z) e^{\mp i n \theta} 
\]
With these expansions
\beqrs
T_{\phi}(z) & = &  {\rm Re} \left\{  \sum_{k_1}\sum_{k_2}\sum_{k_3}\sum_{k_4}  i^{k_1 + k_2}(-1)^{k_4} 
 I_{k_1}(b_1z) I_{k_2}(b_2\sqrt{z}) I_{k_3}( b_3\sqrt{z}) I_{k_4}(b_3\sqrt{z}) 
\right. \\
& &  \int d\phi e^{i\phi} \exp\left[ i( - k_1(2\phi_{-\Dl \phi}- q/2) + k_2( \phi - \Dl\phi - \tau \om_+ - q + Q z \sin 2\phi_{-\Dl \phi} ) \right. \\
&  &  + k_3( \phi - \om(t - \tau) + \tau\om_+ - Q z \sin 2\phi_{-\Dl \phi}+ q/2) \\
&  &  \left. \left.  + k_4(3\phi - 3\om(t - \tau) - \tau\om_+ + Q z \sin 2\phi_{-\Dl \phi} -  3q/2) )  \right] \right\} \\
 & = &  {\rm Re} \left\{  \sum_{k_1}\sum_{k_2}\sum_{k_3}\sum_{k_4}  i^{k_1 + k_2} (-1)^{k_4}  
 I_{k_1}(b_1z) I_{k_2}(b_2\sqrt{z}) I_{k_3}( b_3\sqrt{z}) I_{k_4}(b_3\sqrt{z}) \right. \\
& &  \times \exp[i( k_1(2\om(t - \tau) + q/2)- k_2(\om( t-\tau)- \tau\om_++q) - k_3(\om(t - \tau) -  \tau\om_+ - q/2) \\
&  &  - k_4(3\om(t - \tau) +  \tau\om_+ + 3q/2))] \\
& & \left.  \int d\phi  \exp\left[ i \left( [1 - 2 k_1 + k_2  + k_3 + 3 k_4] \phi + (k_2+k_4 - k_3)Q z\sin  2\phi_{-\Dl \phi}\right) \right]  \right\}
\eeqrs
To do the $\phi$ integration, I have to expand into a Bessel function
\[ \exp[i(k_2+k_4 - k_3)Q z\sin  2\phi_{-\Dl \phi}] = \sum_{l=-\infty}^{infty} J_l((k_2+k_4-k_3)Q z)\exp[i l 2(\phi - \Dl\phi)] \]
We have therefore 
\beqrs
T_{\phi}(z) & = & 
 {\rm Re} \left\{  \sum_{k_1}\sum_{k_2}\sum_{k_3}\sum_{k_4}  \sum_l  i^{k_1 + k_2} (-1)^{k_4}  
 I_{k_1}(b_1z) I_{k_2}(b_2\sqrt{z}) I_{k_3}( b_3\sqrt{z}) I_{k_4}(b_3\sqrt{z}) J_l((k_2+k_4-k_3)Q z)  \right. \\
& &  \times \exp[i( k_1(2\om(t - \tau) + q/2)- k_2(\om( t-\tau)- \tau\om_+ +q) - k_3(\om(t - \tau) -  \tau\om_+ - q/2) \\
& & - k_4(3\om(t - \tau) +  \tau\om_+ + 3q/2)) - 2 l\Dl\phi ]  \\
& & \left.  \int d\phi  \exp\left[ i (1 - 2 k_1 + k_2  + k_3 + 3 k_4 + 2l ) \phi \right]  \right\} \\
& = &  2\pi {\rm Re} \left\{  \sum_{k_1}\sum_{k_2}\sum_{k_3}\sum_{k_4}  \sum_l  i^{k_1 + k_2} (-1)^{k_4}  
 I_{k_1}(b_1z) I_{k_2}(b_2\sqrt{z}) I_{k_3}( b_3\sqrt{z}) I_{k_4}(b_3\sqrt{z}) J_l((k_2 + k_4-k_3)Q z)  \right. \\
& &  \times \exp[i( k_1(2\om(t - \tau) + q/2)- k_2(\om( t-\tau)- \tau\om_+ +q) - k_3(\om(t - \tau) -  \tau\om_+ - q/2) \\
&  &  \left.  - k_4(3\om(t - \tau) +  \tau\om_+ + 3q/2)) - 2 l\Dl\phi ] \dl(1 - 2 k_1 + k_2  + k_3 + 3 k_4 + 2l )  \right\}
\eeqrs
We can replace $k_3$ by $2k_1 - k_2 - 3k_4 - 2l - 1$, drop the sum over $k_3$ and for convenience, replace $k_4$ by $k_3$. 
Writing
\[ \tau \om_+ = \om \tau + \half q^2 \om' \eps_o \tau  z = \om \tau + Q_2 z, \;\;\;\; Q_2 = \half q^2 \om' \eps_0 \tau \]
where $Q_2 \sim O(q^2)$ and dimensionless. The phase factor simplifies to 
\beqrs
\mbox{} & = &  k_1(2\om(t - \tau) + q/2) - k_2(\om( t-\tau)- \tau\om_+ +q)  
-(2k_1 - k_2 - 3 k_3 - 2l - 1)(\om(t - \tau) -  \tau\om_+ - q/2) \\
& &  - k_3(3\om(t - \tau) +  \tau\om_+ + 3q/2)) - 2l \om(t-\tau)\\
& = & \om \left[ t - 2\tau(-k_1 + 2 k_3 + l + 1 1)\right] + Q_2 z(2k_1 - 4k_3 - 2l- 1)  +  q((3/2)k_1 - (3/2)k_2 - 3k_3 - 1/2)
\eeqrs
while the argument of  the Bessel function is
\beqrs
 (k_2 + k_4 - k_3)Q z & = &  (k_2 + k_4 - (2k_1 - k_2 - 3k_4 - 2l - 1))Q z = (2(l + k_2 + 2k_4 - k_1) + 1)Q z \\
&  \rarw  & (2(l + k_2 + 2k_3 - k_1) + 1)Q z 
\eeqrs
Hence
\beqrs
T_{\phi}(J) & = &   2\pi {\rm Re} \left\{  \sum_{k_1}\sum_{k_2}\sum_{k_3} \sum_l  i^{k_1 + k_2} (-1)^{k_3}  
\exp[ i  q((3/2)k_1 - (3/2)k_2 - 3k_3 - 1/2)] \right.  \\
&  &  I_{k_1}(b_1z) I_{k_2}(b_2\sqrt{z}) I_{k_3}( b_3\sqrt{z}) I_{2k_1 - k_2 - 3 k_3 - 1}(b_3\sqrt{z}) J_l([2(l + k_2 + 2k_3 - k_1) + 1]Q z ) \\
& & \left.  \times \exp\left( i\left[ \om( t - 2\tau(-k_1 + 2 k_3 + l + 1))  + Q_2 z(2k_1 - 4k_3 - 2l- 1) \right]\right) \right\}
\eeqrs

\noi Comments
\bit
\item The form of the phase factor  shows the possibility of multiple echoes. Since the amplitude is
locally maximum when the phase factor vanishes, the form above shows that echoes occur at (dropping the small contribution 
from $q$) when
\[ t - 2\tau(2k_3  + l  - k_1 + 1) = 0 \]
This predicts echoes only at  multiples of 2$\tau$
\eit
1
Since we want to identify the amplitudes of the lowest order echoes at $2\tau, 4\tau$, we replace 
\[ -k_1 + 2 k_3 + l = m , \;\;\; k_1 =  2 k_3 + l - m \]
\beqrs
(3/2)k_1 - (3/2)k_2 - 3k_3 - 1/2 & = & \half( 3 l - 3 m - k_2 - 1) \\
2k_1 - k_2 - 3 k_3 - 1 & = & 4k_3 + 2l - 2m - k_2 - 3k_3 - 1 = k_3 + 2l - k_2 - 2m - 1 \\
l + k_2 + 2k_3 - k_1 & = & l + k_2 + 2k_3 - 2k_3 - l + m = k_2 + m  \\
2k_1 - 4k_3 - 2l- 1 & = & 4k_3 + 2l - 2m - 4k_3 - 2l - 1 = - 2m - 1
\eeqrs

Subsequently we can replace $k_2$ by $k_1$ and $k_3$ by $k_2$. Hence
\beqrs
T_{\phi}(J) & = &   2\pi {\rm Re} \left\{  \sum_{m}\sum_{k_2}\sum_{k_3} \sum_l  i^{ k_2 + 2k_3 + l - m} (-1)^{k_3}  
\exp[ i \half q( 3 l - 3 m - k_2 - 1)] \right.  \\
&  &  I_{2k_3 + l - m }(b_1z) I_{k_2}(b_2\sqrt{z}) I_{k_3}( b_3\sqrt{z}) I_{k_3 + 2l - k_2 - 2m - 1}(b_3\sqrt{z}) 
J_l([2(k_2 + m) + 1]Q z ) \\
& & \left.  \times \exp\left( i\left[ \om( t - 2\tau(m + 1))  - Q_2 z(2m + 1) \right]\right) \right\} \\
& = &   2\pi {\rm Re} \left\{  \sum_{m}\sum_{k_1}\sum_{k_2} \sum_l  i^{ k_1 + l - m} 
\exp\left( i\left[ \om( t - 2\tau(m + 1))  - Q_2 z(2m + 1) \right]\right) \exp[ i \half q( 3 l - 3 m - k_1 - 1)] \right.  \\
&  &  \left.  I_{2k_2 + l - m }(b_1z) I_{k_1}(b_2\sqrt{z}) I_{k_2}( b_3\sqrt{z})  
 I_{k_2 + 2l - k_1 - 2m - 1}(b_3\sqrt{z}) J_l([2(k_1 + m) + 1]Q z ) \right\} \\
& = &   2\pi {\rm Re} \left\{ e^{i[\om(t-2\tau) - Q_2 z]} \sum_{k_1}\sum_{k_2} \sum_l  i^{ k_1 + l} 
e^{ i \half q( 3 l - k_1 - 1)}  I_{2k_2 + l }(b_1z) I_{k_1}(b_2\sqrt{z}) I_{k_2}( b_3\sqrt{z}) \right.  \\
&  &  \times I_{k_2 + 2l - k_1 - 1}(b_3\sqrt{z}) J_l([2k_1  + 1]Q z )  \\
&   &  + e^{i[\om(t-4\tau) - 3Q_2 z]} \sum_{k_1}\sum_{k_2} \sum_l  i^{ k_1 + l - 1} e^{ i \half q( 3 l - k_1 - 4)}
 I_{2k_2 + l -1 }(b_1z) I_{k_1}(b_2\sqrt{z}) I_{k_2}( b_3\sqrt{z}) \\
&  &  \times I_{k_2 + 2l - k_1 - 3}(b_3\sqrt{z}) J_l([2k_1  + 3]Q z ) \\
& & \left.   + ...  \right\} \\
\eeqrs
where in the last form we identified the first two terms as those contributing to the echoes at times $2\tau$ and $4\tau$.

Comment
\bit
\item I would like to use a summation formula for the Bessel functions to remove the sum over $k_1$ if possible. I know of these
Neumann summation formulas
\[
\sum_k (-1)^k I_{p + k}(z_1) I_k(z_2) = I_p(z_1 - z_2), \;\;\;  \sum_k  I_{p - k}(z_1) I_k(z_2) = I_p(z_1 + z_2)
\]
We can attempt a small amplitude approximation for the $I_n(z)$ as
\[ I_n(z) \simeq_{\lim z \rarw 0)} \frac{(z/2)^n}{\Gm(n+1)}, \;\;\; n \ne -1, -2, ...   \]
This could be applied to the argument $b_1 J/J_0$ for small actions $J/J_0 \ll 1$. This single power expansion is questionable
for the other arguments $b_2 \sqrt{J/J_0}, b_3\sqrt{J/J_0}$. 
\eit

If we consider the dominant terms contributing to the 1st and 2nd echoes at 2$\tau$ and 4$\tau$ respectively,
\beqr
\lan x(2\tau) \ran & = & \sqrt{2\bt J_0} \exp[-\frac{\bt_k\theta^2}{2 J_0}] \int dz \; \sqrt{z} \exp[-(1 + \half q^2) z] \bar{T}_{1, \phi}(z) \\
T_{1, \phi}(z)  & = &  {\rm Re} \left\{ e^{i[\om(t-2\tau) - Q_2 z]} \sum_{k_1}\sum_{k_2} \sum_l  i^{ k_1 + l} 
e^{ i \half q( 3 l - k_1 - 1)}  I_{2k_2 + l }(b_1z) I_{k_1}(b_2\sqrt{z}) I_{k_2}( b_3\sqrt{z})  \right. \\
&  & \left.  \times I_{k_2 + 2l - k_1 - 1}(b_3\sqrt{z}) J_l([2k_1  + 1]Q z )  \right\} \\
\lan x(4\tau) \ran & = & \sqrt{2\bt J_0} \exp[-\frac{\bt_k\theta^2}{2 J_0}] \int dz \; \sqrt{z} \exp[-(1 + \half q^2) z] \bar{T}_{2, \phi}(z) \\
T_{2, \phi}(z)  & = &  {\rm Re} \left\{ e^{i[\om(t-4\tau) - 3Q_2 z]} \sum_{k_1}\sum_{k_2} \sum_l  i^{ k_1 + l - 1} e^{ i \half q( 3 l - k_1 - 4)}
 I_{2k_2 + l -1 }(b_1z) I_{k_1}(b_2\sqrt{z}) I_{k_2}( b_3\sqrt{z}) \ \right. \\
&  & \left. \times I_{k_2 + 2l - k_1 - 3}(b_3\sqrt{z}) J_l([2k_1  + 3]Q z ) \right\}\\
\eeqr
We can write
\beqrs
 \om(t - 2m\tau) & = &  \om_{\bt}(t- 2m\tau) + \om' \eps_0 (t - 2m\tau) z \equiv \Phi_m + \xi_m z \\
\Phi_m & = & \om_{\bt}(t- 2m\tau),  \;\;\;  \xi_m = \om' \eps_0 (t - 2m\tau)
\eeqrs
Thus the echo at $2\tau$ can be written as
\beqrs
\lan x(2\tau) \ran & = & \sqrt{2\bt J_0} \exp[-\frac{\bt_k\theta^2}{2 J_0}] \int dz \; \sqrt{z} \exp[-(1 + \half q^2) z] \bar{T}_{1, \phi}(z) \\
T_{1, \phi}(z)  & = &  {\rm Re} \left\{ e^{i[\Phi_1 - Q_2 z]} \sum_{k_1}\sum_{k_2} \sum_l  i^{ k_1 + l} 
e^{ i \half q( 3 l - k_1 - 1)}  I_{2k_2 + l }(B(q)z) I_{k_1}(\sqrt{2}C_1a_{\theta} \sqrt{z}) I_{k_2}( \frac{\sqrt{2}}{4}B(q)a_{\theta}\sqrt{z}) 
 \right. \\
&  & \left.  \times I_{k_2 + 2l - k_1 - 1}(\frac{\sqrt{2}}{4}B(q)a_{\theta}\sqrt{z}) J_l([2k_1  + 1]Q z )  \right\} \\
\eeqrs

\noi Limiting cases
We use the properties
\[ I_0(0)  = 1 = J_0(0),  I_{n, n\ne 0}(0) = 0 = J_{n, n\ne 0}(0)  \]

\bit
\item No dipole kick $a_{\theta} = 0$  \\
This implies $k_1 = 0 = k_2 = k_2 + 2l - k_1 - 1$. The last has no solution for integer $l$, hence in this limit the Bessel functions
depending on $a_{\theta}$ all vanish and hence so does the echo as expected.

\item No quadrupole kick \\
Consider the limiting case of no quadrupole kick, then $b_1 = 0 = b_3$. The only terms
contributing to the dipole moment at $t=2\tau$ are those with $2k_2 + l = 0 =  k_2 = k_2 + 2l - k_1 + 1 = l$, we have
as the only solution $k_2 = 0 = l$ and $k_1 = 0$. \\
TBC

\item 1st order in the dipole kick, all orders in the quad kick \\
We can use the ascending series expansion for the Bessel functions \cite{Abram_Steg}
\beqr
I_n(z) &  =  & (\frac{z}{2})^n \sum_{k=0}^{\infty} \frac{1}{k! (n+k)!} (\frac{z}{2})^{2k}, \;\;\; I_{-n}(z) = I_n(z) \\
J_n(z) & = & (\frac{z}{2})^n \sum_{k=0}^{\infty} (-1)^k\frac{1}{k! (n+k)!} (\frac{z}{2})^{2k}, \;\;\; J_{-n}(z) = (-1)^n J_n(z) \\
\eeqr
where $n$ is an integer. 
This shows for example that $I_{\pm 1}(z)$ are the only Bessel functions to have a term linear in $z$. Since we have the product of
three Bessel functions  each of whose arguments is $\propto a_{\theta}$, 
\[
I_{k_1}(\sqrt{2}C_1a_{\theta} \sqrt{z}) I_{k_2}(\frac{\sqrt{2}}{4}B(q)a_{\theta}\sqrt{z})I_{k_2 + 2l - k_1 - 1}(\frac{\sqrt{2}}{4}B(q)a_{\theta}\sqrt{z})
\]
we can have the following
combinations of terms that can result in a term linear in $a_{\theta}$.
\beqrs
k_1  & = & \pm 1,  \;\;\; k_2 = 0, \;\;\; k_2 + 2l - k_1 - 1 = 0; \;\;\; \Rarw 2l = k_1 + 1 = 0, 2 \\
k_1 & = & 0, \;\;\; k_2 = \pm 1, \;\;\; k_2 + 2l - k_1 - 1  = 0; \;\;\; \Rarw  2l = -k_2 + 1 = 0, 2 \\
k_1  & = & 0, \;\;\; k_2 = 0, \;\;\;  k_2 + 2l - k_1 - 1  = \pm 1; \;\;\; \Rarw  2l = 0, 2
\eeqrs
Hence we have these  possibilities: $(k_1, k_2, l) = (\pm 1, 0, 0), (\pm 1, 0, 1), (0, \pm 1, 0),  (0, \pm 1, 1), (0, 0, 0),
(0, 0, 1)$

Hence the following terms contribute to terms of $O(a_{\theta})$ using $\lim_{z\rarw 0} I_0(z)= 1$ and $\lim_{z\rarw 0} I_1(z)= z/2$
\beqrs
T_{1, \phi}(z)  & = &  {\rm Re} \left\{ e^{i[\Phi_1 - Q_2 z]} \left(  \half \sqrt{2}C_1a_{\theta} \sqrt{z}
\sum_{k_1 = \pm 1}  i^{ k_1}  \sum_{l=0,1}  e^{ i \half q(3l - k_1 - 1)} I_{l }(B(q)z) J_l([2k_1  + 1]Q z )   \right. \right.   \\ 
&  & + \half \frac{\sqrt{2}}{4}B(q)a_{\theta}\sqrt{z} \sum_{k_2= \pm 1}\sum_{l=0,1} i^{ l} e^{ i \half q( 3 l - 1)} I_{2 k_2+ l}(B(q)z)  J_l(Q z ) \\
& & \left. \left. +  \half \frac{\sqrt{2}}{4}B(q)a_{\theta}\sqrt{z}\sum_{l= 0, 1}  i^{ l} e^{ i \half q( 3 l - 1)}  I_{l }(B(q)z) J_l(Q z ) 
 \right) \right\}
\eeqrs
\eit

\subsection{A more approximate but simpler calculation}

While doing the nonlinear quad kick only calculation, a good approximation to the final result was initially obtained by
not changing the action $J$ in the distribution function but only the phase. Here I'll do the same to obtain a simpler
result and check if that also gives the multiple echoes.

Bessel function property
\[ I_{-n}(z) = I_n(z), \;\;\;\;  J_n(-z) = (-1)^n J_n(z), \;\;\;  J_{-n}(z) = (-1)^n J_n(z) \]

In this approximation, we have
\beqrs
\psi_5(J, \phi,t) & = & \psi_0(J + \bt_k\theta\sqrt{2 J/\bt}\sin(\phi - D(z) + Qz\sin 2\phi_{-\Dl\phi}) + \half \bt_k\theta^2 ) \\
  &  = & \frac{1}{2\pi J_0}\exp[-\frac{\bt_k\theta^2}{2J_0}]\exp[-\fr{1}{J_0}[J + \bt_k\theta\sqrt{2 J/\bt}
\sin(\phi - D(z) + Qz\sin 2\phi_{-\Dl\phi})]]
\eeqrs
In this case, the simplified expressions can be obtained from the previous section by replacing
\beq
b_1 = 0 = b_3, \;\;\; b_2 = \sqrt{2}a_{\theta}, \;\;\;  C_1 = 1, \;\;\; C_3 = 0
\eeq
We have for the dipole moment
\beqr
\lan x(t) \ran & = & \frac{\sqrt{2\bt J_0}}{2\pi} \exp[-\frac{\bt_k\theta^2}{2 J_0}]\int dz \; \sqrt{z} \exp[- z] 
T_{\phi}(z) \\
T_{\phi}(z)  & = & \int d\phi \cos\phi \exp\left[ - b_2\sqrt{z} \sin( \phi - \Dl\phi - \tau \om_+ - q + Q z\sin 2\phi_{-\Dl \phi}) \right] 
 \nonumber \\
& = &  {\rm Re} \left\{  \int d\phi e^{i\phi} \exp\left[ - b_2\sqrt{z} \sin( \phi - \Dl\phi - \tau \om_+ - q + Q z 
\sin 2\phi_{-\Dl \phi} ) \right] \right\}
\eeqr
Using the expansion: $e^{-z\sin\theta} = \sum_{n}i^n I_n(z) e^{i n \theta}$
\beqrs
T_{\phi}(z) & = &  {\rm Re} \left\{  \sum_{k}i^{k}
 I_{k}(b_2\sqrt{z})   \int d\phi e^{i\phi} \exp[ i( k( \phi - \Dl\phi - \tau \om_+ - q + Q z \sin 2\phi_{-\Dl \phi}))]  \right\} \\
 & = &  {\rm Re} \left\{  \sum_{k} i^{k}  I_{k}(b_2\sqrt{z}) \exp[i( -k(\Dl\phi + \tau\om_++q) ] \right. \\
& & \left.  \int d\phi  \exp\left[ i \left( (1 + k ) \phi + k Q z\sin  2\phi_{-\Dl \phi}\right) \right]  \right\}
\eeqrs
To do the $\phi$ integration, I have to expand into a Bessel function
\[ \exp[i k Q z\sin  2\phi_{-\Dl \phi}] = \sum_{l=-\infty}^{\infty} J_l (k Q z)\exp[2 i l (\phi - \Dl\phi)] \]
Then
\beqrs
T_{\phi}(z) & = &  {\rm Re} \left\{  \sum_{k} \sum_l i^{k}  I_{k}(b_2\sqrt{z}) J_l(k Qz)\exp[i(- k(\Dl\phi + \tau\om_++q) 
- 2l\Dl\phi) ]   \int d\phi  \exp\left[ i ( [1 + k  + 2l ] \phi)  \right]  \right\} \\
& =  &  2\pi {\rm Re} \left\{  \sum_{l} i^{-(2l+1)}  I_{-(2l+1)}(b_2\sqrt{z}) J_{l}(-(2l+1) Qz)
\exp[i((2l+1) (\Dl\phi + \tau\om_++q) - 2l \Dl\phi) ] \right\}
\eeqrs
where we replaced $k \rarw -(2l+1)$
The phase factor simplifies to
\beqrs
 (2l+1)[\Dl\phi + \tau(\om + Q_2 z +q) - 2l\Dl\phi & = &  \om(t - \tau) + (2l+1)\tau(\om + Q_2 z + q) \\
& = &  \om t + 2l \om \tau + (2l+1)(Q_2 z + q) 
 \eeqrs
Since the sum extends over positive and negative values of $l$, I can replace $l = -n$ and write
\[
T_{\phi}(z)  =   {\rm Re} \left\{  \sum_{n=-\infty}^{\infty} i^{2n - 1)}  I_{2n - 1)}(b_2\sqrt{z}) J_{-n}(2n - 1) Qz)
\exp[i\left\{\om(t - 2n\tau) - (2n - 1)(Q_2 z + q) \right)]   \right\}
\]
This form predicts echoes at multiples of 2$\tau$. 
In this approximation, the amplitude of the echo at $2tau$ corresponds to the term with $n=1$. Writing
\[ \om(t - 2 n \tau) = \om_{\bt}(t - 2 n \tau) + \om'\eps_0(t - 2n \tau)z \equiv \Phi_n + \xi_n z \]
and using $J_{-n}(z) = (-1)^n J_n(z)$, we have
\beq
T_{\phi}(z)  =   {\rm Re} \left\{  \sum_{n=-\infty}^{\infty} i^{2n - 1)} (-1)^n  I_{2n - 1)}(b_2\sqrt{z}) J_{n}(2n - 1) Qz)
\exp[i( \Phi_n - (2n - 1) q )]  \exp[i z(\xi_n - (2n-1)Q_2 )]  \right\}
\eeq
 Hence
\beqr
\lan x(t=2\tau) \ran  &  =  &  - \sqrt{2\bt J_0} \exp[-\frac{\bt_k\theta^2}{2 J_0}]
{\rm Re} \left\{ i e^{i(\Phi_1 - q)}\int dz \;  \sqrt{z} \exp[- z\{1 - i (\xi_1 - Q_2) \} ] I_1(b_2\sqrt{z}) J_{1}( Qz) \right\} \nonumber
\\
& = & \sqrt{2\bt J_0} \exp[-\frac{\bt_k\theta^2}{2 J_0}]
{\rm Im} \left\{ e^{i(\Phi_1 - q)}\int dz \;  \sqrt{z} \exp[- z\{1 - i (\xi_1 - Q_2) \} ] I_1(b_2\sqrt{z}) J_{1}( Qz) \right\}
\label{eq: 2tau_NLDip_1}
\eeqr
where in the last form we used the fact that $-{\rm Re}[i f(z)] = {\rm Im}[f(z)]$ for a complex function $f(z)$.  In this form
it has nearly the same form obtained as obtained with the linearized dipole kick approximation, except that
$\bt_k\theta$ is replaced by $I_1(\sqrt{2}a_{\theta}\sqrt{z})$ and there is a factor of $\exp[-\bt_k\theta^2/(2 J_0)]$. 
The presence of $I_1(\sqrt{2}a_{\theta}\sqrt{z})$ shows that without a dipole kick 
$a_{\theta}=0$, the dipole moment also vanishes, as it does in the linearized dipole approximation. 
For ease of comparison,  the form obtained in the linear dipole approximation for $I_1$ in the linearized approximation was
\beq
\lan x(t=2\tau) \ran  = \bt_k\theta {\rm Im}\left\{ e^{i[\Phi_1- q]}\int dz \; \exp[-(1 - i(\xi_1 - Q_2)z)] J_1(Qz)  \right\}
\eeq
If in Eq.(\ref{eq: 2tau_NLDip_1}) we replace $I_1(b_2\sqrt{z})$ by its 1st order approximation
 $\half b_2\sqrt{z} = (\sqrt{2}/2) (\bt_k \theta/\sqrt{\bt J_0}) z$ and $\exp[-\bt_k\theta^2/(2 J_0)]$ by 1, then it
reduces to exactly the same equation above.

The amplitude of the echo at 4$\tau$ corresponds to the term with $n=2$. Hence
\beqr
\lan x(t=4\tau) \ran  & =  & \sqrt{2\bt J_0} \exp[-\frac{\bt_k\theta^2}{2 J_0}] 
{\rm Re} \left\{ i^{3} e^{i(\Phi_2 - 3q)}\int dz \;  \sqrt{z} \exp[-  z \{1 - i(\xi_2 - 3 Q_2) \}] I_3(b_2\sqrt{z}) J_{2}( Qz) \right\}
\nonumber \\
&   =  &  \sqrt{2\bt J_0} \exp[-\frac{\bt_k\theta^2}{2 J_0}] 
{\rm Im} \left\{ e^{i(\Phi_2 - 3q)}\int dz \;  \sqrt{z} \exp[-  z \{1 - i(\xi_2 - 3 Q_2) \}] I_3(b_2\sqrt{z}) J_{2}( Qz) \right\}
\eeqr
Note that since the lowest order term in $I_3(b_2\sqrt{z})$ is $(b_2 \sqrt{z})^3 = (2z)^{3/2} a_{\theta}^3$, there is no
echo at 4$\tau$ in the linearized dipole kick approximation. Note also the phase factor $\exp[i(\Phi_2 - 3q)]$ shows that
the $4\tau$ echo will be shifted slightly from the time $t=4\tau$. 

Neither of these integrations can be done analytically by Mathematica. These integrals also do not seem to appear in the
tables of integrals in \cite{Grad_Ryz}.  However they can be evaluated numerically. 
Putting the phase factors to zero, the ratio of the amplitudes can be found from
\beq
\fr{\lan x(t=4\tau) \ran^{amp}}{\lan x(t=2\tau) \ran^{amp}} =  
\fr{\int dz \;\sqrt{z} \exp[- z] I_3(b_2\sqrt{z}) J_{2}( Qz)}{\int dz \;\sqrt{z} \exp[- z] I_1(b_2\sqrt{z}) J_{1}( Qz)}
\eeq

\section{Multiple Echoes }

Multiple echoes at 4$\tau$, 6$\tau$ with calculated echo amplitudes

\bit
\item What additional information is present in these multiple echoes?
\item How do these multiple echoes help confirm / validate information from the echo at $2\tau$?
\item What can we say if no multiple echoes are observed, besides the obvious fact that diffusion may be too strong? In the RHIC studies, why were multiple echoes not observed in most cases? For example, consider the dipole kick amplitudes: does this kick 
have to be larger (or smaller) than some critical value for multiple echoes to be seen?
\eit

\clearpage

\section{Vlasov equation solution for the echo response}

      Consider the case where the external force is an impulse, i.e. dipole shock excitation at $t=0$.
      For simplicity, consider the 1D transverse case with a betatron tune spread. Thus,
      consider the betatron frequency to be $ \om_x(J_x)$.
      The transformation from the variables $(x, x')$ to $(J_x. \phi_x)$ are
 \beqr
      x & = & \sqrt{2\bt_x J_x}\cos \phi_x, \;\;\;  p = \al_x x + \bt_x x' = - \sqrt{2 \bt_x J_x} \sin\phi_x \\
      J_x & = &  \fr{1}{2\bt_x}[x^2 + p^2] \;\;\; \phi_x = {\rm Arctan}[\fr{-p}{x}]
 \eeqr

\subsection{Linearized Vlasov solution with a dipole kick and quadrupole kick.}

The Hamiltonian and equations of motion following a dipole kick are 
\beqr
 H(J_x,\phi_x) & = & H_0(J_x) - \eps x f_x(t) =  H_0(J_x) - \eps  \sqrt{2 \bt_x J_x} \cos \phi_x f_x(t) \\
  \dot{\phi_x} & = & \fr{\del H}{\del J_x} =\om_x(J_x) - \eps \sqrt{\bt_x} \fr{\cos\phi_x}{\sqrt{2  J_x}} f_x(t) \\
      \dot{J_x} & = & - \fr{\del H}{\del \phi_x} = - \eps \sqrt{2  \bt_x J_x} \sin \phi_x f_x(t)
      \eeqr
where $\eps$ is a suitably small parameter. 

The Vlasov equation is
 \beq
      \fr{\del \psi}{\del t} + \dot{\phi_x} \fr{\del \psi}{\del \phi_x} + \dot{J_x} \fr{\del \psi}{\del J_x}= 0
 \eeq
      Expanding
      \[ \psi(J_x, \phi_x)  = \psi_0(J_x)  + \eps \psi_1(J_x, \phi_x)
      \]
      Hence the Vlasov equation reduces to
      \[       \fr{\del \psi_1}{\del t} + ( \om_x(J_x) - \eps \sqrt{\bt_x}  \fr{\cos\phi_x}{\sqrt{2   J_x}} f_x(t) )\eps
      \fr{\del \psi_1}{\del \phi_x} - \eps \sqrt{2  \bt_x J_x} \sin \phi_x f_x(t)[\fr{\del \psi_0}{\del J_x} +
     \eps   \fr{\del \psi_1}{\del J_x} ] = 0
      \]
      Keeping terms to $O(\eps)$, we have
 \beq
  \fr{\del \psi_1}{\del t} +  \om_x(J_x) \fr{\del \psi_1}{\del \phi_x} -
  \sqrt{2  \bt_x J_x} \sin \phi_x f_x(t)\fr{\del \psi_0}{\del J_x}  = 0
\eeq
        The perturbation is of the form
        \[ f_x(t) = B \dl(t) \]
 Unlike a harmonic perturbation, we cannot make the ansatz of the density responding at the single
 frequency of the driving force.

\subsubsection{Solution by Fourier transform}

        One possibility is to take the Fourier transform w.r.t  time  and write
          \beq
          \psi_1(J_x, \phi_x, t) = e^{i \phi_x}   \int d\om e^{- i \om t} \tilde{\psi_1}(J_x, \phi_x, \om), \;\;\;
          \dl(t) = \fr{1}{2\pi}\int d\om e^{- i \om t}
          \eeq

          Question: Are there subtleties associated with taking the Fourier transform  here? Why is the Laplace
          transform used in some cases?

  Then we have from the Vlasov equation
          \beq
          i e^{i \phi_x}  \int d\om  e^{i \om t} \tilde{\psi_1}[ - \om +  \om_x(J_x)] = 
          B \sqrt{2 \bt_x  J_x} \fr{e^{i \phi_x} - e^{-i \phi_x}}{2i} \int \fr{d\om}{2\pi} \exp[- i \om t]  \fr{\del \psi_0}{\del J_x}
          \eeq
          Equating the integrands
          \[   [  \om -  \om_x(J_x)]\tilde{\psi_1} = \fr{1}{4\pi} B (1 - e^{-2i \phi_x} )  \fr{\del \psi_0}{\del J_x} \]
          Averaging this equation over the phase $\phi_x$ removes the $\phi_x$ dependent term on the RHS
          leaving
          \beq
          \tilde{\psi_1} = \fr{B}{4\pi}  \fr{1}{\om - \om_x(J_x)} \sqrt{2  \bt_x J_x} \fr{\del \psi_0}{\del J_x}
          \eeq
          and the complete solution for the time dependent perturbed density is
          \beq
          \psi_1(J_x, \phi_x, t) = \fr{B}{4\pi} e^{i \phi_x} \sqrt{2  \bt_x J_x} \fr{\del \psi_0}{\del J_x} \int d\om e^{-i \om t}
          \fr{1}{\om - \om_x(J_x)}
          \eeq

          From Mathematica,
          \beq
          \int_{-\infty}^{\infty} d\om \fr{e^{ - i \om t}}{\om - \om_x} = - i \pi  e^{-i \om_x t} {\rm Sign}(t)
          \eeq
          In our case $t > 0$ and we have
          \beq
          \psi_1(J_x, \phi_x, t) = -i \fr{1}{4} B e^{i (\phi_x - \om_x(J_x)t)} \sqrt{2  \bt_x J_x} \fr{\del \psi_0}{\del J_x}
          \eeq
          Comment: The Sign($t$) function could be a source of  a problem suggesting a lack of causality.
          Prior to the kick at $t=0$, there is no perturbation and so we should have $\psi_1(J_x, \phi_x, t < 0) = 0$.
          The solution obtained here does not obey that. This could be corrected by multiplying the above
          expression by the Heavyside theta function $\Theta(t)$ to take this into account. Thus,
          \beq
          \psi_1(J_x, \phi_x, t) = -i \fr{B}{4}  e^{i (\phi_x - \om_x(J_x)t)} \sqrt{2  \bt_x J_x} \fr{\del \psi_0}{\del J_x} \Theta(t)
          \eeq
          
         The inverse Laplace transform  of an arbitrary function $f(s)$ is
          \[ {\cal L}^{-1}[f(s)] = \fr{1}{2\pi i} \int_{\gm -i \infty}^{\gm + i \infty} f(s) e^{st} ds \]
          where $\gm $ is an arbitrary positive constant so that all singularities of $f(s)$ lie to the right of the
          contour. If $f(s)=1$, we can choose $\gm=0$ and we have the
the inverse Laplace transform representation of the delta function as
          \[ \dl(t) = \fr{1}{2\pi i} \int_{-i \infty}^{i \infty} e^{st} ds \]
           This becomes a Fourier transform on replacing $s i \om$. 

The dipole moment is after averaging over $(J_x, \phi_x)$
\beqr
\lan x \ran(t) & = & \int dJ_x \int \fr{d\phi_x}{2\pi}  (\psi_0 + \eps \psi_1) \sqrt{2  \bt_x J_x} \cos \phi_x \\
& = & -i \fr{}{4} (\eps  \bt_x B) \int dJ_x e^{-i \om_x(J_x)t} 2 J_x \fr{\del \psi_0}{\del J_x}
\int \fr{d\phi_x}{2\pi} e^{i \phi_x} \cos\phi_x \\
 & = & -i \fr{1}{4} (\eps  \bt_x  B) \int dJ_x e^{-i \om_x(J_x)t}  J_x \fr{\del \psi_0}{\del J_x}
\eeqr

\subsubsection{Solution by Laplace transform}

Usually problems with delta function impulses are solved by the Laplace transform

First write
 \beq
 \psi_1(J_x, \phi_x, t) = e^{i \phi_x} g(J_x, t)
 \eeq
 Substituting into the Vlasov equation, we have
 \beq
 e^{i\phi_x}[\fr{\del g}{\del t} + i \om_x(J_x) g(J_x, t)]  = \sqrt{2 \bt_x  J_x}\fr{e^{i\phi_x}- e^{-i\phi_x}}{2i}
 B\dl(t)\fr{\del \psi_0}{\del J_x}
 \eeq
 Averaging over $\phi_x$, we have
 \beq
 i \fr{\del g}{\del t} - \om_x(J_x)  g(J_x, t)] = \half \sqrt{2  \bt_x J_x} B\dl(t)\fr{\del \psi_0}{\del J_x}    \label{eq: g_J_t}
\eeq
The Laplace transform of a function $f(t)$ is defined as
\[ F(s) \equiv {\cal L}(f) = \int_0^{\infty} f(t) e^{-st} dt , \;\;\;  {\cal L}(\dl(t)) = 1  \]
while the Laplace transforms of derivatives are
\[ {\cal L}(\dot{f}) = s {\cal L}(f) - f(0), \;\;\; {\cal L}(\ddot{f}) = s^2 {\cal L}(f) - s f(0) - \dot{f}(0) \]
These can be shown by integrating by parts, e.g.
\beqrs
    {\cal L}(\dot{f}) &  =  & \int_0^{\infty} \dot{f(t)} e^{-st} dt = \int_0^{\infty} \fr{d}{dt}(f(t) e^{-st}) -
\int_0^{\infty} f(t) [-s e^{-st}] dt \\
 & = &  s {\cal L}(f) - f(0)
\eeqrs

Define the Laplace transform of $g(J_x, t)$ w.r.t time as
\[ G(J_x, s) = \int g(J_x, t) e^{- st} dt \]
Now we take the Laplace transform of Eq. (\ref{eq: g_J_t})  and we use $g(J_x,t=0) = 0$  to obtain
\beqr
G(J_x, s) (i s - \om_x(J_x) ) &  = &  \half \sqrt{2  \bt_x J_x} B \fr{\del \psi_0}{\del J_x} \\
\Rarw g(J_x, t) = -i {\cal L}^{-1}[\fr{1}{s + i\om_x(J_x)}] \half \sqrt{2  \bt_x J_x} B \fr{\del \psi_0}{\del J_x} \\
\eeqr
Using the inverse Laplace transform 
\[   {\cal L}^{-1}[\fr{1}{s + a}] = e^{-a t} \Theta(t) \]
where $\Theta(t)$ is the Heavyside step function.

Hence, the perturbed density distribution is
\beq
\psi_1(J_x, t) = -i \fr{ B}{2} \sqrt{2  \bt_x J_x} e^{i[\phi_x - \om_x(J_x) t]} \fr{\del \psi_0}{\del J_x} \Theta(t) 
\eeq
which is the same as the solution from the Fourier transform, except for the factor of 2 instead of 4 in the
denominator.

Consequently, the dipole moment is
\beqr
\lan x \ran(t) & = & \int dJ_x \int \fr{d\phi_x}{2\pi}  (\psi_0 + \eps \psi_1) \sqrt{2  \bt_x J_x} \cos \phi_x \\
 & = & -i \fr{(\eps  \bt_x B)}{2}  \int dJ_x e^{-i \om_x(J_x)t}  J_x \fr{\del \psi_0}{\del J_x} \Theta(t) 
\eeqr
Taking the real part, we have
\beq
\lan x \ran(t)  = \fr{(\eps  \bt_x B)}{2}   \int dJ_x \sin[ \om_x(J_x)t]  J_x \fr{\del \psi_0}{\del J_x}\Theta(t) 
\eeq
However with an exponential distribution in the action $\psi_0(J_x) = \fr{1}{2\pi \eps_x}\exp[- \fr{J_x}{\eps_x}]$
it is better to keep the complex form for now.

Compare this solution with that in Chao's lecture notes with the specific choices of $\om_x)J_x), \psi_0$.

\subsection{Linearized Vlasov solution following  dipole kick and quadrupole kicks}

With a dipole kick at $t=0$ and a quadrupole kick at $t= \tau$, the Hamiltonian and the equations of motion
are of the form
\beqr
H(J_x,\phi_x) & = & H_0(J_x) - \eps B x  \dl(t) - \eps K_Q x^2 \dl(t-\tau)  \nonumber \\
& =  & H_0(J_x) - \eps  B \dl(t) \sqrt{2 \bt_x J_x} \cos \phi_x  - \eps K_Q\dl(t-\tau)  (2 \bt_x J_x) \cos^2 \phi_x \\
\dot{\phi_x} & = & \fr{\del H}{\del J_x} = \om_x(J_x) - \eps B \dl(t) \sqrt{\bt_x} \fr{\cos\phi_x}{\sqrt{2  J_x}} -
\eps 2 K_Q  \bt_x \dl(t-\tau)   \cos^2 \phi_x  \\
  \dot{J_x} & = & -\fr{\del H}{\del \phi_x} = - \eps B \dl(t) \sqrt{2  \bt_x J_x} \sin \phi_x - \eps K_Q\dl(t-\tau)  (2 \bt_x J_x) \sin 2 \phi_x
  \eeqr
  Here we assumed the same order of smallness $\eps$ for the dipole and quadrupole kicks.
  The parameter $K_Q$ is related to the quad strength parameter $q$. 

The Vlasov equation after keeping  terms to $O(\eps)$ is
 \beq
 \fr{\del \psi_1}{\del t} +  \om_x(J_x) \fr{\del \psi_1}{\del \phi_x} -
[B \dl(t) \sqrt{2  \bt_x J_x} \sin \phi_x  + K_Q\dl(t-\tau)  (2 \bt_x J_x) \sin 2 \phi_x] \fr{\del \psi_0}{\del J_x}  = 0
\eeq
The dipole kick introduces changes to the phase $\phi_x$ while the quadrupole kick introduces changes
to twice the phase $\phi_x$.

Since the distribution function is periodic in the phase $\phi_x$, it can be expanded as a Fourier series in $\phi_x$.
One possibility is to truncate the expansion at two terms and make an ansatz of the form
\[
\psi_1(J_x, \phi_x, t) = e^{i \phi_x}g_1(J_x, t) + e^{2i \phi_x}g_2(J_x, t)
\]
However this does not generate a real distribution, so the complete expansion is of the form
\beq
\psi_1(J_x, \phi_x, t) = e^{i \phi_x}g_1(J_x, t) + e^{-i \phi_x}g_{-1}(J_x, t) + e^{2i \phi_x}g_2(J_x, t) +
e^{-2i \phi_x}g_{-2}(J_x, t)
\eeq
with the constraints $g_{-1} = g_1^*$,  $g_{-2} = g_2^*$ so that $\psi_1$ is real. 

Constraint on the perturbation is
\[ \int dJ_x d\phi_x \psi_1(J_x, \phi_x, t) = 0 \]
This does not introduce constraints into the functions $g_1, g_2$ since the integrals over the phase vanish for
both terms.

Substituting back into the Vlasov equation,
\beqr
& \mbox{} &  e^{i \phi_x}\fr{\del g_1}{\del t} + e^{2i \phi_x}\fr{\del g_2}{\del t} +
e^{-i \phi_x}\fr{\del g_1^*}{\del t} + e^{-2i \phi_x}\fr{\del g_2^*}{\del t}  + 
i \om_x(J_x)[ e^{i \phi_x} g_1 -  e^{-i \phi_x} g_1^* + 2 e^{2i \phi_x}g_2 - 2 e^{- 2i \phi_x}g_2^*]  \nonumber \\
& = & [B \dl(t) \sqrt{2  \bt_x J_x} \fr{e^{i \phi_x} - e^{-i \phi_x}}{2i}  + K_Q\dl(t-\tau)  (2 \bt_x J_x)
  \fr{e^{2i \phi_x} - e^{-2i \phi_x}}{2i}] \fr{\del \psi_0}{\del J_x}   \label{eq: Vlasov_g1_g2}
\eeqr

We can equate coefficients of $e^{i k \phi_x}, k = \pm 1, \pm 2$ on both sides of this equation or alternatively do
the following. 

Multiplying both sides by $e^{-i \phi_x}$ and averaging over $\phi_x$ leads to
  \beq
  \fr{\del g_1}{\del t} + i \om_x(J_x) g_1  = - \fr{i}{2}B \dl(t) \sqrt{2  \bt_x J_x}  \fr{\del \psi_0}{\del J_x}
  \eeq
  which is the same as the equation for $g$ in the last sub-section, leading to the same solution
  \beq
  g_1(J_x, t) = - \fr{i}{2}B \sqrt{2  \bt_x J_x}e^{-i \om_x(J_x)t}  \fr{\del \psi_0}{\del J_x}
  \eeq
  Now multiplying both sides of Eq.(\ref{eq: Vlasov_g1_g2}) by $e^{-2i \phi_x}$ and averaging over $\phi_x$ leads to
  \beq
  \fr{\del g_2}{\del t} + 2 i \om_x(J_x) g_2  = - \fr{i}{2}K_Q\dl(t-\tau)  (2 \bt_x J_x)\fr{\del \psi_0}{\del J_x}
  \eeq
  Let $G_2(s)$ be the Laplace transform of $g_2$ and we use ${\cal L}[\dl(t - \tau)] = e^{- s \tau}$ to obtain
  \beq
  G_2(J_x, s) = - \fr{i}{2}K_Q \fr{e^{-s \tau}}{s + 2i \om_x(J_x)}  (2 \bt_x J_x)\fr{\del \psi_0}{\del J_x}
    \eeq
    Using
    \[ {\cal L}^{-1}[ \fr{e^{-s \tau}}{s + 2i \om_x(J_x)}] = \exp[-2i\om_x(J_x)(t - \tau)] \Theta(t - \tau) \]
    we have
    \beq
    g_2(J_x, t) = - \fr{i}{2}K_Q e^{[-2i\om_x(J_x)(t - \tau)]} (2 \bt_x J_x)\fr{\del \psi_0}{\del J_x}  \Theta(t - \tau) 
    \eeq
    With the ansatz we made for the perturbed density, we have
    \beqr
    \psi_1(J_x,\phi_x,t) & = & \fr{ 1}{2} B\sqrt{2  \bt_x J_x} [ -i e^{i[\phi_x - \om_x(J_x) t]}
      + i e^{- i[\phi_x - \om_x(J_x) t]} ] \fr{\del \psi_0}{\del J_x} \Theta(t)
    \nonumber \\
    & & -\fr{1}{2}K_Q  (2 \bt_x J_x)[-i e^{2i[\phi_x - \om_x(J_x)(t - \tau)]} + i e^{-2i[\phi_x - \om_x(J_x)(t - \tau)]}] \fr{\del \psi_0}{\del J_x}  \Theta(t - \tau)     \nonumber \\
    &  = &  B\sqrt{2  \bt_x J_x} \sin [\phi_x - \om_x(J_x) t] + K_Q  (2 \bt_x J_x) \sin 2[\phi_x - \om_x(J_x)(t - \tau)]
    \eeqr

    Problem with this solution: The second term from the quadrupole kick is even in $\phi_x$ and will give
    a zero dipole moment. From the Eulerian method of solution for the density distribution, we know that the
    term giving a non-zero contribution to the dipole moment was linear in both the dipole and quadrupole
    parameters. In other words, the non-zero contribution came from a term of $O(\eps^2)$.

    So, I have to solve the Vlasov equation to second order. This should not be a surprise, the theory of
    the plasma wave echo had shown the echo to be a nonlinear process.

    \subsection{Nonlinear Vlasov solution of the echo response}

    Starting with the perturbative solution but setting the small parameter $\eps = 1$ [we are not doing a
      perturbative order by order solution]
    \[ \psi(J_x,\phi_x,t) = \psi_0(J_x) +  \psi_1(J_x, \phi_x, t) \]
and substituting into the complete Vlasov equation
\beqrs
 & \mbox{} &      \fr{\del \psi_1}{\del t} +     \fr{\del \psi_1}{\del \phi_x}
  [  \om_x(J_x) -  B \dl(t) \sqrt{\bt_x}  \fr{\cos\phi_x}{\sqrt{2   J_x}} -
          2   K_Q  \bt_x \dl(t-\tau)  \cos^2 \phi_x ]  \\
&  &  +  ( \fr{\del \psi_0}{\del J_x} +     \fr{\del \psi_1}{\del J_x})
[ -  B \dl(t) \sqrt{2  \bt_x J_x} \sin \phi_x -  K_Q\dl(t-\tau) (2 \bt_x J_x) \sin 2 \phi_x ]  = 0
\eeqrs
Introduce the notation
\beq
d_x = \sqrt{\fr{\bt_x}{2 J_x}}, \;\;\; A_x = \sqrt{2 \bt_x J_x}
\eeq
We can rewrite the Vlasov equation as
\beqrs
 & \mbox{} &      \fr{\del \psi_1}{\del t} + 
  [  \om_x(J_x) -  B \dl(t) d_x \cos\phi_x -
          2   K_Q  \bt_x \dl(t-\tau)  \cos^2 \phi_x ]   \fr{\del \psi_1}{\del \phi_x} \\
&  &  - [ B \dl(t) A_x \sin \phi_x +  K_Q\dl(t-\tau) A_x^2 \sin 2 \phi_x ]    \fr{\del \psi_1}{\del J_x} 
\\
 & = & [ B \dl(t) A_x \sin \phi_x +  K_Q\dl(t-\tau) A_x^2 \sin 2 \phi_x] \fr{\del \psi_0}{\del J_x}
\eeqrs
Since the distribution function is periodic in $\phi_x$, we can expand it in a Fourier series in $\phi_x$ as
\beq
\psi_1(J_x, \phi_x, t) = \sum_{k=-\infty}^{\infty} g_k(J_x, t) e^{i k \phi_x}
\eeq
Since we want a real distribution function   it follows that
\beqrs
& \mbox{} & \sum_k  g_k(J_x, t) e^{i k \phi_x}= \sum_k  g_k^*(J_x, t) e^{- i k \phi_x} = \sum_k  g_{-k}^*(J_x, t) e^{ i k \phi_x} \\
& \Rarw & g_k = g_{-k}^*, \;\;\; {\rm or} \;\;\; g_{-k} = g_k^*
\eeqrs
I expect that the only the components $g_1, g_2, g_3$ will be necessary. It follows that
\beqrs
 & \mbox{} &    \sum_k  \fr{\del g_k}{\del t}  e^{i k \phi_x}  + 
 \left[  \om_x(J_x) -  \half B \dl(t) d_x (e^{i \phi_x} + e^{-i \phi_x})  -
           K_Q  \bt_x \dl(t-\tau)[1 + \half(e^{2i \phi_x} + e^{-2i \phi_x})] \right] \sum_{k} i k  g_k(J_x, t) e^{i k \phi_x} \\
  &  &  - \left[ \fr{B \dl(t) }{2i} A_x [e^{i \phi_x} - e^{-i \phi_x}]  +  \fr{K_Q\dl(t-\tau)}{2i} A_x^2
    (e^{2i \phi_x} - e^{-2i \phi_x})  \right]  \sum_k  \fr{\del g_k}{\del J_x} e^{i k \phi_x} \\
  & = & [ \fr{B \dl(t) }{2i} A_x  [ e^{i \phi_x} - e^{-i \phi_x}] +
    \fr{K_Q\dl(t-\tau)}{2i} A_x^2  (e^{2i \phi_x} - e^{-2i \phi_x})] \fr{\del \psi_0}{\del J_x}
  \eeqrs
  Writing out the LHS to isolate the $k$th  harmonic
  \beqrs
& \mbox{} &  \sum_k \left\{      \fr{\del g_k}{\del t}  e^{i k \phi_x}  + 
  i k g_k \left[ \om_x(J_x) e^{i k \phi_x} -  \half B \dl(t) d_x [e^{i (k+1) \phi_x} + e^{i (k - 1) \phi_x}]
    \right.   \right. \\
& & \left. \left.  -   K_Q  \bt_x \dl(t-\tau)(e^{i k \phi_x} + \half(e^{i(k+2) \phi_x} + e^{i(k- 2) \phi_x})   ]  \right]  \right.  \\
 &  & \left.  -  \fr{\del g_k}{\del J_x} [ \fr{B \dl(t) }{2i} A_x  [ e^{i (k+1)\phi_x} - e^{i(k- 1) \phi_x}]  +
   \fr{K_Q\dl(t-\tau)}{2i} A_x^2 (e^{i(k+2) \phi_x} - e^{i(k-2 ) \phi_x})  ]  \right\}   \\
 & = & \sum_k  e^{i k \phi_x} \left\{\fr{\del g_k}{\del t}  + i k\om_x(J_x) g_k
 -  \fr{i}{2} B \dl(t) d_x  [ (k-1)g_{k-1} + (k+1)g_{k+1}] \right. \\
& &  - i K_Q  \bt_x \dl(t-\tau)[ k g_k + \half( (k-2)g_{k-2} + (k+2)g_{k+2}) ]   \\
 & & \left. - \fr{B \dl(t) }{2i} A_x [ \fr{\del g_{k-1}}{\del J_x}  -  \fr{\del g_{k+1}}{\del J_x} ]
 - \fr{K_Q\dl(t-\tau)}{2i} A_x^2  [\fr{\del g_{k-2}}{\del J_x}  -  \fr{\del g_{k+2 }}{\del J_x} ]  \right\}
 \eeqrs
 Hence the Vlasov equation in terms of the harmonics is
 \beqr
 & \mbox{} & \sum_k  e^{i k \phi_x} \left\{\fr{\del g_k}{\del t}  + i k\om_x(J_x) g_k
 -  \fr{i}{2} B \dl(t) d_x  [ (k-1)g_{k-1} + (k+1)g_{k+1}] \right.  \nonumber \\
& &  - i K_Q  \bt_x \dl(t-\tau)[ k g_k + \half( (k-2)g_{k-2} + (k+2)g_{k+2}) ]  \nonumber  \\
 & & \left. + \fr{B \dl(t) }{2i} A_x \fr{\del }{\del J_x}[ g_{k+1} - g_{k-1} ]
 + \fr{K_Q\dl(t-\tau)}{2i} A_x^2  \fr{\del}{\del J_x}[  g_{k+2} -   g_{k- 2 } ]  \right\} \nonumber \\
& = & [ \fr{B \dl(t) }{2i} A_x  [ e^{i \phi_x} - e^{-i \phi_x}] +
    \fr{K_Q\dl(t-\tau)}{2i} A_x^2  (e^{2i \phi_x} - e^{-2i \phi_x})] \fr{\del \psi_0}{\del J_x}  \label{eq: Vlasov_all_k}
 \eeqr
 This equation shows that the time evolution of the $k$th mode is coupled to the neighboring 4 modes:
 $g_{k-2}, g_{k - 1}, g_{k+1}, g_{k+2}$.

 Equating the $k=0, \pm 1, \pm 2$ and the $|k| > 2$ harmonics on the two sides, we have
first for the $k=0$ harmonic
 \beqrs
 & \mbox{} & \fr{\del g_0}{\del t}   - \fr{i}{2} B \dl(t)  d_x [ -g_{-1} + g_{1}]
 - i K_Q  \bt_x \dl(t-\tau)[ \half (-2g_{-2} + 2g_{2}) ]   \\
&  &  +  \fr{B \dl(t) }{2i} A_x \fr{\del }{\del J_x} [ g_{1}  -  g_{-1} ]
 + \fr{K_Q\dl(t-\tau)}{2i} A_x^2 \fr{\del }{\del J_x} [ g_{2}-  g_{-2 }]   = 0
 \eeqrs
 Using and introducing notation for the real and imaginary parts
   \[ g_{-k} = g_k^*, \;\;\;  g_k + g_k^* = 2 {\rm Re}[g_k] \equiv g_{k, R},  \;\;\;  g_k - g_k^* = 2 i {\rm Im}[g_k]  \equiv g_{k, I} \]
 the above implies 
 \beqrs
& \mbox{} & \fr{\del g_0}{\del t}  +   B \dl(t)  d_x  {\rm Im}[ g_{1}]
 + 2  K_Q  \bt_x \dl(t-\tau){\rm Im}[ g_{2} ]  \\
&  &  +  B \dl(t) A_x \fr{\del}{\del J_x}  {\rm Im}[ g_{1}]
+ K_Q\dl(t-\tau) A_x^2  \fr{\del}{\del J_x}{\rm Im}[ g_{2}]   = 0
\eeqrs

WLOG we can put $g_0=0$  since the zeroth harmonic time independent function is already in $\psi_0$,
Then the above equation can be written as one linking the imaginary parts
\beq
B \dl(t) [ A_x  \fr{\del}{\del J_x}g_{1, I} +  d_x g_{1, I}]
+ K_Q\dl(t-\tau)[ A_x^2  \fr{\del}{\del J_x}g_{2, I} + 2 \bt_x  g_{2, I}] = 0
\label{eq: k=0}
\eeq
Integrating this equation over time leads to
\[ B  [ A_x  \fr{\del}{\del J_x}g_{1, I}(J_x, 0) +  d_x g_{1, I}(J_x, 0)]
+ K_Q[ A_x^2  \fr{\del}{\del J_x}g_{2, I }(J_x, \tau) + 2 \bt_x  g_{2, I}(J_x, \tau)] = 0
\]

Using the initial condition $g_k(J_x, 0) = 0$ for all $k$, this reduces to
\beq
J_x \fr{\del}{\del J_x}g_{2, I }(J_x, \tau) +   g_{2, I}(J_x, \tau) = 0 \label{eq: g2I_1}
\eeq
This has the solution
\beq
g_{2, I}(J_x, \tau) = c_{2, I, \tau}\fr{1}{J_x}
\eeq
where $ c_{2, I, \tau}$ is a real constant. 

 Equation for the $k=1$ harmonic and using $g_0 = 0, g_{-k} = g_k^*$: 
 \beqrs
& \mbox{} & \fr{\del g_1}{\del t}  + i \om_x(J_x)g_1  -  i B \dl(t) d_x  g_2
 - i K_Q  \bt_x \dl(t-\tau)[  g_1 + \half( -g_1^* + 3g_{3}) ]  \\
 & &  + \fr{B \dl(t) }{2i} A_x \fr{\del }{\del J_x} [  g_{2} ]
 + \fr{K_Q\dl(t-\tau)}{2i} A_x^2  \fr{\del }{\del J_x} [g_{3 } -  g_{1}^*     ]  \\
 & = &  \fr{B \dl(t) }{2i} A_x  \fr{\del\psi_0}{\del J_x}
 \eeqrs

 Equation for the $k=-1$ harmonic and using $g_0 = 0, g_{-k} = g_k^*$:
 \beqrs
 & \mbox{}&  \fr{\del g_{1}^*}{\del t}  - i \om_x(J_x) g_{1}^* -  i B \dl(t) d_x  [ -g_{2}^* ] 
  - i K_Q  \bt_x \dl(t-\tau)[ - g_{1}^* + \half( -3 g_{3}^* + g_{1}) ]   \\
 & &  + \fr{B \dl(t) }{2i} A_x \fr{\del }{\del J_x} [ - g_{2}^*  ]
 + \fr{K_Q\dl(t-\tau)}{2i} A_x^2  \fr{\del }{\del J_x} [g_{1} -  g_{3}^*   ]   \\
& = & - \fr{B \dl(t) }{2i} A_x \fr{\del\psi_0}{\del J_x}
 \eeqrs

 Adding the equations for $k=1$ and $k= -1$, we obtain 
 \beqr
 & \mbox{} &  2\fr{\del g_{1,R}}{\del t}  - 2 \om_x(J_x) g_{1,I} + 2  B\dl(t) d_x g_{2, I}  + 3  K_Q  \bt_x \dl(t-\tau)[g_{1,I} +  g_{3,I} ]
 \nonumber \\
& &  + B \dl(t)A_x \fr{\del }{\del J_x} [g_{2,I}]
 + K_Q\dl(t-\tau) A_x^2  \fr{\del }{\del J_x} [ g_{1, I} +  g_{3, I } ]
 = 0
 \eeqr

 Note that the time derivative of the real part $g_{1,R}$ is determined by the imaginary parts of $g_1, g_2, g_3$
 and it does not depend on the unperturbed density $\psi_0$.

Subtracting the equations for $k=- 1$ from that for $k= 1$, we obtain after multiplying by $i$, 
 \beqr
 & \mbox{} &  - 2\fr{\del g_{1,I}}{\del t} - 2 \om_x(J_x) g_{1,R} + 2  B\dl(t) d_x g_{2, R}  +
   K_Q  \bt_x \dl(t-\tau)[ g_{1,R} + 3 g_{3, R} ]   \nonumber \\
& + &  B \dl(t)A_x \fr{\del }{\del J_x} [g_{2, R}] -  K_Q\dl(t-\tau) A_x^2  \fr{\del }{\del J_x} [ g_{1, R} -  g_{3, R } ]
 = B \dl(t)  A_x \fr{\del\psi_0}{\del J_x}
 \eeqr

 This shows the time derivative of the imaginary part $g_{i,I}$ depends on the real parts of $g_1, g_2, g_3$
 and also on $\psi_0$. 

 Assumption: We will assume that modes higher than 3 are relatively small so we put
 \[ g_{|k|} = 0  , \;\;\;  |k| > 3  \]
 With this assumption, we have the equation for the $k=2$ mode
\beqrs
 & \mbox{} &  \fr{\del g_2}{\del t}  + 2 i \om_x(J_x) g_2
 -  \fr{i}{2} B \dl(t) d_x  [ g_{1} + 3 g_{3}]  - i K_Q  \bt_x \dl(t-\tau)[ 2 g_2 ]   \\
 & & + \fr{B \dl(t) }{2i} A_x \fr{\del }{\del J_x}[ g_{3} -  g_{1} ]  = \fr{K_Q\dl(t-\tau)}{2i} A_x^2   \fr{\del \psi_0}{\del J_x}  
 \eeqrs

 and with $k=-2$,
\beqrs
 & \mbox{} & \fr{\del g_2^*}{\del t}  - 2 i \om_x(J_x) g_2^*
 -  \fr{i}{2} B \dl(t) d_x  [ -3 g_{3}^* - g_{1}^*]   - i K_Q  \bt_x \dl(t-\tau)[ -2 g_2^*  ]   \\
 & &  - \fr{B \dl(t) }{2i} A_x \fr{\del }{\del J_x}[ g_{3}^* -  g_{1}^* ] =-  \fr{K_Q\dl(t-\tau)}{2i} A_x^2   \fr{\del \psi_0}{\del J_x}  
 \eeqrs

Adding we have,
 \beqr
2\fr{\del g_{2,R}}{\del t}  - 4 \om_x(J_x) g_{2,I} +   B\dl(t) d_x[ g_{1, I} + 3 g_{3, I}] +
 4  K_Q  \bt_x \dl(t-\tau) g_{2,I}   + B \dl(t)A_x \fr{\del }{\del J_x} [ g_{3,I} - g_{1, I}]   = 0  \nonumber \\
 \eeqr

and subtracting we have, after multiplying by $i$
  \beqr
& \mbox{} &  -2\fr{\del g_{2,I}}{\del t}  - 4 \om_x(J_x) g_{2,R} +   B\dl(t) d_x[ g_{1, R} + 3 g_{3, R}] +
  4  K_Q  \bt_x \dl(t-\tau) g_{2,R}   + B \dl(t)A_x \fr{\del }{\del J_x} [ g_{3,R} - g_{1, R}]  \nonumber \\
&  = & K_Q\dl(t-\tau) A_x^2     \fr{\del \psi_0}{\del J_x}  
 \eeqr

 Next, setting $k=3$, we have
 \beqrs
 & \mbox{} & \fr{\del g_3}{\del t}  + i 3\om_x(J_x) g_3 -  \fr{i}{2} B \dl(t) d_x  [ 2g_{2} ]  
 - i K_Q  \bt_x \dl(t-\tau)[ 3 g_3 + \half  g_1 ]   \\
 & - &  \fr{B \dl(t) }{2i} A_x \fr{\del }{\del J_x}[ g_{2} ] -
 \fr{K_Q\dl(t-\tau)}{2i} A_x^2  \fr{\del}{\del J_x}[  g_{1}  ]   =  0
 \eeqrs

 and $k = -3$, we have
\beqrs
  & \mbox{} & \fr{\del g_3^*}{\del t}  - 3 i \om_x(J_x) g_3^* -  \fr{i}{2} B \dl(t) d_x  [ -2 g_{2}^*]  
- i K_Q  \bt_x \dl(t-\tau)[ -3 g_3^* + \half( -g_{1}^*)]   \\
 & &  + \fr{B \dl(t) }{2i} A_x \fr{\del }{\del J_x}[ g_{2}^* ]
 + \fr{K_Q\dl(t-\tau)}{2i} A_x^2  \fr{\del}{\del J_x}[ g_{1 }^* ]   = 0
 \eeqrs

 Adding the equations, we have
 \beqr
& \mbox{} & 2\fr{\del g_{3,R}}{\del t}  - 6 \om_x(J_x) g_{3,I} +   2 B\dl(t) d_x g_{2, I} +
 K_Q  \bt_x \dl(t-\tau) [ 6 g_{3,I} + g_{1, I}]   \nonumber \\
 & &  - B \dl(t)A_x \fr{\del }{\del J_x}  g_{2,I} - K_Q \dl(t-\tau )A_x^2 \fr{\del }{\del J_x}  g_{1,I}= 0 
 \eeqr

 while subtracting gives
 \beqr
& \mbox{} & - 2\fr{\del g_{3,I}}{\del t}  - 6 \om_x(J_x) g_{3,R} +   2 B\dl(t) d_x g_{2, R} +
K_Q  \bt_x \dl(t-\tau) [ 6 g_{3,R} + g_{1, R}]    \nonumber \\
&  &  - B \dl(t)A_x \fr{\del }{\del J_x}  g_{2,R}
- K_Q \dl(t-\tau )A_x^2 \fr{\del }{\del J_x}  g_{1,R} = 0  
 \eeqr

 The $k = \pm 4, \pm 5$ equations contain $g_2, g_3$ so may have useful relations on them.
 Caution: Is it consistent or self-consistent to use the equations for the higher modes $k \ge 4$ which are ignored in the rest of
 the analysis?
 
 With $k= 4$, we have after assuming $g_k = 0, |k| \ge 4$,
\beqrs
 & \mbox{} &  -  \fr{i}{2} B \dl(t) d_x  ( 3 g_{3})   - i K_Q  \bt_x \dl(t-\tau)[ g_{2}  ]   \\
 & &  - \fr{B \dl(t) }{2i} A_x \fr{\del }{\del J_x}[ g_3  ] - \fr{K_Q\dl(t-\tau)}{2i} A_x^2  \fr{\del}{\del J_x}[  g_{2} ]  
 =  0 
 \eeqrs

 while with $k=-4$, we have
 \beqrs
 & \mbox{} & 
 -  \fr{i}{2} B \dl(t) d_x  [ (-3)g_{3}^*]  - i K_Q  \bt_x \dl(t-\tau)[ - g_{2}^* ]   \\
& &  - \fr{B \dl(t) }{2i} A_x \fr{\del }{\del J_x}[  -  g_{3}^* ] 
 - \fr{K_Q\dl(t-\tau)}{2i} A_x^2  \fr{\del}{\del J_x}[  -   g_{2 }^* ]  = 0
 \eeqrs

 Adding the two equations, we have
\beqr
 3 B \dl(t) d_x  g_{3, I}   + 2 K_Q  \bt_x \dl(t-\tau) g_{2, I}    
 - B \dl(t)  A_x \fr{\del }{\del J_x}[ g_{3, I}  ] - K_Q\dl(t-\tau) A_x^2  \fr{\del}{\del J_x}  g_{2, I}  =  0  \nonumber \\
 \label{eq: k=4_add}
\eeqr

 Subtracting gives us
 \beqr
 & \mbox{} &  3 B \dl(t) d_x  g_{3, R}   + 2 K_Q  \bt_x \dl(t-\tau) g_{2, R}    
 - B \dl(t)  A_x \fr{\del }{\del J_x}[ g_{3, R} ] - K_Q\dl(t-\tau) A_x^2  \fr{\del}{\del J_x}  g_{2, R}  =  0  \nonumber \\
\label{eq: k=4_sub}
 \eeqr

Setting $k=5$ yields 
 \beqrs
  - i K_Q  \bt_x \dl(t-\tau)[ \half( 3 g_{3} ) ]   
 - \fr{K_Q\dl(t-\tau)}{2i} A_x^2  \fr{\del}{\del J_x}[  g_{3}  ]  = 0
 \eeqrs

 and $k=-5$ yields
  \beqrs
 - i K_Q  \bt_x \dl(t-\tau)[  \half( (-3)g_{3}^*) ]   
 - \fr{K_Q\dl(t-\tau)}{2i} A_x^2  \fr{\del}{\del J_x}[   -  g_{3}^* ]  =  0
 \eeqrs

 The equation for $g_3, g_3^*$ can be written as (using $A_x^2 = 2 \bt_x J_x$)
 \beqr
 \dl(t - \tau) \left\{ \fr{3}{2} g_3 -  J_x \fr{\del}{\del J_x}g_3 \right\} & = &  0 \\
 \dl(t - \tau) \left\{ \fr{3}{2} g_3^* -  J_x \fr{\del}{\del J_x}g_3^* \right\} & = &  0
 \eeqr

Integrating these over time yields the same equation for $g_3, g_3^*$
 \beqr
 \fr{3}{2} g_3(J_x, \tau) & -  &  J_x \fr{\del}{\del J_x}g_3(J_x, \tau) = 0  \label{eq: g3} \\
\Rarw g_3(J_x,\tau) &  = &  C_{3,\tau}J_x^{3/2}
 \eeqr
where $C_{3,\tau}$ is a constant, possibly complex. Eq. (\ref{eq: g3}) shows that $g_{3,R}, g_{3,I}$ obey the same equation. 

 Integrating Eq.(\ref{eq: k=4_add}) over time yields
 \beqrs
3 B  d_x  g_{3, I}(J_x,0)   + 2 K_Q  \bt_x  g_{2, I}(J_x,\tau)
 - B  A_x \fr{\del }{\del J_x}[ g_{3, I}(J_x, 0)  ] - K_Q A_x^2  \fr{\del}{\del J_x}  g_{2, I}(J_x,\tau)  =  0 
 \eeqrs
 With the initial condition that $\psi_1(t=0) = 0$, we also have $g_k(J_x, 0) = 0$. Hence the above equation
 reduces to, after using $A_x^2 = 2 \bt_x J_x$, 
 \beq
g_{2, I}(J_x,\tau) -  J_x \fr{\del}{\del J_x}  g_{2, I}(J_x,\tau)  = 0   \label{eq_g2I_2}
 \eeq
 Adding Equations (\ref{eq: g2I_1}) and (\ref{eq_g2I_2}), we have
 \beq
 g_{2, I}(J_x, \tau) = 0
 \eeq
 Since we now have $g_{2,I}(J_x, 0) = 0 = g_{2,I}(J_x, \tau)$, this suggests we can set
 \beq
 g_{2, I}(J_x, t) = 0
 \eeq
 
Integrating Eq.(\ref{eq: k=4_sub}) over time yields the   solution  for $g_{2, R}(J_x, \tau)$
 \[
 g_{2, R}(J_x,\tau) = c_{2, R, \tau}J_x
 \]
 where $c_{2, R, \tau}$ is a real constant.  Thus the equations at $k= \pm 4, \pm 5$ yield
 \beq
 g_{2}(J_x, \tau) = c_{2,R}J_x, \;\;\;  g_{3}(J_x, \tau) = C_{3,\tau} J_x^{3/2}
 \eeq
Notation: Lower case constants $c_k$ are real while upper case constants $C_k$ are in general complex. 

Setting $g_{2,I}=0$, the remaining equations from $k = \pm 1, \pm 2, \pm 3$ are 
 \beqr
 & \mbox{} &  2\fr{\del g_{1,R}}{\del t}  - 2 \om_x(J_x) g_{1,I} + 
 3  K_Q  \bt_x \dl(t-\tau)[g_{1,I} +  g_{3,I} ]  
 + K_Q A_x^2 \dl(t-\tau) \fr{\del }{\del J_x} [ g_{1, I} +  g_{3, I } ] = 0 \\
 & \mbox{} &  - 2\fr{\del g_{1,I}}{\del t} - 2 \om_x(J_x) g_{1,R} + 2  B d_x \dl(t)  g_{2, R}  +
   K_Q  \bt_x \dl(t-\tau)[ g_{1,R} + 3 g_{3, R} ]   \nonumber \\
& + &  B A_x  \dl(t) \fr{\del }{\del J_x} [g_{2, R}] -  K_QA_x^2 \dl(t-\tau)   \fr{\del }{\del J_x} [ g_{1, R} -  g_{3, R } ]
 = B   A_x \dl(t) \fr{\del\psi_0}{\del J_x} \\
 & \mbox{} & 2\fr{\del g_{2,R}}{\del t}   +   B  d_x \dl(t)[ g_{1, I} + 3 g_{3, I}] 
   - B A_x \dl(t) \fr{\del }{\del J_x} [ g_{1,I} - g_{3, I}]   = 0  \label{eq: g2R}\\
& \mbox{} &    - 4 \om_x(J_x) g_{2,R} +   B d_x \dl(t) [ g_{1, R} + 3 g_{3, R}] +
   4  K_Q  \bt_x \dl(t-\tau) g_{2,R}   - B A_x \dl(t) \fr{\del }{\del J_x} [ g_{1,R} - g_{3, R}]  \nonumber \\
   & = &  K_Q A_x^2  \dl(t-\tau)  \fr{\del \psi_0}{\del J_x}  \\
&  & 2\fr{\del g_{3,R}}{\del t}  - 6 \om_x(J_x) g_{3,I} +   K_Q  \bt_x \dl(t-\tau) [ 6 g_{3,I} + g_{1, I}]   
 - K_Q A_x^2 \dl(t-\tau )  \fr{\del }{\del J_x}  g_{1,I}= 0  \\
& \mbox{} & - 2\fr{\del g_{3,I}}{\del t}  - 6 \om_x(J_x) g_{3,R} +   2 B d_x \dl(t) g_{2, R} +
K_Q  \bt_x \dl(t-\tau) [ 6 g_{3,R} + g_{1, R}]    \nonumber \\
&  &  - B A_x \dl(t)  \fr{\del }{\del J_x}  g_{2,R} - K_Q A_x^2 \dl(t-\tau ) \fr{\del }{\del J_x}  g_{1,R} = 0  
 \eeqr

 We take the Laplace transform of these equations w.r.t time to solve them. We define
 \[ G_{k, R}(J_x, s) = {\cal L}[g_{k, R}(J_x, t)] , \;\;\; G_{k, I}(J_x, s) = {\cal L}[g_{k, I}(J_x, t)]  \]
 and using
 \[ \int dt e^{-s t}\dl(t - \tau) g(t)  = e^{- s \tau} g(\tau), \;\;\; {\cal L}\fr{\del g_k}{\del t} = s  G_k(s) , \;\;\;
 g_k(J_x, t=0) = 0  \]

 The Laplace transform of Eq.(\ref{eq: g2R}) yields
 \beq
 s G_{2, R}(J_x, s) = 0
 \eeq
 which yields identically $G_{2,R}=0$ which in turn implies $g_{2, R}(J_x, t) = 0$. Hence we have no contribution at the
 2nd harmonic, $g_2= 0$. With this the remaining equations for $g_1, g_3$ simplify to
 \beqr
& \mbox{} & 2\fr{\del g_{1,R}}{\del t}   - 2 \om_x(J_x) g_{1,I} + 
K_Q   \dl(t-\tau) \left[ ( 3 \bt_x g_{1,I} + A_x^2  \fr{\del }{\del J_x}g_{1, I}) +  ( 3 \bt_x g_{3,I} + A_x^2  \fr{\del }{\del J_x}g_{3, I})
  \right] = 0 \nonumber  \\
\mbox{} \\
&\mbox{} &  - 2\fr{\del g_{1,I}}{\del t}   - 2 \om_x(J_x) g_{1,R} + 
K_Q   \dl(t-\tau) \left[ ( \bt_x g_{1,R} - A_x^2  \fr{\del }{\del J_x}g_{1,R})  + ( 3 \bt_x  g_{3, R} + A_x^2  \fr{\del }{\del J_x} g_{3, R })
\right]  \nonumber  \\
& = &  B \dl(t)  A_x \fr{\del\psi_0}{\del J_x} \\
& \mbox{} &      B\dl(t)  \left\{ d_x[ g_{1, R} + 3 g_{3, R}] 
- A_x \fr{\del }{\del J_x} [ g_{1,R} - g_{3, R}] \right\}  =  K_Q A_x^2 \dl(t- \tau) \fr{\del\psi_0}{\del J_x}  \\
& \mbox{}  &  2\fr{\del g_{3,R}}{\del t}    - 6 \om_x(J_x) g_{3,I} +
K_Q  \dl(t-\tau) \left [ 6 \bt_x g_{3,I} +  \bt_x g_{1, I} - A_x^2 \fr{\del }{\del J_x}  g_{1,I} \right]  = 0    \\
& \mbox{} &  - 2\fr{\del g_{3,I}}{\del t}  - 6 \om_x(J_x) g_{3,R} +
K_Q  \bt_x \dl(t-\tau) \left[ 6 \bt_x  g_{3,R} + \bt_x  g_{1, R} - A_x^2 \fr{\del }{\del J_x}  g_{1,R} \right] = 0  
 \eeqr
 
Laplace transforms of  the first of the above two equations yield
 \beqrs
& \mbox{} &  2 s G_{1,R}  - 2 \om_x(J_x) G_{1,I} +
 K_Q e^{ - s \tau} [3 \bt_x g_{1,I}(J_x,\tau) +  A_x^2  \fr{\del }{\del J_x} g_{1, I}(J_x,\tau) ]  \nonumber \\
& & + K_Q e^{ - s \tau}  [ 3  \bt_x g_{3, I }(J_x,\tau) +  A_x^2 \fr{\del }{\del J_x}  g_{3, I }(J_x,\tau) ] = 0  \\
 & \mbox{} &  - 2 s G_{1,I} - 2 \om_x(J_x) G_{1,R} + 
   K_Q  e^{ - s \tau} [ \bt_x  g_{1,R}(J_x,\tau)  - A_x^2 \fr{\del }{\del J_x} g_{1, R}(J_x,\tau) ]    \nonumber \\
&  &  -  K_Q e^{ - s \tau}  [ 3 \bt_x  g_{3, R}(J_x,\tau)  + A_x^2 \fr{\del }{\del J_x} g_{3, R }(J_x,\tau) ]
 = B   A_x \fr{\del\psi_0}{\del J_x} 
\eeqrs
We have using Eq. (\ref{eq: g3}), that
\[  A_x^2 \fr{\del }{\del J_x}  \left\{ \begin{array}{c} g_{3, I }(J_x,\tau) \\ g_{3, I }(J_x,\tau) \end{array} \right.
 = 3  \bt_x  \left\{ \begin{array}{c} g_{3, I }(J_x,\tau) \\ g_{3, I }(J_x,\tau) \end{array} \right.
 \]
Hence
\beqrs
 2 s G_{1,R}  - 2 \om_x(J_x) G_{1,I} & = & 
- K_Q   e^{ - s \tau} [3 \bt_x (g_{1,I}(J_x,\tau) + 2 g_{3, I }(J_x,\tau)) -  A_x^2  \fr{\del }{\del J_x} g_{1, I}(J_x,\tau) ]   \\ 
 - 2 s G_{1,I} - 2 \om_x(J_x) G_{1,R} & = &
-   K_Q  e^{ - s \tau} [ \bt_x  g_{1,R}(J_x,\tau) - A_x^2 \fr{\del }{\del J_x} g_{1, R}(J_x,\tau) ]   
 + B   A_x \fr{\del\psi_0}{\del J_x} 
\eeqrs

\clearpage

\section{Theory of Nonlinear Dipole Kick from a different approach}

\bit
\item Extend the present approach to nonlinear dipole kicks

\item Use the Vlasov equation to find the solution with nonlinear dipole kicks if the approach above
does not work
\eit

\subsection{Integrating the equations of motion}

We assume that the Hamiltonian in the absence of the dipole and quadrupole kicks leads to motion 
where the linear action $J$ is conserved but the betatron frequency depends on the action. In terms
of the usual canonical coordinates $(x,x')$, this Hamiltonian is $\bar{H}(x,x')$. 
\[ \frac{dx}{ds} = \frac{\del \bar{H}}{\del x'}, \;\;\;\; \frac{dx'}{ds} = 
-\frac{\del \bar{H}}{\del x'}
\]

Now consider the Hamiltonian $H_d$ which describes the impulsive dipole kick with angle $\theta$ at $t=0$.
\[ \theta = \Dl x' \equiv \int \frac{dx'}{dt}dt = (\bt_{kin}c) \int \frac{dx'}{ds} dt = 
-(\bt_{kin}c) \int \frac{\del H_d}{\del x} dt \]
where we used the longitudinal distance variable $s = \bt_{kin}c t$ where $\bt_{kin}$ is the kinematic 
$\bt$. This implies that 
\beq
H_d(x,x') = -\theta x \frac{\dl(t)}{\bt_{kin}c}
\eeq
Next, the quadrupole kick at $t=\tau$ leads to a kick $\Dl x' = - x/f$ and is given by a Hamiltonian 
$H_q$ which obeys at time $t=\tau$,
\[ -\frac{x}{f} = \Dl x' = -(\bt_{kin}c) \int \frac{\del H_q}{\del x} dt \]
This leads to
\beq
H_q(x,x') = \half \frac{x^2}{f} \frac{\dl(t-\tau)}{\bt_{kin}c}
\eeq
The complete Hamiltonian is $\bar{H} = \bar{H} + H_d + H_q$. Note that each of these Hamiltonians is
dimensionless. 

We now make a canonical transformation from $(x,x')$ to action angle variables $(J,\phi)$ where
\beqr
x & = & \sqrt{2\bt J}\cos \phi , \;\;\; x' = -\sqrt{\frac{2 J}{\bt}}[\sin\phi + \al \cos\phi] \\
J & = & \frac{1}{2\bt}[x^2 + (\bt x' + \al x)^2], \;\;\; \tan\phi = -\frac{\bt x' + \al x}{x}
\eeqr
This can be done by a generating function of the type, say $F_1$. Then the transformed Hamiltonian
is
\[ \bar{H} \rarw H = \bar{H} + \frac{\del F_1}{\del s} = H_0(J) + H_1(J,\phi) + H_d + H_q \]
The term $H_1(J,\phi)$ drives resonances but we assume in the following discussion that we are
sufficiently far from resonances that this term has a negligible impact and can be dropped.
The Hamiltonian we consider in the following has the form
\beq
H(J,\phi) = H_0(J) - \theta \sqrt{2 \bt_K J}\cos\phi \frac{\dl(t)}{\bt_{kin}c} 
+  \frac{\bt_Q}{f} J \cos^2 \phi \frac{\dl(t-\tau)}{\bt_{kin}c}
\eeq
I have used $\bt_K$ (beta function at dipole kicker) with the $\theta$ term and defined $q = \bt_Q/f$. 
The equations of motion are
\[ \frac{d\phi}{ds} = \frac{\del H}{\del J}, \;\;\; \frac{d J}{ds} = -\frac{\del H}{\del \phi} \]
Define
\[ \om(J) = \bt_{kin}c \frac{\del H_0}{\del J} \]
Note that since $H_0$ is dimensionless, the RHS has the dimensions of frequency.
We have
\beqr
\frac{d\phi}{d t} & = & \om(J) - \theta \sqrt{\frac{\bt_K}{2 J}}\cos\phi \dl(t) 
+  q \cos^2 \phi \dl(t-\tau) \\
\frac{d J}{d t} & = & -\theta \sqrt{2 \bt_K J}\sin\phi \dl(t) + q J \sin 2\phi \dl(t-\tau)
\eeqr
where we set $q = \bt_Q/f$. 

Integrating the equations of motion from $0 < t < \tau$, 
\beqr
J(t) & = & J(0) + \int_0^t dt[-\theta \sqrt{2 \bt_K J}\sin\phi \dl(t) + q J \sin 2\phi \dl(t-\tau)]
\nonumber \\
J(t) & \equiv J_{0+} & = J_i - \Dl J_{\theta} , \;\;\;\; 0 < t < \tau \\
\Dl J_{\theta} & = & \theta \sqrt{2 \bt_K J_i}\sin\phi_i
\eeqr
where we set the initial values $J_i = J(0), \phi_i = \phi(0)$ and we used the fact that 
$J(0 < t < \tau) = J_{0+} $ stays constant over this time interval. Note that in the 1st quadrant where $(x,x') > 0$
and provided that $\bt x' + \al x > 0$, we have $\phi < 0$ and hence $\Dl J_{\theta} < 0$, so that $J_{0+} > J_i$. 

Now integrating the equation for $\phi$, we have over this same time interval
\beqr
\phi(t) & = & \phi_i + \om(J_{0+})t - \Dl \phi_{\theta} , \;\;\;\; 0 < t < \tau \\
\Dl \phi_{\theta} & = & \theta \sqrt{\frac{\bt_K}{2 J_i}}\cos\phi_i
\eeqr

Integrating over times $t > \tau$, we have
\beqr
J(t) & = & J(\tau) + q J(\tau) \sin 2\phi(\tau) =  J_{0+} + \Dl J_q, \;\;\; t > \tau \\
\Dl J_q & = & q J_{0+} \sin 2\phi(\tau)
\eeqr
Again, this is time independent for $t > \tau$, so we set $J_{\tau +} = J(t>\tau)$. For the $\phi$ variable
we obtain
\beqr
\phi(t) & = & \phi(\tau) + \int_{\tau}^t  dt [\om(J) +  q \cos^2 \phi \dl(t-\tau)] = \phi(\tau) + 
\om(J_{\tau +})(t-\tau) + \Dl \phi_q \\
\Dl\phi_q & = &  q \cos^2 \phi(\tau) 
\eeqr

In order to do the phase space integration to find the dipole moment, I need to express 
$J_{\tau +}, \phi(t)$ in terms of the initial values $J_i, \phi_i$. 
We have
\beqr
J_{\tau +} & = & J_i - \Dl J_{\theta} + q(J_i - \Dl J_{\theta}) \sin 2\phi(\tau) \nonumber \\
& = & J_i - \Dl J_{\theta} 
 + q (J_i - \Dl J_{\theta})\sin 2\left[\phi_i + \om(J_{0+})\tau - \Dl\phi_{\theta} \right]\nonumber \\
& \approx & J_i - \Dl J_{\theta}  + q(J_i- \Dl J_{\theta}) \sin 2\left[\phi_i + \om(J_{0+})\tau \right]
\eeqr
where we dropped the phase correction $\Dl\phi_{\theta}$ that does not grow with either $t$ or $\tau$. 
We now do a Taylor expansion of $\om(J_{0+})$ about $J_i$. Keeping terms to first order in $\theta$, we
have
\[ \om(J_{0+}) = \om(J_i) - \Dl J_{\theta} \om'(J_i) \]
Similarly we do a Taylor expansion to 1st order in $q$,
\beqrs
\om(J_{\tau +}) & = & \om(J_{0+}) + q J_{0+} \sin 2\phi(\tau) \om'(J_{0+}) \\
& = & \om(J_i) - \Dl J_{\theta} \om'(J_i) 
+ q(J_i - \Dl J_{\theta})
\sin 2\left[\phi_i +  \om(J_{0+})\tau - \Dl\phi_{\theta} \right] \\
& & \times \left(\om'(J_i) - \Dl J_{\theta} \om''(J_i)\right)
\eeqrs
We drop the phase correction $\Dl\phi_{\theta}$ as before and also the correction 
$\Dl J_{\theta} \om''(J_i)$ for reasons of smallness. Later we will explicitly assume that $\om''=0$.
\beqr
\om(J_{\tau +}) & = & \om(J_i) - \Dl J_{\theta} \om'(J_i) 
+ q (J_i - \Dl J_{\theta}) \om'(J_i) \sin 2\left[\phi_i + \om(J_{0+})\tau\right] \nonumber \\
& = & \om(J_i) - \Dl J_{\theta}\om'(J_i) 
+ q (J_i - \Dl J_{\theta})\om'(J_i) \sin 2\left[\phi_i + \left(\om(J_i) - \Dl J_{\theta} \om'(J_i)\right)\tau\right] \nonumber \\
\eeqr
Using this form, we have 
\beqrs
\phi(t) & =  & \phi(\tau) + \Dl\phi_q  + (t-\tau)\left\{\om(J_i) - \Dl J_{\theta} \om'(J_i) \right. 
\nonumber \\ 
& & \left. + q (J_i - \Dl J_{\theta}) \om'(J_i) \sin 2\left[\phi_i + \left(\om(J_i) - \Dl J_{\theta} \om'(J_i)\right)\tau\right] \right\} \nonumber \\ 
& = &  \phi_i + \om(J_{0+})\tau - \Dl\phi_{\theta} + \Dl\phi_q  
 + (t-\tau)\left\{ \om(J_i) - \Dl J_{\theta} \om'(J_i) \right. \nonumber \\
& & \left. + q (J_i - \Dl J_{\theta}) \om'(J_i) \sin 2\left[\phi_i + \left(\om(J_i) - \Dl J_{\theta} \om'(J_i)
\right)\tau\right] \right\} \nonumber \\ 
& = & \phi_i + [\om(J_i) - \Dl J_{\theta} \om'(J_i)]\tau  - \Dl\phi_{\theta} + \Dl\phi_q  
  + (t-\tau)\left\{ \om(J_i) - \Dl J_{\theta} \om'(J_i) \right. \nonumber \\
& & \left. + q (J_i - \Dl J_{\theta}) \om'(J_i) \sin 2\left[\phi_i + \left(\om(J_i) - \Dl J_{\theta} \om'(J_i)\right)\tau\right] \right\} 
\eeqrs
Both $\Dl \phi_{\theta}$ and $\Dl\phi_q$ are small terms that do not grow with time and will be dropped from here on. 
With this approximation,
\beqr
\phi(t) & =  & \phi_i + [\om(J_i) - \Dl J_{\theta} \om'(J_i) ]t  \nonumber \\
 &  &  + (t-\tau)
 q (J_i - \Dl J_{\theta}) \om'(J_i) \sin 2\left[\phi_i + \left(\om(J_i) - \Dl J_{\theta}\om'(J_i)\right)\tau\right]  \nonumber \\
 & = &  \phi_i + t(\om(J_i)  - a_{\theta}  \sin\phi_i) + a_q(t-\tau)(1  - \frac{\Dl J_{\theta}}{J_i})\sin 2[\phi_i + 
\om(J_i)\tau - a_{\theta} \tau \sin\phi_i]   \label{eq: phi_Ji_phii}  \\
a_{\theta} & = & \theta  \sqrt{2 \bt_K J_i} \om'(J_i), \;\;\;  a_q = q J_i\om'(J_i)
\eeqr
Note that both $a_{\theta}$ and $a_q$  depend on $J_i$ and have dimension of $[1/T]$.
In this form, Eq.(\ref{eq: phi_Ji_phii}) shows the dependence of $\phi(t)$ on $(J_i,\phi_i,t)$. 

The time dependent dipole moment is
\beq
\lan x(t) \ran = \int dJ d\phi \sqrt{2\bt J(t)}\cos\phi(t) \psi(J,\phi)
\eeq
By the conservation of phase space, we have
\beq
dJ d\phi \psi(J,\phi) = dJ_i d\phi_i \psi_i(J_i)
\eeq
where the initial distribution function depends only on $J_i$. Thus
\beqr
\lan x(t) \ran & = & \sqrt{2\bt} \int dJ_i d\phi \psi_i(J_i) \sqrt{J_{\tau +}(J_i,\phi_i)}\cos\phi(J_i,\phi_i,t)  \nonumber \\
& = & \sqrt{2\bt} \int dJ_i d\phi \psi_i(J_i) \sqrt{ J_i - \Dl J_{\theta} + q (J_i - \Dl J_{\theta}) 
\sin 2\left(\phi_i + \om(J_{0+})\tau\right) }  \nonumber \\
& & \times \cos[\phi_i + t(\om(J_i)  - a_{\theta}  \sin\phi_i) + a_q(t-\tau)(1  - \frac{\Dl J_{\theta}}{J_i})\sin 2[\phi_i + 
\om(J_i)\tau - a_{\theta} \tau \sin\phi_i]    \label{eq: xdip_exact}  \nonumber \\
\eeqr
In arriving at this form, we have dropped the small phase corrections $\Dl\phi_{\theta}, \Dl\phi_q$ but not made
assumptions on the smallness of $\Dl J_{\theta}/J_i$. This Eq.(\ref{eq: xdip_exact}) could therefore be useful for
\underline{numerical integration} to arrive at a relatively exact result. 

To make analytical progress, we have to make further approximations. 

\noi \underline{Approximation of small $\Dl J_{\theta}/J_i$}: We now approximate
\beqrs
 \sqrt{J_{\tau +}} & = & \left[ J_i - \Dl J_{\theta} + q (J_i - \Dl J_{\theta}) \sin 2\left(\phi_i + \om(J_{0+})\tau  \right)\right]^{1/2} \\
& = & \sqrt{J_i}\left[ 1 - \frac{\Dl J_{\theta}}{J_i}
+ q (1  - \frac{\Dl J_{\theta}}{J_i})\sin 2\left(\phi_i + \om(J_{0+})\tau \right)\right]^{1/2} \\
& \approx & \left[ \sqrt{J_i} - \half \theta \sqrt{2 \bt_K}\sin\phi_i 
 + \half q  \sqrt{J_i}(1  - \frac{\Dl J_{\theta}}{J_i}) \sin 2\left(\phi_i + \om(J_{0+})\tau \right)\right] \\
& \approx & \sqrt{J_i} - \half \theta \sqrt{2 \bt_K}\sin\phi_i 
 + \half q  \sqrt{J_i}(1  - \frac{\Dl J_{\theta}}{J_i}) \sin 2\left[\phi_i + \left(\om(J_i) - \Dl J_{\theta} \om'(J_i)\right)\tau \right]
\eeqrs
Here we assumed that $\Dl J_{\theta}/J_i = \theta\sqrt{2\bt_K/J_i}\sin\theta_i \ll 1$. 
\begin{quote}
If the above approximation is not a good one, then an alternative could be to instead
\beqrs
 \sqrt{J_{\tau +}} & = & \sqrt{J_i - \Dl J_{\theta} }\left[ 1 + q  \sin 2\left(\phi_i + \om(J_{0+})\tau  \right)\right]^{1/2} \\
& \approx & \sqrt{J_i - \Dl J_{\theta} }\left[ 1 + \half q \sin 2\left(\phi_i + \om(J_{0+})\tau  \right)\right] 
\eeqrs
The problems is that $\sqrt{J_i - \Dl J_{\theta} } = \sqrt{J_i - \theta\sqrt{2\bt_K J_i}\om'(J_i)\sin\phi_i}$ and it's not clear
how to do the $\phi_i$ integration.
\end{quote}

Returning to the first approximation above,  the integration over phase space can be written as
\beq
\lan x(t) \ran = \sqrt{2\bt}\int dJ_i \psi_i(J_i) \left[ \sqrt{J_i} T_1 - \half \theta \sqrt{2 \bt_K} T_2
+ \half q  \sqrt{J_i} T_3 \right]
\eeq
where
\beqr
T_1 & = & \int d\phi_i \cos\phi(t) = {\rm Re}\left[\int d\phi_i \exp[i \phi(t)]\right] \\
T_2 & = &\ \int d\phi_i \cos\phi(t) \sin\phi_i = \half\int d\phi_i [\sin(\phi(t) + \phi_i) - \sin(\phi(t) - \phi_i) ]\\
T_3 & = & \int d\phi_i (1  - \frac{\Dl J_{\theta}}{J_i})\cos\phi(t) \sin 2\left[\phi_i + \left(\om(J_i) - \theta \sqrt{2 \bt_K J_i}\sin\phi_i \om'(J_i)\right)\tau \right]
\eeqr
Hence
\beqrs
 T_1 & = & {\rm Re}\left[e^{i\om(J_i)t}\int d\phi_i \exp\left(i\{ \phi_i - a_{\theta} t \sin\phi_i  \right. \right. \\ 
& & \left.\left. + a_q(t - \tau)(1  - \frac{\Dl J_{\theta}}{J_i})\sin 2[\phi_i + \om(J_i)\tau - a_{\theta} \tau \sin\phi_i]\}\right) \right]
\eeqrs
We decompose the product
\beqrs
\frac{\Dl J_{\theta}}{J_i}\sin 2[\phi_i + \om(J_i)\tau - a_{\theta} \tau \sin\phi_i] & = & 
\theta \sqrt{\frac{2\bt_K}{J_i}}\sin\phi_i \sin 2[\phi_i + \om(J_i)\tau - a_{\theta} \tau \sin\phi_i]  \\
& = & \half \theta \sqrt{\frac{2\bt_K}{J_i}}\left\{\cos [\phi_i + 2\om(J_i)\tau - 2a_{\theta} \tau \sin\phi_i] \right. \\
& & \left. - \cos [3\phi_i + 2\om(J_i)\tau - 2a_{\theta} \tau \sin\phi_i] \right\}
\eeqrs
Define
\beq
a_{\theta q} = a_q \theta \sqrt{\frac{2\bt_K}{J_i}} = q \theta  \sqrt{2\bt_K J_i}\om'(J_i) = q a_{\theta} \ll a_{\theta}
\eeq
Like $a_{\theta}, a_q$,  $a_{\theta q}$ also  has dimension [1/T]. Hence
\beqrs
 T_1 & = & {\rm Re}\left[e^{i\om(J_i)t}\int d\phi_i \exp\left(
i\left\{ \phi_i - a_{\theta} t \sin\phi_i + 
a_q(t - \tau)\sin 2[\phi_i + \om(J_i)\tau - a_{\theta} \tau \sin\phi_i]\} \right. \right. \right. \\
& & \left. \left.\left. - \half a_{\theta q}(t-\tau)\cos [\phi_i + 2\om(J_i)\tau - 2a_{\theta} \tau \sin\phi_i] 
+\half  a_{\theta q}(t-\tau)\cos [3\phi_i + 2\om(J_i)\tau - 2a_{\theta} \tau \sin\phi_i] 
 \right\} \right) \right]
\eeqrs
The parameter $a_{\theta q}$ is an order of magnitude smaller than the leading parameter
$a_{\theta}$, so I drop it in the following. This also implies that the factor
$q(1 - \Dl J_{\theta}/J_i)$ will be approximated as $q$ in the following. 

\noi \underline{Approximation of dropping $a_{\theta q}$} \newline
With this approximation, we have
\beqr
\phi(t)  & = &  \phi_i + t(\om(J_i)  - a_{\theta}  \sin\phi_i) + a_q(t-\tau)\sin 2[\phi_i + 
\om(J_i)\tau - a_{\theta} \tau \sin\phi_i]   \label{eq: phi_Ji_phii_1}  \\
 T_1 & = & {\rm Re}\left[e^{i\om(J_i)t}\int d\phi_i \exp\left(
i\left\{ \phi_i - a_{\theta} t \sin\phi_i + 
a_q(t - \tau)\sin 2[\phi_i + \om(J_i)\tau - a_{\theta} \tau \sin\phi_i]\}  \right\} \right) \right] \\
T_3 & = & \int d\phi_i \cos\phi(t) \sin 2\left[\phi_i + \left(\om(J_i) - \theta \sqrt{2 \bt_K J_i}\sin\phi_i \om'(J_i)\right)\tau \right]
\eeqr

Now use the expansion
\[ \exp[i z \sin\theta ] = \sum_{l=-\infty}^{\infty} J_l(z) e^{i l\theta}
\]
where $J_l$ is the Bessel function of order $l$. Hence
\beqrs
e^{-i a_{\theta} t \sin\phi_i} & = & \sum_{l_1}J_{l_1}(a_{\theta} t)\exp[-i l_1 \phi_i] \\
e^{i a_q(t - \tau)\sin 2[\phi_i + \om(J_i)\tau - a_{\theta} \tau \sin\phi_i]} & = & \sum_{l_2}J_{l_2}(a_q(t - \tau))
\exp[2 i l_2 (\phi_i + \om(J_i)\tau - a_{\theta} \tau \sin\phi_i)] \\
& = &  \sum_{l_2} \sum_{l_3} J_{l_2}(a_q(t - \tau)) J_{l_3}(2 l_2 a_{\theta}\tau)\exp[2 i l_2 (\phi_i + \om(J_i)\tau]
\exp[-i l_3\phi_i]
\eeqrs

Combining all terms,
\beqrs
T_1 & = &  {\rm Re}\left[e^{i\om(J)t} \sum_{l_1}\sum_{l_2} \sum_{l_3} \exp[2 i l_2 \om(J_i)\tau] 
J_{l_1}(a_{\theta} t) J_{l_2}(a_q(t - \tau)) J_{l_3}(2 l_2 a_{\theta}\tau) \right.  \\
& & \left. \times \int d\phi_i  \exp[i\{1 - l_1 + 2 l_2 - l_3\}\phi_i] \right] \\
& = & 2\pi {\rm Re}\left[e^{i\om(J)t} \sum_{l_1}\sum_{l_2} \sum_{l_3} \exp[2 i l_2 \om(J_i)\tau] 
J_{l_1}(a_{\theta} t) J_{l_2}(a_q(t - \tau)) J_{l_3}(2 l_2 a_{\theta}\tau) \dl_{1 - l_1 + 2l_2 - l_3} \right] \\
& = & 2\pi {\rm Re}\left[e^{i\om(J)t} \sum_{l_1}\sum_{l_2}  \exp[2 i l_2 \om(J_i)\tau] 
J_{l_1}(a_{\theta} t) J_{l_2}(a_q(t - \tau)) J_{2l_2 +1 - l_1}(2 l_2 a_{\theta}\tau) \right]
\eeqrs
Now we use the identity \cite{Abram_Steg}
\[ \sum_{k=-\infty}^{\infty}J_{p \mp k}(u) J_k(v) = J_p(u \pm  v)  \]
Hence
\[ \sum_{l_1}J_{l_1}(a_{\theta} t) J_{2l_2 +1 - l_1}(2 l_2 a_{\theta}\tau) = J_{2l_2 + 1}(a_{\theta}(t + 2 l_2\tau)) \]
which leads to
\beq
T_1  = 2\pi {\rm Re}\left[\sum_{l}\exp[i\om(J)(t + 2 l\tau)]J_{l}(a_q(t - \tau)) J_{2l + 1}(a_{\theta}(t + 2 l\tau))\right]
\label{eq: T1_sol}
\eeq
\underline{CHECK}: Consider the expression for $T_1$ with $a_{\theta} = 0 = a_q$.  From the definition of $\phi(t)$
it follows that in this case $\phi(t) = \phi_i + \om(I)t$, then the original expression for $T_1$ yields
\[
T_1  =  \int d\phi_i \cos\phi(t) = \int d\phi_i \cos[\phi_i + \om(J)t] = 0
\]
while if I substitute $a_{\theta} = 0 = a_q$ into Eq. (\ref{eq: T1_sol}), then using that only $J_l(0) = 0$ for all $l \ne 0$ 
and $J_0(0) = 1$, 
\[
T_1  =  2\pi {\rm Re}\left[\sum_{l}\exp[i\om(J)(t + 2 l\tau)]J_{l}(0) J_{2l + 1}(0)\right] 
\]
In order for both Bessel functions not to vanish, we must have $l = 0 = 2l + 1$ which has no solution. Hence
the integrated result  vanishes as it should. This shows that the integrated result in Eq.(\ref{eq: T1_sol}) is correct in this
simple limiting case.

Consider the three lowest order terms: $l=0, \pm 1$
\beqr
T_1 & = & 2\pi {\rm Re}\left[ e^{i\om(J)t} J_0(a_q(t - \tau))J_1(a_{\theta} t)+ 
e^{i\om(J)(t - 2\tau)} J_{1}(a_q(t - \tau))J_{1}(a_{\theta}(t-2\tau)) \right. \nonumber\\
& & \left.  + e^{i\om(J)(t + 2\tau)} J_{1}(a_q(t - \tau))J_{3}(a_{\theta}(t+2\tau)) + ... \right]
\eeqr
where we used the relation $ J_{-n}(z) = (-1)^n J_n(z)$ to reverse the index on the $l=-1$ factors. 

The complete contribution to the dipole moment from this term is 
\beqr
\lan x(t) \ran_{T_1} &  = & \sqrt{2\bt}\int dJ_i \psi_i(J_i)\sqrt{J_i} T_1 \\
& = &  \frac{\sqrt{2\bt}}{J_0}{\rm Re}\left[ \int dJ_i e^{-J_i/J_0}\sqrt{J_i} \right. \nonumber \\
& & \times \left\{ e^{i\om(J)t} J_0(a_q(t - \tau))J_1(a_{\theta} t)+e^{i\om(J)(t - 2\tau)} J_{1}(a_q(t - \tau))J_{1}(a_{\theta}(t-2\tau)) 
\right. \nonumber\\
& & \left. \left.  + e^{i\om(J)(t + 2\tau)} J_{1}(a_q(t - \tau))J_{3}(a_{\theta}(t+2\tau)) + ... \right\} \right]
\eeqr
The exponents in the first and third terms never vanish since we have $t > \tau$, while the second
term has a vanishing exponent at $t=2\tau$, the time of the echo. At exactly $t=2\tau$, the second term
vanishes because of the factor $J_1(a_{\theta}(t-2\tau))$.  The dominant contribution to the echo at $t$ close to $2\tau$ is
\beq
\lan x(t) \ran_{T_1}(t \approx 2\tau) = \frac{\sqrt{2\bt}}{J_0}{\rm Re}\left[ \int dJ_i e^{-J_i/J_0}\sqrt{J_i}
e^{i\om(J)(t - 2\tau)} J_{1}(a_q(t - \tau))J_{1}(a_{\theta}(t-2\tau)) \right]
\eeq

\noi \underline{Linearized approximation for $T_1$} \newline 
For small arguments
\[ J_n(z) \simeq \frac{(z/2)^n}{\Gm(n+1)} \Rarw \lim_{z\rarw 0} J_0(z) = 1, \;\;\; 
\lim_{z\rarw 0} J_1(z) =  \frac{z}{2}, \;\;\;   J_2(z) = \frac{z^2}{8} \]

Considering only the second term and taking small arguments for both Bessel functions, we have
\beqrs
\lan x(t) \ran_{T_1} &  \approx & 
\frac{\sqrt{2\bt}}{J_0}{\rm Re}\left\{ \int dJ_i e^{-J_i/J_0}\sqrt{J_i} \right. \nonumber \\
& & \left. \times e^{i\om(J_i)(t - 2\tau)} \left[\half \theta \sqrt{2 \bt_K J_i}\om'(J_i) (t-2\tau)\right]
\left[ \half q J_i\om'(J_i)(t-\tau)\right] \right\} \\
& = & \frac{1}{2J_0} \sqrt{\bt \bt_K} \theta q (t-2\tau)(t-\tau){\rm Re}\left[\int dJ_i e^{-J_i/J_0} J_i^2 
(\om'(J_i))^2 e^{i\om(J_i)(t - 2\tau)} \right]
\eeqrs
Assuming now that 
\[ \om(J) = \om_{\bt} + \om' J_i \]
Let as before
\[ \Phi = \om_{\bt}(t - 2\tau), \;\;\; \xi = \om' J_0(t- 2\tau) \]
Then using the result
\[ \int dz \;  z^2 \exp[-a z] = \frac{2}{a^3} \]
we have with $a = 1 - i\xi$
\beqrs
\lan x(t) \ran_{T_1} &  \approx & 
\sqrt{\bt \bt_K} \theta q (t-2\tau)(t-\tau) (\om' J_0)^2 {\rm Re}\left[ \frac{e^{i\Phi}}{(1 -i\xi)^3}\right] \\
& = & \sqrt{\bt \bt_K} \theta q \xi (\om' J_0)(t-\tau) {\rm Re}\left[ \frac{e^{i\Phi}}{(1-i\xi)^3}\right] \\
& = & \sqrt{\bt \bt_K} \theta q \xi (\om' J_0)(t-\tau)
\frac{(1 - 3\xi^2)\cos\Phi - \xi(3 - \xi^2)\sin\Phi}{(1+\xi^2)^3}
\eeqrs
Letting 
\[ \chi = \tan^{-1}[\frac{\xi(3 - \xi^2)}{1 - 3\xi^2}] = 3 \tan^{--1}\xi \]
we have in the completely linearized approximation
\beq
\lan x(t) \ran_{T_1} = \sqrt{\bt \bt_K} \theta q \frac{\om' J_0 \xi (t-\tau)}{(1+\xi^2)^{3/2}} \cos[\Phi + \chi]
\eeq
This has the features that the moment amplitude vanishes at $t=\tau$ and also at $t=2\tau$.

\noi \underline{Contribution of $T_2$}

Evaluating the contribution from $T_2$. We have
\beqrs
T_2 & = & \half {\rm Im}\left[
e^{i\om(J_i)t}\int d\phi_i \left( \exp[i(\phi(t)+\phi_i)]  - \exp[i(\phi(t)-\phi_i)]\right)\right] \\
& = & \half {\rm Im}\left[
e^{i\om(J_i)t}\int d\phi_i \exp\left(i\{ 2\phi_i - a_{\theta} t \sin\phi_i + 
a_q(t - \tau)\sin 2[\phi_i + \om(J_i)\tau - a_{\theta} \tau \sin\phi_i]\}\right) \right. \\ 
& & \left. - e^{i\om(J_i)t}\int d\phi_i \exp\left(i\{- a_{\theta} t \sin\phi_i + 
a_q(t - \tau)\sin 2[\phi_i + \om(J_i)\tau - a_{\theta} \tau \sin\phi_i]\}\right) \right] \\
& = & 
\pi {\rm Im}\left[e^{i\om(J)t} \sum_{l_1}\sum_{l_2} \sum_{l_3} \exp[2 i l_2 \om(J_i)\tau] 
J_{l_1}(a_{\theta} t) J_{l_2}(a_q(t - \tau)) J_{l_3}(2 l_2 a_{\theta}\tau) \right. \\
& & \left. \times \left(\dl_{2 - l_1 + 2l_2 - l_3} - \dl_{-l_1 + 2l_2 - l_3}\right) \right] \\
& = & 
\pi {\rm Im}\left[e^{i\om(J)t} \sum_{l_1}\sum_{l_2} \exp[2 i l_2 \om(J_i)\tau] 
J_{l_1}(a_{\theta} t) J_{l_2}(a_q(t - \tau)) 
\left(J_{2l_2 +2  - l_1}(2 l_2 a_{\theta}\tau) - J_{2l_2 - l_1}(2 l_2 a_{\theta}\tau)
\right) \right] \\
& = & 
\pi {\rm Im}\left[e^{i\om(J)t} \sum_{l} \exp[2 i l \om(J_i)\tau] 
J_{l}(a_q(t - \tau))  \left(J_{2l +2 }(a_{\theta}(t + 2 l \tau)) - J_{2l}(a_{\theta}(t + 2 l \tau)) \right) \right]
\eeqrs
\underline{CHECK}: Case when $a_{\theta} = 0 = a_q$. Using the original expression for $T_2$ with $\phi(t) = \phi_i + \om(J)t$,
\beqrs
T_2 & = &  \half\int d\phi_i [\sin(\phi(t) + \phi_i) - \sin(\phi(t) - \phi_i) ] =  \half\int d\phi_i [\sin(2\phi_i + \om(J)t) - \sin(\om(J)t) ]\\
 &  = & -\pi \sin(\om(J)t)
\eeqrs
Substituting $a_{\theta} = 0 = a_q$ in the sum over Bessel functions above, we have
\beqrs
T_2 & = & 
\pi {\rm Im}\left[e^{i\om(J)t} \sum_{l} \exp[2 i l \om(J_i)\tau] 
J_{l}(0)  \left\{J_{2l +2 }(0) - J_{2l}(0) \right\} \right] \\
& = & -\pi {\rm Im}[e^{i\om(J)t} = -\pi \sin(\om(J)t) ]
\eeqrs
where in the second line we used the result that only the second term within the curly braces $\{ \}$ contributes with $l=0$.
This agrees with the exact integrated result. 

Keeping the lowest order terms with $l=0,\pm 1$, the contribution of this term to the dipole moment is
\beqrs
\lan x (t) \ran_{T2} & = & - \sqrt{\bt \bt_K}\theta  \int dJ_i \; \psi_i(J_i) T_2 \nonumber \\
& = & -\pi \sqrt{\bt \bt_K}\theta {\rm Im}\left[  \int dJ_i \; \psi_i(J_i) e^{i\om(J)t}  \right. \nonumber \\
& & \times \left( e^{- 2 i \om(J_i)\tau}J_{-1}(a_q(t - \tau))\left\{J_0(a_{\theta}(t - 2 \tau)) - J_{-2}(a_{\theta}(t - 2 \tau)) \right\}  \right. \\
& &  + J_0(a_q(t - \tau))\left\{J_2(a_{\theta}t) - J_0(a_{\theta}t) \right\}   \\
 & &  \left. \left. e^{ 2 i \om(J_i)\tau}J_{1}(a_q(t - \tau))\left\{J_4(a_{\theta}(t+2\tau)) - J_2(a_{\theta}(t+2\tau)) \right\} + \ldots
 \right) \right]
\eeqrs
Considering the terms with phase factor close to zero around $t = 2\tau$, we have as an approximation
\beqr
\lan x (t) \ran_{T2} & = & 
-\pi \sqrt{\bt \bt_K}\theta {\rm Im}\left[  \int dJ_i \; \psi_i(J_i) e^{i\om(J)(t-2\tau)}  \right. \nonumber \\
& & \left. \times J_{-1}(a_q(t - \tau))\left\{J_0(a_{\theta}(t - 2 \tau)) - J_{-2}(a_{\theta}(t - 2 \tau)) \right\} \right]  \nonumber \\
& = & \pi \sqrt{\bt \bt_K}\theta {\rm Im}\left[  \int dJ_i \; \psi_i(J_i) e^{i\om(J)(t-2\tau)}  \right. \nonumber \\
& & \left. \times J_{1}(a_q(t - \tau))\left\{J_0(a_{\theta}(t - 2 \tau)) - J_{2}(a_{\theta}(t - 2 \tau)) \right\} \right]  
\eeqr
where in the last line we used the relation $J_{-n}(z) = (-1)^n J_n(z)$ for integer $n$. 
Now substitute $z= J_i/J_0$ and $\psi(J_i) = (1/(2\pi J_0))\exp[-z]$, we have 
\beqr
\lan x (t) \ran_{T2} & = & \half \sqrt{\bt \bt_K}\theta {\rm Im}\left[   e^{i\Phi} \int dz \;  e^{-a_1 z}  \right. \nonumber \\
& & \left. \times J_{1}(a_q(t - \tau))\left\{J_0(a_{\theta}(t - 2 \tau)) - J_{2}(a_{\theta}(t - 2 \tau)) \right\} \right]  
\eeqr

Consider the approximation of this equation above, replacing
\[ J_0(z) \simeq 1 -  \frac{1}{2}z^2 \;\;\; J_1(z) = \frac{z}{2}, \;\;\; J_z(z) = \frac{z^2}{8} \]
Then
\beqrs
\lan x(t) \ran_{T2} & \approx & 
\pi \sqrt{\bt \bt_K}\theta {\rm Im}\left[  \int dJ_i \; \psi_i(J_i) e^{i\om(J)(t-2\tau)}  \right. \nonumber \\
& & \left. \times \half q J_i\om'(t-\tau)\{1 - \frac{3}{8}(\theta\om'(t - 2 \tau))^2 (2\bt_K J_i)\} 
\right] \\
& \approx & \frac{1}{4} \sqrt{\bt \bt_K}\theta q \om'J_0 (t-\tau){\rm Im}\left[ e^{i\Phi} 
 \int dz e^{- a_1 z} z \{1 - \frac{3}{4}\frac{\bt_K}{J_0} (\theta\xi)^2 z\} \right]
\eeqrs
Using the result 
\[ \int dz z \exp[-a_1 z] =    \frac{1}{a^2}   \] 
we have 
\beqr
\lan x(t) \ran_{T2} & \approx & \frac{1}{4} \sqrt{\bt \bt_K}\theta q \om'J_0 (t-\tau)
{\rm Im}\left[ e^{i\Phi} \left\{ \frac{1}{a_1^2} - \frac{3}{2}\frac{\bt_K}{J_0} (\theta\xi)^2 
\frac{2}{a_1^3} \right\} \right]
\eeqr
Keeping only the leading term,
\beq
\lan x(t) \ran_{T2} \approx  \frac{1}{4} \sqrt{\bt \bt_K}\theta q \om'J_0 (t-\tau)
{\rm Im}\left[ \frac{e^{i\Phi}}{(1- i\xi)^2} \right]
\eeq
This is larger than the approximate $\lan x(t) \ran_{T1}$ term because it does not have the
multiplying factor $\xi$. It also does not agree with the linear result from Chao because the 
denominator has $(1 - i\xi)^2$ instead of $(1 - i\xi)^3$ as in Chao. 

\noi \underline{Consider the $T_3$ term}
\beqrs
T_3 & = &  \half\int d\phi_i  \left[\sin \left(\phi(t) + 2\phi_i +2\tau(\om(J_i)- a_{\theta}\sin\phi_i) \right) \right. \\
& & \left. - \sin\left( \phi(t) - 2\phi_i  - 2\tau(\om(J_i)- a_{\theta}\sin\phi_i) \right)  \right] 
\eeqrs

\beqrs
T_3 & = & 
\half\int d\phi_i \left[\sin \left(\phi(t) + 2\phi_i +2\tau(\om(J_i)- a_{\theta}\sin\phi_i) \right) - 
\sin\left( \phi(t) - 2\phi_i  - 2\tau(\om(J_i)- a_{\theta}\sin\phi_i) \right)  \right] \\
& = &  \half {\rm Im}\left[ e^{i\om(J_i)(t+2\tau)}\int d\phi_i   \right. \\
& &  \times \exp\left( 
i\left\{ 3\phi_i - a_{\theta} (t + 2\tau) \sin\phi_i 
 + a_q(t - \tau)\sin 2[\phi_i + \om(J_i)\tau - a_{\theta} \tau \sin\phi_i]\right\}\right) \\ 
& & - e^{i\om(J_i)(t-2\tau)}\int d\phi_i  \\ 
& & \left. \times \exp\left(i\left\{-\phi_i - a_{\theta}( t-2\tau) \sin\phi_i + 
 a_q(t - \tau)\sin 2[\phi_i + \om(J_i)\tau - a_{\theta} \tau \sin\phi_i] \right\}\right) \right] \\
& = & \pi {\rm Im}\left[e^{i\om(J_i)t}
 \sum_{l_1}\sum_{l_2} \sum_{l_3} \exp[2 i l_2 \om(J_i)\tau]  J_{l_2}(a_q(t - \tau)) J_{l_3}(2 l_2 a_{\theta}\tau) \right.  \\
& & \left. \times \left(e^{i\om(J_i)2\tau}J_{l_1}(a_{\theta}( t+2\tau))\dl_{3 - l_1 + 2l_2 - l_3} -
 (e^{-i\om(J_i)2\tau}J_{l_1}(a_{\theta}( t - 2\tau)) \dl_{-1 - l_1 + 2l_2 - l_3}) \right) \right] \\
& = & \pi {\rm Im}\left[e^{i\om(J_i)t}  \sum_{l_1}\sum_{l_2}  \exp[2 i l_2 \om(J_i)\tau]  J_{l_2}(a_q(t - \tau))  \right. \\
& & \left. \times \left(e^{i\om(J_i) 2\tau}J_{l_1}(a_{\theta}( t+2\tau))  J_{2l_2 +3 - l_1}(2 l_2 a_{\theta}\tau) -  
e^{-i\om(J_i) 2\tau} J_{l_1}(a_{\theta}( t - 2\tau)) J_{2l_2 -1 - l_1}(2 l_2 a_{\theta}\tau) \right) \right] \\
& = & \pi {\rm Im}\left[e^{i\om(J_i)t}  \sum_{l}  \exp[2 i l \om(J_i)\tau]  J_{l}(a_q(t - \tau))  \right. \\
& & \left. \times \left(e^{i2\om(J)\tau}J_{2l+3}(a_{\theta}( t+2(l+1)\tau)) - 
e^{-i2\om(J) \tau} J_{2l-1}(a_{\theta}( t+2(l-1)\tau)) 
\right) \right] 
\eeqrs

Now considering the three lowest order terms $l = 0, \pm 1$, the contribution to the dipole moment from $T_3(1)$ is
\beqrs
\lan x(t) \ran_{T_3} & = & \frac{\pi}{2} q\sqrt{2\bt} \int dJ_i \; \psi_i(J_i)\sqrt{J_i} T_3 \\
& = &  \frac{\pi}{2} q\sqrt{2\bt} \int dJ_i \; \psi_i(J_i)\sqrt{J_i} {\rm Im}\left[
e^{i\om(J_i)t} \left(   \right. \right. \\ 
& & \times e^{-2 i \om(J_i)\tau} J_{-1}(a_q(t - \tau)) \left\{ e^{2 i \om(J_i)\tau}J_1(a_{\theta}t) 
 - e^{-2 i \om(J_i)\tau} J_{-3}(a_{\theta}(t - 4\tau)) \right\}   \\
& & + J_0(a_q(t - \tau))\left\{ e^{2 i \om(J_i)\tau}J_3(a_{\theta}(t + 2\tau)) - e^{-2 i \om(J_i)\tau} J_{-1}(a_{\theta}(t -2\tau))\right\}
  \\
& & \left. \left. + e^{2 i \om(J_i)\tau} J_{1}(a_q(t - \tau)) \left\{ e^{2 i \om(J_i)\tau}J_5(a_{\theta}(t + 4\tau))
 - e^{-2 i \om(J_i)\tau} J_1(a_{\theta}t) \right\} + \ldots   \right) \right]
\eeqrs
If we further  approximate this by keeping only those phase factors which will be small around $t=2\tau$, 
 we have the approximate result
\beq
\lan x(t) \ran_{T_3}    = 
 \frac{\pi}{2} q\sqrt{2\bt} {\rm Im}\left[ \int dJ_i \; \psi_i(J_i)\sqrt{J_i} e^{i\om(J_i)(t-2\tau)} 
J_0(a_q(t - \tau)) J_{1}(a_{\theta}(t -2\tau)) \right]
\eeq

Consider the small amplitude approximation with $J_0(a_q(t - \tau)) \approx 1$, 
$J_1(a_{\theta}(t-2\tau)) = a_{\theta}(t-2\tau)/2$, we have
\beqr
\lan x(t) \ran_{T_3}  &  \approx & 
 \frac{\pi}{2} q\sqrt{2\bt} \theta (\half\om'\sqrt{2\bt_K}(t-2\tau)) {\rm Im}\left[e^{i\Phi} 
\int dJ_i (\frac{1}{2\pi J_0}e^{-J_i/J_0}) e^{i\xi J_i/J_0}(\sqrt{J_i})^2 \right] \nonumber \\
& \approx  & \frac{1}{4}\sqrt{\bt \bt_K} J_0\theta q \om'(t- 2\tau){\rm Im}\left[e^{i\Phi} \int dz \;
z e^{-a_1 z} \right] \nonumber \\
& \approx  & \frac{1}{4}\sqrt{\bt \bt_K} J_0\theta q \om'(t- 2\tau){\rm Im}\left[\frac{e^{i\Phi}}{(1- i\xi)^2}
\right] \nonumber \\
& \approx  & \frac{1}{4}\sqrt{\bt \bt_K} J_0\theta q \om'(t- 2\tau)\frac{2\xi\cos\Phi + (1 - \xi^2) \sin\Phi}{(1+\xi^2)^2} \nonumber \\
& \approx  & \frac{1}{4}\sqrt{\bt \bt_K} J_0\theta q \om'(t- 2\tau)\frac{1}{(1+\xi^2)}\sin(\Phi + 2\tan^{-1}\xi)
\eeqr

In the completely linearized approximation, we have the contributions from $T_1, T_2, T_3 $ as the dipole moment
\beqr
\lan x(t) \ran_{lin} & = & \sqrt{\bt \bt_K} \theta q \frac{\om' J_0 \xi (t-\tau)}{(1+\xi^2)^{3/2}} \cos[\Phi + \chi]  \nonumber \\
& & + \frac{1}{4} \sqrt{\bt \bt_K}\theta q \om'J_0 (t-\tau)\frac{1}{(1+\xi^2)}\sin(\Phi + 2\tan^{-1}\xi) \nonumber \\
& & + \frac{1}{4}\sqrt{\bt \bt_K} \theta q \om' J_0 (t- 2\tau)\frac{1}{(1+\xi^2)}\sin(\Phi + 2\tan^{-1}\xi) \nonumber \\
& = & \sqrt{\bt \bt_K} \theta q \om' J_0 \left[ \frac{\xi (t-\tau)}{(1+\xi^2)^{3/2}} \cos[\Phi + \chi] \right.  \nonumber \\
& & \left. + \frac{1}{4}(2t - 3\tau) \frac{1}{(1+\xi^2)}\sin(\Phi + 2\tan^{-1}\xi) \right] 
\eeqr
This is WRONG.

Combining the contributions from $T_1, T_2, T3$ for the echo around $t=2\tau$, we have
\beqr
\lan x(t) \ran & = & 2\pi\sqrt{2\bt}{\rm Re}\left[\int dJ_i \psi_i(J_i)\sqrt{J_i}e^{i\om(J_i)(t-2\tau)} 
J_1(a_q(t - \tau))J_1(a_{\theta}(t -2\tau))\right] \nonumber \\
&  & + \pi \sqrt{\bt \bt_K}\theta {\rm Im}\left[  \int dJ_i \; \psi_i(J_i) e^{i\om(J)(t-2\tau)}  \right. \nonumber \\
& & \left. \times J_{1}(a_q(t - \tau))\left\{J_0(a_{\theta}(t - 2 \tau)) - J_{2}(a_{\theta}(t - 2 \tau)) \right\} \right]  \nonumber \\
& & +  \frac{\pi}{2} q\sqrt{2\bt} {\rm Im}\left[ \int dJ_i \; \psi_i(J_i)\sqrt{J_i} e^{i\om(J_i)(t-2\tau)} 
J_0(a_q(t - \tau)) J_{1}(a_{\theta}(t -2\tau)) \right]  \nonumber \\
& = & 2\pi\sqrt{2\bt} \left\{ {\rm Re}\left[\int dJ_i \psi_i(J_i)\sqrt{J_i}e^{i\om(J_i)(t-2\tau)} 
J_1(a_q(t - \tau))J_1(a_{\theta}(t -2\tau))\right] \right. \nonumber \\
& & + \frac{\sqrt{2\bt_K}\; \theta}{4} {\rm Im}\left[  \int dJ_i \; \psi_i(J_i) e^{i\om(J)(t-2\tau)}  
 J_{1}(a_q(t - \tau))\left\{J_0(a_{\theta}(t - 2 \tau)) - J_{2}(a_{\theta}(t - 2 \tau)) \right\} \right]  \nonumber \\
& & + \left. \frac{q}{2} {\rm Im}\left[ \int dJ_i \; \psi_i(J_i)\sqrt{J_i} e^{i\om(J_i)(t-2\tau)} 
J_0(a_q(t - \tau)) J_{1}(a_{\theta}(t -2\tau)) \right] \right\}
\label{eq: echo_T1T2T3}
\eeqr
With the usual assumption that $\om(J) = \om_{\bt} + \om' J$, and using $z= J_i/J_0$ as the independent
variable, we have
\beqrs
 \om(J_i)(t - 2\tau) & = &  \om_{\bt}(t-2\tau) + \om'(t-2\tau) J_i = \Phi + \om'(t-2\tau)  J_0 z  = \Phi + \xi z \\
a_{\theta}(t-2\tau) & = & \theta\om'(J_i)\sqrt{2\bt_K J_i}(t-2\tau) = \sqrt{2\bt_K}\theta (\xi/J_0) \sqrt{J_i} \equiv 
a_2\sqrt{z} , \;\;\;  a_2 = \sqrt{\frac{2\bt_K}{ J_0}} \theta \xi \\ 
a_q(t - \tau) & = & q\om'(J_i )(t - \tau)J_i = q\om' (t - \tau)J_0 z  \equiv a_3 z , \;\;\;\; a_3 = q \xi_1, \;\;\; \xi_1 = \om'(t - \tau) J_0  \\
\psi_i(J_i)e^{i\om(J_i)(t-2\tau)}  & = & \frac{1}{2\pi J_0}\exp[i\Phi]\exp[- a_1 z] , \;\;\; a_1 = 1 - i \xi 
\eeqrs
The parameters $a_1, a_2, a_3$ are all dimensionless. 

We can now rewrite Eq. (\ref{eq: echo_T1T2T3}) in the form, after changing the integration variable to 
$z=J_i/J_0$, 
\beqr
\lan x(t) \ran & = &  \sqrt{2\bt}\left\{ \sqrt{J_0} {\rm Re}[e^{i\Phi} I_1] + 
\frac{\sqrt{2\bt_K}\; \theta}{4} {\rm Im}[e^{i\Phi} I_2] + \frac{q}{2}\sqrt{J_0}  {\rm Im}[e^{i\Phi} I_3] \right\} \\
 I_1 & = &  \int dz \exp(-a_1 z) \sqrt{z} J_1(a_2 \sqrt{z}) J_1(a_3 z)  \label{eq: I1_integ} \\
 I_2 & =  & \int dz \exp[-a_1 z]J_1(a_3 z)[J_0(a_2\sqrt{z}) - J_2(a_2\sqrt{z})]  
\label{eq: I2_integ} \\
I_3 & = &  \int dz \exp(-a_1 z) \sqrt{z} J_1(a_2 \sqrt{z}) J_0(a_3 z)  
\label{eq: I3_integ}
\eeqr
The approximations made in deriving this result have been in dropping the phase corrections
$\Dl\phi_{\theta}, \Dl \phi_q$ and approximating $q(1 - \Dl J{\theta}/J_i)$ by $q$. 
None of the integrals $I_1, I_2, I_3$ can be done by Mathematica. 
These integrals can of course be evaluated numerically. 

One analytic possibility is to expand the Bessel function with $\theta$ in
its argument in powers of the argument and integrate term by term. Using
\[ J_n(w) = \frac{w^n}{2^n} \sum_{p=0}^{\infty} (-1)^p \frac{1}{2^{2p} p! \Gm(n+p+1)} w^{2p}  \equiv
\frac{w^n}{2^n} \sum_{p=0}^{\infty} (-1)^p \frac{c_p}{\Gm(n+p+1)} w^{2p}
\]
Then we have
\beqrs
I_1 & = & \frac{a_2}{2} \int dz \exp(-a_1 z) J_1(a_3 z) z \sum_{p=0} (-1)^p \frac{1}{2^{2p} p! \Gm(p+2)} a_2^{2p} z^p
 \\
& = & \frac{a_2}{2} \int dz \exp(-a_1 z) J_1(a_3 z) z \left[ 1 - \frac{1}{2^3}a_2^2 z + 
\frac{1}{2^4 \times 2  \times 6} a_2^4 z^2 + \ldots  \right] 
\eeqrs
This can be integrated term by term using 
\[
\int dz \exp(-a_1 z) z^m J_1(a_3 z) = \half {a_3}{a_1^{m+2}}\Gm(m+2)
 \mbox{}_2F_1\left[\frac{m+2}{2},\frac{m+3}{2},2, -(\frac{a_3}{a_1})^2\right] 
\]
where $ \mbox{}_2F_1$ is the hypergeometric function.  For $m=0, 1, 2$ we have explicitly, 
\beqrs
\int dz \exp(-a_1 z)  J_1(a_3 z) & = & \frac{1}{a_3}[1 - \frac{a_1}{(a_1^2 + a_3^2)^{1/2}}] \\
\int dz \exp(-a_1 z) z J_1(a_3 z) & = & \frac{a_3}{(a_1^2 + a_3^2)^{3/2}} \\
\int dz \exp(-a_1 z) z^2 J_1(a_3 z) & = & \frac{3 a_1 a_3}{(a_1^2 + a_3^2)^{5/2}} 
\eeqrs
Writing out the 1st two terms of $I_1$ explicitly we have
\beq
I_1  =  \frac{a_2}{2}\left[ \frac{a_3}{(a_1^2 + a_3^2)^{3/2}} - \frac{a_2^2}{8} \frac{3 a_1 a_3}{(a_1^2 + a_3^2)^{5/2}} + \ldots  \right]
\eeq

$I_2$ also has to be integrated term by term by expanding $J_0(a_2\sqrt{z}) , J_2(a_2\sqrt{z})$. First we note
that using the series expansion above, we can write 
\[ J_0(a_2\sqrt{z}) - J_2(a_2\sqrt{z}) = 1 - (c_1 + \frac{1}{2^2 \Gm(3)} ) a_2^2 z + 
(\frac{1}{\Gm(3)} c_2 + \frac{1}{2^2 \Gm(5)} c_1)a_2^4 z^2 + \ldots \]
where as defined above, $c_1 = 1/2^2$, $c_2 = 1/2^5$ etc. 
Again, writing out the first two terms in the term by term integration, we have
\beq
I_2 = \frac{1}{a_3}(1 - \frac{a_1}{(a_1^2 + a_3^2)^{1/2}}) - \frac{3}{8} a_2^2 \frac{a_3}{(a_1^2 + a_3^2)^{3/2}}
 + \ldots
\eeq
In the limit of small $a_3$, the first term behaves as
\beqrs
\lim_{a_3 \rarw 0} \frac{1}{a_3}(1 - \frac{a_1}{(a_1^2 + a_3^2)^{1/2}}) & = &
 \lim_{a_3 \rarw 0} \frac{1}{a_3}(1 - [1 + (\frac{a_3}{a_1})^2 ]^{-1/2} \\
& \approx & \frac{1}{a_3}[\half (\frac{a_3}{a_1})^2 + {\rm HOTs}] = \frac{a_3}{2 a_1^2} + \dots
\eeqrs
so it vanishes linearly as $a_3\rarw 0$. 

Finally for $I_3$ we have
\beqr
I_3 & = & \int dz \exp(-a_1 z) \sqrt{z} J_0(a_3 z)  [\frac{a_2 \sqrt{z}}{2} - \frac{1}{16}a_2^3 z^{3/2} + \ldots ]
\nonumber \\
& = & \frac{a_2}{2}\frac{a_1}{(a_1^2 + a_3^2)^{3/2}} - \frac{a_2^3}{16}\frac{2a_1^2 - a_3^2}{(a_1^2 + a_3^2)^{5/2}}
\eeqr
Hence,
\beqr
\lan x(t) \ran & = &  \sqrt{2\bt}\left\{ \sqrt{J_0} {\rm Re}\left[ e^{i\Phi} \frac{a_2}{2}\left( \frac{a_3}{(a_1^2 + a_3^2)^{3/2}} - 
\frac{a_2^2}{8} \frac{3 a_1 a_3}{(a_1^2 + a_3^2)^{5/2}} + \ldots  \right) \right]  \right.  \nonumber \\ 
& & + \frac{\sqrt{2\bt_K}\; \theta}{2} {\rm Im}\left[ e^{i\Phi} \left( \frac{1}{a_3}(1 - \frac{a_1}{(a_1^2 + a_3^2)^{1/2}}) - 
\frac{3}{8} a_2^2 \frac{a_3}{(a_1^2 + a_3^2)^{3/2}} \right) \right]   \nonumber \\
& & \left. + \frac{q}{2}\sqrt{J_0}  {\rm Im}\left[ e^{i\Phi} \left( \frac{a_2}{2}\frac{a_1}{(a_1^2 + a_3^2)^{3/2}} - 
\frac{a_2^3}{16}\frac{2a_1^2 - a_3^2}{(a_1^2 + a_3^2)^{5/2}}\right) \right] \right\} 
\eeqr

Keeping only the 1st term in the series from each contribution, we have from the $I_1$ term,
\beqrs
 {\rm Re}[ e^{i\Phi} \frac{a_2 a_3}{(a_1^2 + a_3^2)^{3/2}}] & = & a_2 a_3 {\rm Re}[ e^{i\Phi} \frac{1}{[(1-i\xi)^2 + a_3^2]^{3/2}} ] \\
& & a_2 a_3 {\rm Re}[ e^{i\Phi} \frac{1}{[1-\xi^2 - 2 i\xi + a_3^2]^{3/2}}] \\
&= &  \frac{a_2 a_3}{[(1 - \xi^2 + a_3^2)^2 + 4\xi^2]^{3/2}}{\rm Re}[ e^{i\Phi} (1-\xi^2 + 2 i\xi + a_3^2)^{3/2}]  \\
& = &  \frac{a_2 a_3}{[(1 - \xi^2 + a_3^2)^2 + 4\xi^2]^{3/4}}{\rm Re}[ e^{i\Phi} (\exp[i\Theta] )^{3/2}]  \\
& = & \frac{a_2 a_3}{[(1 - \xi^2 + a_3^2)^2 + 4\xi^2]^{3/4}}\cos[\Phi + \frac{3}{2}\Theta ] \\
\Theta & = & \tan^{-1}\left[ \frac{2\xi}{1 - \xi^2 + a_3^2}\right] 
\eeqrs
Here we used
\[ 1-\xi^2 + 2 i\xi + a_3^2 = [(1 - \xi^2 + a_3^2)^2 + 4\xi^2]^{1/2} \exp[i\Theta]  \]

The contribution from the $I_2$ term is
\beqrs
{\rm Im}[ e^{i\Phi} \left( \frac{1}{a_3}(1 - \frac{a_1}{(a_1^2 + a_3^2)^{1/2}})\right) ]  & = & 
\frac{1}{a_3}{\rm Im}\left[  e^{i\Phi}  - e^{i\Phi} \frac{1 - i\xi}{[(1- i\xi)^2 +a_3^2]^{1/2}} \right] \\
& = & \frac{1}{a_3}{\rm Im}\left[  e^{i\Phi}  - \frac{e^{i\Phi} (1 - i\xi)e^{i\Theta/2}}{[(1-\xi^2 + a_3^2)^2 + 4\xi^2]^{1/4}}  \right] \\
& = & \frac{1}{a_3}\left[ \sin\Phi - \frac{\left\{\sin(\Phi+\Theta/2) - \xi\cos(\Phi+\Theta/2)\right\}}{[(1-\xi^2 + a_3^2)^2 +  4\xi^2]^{1/4}} \right]
\eeqrs

Similarly the contribution from the 1st term in $I_3$ is 
\beqrs
{\rm Im}\left[ e^{i\Phi} \frac{a_2}{2}\frac{a_1}{(a_1^2 + a_3^2)^{3/2}} \right]  &  = & 
 \frac{a_2}{2}\frac{\left\{\sin(\Phi+\Theta/2) - \xi\cos(\Phi+\Theta/2)\right\}}{[(1-\xi^2 + a_3^2)^2 +  4\xi^2]^{1/4}}
\eeqrs

Combining the leading terms from $I_1, I_2, I_3$, we have
\beqr
\lan x(t \ge \tau) \ran & = & \sqrt{2\bt}\left\{ \sqrt{J_0}
\left( \frac{a_2 a_3}{[(1 - \xi^2 + a_3^2)^2 + 4\xi^2]^{3/4}}\cos(\Phi + \frac{3}{2}\Theta) \nonumber  \right) \right.\\ 
& & + \half\sqrt{2\bt_K}\theta \left(
\frac{1}{a_3}\left[ \sin\Phi - \frac{\left\{\sin(\Phi+\Theta/2) - \xi\cos(\Phi+\Theta/2)\right\}}{[(1-\xi^2 + a_3^2)^2 +  4\xi^2]^{1/4}} \right] \right) \nonumber \\
& & \left. + \half q\sqrt{J_0}
\left( \frac{a_2}{2}\frac{\left\{\sin(\Phi+\Theta/2) - \xi\cos(\Phi+\Theta/2)\right\}}{[(1-\xi^2 + a_3^2)^2 +  4\xi^2]^{1/4}} \right) \right\} 
\eeqr
where $ a_2 = \sqrt{\frac{2\bt_K}{ J_0}} \theta \xi$, $a_3 = q\xi_1$, $\xi = \om'J_0(t-2\tau)$,
$\xi_1 = \om' J_0(t-\tau)$, $\Phi = \om_{\bt}(t-2\tau)$, 
$\Theta = \tan^{-1}[2\xi/(1 - \xi^2 + a_3^2)]$. 

Questions
\bit
\item At $t=2\tau$, the dipole moment vanishes because $\xi = a_2 = \Phi = \Theta = 0$.
At what time does this moment have its largest value? It is expected to be at $t=2\tau \pm T_{rev}$
Evaluate the above perturbative result numerically to find the times at which amplitude is
largest. Compare with Chao's linearized theory for the time evolution of the echo.

\item At what value of $q$ does the amplitude have its largest value?

\item Is there a value of $\theta$ at which the amplitude have its largest value or does this
amplitude increase monotonically with $\theta$?

\item Can the integral forms for $I_1, I_2, I_3$ in Eqs. (\ref{eq: I1_integ}), (\ref{eq: I2_integ}), (\ref{eq: I3_integ}) respectively be used to obtain the values of $q, \theta$ where the amplitude
reaches maxima?

\eit

\noi {\bf Problems with the above calculation}
\bit
\item The expansion for $\sqrt{J_{\tau, +}}$ in Eq.() is only valid for small $\theta$ such that
\[ \frac{\Dl J_{\theta}}{J_i} = \theta \sqrt{\frac{2\bt_K}{J_i}}\sin\phi_i \ll 1 \]
If we write the initial action as $J_i = (a_i^2/2)\eps_0$ where $a_i$ is an amplitude factor as
in $x_i = a_i \sg_0$ where $\sg_0 = \sqrt{\bt \eps}$ is the initial beam size at the BPM, then
\beqrs
\frac{\Dl J_{\theta}}{J_i} & = & (\bt\theta)\sqrt{\frac{4\bt_K}{a_i^2 \bt \sg_0^2}}\sin\phi_i
 =  2 \frac{x_{\theta}}{a_i \sg_o}\sqrt{\frac{\bt_K}{\bt}}\sin\phi_i 
\eeqrs
where $x_{\theta} = \bt\theta$ is the amplitude of the centroid at the BPM after the dipole kick.
Assuming as is usually the case
that the two beta functions are comparable, i.e. $\bt_K/\bt \simeq O(1)$, we have the requirement
for the expansion to be valid for all $\phi_i$ is
\beq
x_{\theta} \ll \half a_i \sg_o
\eeq
If we pick $a_i = 1$ for a particle initially at an amplitude $\sg_0$, then the expansion is only
for small enough kick angles $\theta$ such that $x_{\theta} \ll \sg_0/2$. 
\eit

The amplitude of these echoes at later times can be found by identifying the dominant terms with 
phase factors vanishing at these times.

\clearpage

\section{Echoes in 2D transverse motion }

Calculations to be done
\bit
\item Nonlinear quad theory (QT) and the moments $\lan x \ran, \lan y \ran$
\item QT and the 2nd order moments $\lan x^2\ran, \eps_x$ and the quadrupole moment $Q_2 = \lan x^2 - y^2\ran $.
  Find the decoherence (or damping) time and the initial frequency of ringing oscillations of  $Q_2$
\item Nonlinear dipole and quad theory (DQT) and similar calculations as above. 

\item Decoherence time in 2D after the initial dipole kick

\item Can this be used to measure the strength of the coupling?  Diffusion rates in
the transverse planes separately, or a mix of the diffusion coefficients 
$D_{xx},D_{xy}, D_{yy}$.

\item Options with a quad kick: kick in the same plane as the dipole, i.e follow with
$p_x \rarw p_x + k_q x$ or in the other plane as $p_y \rarw p_y + k_q y$. 
In the 1st option, is the echo dynamics in the $x$ plane the same as without coupling?
In the 2nd option, does the coupling lead to an echo in $y$ and/or in $x$ ?

If the quad kick is in the complementary plane, the echo may be quite weak.
\eit

\subsection{RMS Tune Width}

\noi {\tt Notation needs changing}

Initial action
\beq
\rho(J_x, J_y) = \frac{1}{(2\pi)^2 \eps_x \eps_y} \exp[-\frac{J_x}{\eps_x} - \frac{J_y}{\eps_y}]
\eeq
which is normalized to 1 as
\[ \int dJ_x dJ_y d\phi_x d\phi_y  \rho(J_x, J_y) = 1 \]

Tunes as a function of $J_x, J_y$
\beqr
\Dl\nu_x & = & \al_{xx} J_x + \al_{xy} J_y \nonumber \\
\Dl\nu_y & = & \al_{xy} J_x + \al_{yy} J_y 
\eeqr
Mean tune shift
\beqr
\lan \Dl\nu_x \ran & = & \int dJ_x dJ_y d\phi_x d\phi_y  \rho(J_x, J_y) \Dl\nu_x \nonumber \\
& = & \frac{1}{\eps_x\eps_y}\left(\al_{xx}\int dJ_x J_x \exp[-\frac{J_x}{\eps_x}] \int dJ_y \exp[-\frac{J_y}{\eps_y}]  + 
\al_{xy}\int dJ_x  \exp[-\frac{J_x}{\eps_x}] \int J_y dJ_y \exp[-\frac{J_y}{\eps_y}]\right)  \nonumber \\
& = & \al_{xx}\eps_x + \al_{xy}\eps_y \\
\lan \Dl\nu_y \ran & = &  \al_{xy}\eps_x + \al_{yy}\eps_y 
\eeqr
And for the second moment
\beqr
\lan \Dl\nu_x^2 \ran & = & \int dJ_x dJ_y d\phi_x d\phi_y  \rho(J_x, J_y) (\Dl\nu_x)^2 \nonumber \\
&  =  & \frac{1}{\eps_x\eps_y} \left( \al_{xx}^2\int dJ_x J_x^2 \exp[-\frac{J_x}{\eps_x}] \int dJ_y \exp[-\frac{J_y}{\eps_y}]  +  
\al_{xy}^2 \int dJ_x  \exp[-\frac{J_x}{\eps_x}] \int J_y^2 dJ_y \exp[-\frac{J_y}{\eps_y}]  \right. \nonumber \\
& & \left.  + 2 \al_{xx}\al_{xy}\int dJ_x  J_x\exp[-\frac{J_x}{\eps_x}] \int J_y dJ_y \exp[-\frac{J_y}{\eps_y}]  \right) \nonumber \\
& = & 2\al_{xx}^2 \eps_x^2 + 2\al_{xy}^2 \eps_y^2 + 2 \al_{xx} \al_{xy} \eps_x \eps_y \\
\lan \Dl\nu_y^2 \ran & = & 2\al_{xy}^2 \eps_x^2 + 2\al_{yy}^2 \eps_y^2 + 2 \al_{xy} \al_{yy} \eps_x \eps_y 
\eeqr
where we used
\[ \int_0^{\infty} z \exp[-z] = 1,   \int_0^{\infty} z^2 \exp[-z] = 2 \]
Hence the rms widths of the tune distributions are
\beqr
\sg_{\nu_x} & \equiv & [ \lan \Dl\nu_x^2 \ran - \lan \Dl\nu_x \ran^2]^{1/2} = 
[ 2\al_{xx}^2 \eps_x^2 + 2\al_{xy}^2 \eps_y^2 + 2 \al_{xx} \al_{xy} \eps_x \eps_y - (\al_{xx}\eps_x + \al_{xy}\eps_y)^2]^{1/2} \nonumber \\
& = & [ \al_{xx}^2 \eps_x^2  + \al_{xy}^2 \eps_y^2]^{1/2} \\
\sg_{\nu_y} & =  & [ \al_{xy}^2 \eps_x^2  + \al_{yy}^2 \eps_y^2]^{1/2} 
\eeqr

Special case \\
\noi 1. $\eps_x = \eps_y = \eps$, then
\beq
\sg_{\nu_x}  = [\al_{xx}^2 + \al_{xy}^2 ]^{1/2} \eps, \;\;\;\;  \sg_{\nu_y}  = [\al_{xy}^2 + \al_{yy}^2 ]^{1/2} \eps
\eeq

The coupling may be wither due to linear optics or may only arise due to the nonlinear fields. In the latter case
the coupling may only manifest in the dependence of the dependence of the tunes on both the actions.
We will consider the two cases separately, the latter case first.

\subsection{Only nonlinear coupling}
Follow similar line of argument as in the 1D case

\subsubsection{1st moment}
The phase space coordinates are
\beqr
x & = & \sqrt{2 \bt_y J_x}\cos\phi_x , \;\;\; p_x = \bt_x x'+ \al_x x = -\sqrt{2\bt_x J_x} \sin\phi_x \non \\
y & = & \sqrt{2 \bt_y J_y}\cos\phi_y , \;\;\; p_y = \bt_y y'+ \al_y y = -\sqrt{2\bt_y J_y} \sin\phi_y
\eeqr

The initial density is uncoupled, so it is legitimate to take the initial density as
\beq
\psi_0(J_x, J_y) = \frac{1}{(2\pi)^2 \eps_x \eps_y} \exp[-\frac{J_x}{\eps_x} - \frac{J_y}{\eps_y}]
\eeq
which is normalized to 1 as
\[ \int dJ_x dJ_y d\phi_x d\phi_y  \psi_0(J_x, J_y) = 1 \]
The  motion is coupled via the magnetic forces acting on the beam. Now the angular frequencies are
assumed to depend linearly on the actions
\beq
\om_x(J_x, J_y) = \om_{x0} + w_{xx}J_x + w_{xy}J_y,  \;\;\;\;  \om_y(J_x, J_y) = \om_{y0} + w_{xy}J_x + w_{yy}J_y
\eeq
Here $\om_{x0}, \om_{y0}$ are the zero amplitude frequencies corresponding to the nominal tunes.

Under the action of a horizontal dipole kick $p_x \rarw p_x - \bt_K \theta_x$, the DF evolves to 1st order in the kick as
\beq
\psi_1(J_x, \phi_x, J_y) = \psi_0(J_x, J_y) + \bt_K \theta_x \psi_{0,J_x}\fr{\del J_x}{\del p_x} =
\psi_0(J_x, J_y) + \bt_K \theta_x \psi_{0,J_x}\sqrt{\fr{2 J_x}{\bt_x}}\sin\phi_x
\eeq
where $\psi_{0, J_x} \equiv \del \psi_0/\del J_x$. 
This depends only the horizontal phase $\phi_x$ but not on the vertical phase $\phi_y$. At time $t$ after the
dipole kick, the DF is
\beq
\psi_2(J_x, \phi_x, J_y, t) = \psi_0(J_x, J_y) + \bt_K \theta_x \psi_{0,J_x}\sqrt{\fr{2 J_x}{\bt_x}}\sin(\phi_x - \om_x(J_x, J_y) t)
\eeq
At the time $\tau$ of the quadrupole kick, the DF is $\psi_3(J_x, \phi_x, J_y, \tau) = \psi_2(J_x, \phi_x, J_y, t=\tau))$.
A quadrupole kicker of focal length $f$ applies kicks to both planes, but with opposite signs, as
\beq
\Dl p_x = - q_x x , \;\;\; \Dl p_y =  q_y y , \;\;\; q_x = \fr{\bt_{Q, x}}{f}, \;\;\;   q_y = \fr{\bt_{Q, y}}{f} 
\eeq
The DF following the quad kick is
\beq
\psi_4(x, p_x, y, p_y, \tau) = \psi_3(x, p_x - \Dl p_x, y, p_y - \Dl p_y, \tau) = \psi_3(x, p_x + q_x x, y, p_y - q_y y, \tau)
\eeq
Following the quad kick, the transformed DF depends on the action and angle variables in both transverse planes.

The action and angle variables in the argument of the DF transform to
\beqr
J_x &  \rarw & \bar{J}_x = \fr{1}{2\bt_x}[x^2 + (p_x + q_x x)^2] = J_x(1 + A_x(q_x, \phi_x)), \;\;\;
\phi_x \rarw \bar{\phi}_x = {\rm Arctan}[- \fr{p_x + q_x x}{x}] = {\rm Arctan}[\tan\phi_x - q_x] \non \\
J_y &  \rarw &  \bar{J}_y = \fr{1}{2\bt_y}[y^2 + (p_y - q_y y)^2] = J_y(1 + A_y(q_y, \phi_y)), \;\;\;
\phi_y \rarw \bar{\phi}_y = {\rm Arctan}[- \fr{p_y - q_y y}{y}] = {\rm Arctan}[\tan\phi_y + q_y] \\
A_x & = &  - q_x \sin 2\phi_x + q_x^2 \cos^2\phi_x, \;\;\; A_y  =    q_y \sin 2\phi_y + q_y^2 \cos^2\phi_y    \non
\eeqr
Using Taylor expansions to expand the angle variables in powers of the quad kick as
\beqrs
    {\rm Arctan}[\tan\phi_x - q_x] & = & \phi_x - q_x \cos^2\phi_x  - \qrtr q_x^2(\sin 2\phi_x + \half \sin 2 \phi_x) + O(q_x^3) \\
    {\rm Arctan}[\tan\phi_y + q_y] & = & \phi_y + q_y \cos^2\phi_y  - \qrtr q_y^2(\sin 2\phi_y + \half \sin 2 \phi_y) + O(q_y^3)
 \eeqrs
 Keeping terms to 1st order in $q_x, q_y$, we approximate
 \beqrs
 A_x & \approx & - q_x \sin 2\phi_x , \;\;\; A_y \approx q_y \sin 2\phi_y \\
 {\rm Arctan}[\tan\phi_x - q_x] & \approx & \phi_x - q_x \cos^2\phi_x , \;\;\;
 {\rm Arctan}[\tan\phi_y + q_y]  \approx  \phi_y + q_y \cos^2\phi_y 
 \eeqrs

 Dropping the term $\psi_0$ from the DF, the transformed DF is
 \beqr
 \psi_4(J_x, \phi_x, J_y, \phi_y, \tau) &  =  & 
 \bt_K \theta_x \psi_{0,J_x}\left(J_x[1- q_x \sin 2\phi_x],  J_y[1 +  q_y \sin 2\phi_y]\right)
 \sqrt{\fr{2 J_x[1- q_x \sin 2\phi_x]}{\bt_x}}  \non  \\
&  &  \times \sin (\phi_{x,-\tau} - q_x \cos^2\phi_x)   \\
 \phi_{x,-\tau}  & = & \phi_x - \om_x\left(J_x[1- q_x \sin 2\phi_x],  J_y[1 +  q_y \sin 2\phi_y]\right)
 \eeqr
 After the quad kick, the DF at time $t$ (from the dipole kick) is
 \beqr
 \psi_5(J_x, \phi_x, J_y, \phi_y, t) &  =  & \psi_4(J_x, \phi_{x,-\Dl \phi_x}, J_y, \phi_{y,-\Dl \phi_y}) \\
 \phi_{x,-\Dl \phi_x} & = & \phi_x - \Dl \phi_x,  \;\;\;  \phi_{y,-\Dl \phi_y}  =  \phi_y - \Dl \phi_y \\
 \Dl \phi_x & = & \om_x(J_x, J_y)(t - \tau) , \;\;\;   \Dl \phi_y  =  \om_y(J_x, J_y)(t - \tau)
 \eeqr

\subsubsection{Decoherence time in 2D}

 We also need to expand the angular betatron frequencies to 1st order in the actions
 \beqr
 \om_x\left(J_x[1- q_x \sin 2\phi_x],  J_y[1 +  q_y \sin 2\phi_y]\right) & = &  \om _{x0} +
 ( 1- q_x \sin 2\phi_x)\om_{x, J_x}J_x + ( 1+ q_y \sin 2\phi_y)\om_{x, J_y}J_y \non \\
& = &  \om_{x0} + ( 1- q_x \sin 2\phi_x) w_{xx} J_x + ( 1+ q_y \sin 2\phi_y) w_{xy} J_y
 \eeqr
 Here the expansion is exact because of the assumed dependence of the betatron frequencies on the actions. 
 
 Under the change $\phi_x \rarw  \phi_{x,-\Dl \phi_x}$
 \[  \phi_{x,-\tau}  \rarw \phi_{x,-\Dl\phi_x,-\tau} \equiv \phi_x - \Dl\phi_x -
     [\om_{x0} + w_{xx}(1- q_x \sin 2(\phi_x - \Dl\phi_x) )J_x + w_{xy}(1 +  q_y \sin 2(\phi_y - \Dl\phi_y) )J_y]\tau
\]
Written out explicitly, 
\beqr
& \mbox{} &  \psi_{0, J_x}\left(J_x[1- q_x \sin 2\phi_x],  J_y[1 +  q_y \sin 2\phi_y]\right) \non \\
& = & - \fr{1}{(2\pi)^2 \eps_x \eps_y} \fr{[1- q_x \sin 2\phi_x]}{\eps_x} 
 \exp[-\frac{J_x[1- q_x \sin 2\phi_x]}{\eps_x} - \frac{J_y[1+ q_y \sin 2\phi_y]}{\eps_y}]  \\
&   & \psi_5(J_x, \phi_x, J_y, \phi_y, t)   \non \\
& = & \bt_K \theta_x \psi_{0,J_x}\left(J_x[1- q_x \sin 2\phi_{x,-\Dl \phi_x}],  J_y[1 +  q_y \sin 2\phi_{y,-\Dl \phi_y}]\right)  \non\\
 &  &  \times  \sqrt{\fr{2 J_x[1- q_x \sin 2\phi_{x,-\Dl \phi_x}]}{\bt_x}}  \sin (\phi_{x,-\Dl\phi_x,-\tau} - q_x \cos^2\phi_{x,-\Dl\phi_x})   \non \\
 & = & -\fr{ \bt_K \theta_x }{(2\pi)^2 \eps_x^2 \eps_y} [1- q_x \sin 2\phi_{x,-\Dl \phi_x}] \non \\
& & \times \exp[-\frac{J_x[1- q_x \sin 2\phi_{x,-\Dl \phi_x}]}{\eps_x} - \frac{J_y[1+ q_y \sin 2\phi_{y,-\Dl \phi_y}]}{\eps_y}]
\non \\
& & \times \sqrt{\fr{2 J_x[1- q_x \sin 2\phi_{x,-\Dl \phi_x}]}{\bt_x}}  \sin (\phi_{x,-\Dl\phi_x,-\tau} - q_x \cos^2\phi_{x,-\Dl\phi_x})
\eeqr
Using
\[ \cos^2\phi_{x,-\Dl\phi_x} = \half(1 + \cos 2\phi_{x,-\Dl\phi_x}) \]
the argument of the last sine function can be written as
\beqrs
& \mbox{} & \phi_{x,-\Dl\phi_x,-\tau} - q_x \cos^2\phi_{x,-\Dl\phi_x}    \\
& = & \phi_x - \Dl\phi_x -
 [\om_{x0} + w_{xx}(1- q_x \sin 2\phi_{x, - \Dl\phi_x}) J_x + w_{xy}(1 +  q_y \sin 2(\phi_{y, - \Dl\phi_y} ) J_y]\tau \\
    & &  - \fr{q_x}{2}(1 + \cos 2\phi_{x,-\Dl\phi_x})  \\
 & = &  \phi_x - \Dl\phi_x - \left[ \om_{x0} + w_{xx}J_x    + w_{xy}(1 +  q_y \sin 2(\phi_{y, - \Dl\phi_y} ) J_y \right] \tau \\
 &  &  + q_x(w_{xx}\sin 2(\phi_{x, - \Dl\phi_x} J_x \tau - \half \cos 2\phi_{x,-\Dl\phi_x} - \half )
 \eeqrs
 We can approximate
 \beqrs
 \sin 2(\phi_{x, - \Dl\phi_x} w_{xx}J_x \tau - \half \cos 2\phi_{x,-\Dl\phi_x} & = & \sqrt{(w_{xx} J_x \tau)^2 + \qrtr}
 \sin [2 \phi_{x, - \Dl\phi_x} - {\rm Arctan}(\fr{1}{2w_{xx}J_x\tau})] \\
 & \approx & w_{xx}J_x\tau \sin 2 \phi_{x, - \Dl\phi_x}
 \eeqrs
 where we used $w_{xx}J_x\tau \simeq \tau/\tau_D \gg 1$.
\beqrs
& \mbox{} & \phi_{x,-\Dl\phi_x,-\tau} - q_x \cos^2\phi_{x,-\Dl\phi_x})    \\
& \simeq & \phi_x - \Dl\phi_x - \left[ \om_{x0} + w_{xx}J_x(1 - q_x \sin 2 \phi_{x, - \Dl\phi_x})   +
  w_{xy}(1 +  q_y \sin 2(\phi_{y, - \Dl\phi_y} ) J_y \right] \tau - \half q_x
\eeqrs
 We expand
 \[  \sqrt{ [1- q_x \sin 2\phi_{x,-\Dl \phi_x}]} \approx  1 - \half q_x \sin 2\phi_{x,-\Dl \phi_x} \]
 \beqr 
\!\!\!\! \!\!\!\! \!\!\!\! & \mbox{} &  \psi_5(J_x, \phi_x, J_y, \phi_y, t)    \non \\
  & = &  -\fr{ \bt_K \theta_x }{(2\pi)^2 \eps_x^2 \eps_y} \sqrt{\fr{2 J_x}{\bt_x}} [1- q_x \sin 2\phi_{x,-\Dl \phi_x}]
 \exp[-\frac{J_x[1- q_x \sin 2\phi_{x,-\Dl \phi_x}]}{\eps_x} - \frac{J_y[1+ q_y \sin 2\phi_{y,-\Dl \phi_y}]}{\eps_y}]
 (1 - \half q_x \sin 2\phi_{x,-\Dl \phi_x})  \non \\
& & \times
\sin \left( \phi_{x,-\Dl\phi_x} - \left[ \om_{x0} + w_{xx}J_x(1 - q_x \sin 2 \phi_{x, - \Dl\phi_x})   +
  w_{xy}(1 +  q_y \sin 2\phi_{y, - \Dl\phi_y} ) J_y \right] \tau - \half q_x \right) \non \\
\label{eq: psi5_2D}
\eeqr

The dipole moments are
\beqr
\lan x(t) \ran & \equiv & \int d J_x dJ_y d\phi_x d\phi_y \sqrt{2\bt_x J_x} \cos \phi_x   \psi_5(J_x, \phi_x, J_y, \phi_y, t)  \\
\lan y(t) \ran & \equiv & \int d J_x dJ_y d\phi_x d\phi_y \sqrt{2\bt_y J_y} \cos \phi_y   \psi_5(J_x, \phi_x, J_y, \phi_y, t)
\eeqr
Using the expression for the DF in Eq.(\ref{eq: psi5_2D}), we have for the horizontal dipole moment
\beqr
\lan x(t) \ran & = & -\fr{2 \bt_K \theta_x }{(2\pi)^2 \eps_x^2 \eps_y} \int d J_x dJ_y d\phi_x d\phi_y \; J_x \cos\phi_x   \non \\
&  & \times  [1- q_x \sin 2\phi_{x,-\Dl \phi_x}]
 \exp[-\frac{J_x[1- q_x \sin 2\phi_{x,-\Dl \phi_x}]}{\eps_x} - \frac{J_y[1+ q_y \sin 2\phi_{y,-\Dl \phi_y}]}{\eps_y}]
 (1 - \half q_x \sin 2\phi_{x,-\Dl \phi_x})  \non \\
& & \times
\sin \left( \phi_{x,-\Dl\phi_x} - \left[ \om_{x0} + w_{xx}J_x(1 - q_x \sin 2 \phi_{x, - \Dl\phi_x})   +
  w_{xy}(1 +  q_y \sin 2\phi_{y, - \Dl\phi_y} ) J_y \right] \tau - \half q_x \right) \non 
\label{eq: xmom_2D_1} \\
\eeqr

CHECK: Setting $q_y = 0 = w_{xy} = 0 = w_{yy}$ should reduce this to the 1D expressions found earlier (in PR-AB paper)
In this case
\[ \Dl \phi_x = [\om_{x,0} + w_{xx}J_x](t - \tau), \;\;\; \Dl\phi_y = \om_{y,0}(t - \tau) \]
These variables do not depend on $J_y$. The dipole moment simplifies to 
\beqrs
\lan x(t) \ran & = & -\fr{2 \bt_K \theta_x }{(2\pi)^2 \eps_x^2 \eps_y} \int d J_x dJ_y d\phi_x d\phi_y \non \\
& = & J_x \cos\phi_x   [1- q_x \sin 2(\phi_{x} - \Dl \phi_x)]
 \exp[-\frac{J_x[1- q_x \sin 2(\phi_{x}-\Dl \phi_x)]}{\eps_x} - \frac{J_y}{\eps_y}]
 (1 - \half q_x \sin 2(\phi_x-\Dl \phi_x))  \non \\
& & \times
\sin \left( \phi_{x,-\Dl\phi_x} - \left[ \om_{x0} + w_{xx}J_x(1 - q_x \sin 2 \phi_{x, - \Dl\phi_x})  \right] \tau - \half q_x \right) \non 
\eeqrs
This does not depend on $\phi_y$ and the integration over $J_y$ is simple, leaving us with 
\beqrs
\lan x(t) \ran & = & -\fr{2 \bt_K \theta_x }{(2\pi) \eps_x^2 } \int d J_x d\phi_x  \non \\
& = & J_x \cos\phi_x   [1- q_x \sin 2(\phi_{x} - \Dl \phi_x)] \exp[-\frac{J_x[1- q_x \sin 2(\phi_{x}-\Dl \phi_x)]}{\eps_x}]
 (1 - \half q_x \sin 2(\phi_x-\Dl \phi_x))  \non \\
& & \times
\sin \left( \phi_{x,-\Dl\phi_x} - \left[ \om_{x0} + w_{xx}J_x(1 - q_x \sin 2 \phi_{x, - \Dl\phi_x})  \right] \tau - \half q_x \right) \non 
\eeqrs
This resembles the simplified version of Eq.(2.13) in the PR-AB 2018 paper. 

Returning to the 2D form of the centroid in Eq.(\ref{eq: xmom_2D_1}), the integrations over the phases $\phi_x, \phi_y$ have to be
done first. We write the last sine function in the integrand 
\beqrs
& \mbox{} & \sin \left( \phi_{x,-\Dl\phi_x} - \left[ \om_{x0} + w_{xx}J_x(1 - q_x \sin 2 \phi_{x, - \Dl\phi_x})   +
  w_{xy}J_y (1 +  q_y \sin 2\phi_{y, - \Dl\phi_y} )  \right] \tau - \half q_x \right) \\
& = & \sin \left( \phi_{x,-\Dl\phi_x} + [q_x w_{xx}J_x \sin 2 \phi_{x, - \Dl\phi_x} - q_y w_{xy}J_y\sin 2 \phi_{y, - \Dl\phi_y}] \tau -
    [ \om_{x0} + w_{xx}J_x + w_{xy}J_y]\tau - \half q_x \right) \\
    & \equiv & \sin( \phi_{x,-\Dl\phi_x} + c_{xx}\sin 2 \phi_{x, - \Dl\phi_x} - c_{xy}\sin 2 \phi_{y, - \Dl\phi_y} - c_{00} ) \\
    c_{xx}(J_x) & = &   q_x w_{xx}J_x \tau, \;\;\; c_{xy}(J_y) = q_y w_{xy}J_y \tau, \;\;\; c_{00} = [ \om_{x0} + w_{xx}J_x + w_{xy}J_y]\tau +
    \half q_x
\eeqrs
Keeping terms to $O(q_x)$, we can write
\[
  [1- q_x \sin 2(\phi_{x} - \Dl \phi_x)] (1 - \half q_x \sin 2(\phi_x-\Dl \phi_x)) \simeq 1 - \fr{3}{2}q_x \sin 2\phi_{x, - \Dl \phi_x}
\]

Combining the phase dependent trigonometric terms
\beqrs
& \mbox{} & \cos\phi_x \sin( \phi_{x,-\Dl\phi_x} + c_{xx}\sin 2 \phi_{x, - \Dl\phi_x} - c_{xy}\sin 2 \phi_{y, - \Dl\phi_y} - c_{00} ) \\
& & \times (1 - \fr{3}{2}q_x \sin 2\phi_{x, - \Dl \phi_x}) \\
& =  & \half( \sin[ 2\phi_x - \Dl\phi_x + c_{xx}\sin 2 \phi_{x, - \Dl\phi_x} - c_{xy}\sin 2 \phi_{y, - \Dl\phi_y} - c_{00} ] \\
& &  + \sin[- \Dl\phi_x + c_{xx}\sin 2 \phi_{x, - \Dl\phi_x} - c_{xy}\sin 2 \phi_{y, - \Dl\phi_y} - c_{00} ] ) \\
&  &  \times (1 - \fr{3}{2}q_x \sin 2\phi_{x, - \Dl \phi_x}) \\
& =  &  \half(\sin\phi_1 + \sin\phi_2) - \fr{3}{4}q_x \left[\cos(\phi_1 - 2 \phi_{x, - \Dl \phi_x}) - \cos(\phi_1 + 2 \phi_{x, - \Dl \phi_x}) \right. \\  &  &  \left. + \cos(\phi_2 - 2 \phi_{x, - \Dl \phi_x}) - \cos(\phi_2 + 2 \phi_{x, - \Dl \phi_x}) \right] 
\eeqrs
where 
\beqrs
 \phi_1 & =  & 2\phi_x - \Dl\phi_x + c_{xx}\sin 2 \phi_{x, - \Dl\phi_x} - c_{xy}\sin 2 \phi_{y, - \Dl\phi_y} - c_{00}  \\
\phi_2 & = & - \Dl\phi_x + c_{xx}\sin 2 \phi_{x, - \Dl\phi_x} - c_{xy}\sin 2 \phi_{y, - \Dl\phi_y} - c_{00} \\
\phi_3 & \equiv & \phi_1 - 2 \phi_{x, - \Dl \phi_x} = \Dl \phi_x + c_{xx}\sin 2 \phi_{x, - \Dl\phi_x} - c_{xy}\sin 2 \phi_{y, - \Dl\phi_y} - c_{00} \\
\phi_4 & \equiv & \phi_1 + 2 \phi_{x, - \Dl \phi_x} = 4\phi_x - 3\Dl \phi_x + c_{xx}\sin 2 \phi_{x, - \Dl\phi_x} - c_{xy}\sin 2 \phi_{y, - \Dl\phi_y} - c_{00} \\
\phi_5 & \equiv & \phi_2 - 2 \phi_{x, - \Dl \phi_x}  =  -2\phi_x + \Dl \phi_x + c_{xx}\sin 2 \phi_{x, - \Dl\phi_x} - c_{xy}\sin 2 \phi_{y, - \Dl\phi_y} - c_{00} \\
\phi_6 & \equiv & \phi_2 + 2 \phi_{x, - \Dl \phi_x}  =  2\phi_x - 3\Dl \phi_x + c_{xx}\sin 2 \phi_{x, - \Dl\phi_x} - c_{xy}\sin 2 \phi_{y, - \Dl\phi_y} - c_{00}
\eeqrs
The coefficients of $\sin 2 \phi_{x, - \Dl\phi_x}$ and $\sin 2 \phi_{y, - \Dl\phi_y}$  are the same for all the phases $\phi_j$.

The dipole moment in 2D is now
\beqr
\lan x(t) \ran & = & -\fr{2 \bt_K \theta_x }{(2\pi)^2 \eps_x^2 \eps_y} \int d J_x dJ_y \; J_x
\exp[-\frac{J_x}{\eps_x} - \frac{J_y}{\eps_y}]  \non \\
&  & \times d\phi_x d\phi_y  \exp[\frac{ q_x  J_x\sin 2\phi_{x,-\Dl \phi_x}}{\eps_x} - \frac{q_y  J_y \sin 2\phi_{y,-\Dl \phi_y}}{\eps_y}] \non \\
& & \times  \left\{ \half(\sin\phi_1 + \sin\phi_2) - \fr{3}{4}q_x \left[\cos\phi_3  - \cos\phi_4 + \cos\phi_5  - \cos\phi_6 \right]  \right\}  \\
& \equiv & -\fr{2 \bt_K \theta_x }{(2\pi)^2 \eps_x^2 \eps_y} \int d J_x dJ_y \;  J_x
\exp[-\frac{J_x}{\eps_x} - \frac{J_y}{\eps_y}]  \left\{ \Phi_1 + \Phi_2 - \Phi_3 + \Phi_4 - \Phi_5 + \Phi_6 \right\}
\eeqr
where
\beqr
\Phi_1(J_x, J_y) & = & \half \int \int d\phi_x d\phi_y  \exp[\frac{ q_x  J_x}{\eps_x} \sin 2\phi_{x,-\Dl \phi_x} -
  \frac{q_y  J_y}{\eps_y} \sin 2\phi_{y,-\Dl \phi_y}] \sin\phi_1  \\
\Phi_2(J_x, J_y) & = & \half \int \int d\phi_x d\phi_y  \exp[\fr{ q_x  J_x}{\eps_x}\sin 2\phi_{x,-\Dl \phi_x} - 
  \frac{q_y  J_y }{\eps_y}\sin 2\phi_{y,-\Dl \phi_y}] \sin\phi_2  \\
\Phi_3(J_x, J_y) & = & \fr{3}{4}q_x \int \int d\phi_x d\phi_y  \exp[\frac{ q_x  J_x}{\eps_x}\sin 2\phi_{x,-\Dl \phi_x} -
  \frac{q_y  J_y}{\eps_y}\sin 2\phi_{y,-\Dl \phi_y} ] \cos\phi_3  \\
\Phi_4(J_x, J_y) & = & \fr{3}{4}q_x \int \int d\phi_x d\phi_y  \exp[\frac{ q_x  J_x}{\eps_x}\sin 2\phi_{x,-\Dl \phi_x} -
  \frac{q_y  J_y }{\eps_y}\sin 2\phi_{y,-\Dl \phi_y}] \cos\phi_4 \\
\Phi_5(J_x, J_y) & = & \fr{3}{4}q_x \int \int d\phi_x d\phi_y  \exp[\frac{ q_x  J_x}{\eps_x } \sin 2\phi_{x,-\Dl \phi_x} -
  \frac{q_y  J_y }{\eps_y}\sin 2\phi_{y,-\Dl \phi_y}] \cos\phi_5  \\
\Phi_6(J_x, J_y) & = & \fr{3}{4}q_x \int \int d\phi_x d\phi_y  \exp[\frac{ q_x  J_x}{\eps_x}\sin 2\phi_{x,-\Dl \phi_x} -
  \frac{q_y  J_y }{\eps_y}\sin 2\phi_{y,-\Dl \phi_y}] \cos\phi_6
\eeqr

Thus the integrands in the terms $\Phi_j$, j=1,... 6 are respectively,
\beqrs 
{\rm Int}(\Phi_1) & = &  \exp[\fr{ q_x  J_x}{\eps_x} \sin 2\phi_{x,-\Dl \phi_x} -  \fr{q_y  J_y}{\eps_y} \sin 2\phi_{y,-\Dl \phi_y}] \sin\phi_1 \\
  & = & {\rm Im}\left(\exp[i \{ \phi_1 - i \fr{ q_x  J_x}{\eps_x} \sin 2\phi_{x,-\Dl \phi_x} - i\fr{q_y  J_y}{\eps_y} \sin 2\phi_{y,-\Dl \phi_y}
    \}] \right) \\
& = & {\rm Im}\left(\exp\left[i\{2\phi_x - \Dl\phi_x  - c_{00}\}\right]
\exp\left[ i\{ z_{xx}J_x\sin 2 \phi_{x, - \Dl\phi_x} -    z_{xy}J_y\sin 2 \phi_{y, - \Dl\phi_y}\} \right] \right)  \\
{\rm Int}(\Phi_2)  & = & \exp[\fr{ q_x  J_x}{\eps_x} \sin 2\phi_{x,-\Dl \phi_x} -  \fr{q_y  J_y}{\eps_y} \sin 2\phi_{y,-\Dl \phi_y}] \sin\phi_2 \\
& = & {\rm Im}\left(\exp[i \{ \phi_2 - i \fr{ q_x  J_x}{\eps_x} \sin 2\phi_{x,-\Dl \phi_x} - i\fr{q_y  J_y}{\eps_y} \sin 2\phi_{y,-\Dl \phi_y} \}
] \right) \\
&  = & {\rm Im}\left(\exp\left[i\{- \Dl\phi_x  - c_{00}\}\right] \exp\left[ i\{ z_{xx}J_x\sin 2 \phi_{x, - \Dl\phi_x} -    z_{xy}J_y\sin 2 \phi_{y, - \Dl\phi_y}\} \right] \right)  \\
{\rm Int}(\Phi_3) & = &  \exp[\fr{ q_x  J_x}{\eps_x} \sin 2\phi_{x,-\Dl \phi_x} -  \fr{q_y  J_y}{\eps_y} \sin 2\phi_{y,-\Dl \phi_y}]\cos\phi_3 \\
& = &{\rm Re}\left(\exp\left[i\{\Dl \phi_x + z_{xx}J_x\sin 2 \phi_{x, - \Dl\phi_x} - z_{xy} J_y \sin 2 \phi_{y, - \Dl\phi_y} - c_{00} \}\right]\right)  \\
& = & {\rm Re}\left(\exp\left[i\{\Dl \phi_x - c_{00} \}\right] 
\exp\left[ i\{ z_{xx}J_x\sin 2 \phi_{x, - \Dl\phi_x} -    z_{xy}J_y\sin 2 \phi_{y, - \Dl\phi_y}\} \right] \right)  \\
{\rm Int}(\Phi_4)  & = &  \exp[\fr{ q_x  J_x}{\eps_x} \sin 2\phi_{x,-\Dl \phi_x} -  \fr{q_y  J_y}{\eps_y} \sin 2\phi_{y,-\Dl \phi_y}] \cos\phi_4 \\
  & = &  {\rm Re}\left(\exp\left[i\{4\phi_x - 3\Dl \phi_x  + z_{xx}J_x\sin 2 \phi_{x, - \Dl\phi_x} 
- z_{xy} J_y \sin 2 \phi_{y, - \Dl\phi_y} - c_{00} \}\right]\right)  \\
{\rm Int}(\Phi_5) & = &   \exp[\fr{ q_x  J_x}{\eps_x} \sin 2\phi_{x,-\Dl \phi_x} -  \fr{q_y  J_y}{\eps_y} \sin 2\phi_{y,-\Dl \phi_y}]\cos\phi_5  \\
& = &  {\rm Re}\left(\exp\left[i\{-2\phi_x + \Dl \phi_x  + z_{xx}J_x\sin 2 \phi_{x, - \Dl\phi_x}
- z_{xy}J_y  \sin 2 \phi_{y, - \Dl\phi_y} - c_{00} \}\right]\right)  \\
{\rm Int}(\Phi_6) & =  & \exp[\fr{ q_x  J_x}{\eps_x} \sin 2\phi_{x,-\Dl \phi_x} -  \fr{q_y  J_y}{\eps_y} \sin 2\phi_{y,-\Dl \phi_y}]\cos\phi_6 \\
    & = &  {\rm Re}\left(\exp\left[i\{ 2\phi_x - 3\Dl \phi_x   + z_{xx}J_x\sin 2 \phi_{x, - \Dl\phi_x} 
- z_{xy}J_y  \sin 2 \phi_{y, - \Dl\phi_y} - c_{00} \}\right]\right) 
\eeqrs
We have defined the constant complex parameters (independent of phase space variables)
\beq
z_{xx} = c_{xx}/J_x - i \fr{ q_x }{\eps_x} = \fr{ q_x }{\eps_x}(w_{xx}\tau\eps_x - i) , \;\;\;
z_{xy} = c_{xy}/J_y - i \fr{ q_y }{\eps_y} = \fr{ q_y }{\eps_y}(w_{xy}\tau\eps_y - i)
\eeq
The parameters $z_{xx}, z_{xy}$ are constant, independent of the actions $J_x, J_y$. We also have $w_{xx}\eps_x\tau \gg 1$
and we must also require that $w_{xy}\tau\eps_y \gg 1$ (the tune spread in $x$ is determined both by $w_{xx}\eps_x$
and  $w_{xy}\eps_y$.  Under these conditions, $|z_{xx}| \simeq q_x w_{xx} \tau$ and $|z_{xy}| \simeq q_y w_{xy} \tau$.

The expansion into Bessel functions is the same in all the phases $\phi_j$, the differences are in the exponential factors in
the front of the form $\exp[i\{a_j \phi_x + b_j \Dl\phi_x\}]$, $j= 1,... 6$. 

The integrations over $\phi_x, \phi_y$ are done using the result
\[ \int d\phi \exp[i m \phi] \exp[i a \sin(2\phi - \theta)] = 2\pi J_{-m/2}(a)\exp[i (m/2) \theta] \]
which follow from the expansion
\[ \exp[i a \sin \theta] = \sum_k J_k(a) \exp[i k \theta] \]

Hence
\beqrs
\Phi_1(J_x, J_y) & = & \half {\rm Im}\left(\int \int d\phi_x d\phi_y  
\left(\exp\left[i\{2\phi_x - \Dl\phi_x  - c_{00}\}\right]
\exp\left[ i\{ z_{xx}J_x\sin 2 \phi_{x, - \Dl\phi_x} -    z_{xy}J_y\sin 2 \phi_{y, - \Dl\phi_y}\} \right] \right)  \right)  
\eeqrs
The $\phi_x$ and $\phi_y$ integrations can be done separately 
\beqrs
& \mbox{} & \int d\phi_x  \exp\left[i\{2\phi_x - \Dl\phi_x  - c_{00}\}\right]
\exp\left[ i\{ z_{xx}J_x\sin 2 \phi_{x, - \Dl\phi_x} \} \right]   \\
&  = & 2\pi \exp[- i(\Dl\phi_x + c_{00})]  J_{-1}(z_{xx}J_x)\exp[i 2\Dl\phi_x] \\
& = &  -2\pi \exp[ i(\Dl\phi_x  - c_{00})] J_{1}(z_{xx}J_x) \\
&  & \int d\phi_y \exp\left[ i\{ - z_{xy}J_y\sin 2 \phi_{y, - \Dl\phi_y} \} \right]  
=  2\pi  J_{0}(-z_{xy}) =  2\pi J_{0}(z_{xy}J_y) 
\eeqrs
where we used $J_0(-z) = J_0(z), J_{-1}(z) = - J_1(z) $. Hence 
\beq
\Phi_1(J_x, J_y) = -2\pi^2 {\rm Im}\left\{ \exp[ i(\Dl\phi_x  - c_{00} )]  J_{1}(z_{xx}J_x)  J_{0}(z_{xy}J_y) \right\}
\eeq
We have
\beqrs
\Dl\phi_x  - c_{00} & = & [\om_{x,0} + w_{xx}J_x + w_{xy}J_y](t - \tau) - [\om_{x,0} + w_{xx}J_x + w_{xy}J_y]\tau - \half q_x  \\
& = & [\om_{x,0} + w_{xx}J_x + w_{xy}J_y ](t - 2 \tau) -  \half q_x
\eeqrs
The dominant term in this phase  $\om_{x,0}(t - 2\tau) \rarw 0$ at the echo time $t = 2\tau$. Hence this term $\Phi_1$
will have a significant contribution to the echo. 

Continuing
\beqrs
\Phi_2(J_x, J_y)  & = & \half {\rm Im}\left(\int \int d\phi_x d\phi_y  
\exp\left[i\{- \Dl\phi_x  - c_{00}\}\right] \exp\left[ i\{ z_{xx}J_x\sin 2 \phi_{x, - \Dl\phi_x} -    z_{xy}J_y\sin 2 \phi_{y, - \Dl\phi_y}\} \right] \right)  \\
& = & 2 \pi^2 {\rm Im}\left( \exp\left[i\{- \Dl\phi_x  - c_{00}\}\right] J_0(z_{xx}J_x) J_0(z_{xy}J_y) \right)
\eeqrs
The phase factor shows that it is a sub-dominant term and $\Phi_2$ can be dropped from the echo amplitude.
\beqrs
\Phi_3(J_x, J_y)  & = & \half {\rm Re}\left(\int \int d\phi_x d\phi_y  
\exp[i\{\Dl \phi_x - c_{00} \}] \exp\left[ i\{ z_{xx}\sin 2 \phi_{x, - \Dl\phi_x} -    z_{xy}\sin 2 \phi_{y, - \Dl\phi_y}\} \right] \right)  \\
& = & 2\pi^2 {\rm Re}\left\{ \exp[ i(\Dl\phi_x  - c_{00} )]  J_{0}(z_{xx}J_x)  J_{0}(z_{xy}J_y) \right\}
\eeqrs
This will contribute to the echo.
\beqrs
\Phi_4(J_x, J_y)  & = & \half {\rm Re}\left(\int \int d\phi_x d\phi_y  
\exp\left[i\{4\phi_x - 3\Dl \phi_x  + z_{xx} J_x \sin 2 \phi_{x, - \Dl\phi_x} 
- z_{xy} J_y \sin 2 \phi_{y, - \Dl\phi_y} - c_{00} \}\right]\right)  \\
& = & 2 \pi^2 {\rm Re}\left( \exp[i(- 3\Dl \phi_x - c_{00})] J_{-2}(z_{xx}J_x)\exp[i 4\Dl\phi_x] J_0(z_{xy}J_y) \right) \\
& = & 2 \pi^2 {\rm Re}\left(\exp[i(\Dl \phi_x - c_{00})] J_{-2}(z_{xx}J_x)  J_0(z_{xy}J_y) \right)
\eeqrs
This will contribute
\beqrs
\Phi_5(J_x, J_y)  & = & \half {\rm Re}\left(\int \int d\phi_x d\phi_y  
\exp\left[i\{-2\phi_x + \Dl \phi_x  + z_{xx}J_x\sin 2 \phi_{x, - \Dl\phi_x}
- z_{xy}J_y  \sin 2 \phi_{y, - \Dl\phi_y} - c_{00} \}\right]\right)  \\
& = & 2\pi^2 {\rm Re}\left( \exp[i( \Dl \phi_x - c_{00}) ] J_1(z_{xx}J_x) \exp[-i 2 \Dl\phi_x] J_0(z_{xy}J_y) \right) \\
& = & 2\pi^2 {\rm Re}\left( \exp[i( -\Dl \phi_x - c_{00}) ] J_1(z_{xx}J_x) J_0(z_{xy}J_y) \right)
\eeqrs
This will be sub-dominant and can be dropped.
\beqrs
\Phi_6(J_x, J_y)  & = & \half {\rm Re}\left(\int \int d\phi_x d\phi_y  
\exp\left[i\{ 2\phi_x - 3\Dl \phi_x   + z_{xx}J_x\sin 2 \phi_{x, - \Dl\phi_x} 
- z_{xy}J_y  \sin 2 \phi_{y, - \Dl\phi_y} - c_{00} \}\right]\right) \\
& = & 2\pi^2 {\rm Re}\left( \exp[i( -3\Dl \phi_x - c_{00}) ] J_{-1}(z_{xx}J_x) \exp[i 2 \Dl\phi_x] J_0(z_{xy}J_y) \right) \\
& = & -2\pi^2 {\rm Re}\left( \exp[i( -\Dl \phi_x - c_{00}) ] J_1(z_{xx}J_x) J_0(z_{xy}J_y) \right)
\eeqrs
This will be sub-dominant and can be dropped. Combining the 3 dominant terms 
\beqr
\lan x(t) \ran & \equiv & -\fr{2 \bt_K \theta_x }{(2\pi)^2 \eps_x^2 \eps_y} \int d J_x dJ_y \;  J_x
\exp[-\frac{J_x}{\eps_x} - \frac{J_y}{\eps_y}]  \left\{ \Phi_1 - \Phi_3 + \Phi_4 \right\}  \non \\
& = & \fr{ \bt_K \theta_x }{\eps_x^2 \eps_y} {\rm Im}\left\{  \int d J_x dJ_y \;  J_x
\exp[-\frac{J_x}{\eps_x} - \frac{J_y}{\eps_y}]  \exp[ i(\Dl\phi_x  - c_{00} )]  J_{1}(z_{xx}J_x)  J_{0}(z_{xy}J_y) \right\}  \non \\
&  &  + \fr{ \bt_K \theta_x }{\eps_x^2 \eps_y} {\rm Re}\left\{  \int d J_x dJ_y \;  J_x \exp[ i(\Dl\phi_x  - c_{00} )]
[  J_{0}(z_{xx}J_x)  - J_{2}(z_{xx}J_x) ] J_0(z_{xy}J_y) \right\}
\eeqr
The 1D theory had shown that the 2nd and 3rd terms, i.e the contributions from $\Phi_3, \Phi_4$ are about 10\% of that
from the 1st term, so they will be dropped.
Hence
\beqr
\lan x(t) \ran & \simeq &  \fr{ \bt_K \theta_x }{\eps_x^2 \eps_y} {\rm Im}\left\{  \int d J_x dJ_y \;  J_x
\exp[-\fr{J_x}{\eps_x} - \fr{J_y}{\eps_y}]  \exp[ i(\Dl\phi_x  - c_{00} )]  J_{1}(z_{xx}J_x)  J_{0}(z_{xy}J_y) \right\}  \non \\
& = &  \fr{ \bt_K \theta_x }{\eps_x^2 \eps_y} {\rm Im}\left\{  e^{i[\om_{x,0} ](t - 2 \tau) -   q_x/2}
\left[ \int d J_x \;  J_x \exp[-\fr{J_x}{\eps_x} + iw_{xx}J_x(t-2\tau)] J_{1}(z_{xx}J_x)   \right]  \right. \\
& &  \times \left. \left[ dJ_y \exp[-\fr{J_y}{\eps_y} + i w_{xy}J_y(t-2\tau)] J_{0}(z_{xy}J_y) \right]  \right\}
\eeqr
Define the dimensionless variables
\beqrs
u_x & = & J_x/\eps_x, \;\;\; u_y = J_y/\eps_y, \;\;\; \xi_x(t) = w_{xx}\eps_x(t-2\tau),\;\;\;  \xi_y(t) = w_{xy}\eps_y(t-2\tau), \\
a_x & = & 1 - i\xi_x, \;\;\; a_y = 1 - i\xi_y, \;\;\;  Q_x = z_{xx}\eps_x, \;\;\; Q_y = z_{xy}\eps_y,  \\
\Phi_x(t) & = & \om_{x, 0}(t - 2 \tau)
\eeqrs
We have
\beqr
\lan x(t) \ran  & = &  \bt_K \theta_x {\rm Im}\left\{  e^{i(\Phi_x - q_x/2)}\left[ \int du_x \; u_x \exp[-a_x u_x] J_1(Q_x u_x) \right]
\left[ \int du_y \; \exp[-a_y u_y] J_0(Q_y u_y) \right] \right\} \non  \\
& = &  \bt_K \theta_x {\rm Im}\left\{  e^{i(\Phi_x - q_x/2)} H_{1,1}(a_x, Q_x) H_{0, 0}(a_y, Q_y) \right\}
\eeqr
where. as in the PR-AB paper, the functions $H_{m, n}(a, Q)$ are defined as
\[ H_{m, n}(a, Q) = \int_0^{\infty} du \; u^m \exp[-a u] J_n(Q u) \]
The expression for the centroid is the same as in Eq.(2.23) in the PR-AB paper apart from the multiplication by the $y$
dependent term $H_{0,0}(a_y, Q_y)$. We have
\beqrs
H_{0, 0}(a, Q) & = &  \fr{1}{[a^2 + Q^2]^{1/2}}, \;\;\;\; H_{1, 1}(a, Q)  =   \fr{Q}{[a^2 + Q^2]^{3/2}}
\eeqrs
Hence
\beqr
\lan x(t) \ran  & = &  \bt_K \theta_x Q_x {\rm Im}\left\{  e^{i(\Phi_x - q_x/2)}\fr{1}{[a_x^2 + Q_x^2]^{3/2}}\fr{1}{[a_y^2 + Q_y^2]^{1/2}}\right\}
\eeqr
Writing
\beqr
 [a_x^2 + Q_x^2]^{3/2} & = &   [(1 - \xi_x)^2 + Q_x^2]^{3/2} \equiv A_{1, x}\exp[ - \fr{3}{2} i \Theta_x] \\
 A_{1, x} & = &  [ (1 - \xi_x^2 + Q_x^2)^2 + 4 \xi_x^2]^{3/4}, \;\;\; \Theta_x(t) = {\rm Arctan}[\fr{2 \xi_x}{1  - \xi_x^2 + Q_x^2}] \\
 (a_y^2 + Q_y^2)^{1/2} & = &  [(1 - \xi_y)^2 + Q_y^2]^{1/2} \equiv A_{1, y} \exp[ - \half i \Theta_y] \\
  A_{1, y} & = &  [ (1 - \xi_y^2 + Q_y^2)^2 + 4 \xi_y^2]^{1/4}, \;\;\; \Theta_y(t) = {\rm Arctan}[\fr{2 \xi_y}{1  - \xi_y^2 + Q_y^2}]
\eeqr
Hence
\beqr
\lan x(t) \ran  & = &  \bt_K \theta_x Q_x \fr{1}{A_{1, x}A_{1, y}}{\rm Im}\left\{  e^{i(\Phi_x - q_x/2)}
\exp[i(\fr{3}{2}\Theta_x + \half \Theta_y)]
\right\} \non\\
& = & \bt_K \theta_x \fr{Q_x }{A_{1, x}A_{1, y}} \sin(\Phi_x(t) + \fr{3}{2}\Theta_x(t) + \half \Theta_y(t) - \half q_x)
\eeqr
The time dependent amplitude of the echo pulse is
\beq
\lan x(t)_{Amp} \ran  = \bt_K \theta_x \fr{Q_x }{A_{1, x}(t) A_{1, y}(t)} =
\bt_K \theta_x \fr{Q_x }{[ (1 - \xi_x(t)^2 + Q_x^2)^2 + 4 \xi_x(t)^2]^{3/4}[ (1 - \xi_y(t)^2 + Q_y^2)^2 + 4 \xi_y(t)^2]^{1/4}}
\eeq
The amplitude has a maximum when $\xi_x(t) = 0 = \xi_y(t)$ or at  $t = 2\tau$, and we have for the echo amplitude
\beq
\lan x(2\tau) \ran^{echo amp} = \bt_K \theta_x \fr{Q_x }{[1 + Q_x^2]^{3/2} [1 + Q_y^2]^{1/2}}
\eeq
This only modifies the 1D expression by the multiplicative factor $ [1 + Q_y^2]^{-1/2}$.
Note that $Q_x, Q_y$ are not truly independent variables, since they both refer to the same quadrupole. 
Unfortunately, this expression does not depend on the tunes $\nu_x, \nu_y$.

\subsubsection{FWHM of the echo pulse}

At a time $t = 2\tau + \Dl t_H$, the echo drops to half its maximum amplitude. At this time, we have
\[ \xi_x(t) = w_{xx} \eps_x \Dl t_H,  \;\;\;  \xi_y(t) = w_{xy} \eps_y \Dl t_H  \]

In the first approximation, I will assume the linear quadrupole theory for the echo amplitude, which is valid as long as the
quad strength $Q_x \le Q_{x, opt}$ where $Q_{x, opt}$ is the strength that maximizes the echo. Initially I will also assume that
the echo falls to half the maximum when the amplitude falls to half the maximum amplitude.
In the linear approximation, we have $ \lan x(2\tau) \ran^{echo amp} = \bt_K \theta_x Q_x$ and the amplitude falls to half maximum
at
\beqrs
\fr{1}{[ (1 - \xi_x(t)^2 )^2 + 4 \xi_x(t)^2]^{3/4}[ (1 - \xi_y(t)^2 )^2 + 4 \xi_y(t)^2]^{1/4}}  & = & \half    \\
\Rarw \fr{1}{[ (1 + \xi_x(t)^2]^{3/2}[ (1 + \xi_y(t)^2]^{1/2}}  & = &  \half  \\
\Rarw [1 +  \xi_x(t)^2 ][ 1 + \xi_y^2]^{1/3} & = & 2^{2/3}
\eeqrs
This equation for $\Dl t_H$ depends only on the emittances $\eps_x, \eps_y$ and the detuning parameter $w_{xx}, w_{xy}$ and is independent of the other parameters such as the delay and the quad strength.

If we have $\xi_y(t) \ll 1$ (either because $\eps_y/\eps_x \ll 1$, true for electron beams or because $w_{xy} \ll w_{xx}$), we have
the approximation
\beqrs
(1 + \xi_x(t= 2\tau + \Dl t_H)^2)[ 1 + \fr{1}{3}\xi_y(t= 2\tau + \Dl t_H)^2 + ...] & = & 2^{2/3} \\
\sqrt{ (w_{xx} \eps_x)^2 + \fr{1}{3}(w_{xy}\eps_y)^2} \Dl t_H &  \approx &  \sqrt{2^{2/3} - 1}
\eeqrs
where we also dropped the term $(w_{xx} \eps_x)^2 (w_{xy} \eps_y)^2 \Dl t_H^4$. 
The FWHM in this (effectively 1D case) is
\beq
\Dl t_{FWHM} = 2 \Dl t_H \approx 2 \sqrt{\fr{(2^{2/3} - 1)}{(w_{xx} \eps_x)^2 + (1/3)(w_{xy} \eps_y)^2}}
\eeq
This approximation could be slightly improved by keeping the term dropped $(w_{xx} \eps_x)^2 (w_{xy} \eps_y)^2 \Dl t_H^4$
and solving a quadratic equation for $\Dl t_H^2$, This would be apple polishing at best and would not introduce dependences on
other parameters.

A more qualitative improvement would be to include the phase dependent part. 
In the linear quadrupole theory, we can simplify the phases as
\beqrs
\Theta_x(t) & = & {\rm Arctan}[\fr{2 \xi_x}{1  - \xi_x^2}] = 2 {\rm Arctan}[\xi_x(t)] \approx 2\xi_x(t) \;\; {\rm if}\; \xi_x \ll 1  \\
\Theta_y(t) & = & {\rm Arctan}[\fr{2 \xi_y}{1  - \xi_y^2}] = 2 {\rm Arctan}[\xi_y(t)] \approx 2\xi_y(t) \;\; {\rm if}\; \xi_y \ll 1  \\
\eeqrs
Let $t_H = 2\tau + \Dl t_H$. Thus at the half max time, we have instead
\beqr
\fr{1}{[(1 + \xi_x(t_H)^2]^{3/2}[ (1 + \xi_y(t_H)^2]^{1/2}}\sin[\om_{x,0}\Dl t_H + \fr{3}{2}\Theta_x(t_H) +  \half \Theta_y(t_H) - \half q_x] & = &  \half
\eeqr
where $\xi_x(t_H) = w_{xx}\eps_x \Dl t_H, \xi_y(t_H) = w_{xy}\eps_y \Dl t_H$. While this equation most likely has to be solved
numerically, this equation also shows that under the approximations made above, the FWHM does not depend on the quadrupole
strength or the delay time but on the detunings $w_{xx}, w_{xy}$ and the emittances $\eps_x, \eps_y$.

\subsubsection{The $\lan y \ran$ moment}

Let me now calculate $\lan y\ran$. 
\beq
\lan y(t) \ran =  \int d J_x dJ_y d\phi_x d\phi_y \sqrt{2\bt_y J_y} \cos \phi_y   \psi_5(J_x, \phi_x, J_y, \phi_y, t)
\eeq
Substituting for $\psi_5$ from Eq. \ref{eq: psi5_2D}, using the same notation and keeping terms to $O(q_x)$, we have
\beqrs
\lan y(t) \ran & \equiv & -\fr{ 2 \bt_K \theta_x }{(2\pi)^2 \eps_x^2 \eps_y} \sqrt{\fr{\bt_y }{\bt_x}}
\int d J_x dJ_y d\phi_x d\phi_y \sqrt{J_x  J_y} \cos \phi_y   \\
&  & \times [1- \fr{3}{2} q_x \sin 2\phi_{x,-\Dl \phi_x}]
 \exp[-\frac{J_x[1- q_x \sin 2\phi_{x,-\Dl \phi_x}]}{\eps_x} - \frac{J_y[1+ q_y \sin 2\phi_{y,-\Dl \phi_y}]}{\eps_y}] \\
& & \times
\sin( \phi_{x,-\Dl\phi_x} + c_{xx}\sin 2 \phi_{x, - \Dl\phi_x} - c_{xy}\sin 2 \phi_{y, - \Dl\phi_y} - c_{00} )
\eeqrs
Combining trigonometric terms,
\beqrs
& \mbox{} & \cos\phi_y \sin( \phi_{x,-\Dl\phi_x} + c_{xx}\sin 2 \phi_{x, - \Dl\phi_x} - c_{xy}\sin 2 \phi_{y, - \Dl\phi_y} - c_{00} )  \\
&  & \times [1- \fr{3}{2} q_x \sin 2\phi_{x,-\Dl \phi_x}]  \\
& =  & \half \left( \sin[ \phi_y + \phi_x - \Dl\phi_x + c_{xx}\sin 2 \phi_{x, - \Dl\phi_x} - c_{xy}\sin 2 \phi_{y, - \Dl\phi_y} - c_{00} ]  \right. \\
& &  \left. + \sin[\phi_x - \phi_y - \Dl\phi_x + c_{xx}\sin 2 \phi_{x, - \Dl\phi_x} - c_{xy}\sin 2 \phi_{y, - \Dl\phi_y} - c_{00} ] \right) \\
&  &  \times (1 - \fr{3}{2}q_x \sin 2\phi_{x, - \Dl \phi_x}) \\
& =  &  \half(\sin\phi_{1y} + \sin\phi_{2y}) - \fr{3}{4}q_x \left[\cos(\phi_{1y} - 2 \phi_{x, - \Dl \phi_x}) - \cos(\phi_{1y} + 2 \phi_{x, - \Dl \phi_x}) \right. \\  &  &  \left. + \cos(\phi_{2y} - 2 \phi_{x, - \Dl \phi_x}) - \cos(\phi_{2y} + 2 \phi_{x, - \Dl \phi_x}) \right] 
\eeqrs
where 
\beqrs
 \phi_{1y} & =  & \phi_x + \phi_y - \Dl\phi_x + c_{xx}\sin 2 \phi_{x, - \Dl\phi_x} - c_{xy}\sin 2 \phi_{y, - \Dl\phi_y} - c_{00}  \\
\phi_{2y} & = & \phi_x - \phi_y - \Dl\phi_x + c_{xx}\sin 2 \phi_{x, - \Dl\phi_x} - c_{xy}\sin 2 \phi_{y, - \Dl\phi_y} - c_{00} \\
\phi_{3y} & \equiv & \phi_{1y} - 2 \phi_{x, - \Dl \phi_x} = \phi_y - \phi_x + \Dl \phi_x + c_{xx}\sin 2 \phi_{x, - \Dl\phi_x} - c_{xy}\sin 2 \phi_{y, - \Dl\phi_y} - c_{00} \\
\phi_{4y} & \equiv & \phi_{1y} + 2 \phi_{x, - \Dl \phi_x} = 3\phi_x + \phi_y - 3\Dl \phi_x + c_{xx}\sin 2 \phi_{x, - \Dl\phi_x} - c_{xy}\sin 2 \phi_{y, - \Dl\phi_y} - c_{00} \\
\phi_{5y} & \equiv & \phi_{2y} - 2 \phi_{x, - \Dl \phi_x}  =  -\phi_x - \phi_y + \Dl \phi_x + c_{xx}\sin 2 \phi_{x, - \Dl\phi_x} - c_{xy}\sin 2 \phi_{y, - \Dl\phi_y} - c_{00} \\
\phi_{6y} & \equiv & \phi_{2y} + 2 \phi_{x, - \Dl \phi_x}  =  3\phi_x - \phi_y  - 3\Dl \phi_x + c_{xx}\sin 2 \phi_{x, - \Dl\phi_x} - c_{xy}\sin 2 \phi_{y, - \Dl\phi_y} - c_{00}
\eeqrs
As in the case for $\lan x\ran $, we now have
\beqr
\lan y(t) \ran & = & -\fr{ 2 \bt_K \theta_x }{(2\pi)^2 \eps_x^2 \eps_y} \sqrt{\fr{\bt_y }{\bt_x}}
\int d J_x dJ_y d\phi_x d\phi_y \sqrt{J_x  J_y} \cos \phi_y   \non \\
&  & \times  \exp[\frac{ q_x  J_x\sin 2\phi_{x,-\Dl \phi_x}}{\eps_x} - \frac{q_y  J_y \sin 2\phi_{y,-\Dl \phi_y}}{\eps_y}] \non \\
& & \times  \left\{ \half(\sin\phi_{1y} + \sin\phi_{2y}) - \fr{3}{4}q_x \left[\cos\phi_{3y}  - \cos\phi_{4y} + \cos\phi_{5y}  - \cos\phi_{6y} \right]  \right\}  \non  \\
& \equiv & -\fr{2 \bt_K \theta_x }{(2\pi)^2 \eps_x^2 \eps_y} \sqrt{\fr{\bt_y }{\bt_x}} \int d J_x dJ_y \;   \sqrt{J_x  J_y}
\exp[-\frac{J_x}{\eps_x} - \frac{J_y}{\eps_y}]  \left\{ \Phi_{1y} + \Phi_{2y} - \Phi_{3y} + \Phi_{4y} - \Phi_{5y} + \Phi_{6y} \right\}
\label{eq: ymom_1}
\eeqr
where
\beqr
\Phi_{1y}(J_x, J_y) & = & \half \int \int d\phi_x d\phi_y  \exp[\frac{ q_x  J_x}{\eps_x} \sin 2\phi_{x,-\Dl \phi_x} -
  \frac{q_y  J_y}{\eps_y} \sin 2\phi_{y,-\Dl \phi_y}] \sin\phi_{1y}  \\
\Phi_{2y}(J_x, J_y) & = & \half \int \int d\phi_x d\phi_y  \exp[\fr{ q_x  J_x}{\eps_x}\sin 2\phi_{x,-\Dl \phi_x} - 
  \frac{q_y  J_y }{\eps_y}\sin 2\phi_{y,-\Dl \phi_y}] \sin\phi_{2y}  \\
\Phi_{3y}(J_x, J_y) & = & \fr{3}{4}q_x \int \int d\phi_x d\phi_y  \exp[\frac{ q_x  J_x}{\eps_x}\sin 2\phi_{x,-\Dl \phi_x} -
  \frac{q_y  J_y}{\eps_y}\sin 2\phi_{y,-\Dl \phi_y} ] \cos\phi_{3y}  \\
\Phi_{4y}(J_x, J_y) & = & \fr{3}{4}q_x \int \int d\phi_x d\phi_y  \exp[\frac{ q_x  J_x}{\eps_x}\sin 2\phi_{x,-\Dl \phi_x} -
  \frac{q_y  J_y }{\eps_y}\sin 2\phi_{y,-\Dl \phi_y}] \cos\phi_{4y} \\
\Phi_{5y}(J_x, J_y) & = & \fr{3}{4}q_x \int \int d\phi_x d\phi_y  \exp[\frac{ q_x  J_x}{\eps_x } \sin 2\phi_{x,-\Dl \phi_x} -
  \frac{q_y  J_y }{\eps_y}\sin 2\phi_{y,-\Dl \phi_y}] \cos\phi_{5y}  \\
\Phi_{6y}(J_x, J_y) & = & \fr{3}{4}q_x \int \int d\phi_x d\phi_y  \exp[\frac{ q_x  J_x}{\eps_x}\sin 2\phi_{x,-\Dl \phi_x} -
  \frac{q_y  J_y }{\eps_y}\sin 2\phi_{y,-\Dl \phi_y}] \cos\phi_{6y}
\eeqr
Thus the integrands in the terms $\Phi_{jy}$, j=1,... 6 are respectively,
\beqrs 
    {\rm Int}(\Phi_{1y}) & = &  \exp[\fr{ q_x  J_x}{\eps_x} \sin 2\phi_{x,-\Dl \phi_x} -  \fr{q_y  J_y}{\eps_y} \sin 2\phi_{y,-\Dl \phi_y}]
    \sin\phi_{1y} \\
  & = & {\rm Im}\left(\exp[i \{ \phi_{1y} - i \fr{ q_x  J_x}{\eps_x} \sin 2\phi_{x,-\Dl \phi_x} - i\fr{q_y  J_y}{\eps_y} \sin 2\phi_{y,-\Dl \phi_y}
    \}] \right) \\
& = & {\rm Im}\left(\exp\left[i\{\phi_x + \phi_y  - \Dl\phi_x  - c_{00}\}\right]
\exp\left[ i\{ z_{xx}J_x\sin 2 \phi_{x, - \Dl\phi_x} -    z_{xy}J_y\sin 2 \phi_{y, - \Dl\phi_y}\} \right] \right)  \\
{\rm Int}(\Phi_{2y})  & = & \exp[\fr{ q_x  J_x}{\eps_x} \sin 2\phi_{x,-\Dl \phi_x} -  \fr{q_y  J_y}{\eps_y} \sin 2\phi_{y,-\Dl \phi_y}] \sin\phi_{2y} \\
& = & {\rm Im}\left(\exp[i \{ \phi_{2y} - i \fr{ q_x  J_x}{\eps_x} \sin 2\phi_{x,-\Dl \phi_x} - i\fr{q_y  J_y}{\eps_y} \sin 2\phi_{y,-\Dl \phi_y} \}
] \right) \\
&  = & {\rm Im}\left(\exp\left[i\{\phi_x - \phi_y - \Dl\phi_x  - c_{00}\}\right] \exp\left[ i\{ z_{xx}J_x\sin 2 \phi_{x, - \Dl\phi_x} -    z_{xy}J_y\sin 2 \phi_{y, - \Dl\phi_y}\} \right] \right)  \\
{\rm Int}(\Phi_{3y}) & = &  \exp[\fr{ q_x  J_x}{\eps_x} \sin 2\phi_{x,-\Dl \phi_x} -  \fr{q_y  J_y}{\eps_y} \sin 2\phi_{y,-\Dl \phi_y}]\cos\phi_{3y} \\
& = &{\rm Re}\left(\exp\left[i\{\phi_y - \phi_x + \Dl \phi_x + z_{xx}J_x\sin 2 \phi_{x, - \Dl\phi_x} - z_{xy} J_y \sin 2 \phi_{y, - \Dl\phi_y} - c_{00} \}\right]\right)  \\
{\rm Int}(\Phi_{4y})  & = &  \exp[\fr{ q_x  J_x}{\eps_x} \sin 2\phi_{x,-\Dl \phi_x} -  \fr{q_y  J_y}{\eps_y} \sin 2\phi_{y,-\Dl \phi_y}]
        \cos\phi_{4y} \\
  & = &  {\rm Re}\left(\exp\left[i\{3\phi_x + \phi_y  - 3\Dl \phi_x  + z_{xx}J_x\sin 2 \phi_{x, - \Dl\phi_x} 
- z_{xy} J_y \sin 2 \phi_{y, - \Dl\phi_y} - c_{00} \}\right]\right)  \\
 {\rm Int}(\Phi_{5y}) & = &   \exp[\fr{ q_x  J_x}{\eps_x} \sin 2\phi_{x,-\Dl \phi_x} -  \fr{q_y  J_y}{\eps_y} \sin 2\phi_{y,-\Dl \phi_y}]
  \cos\phi_{5y}  \\
& = &  {\rm Re}\left(\exp\left[i\{-\phi_x -\phi_y + \Dl \phi_x  + z_{xx}J_x\sin 2 \phi_{x, - \Dl\phi_x}
- z_{xy}J_y  \sin 2 \phi_{y, - \Dl\phi_y} - c_{00} \}\right]\right)  \\
 {\rm Int}(\Phi_{6y}) & =  & \exp[\fr{ q_x  J_x}{\eps_x} \sin 2\phi_{x,-\Dl \phi_x} -  \fr{q_y  J_y}{\eps_y} \sin 2\phi_{y,-\Dl \phi_y}]
            \cos\phi_{6y} \\
    & = &  {\rm Re}\left(\exp\left[i\{ 3\phi_x - \phi_y - 3\Dl \phi_x   + z_{xx}J_x\sin 2 \phi_{x, - \Dl\phi_x} 
- z_{xy}J_y  \sin 2 \phi_{y, - \Dl\phi_y} - c_{00} \}\right]\right) 
\eeqrs
No new additional complex parameters need to be introduced. 

Again using
\[ \int d\phi \exp[i m \phi] \exp[i a \sin(2\phi - \theta)] = 2\pi J_{-m/2}(a)\exp[i (m/2) \theta] \]
The difference with the integrations over $\phi_x, \phi_y$ for the $\lan y \ran$ moment compared to those for the $\lan x \ran$
moment are that the integers $m$ are odd. Thus
\beqr
\Phi_{1y}(J_x, J_y) & = & \half {\rm Im}\left( \int \int d\phi_x d\phi_y  
\exp\left[i\{\phi_x + \phi_y - \Dl\phi_x  - c_{00}\}\right]
\exp\left[ i\{ z_{xx}J_x\sin 2 \phi_{x, - \Dl\phi_x} -    z_{xy}J_y\sin 2 \phi_{y, - \Dl\phi_y}\} \right] \right)  \non \\
& = & \half {\rm Im}\left( \int d\phi_x  \exp\left[i\{\phi_x  - \Dl\phi_x  - c_{00}\}\right]
\exp\left[ i\{ z_{xx}J_x\sin 2 \phi_{x, -\Dl\phi_x} \} \right]  \right.  \non \\
&  & \left. \times \int d\phi_y \exp\left[ i\{\phi_y - z_{xy}J_y\sin 2 \phi_{y, - \Dl\phi_y} \} \right]  \right)  \non \\
&  = & 2\pi^2 {\rm Im}\left( \exp[- i(\Dl\phi_x + c_{00})]  J_{-1/2}(z_{xx}J_x)\exp[i \Dl\phi_x]  J_{-1/2}(-z_{xy}) \exp[i\Dl\phi_y] \right) \non \\
&  = & 2\pi^2{\rm Im}\left( \exp[ i(\Dl\phi_y - c_{00})] J_{-1/2}(z_{xx}J_x) J_{-1/2}(-z_{xy}J_y)  \right)
\eeqr
The phase factor is
\beqrs
\varphi_1(t) & = & \Dl \phi_y - c_{00}  =  [\om_{y0} + w_{xy}J_x + w_{yy}J_y](t - \tau) - [\om_{x0} + w_{xx}J_x + w_{xy}J_y]\tau - \half q_x \\
& = & \om_{y0}(t-\tau) - \om_{x0}\tau - w_{xx}J_x \tau + w_{xy}[J_x(t - \tau) - J_y \tau]  + w_{yy}J_y (t - \tau)  - \half q_x
\eeqrs
This phase factor does not in general become small at any particular time $t$, thus there is no echo in the $y$ plane.
At $t = 2\tau$, this phase factor is
\beq
\varphi_1(2\tau) = \left(\om_{y0} - \om_{x0} - w_{xx}J_x + w_{xy}[J_x - J_y ]  + w_{yy}J_y \right) \tau  - \half q_x
\eeq
If the detuning terms $w_{xx}J_x, ....$ are small compared to the nominal tunes, then we can approximate
\[ \varphi_1(2\tau) \approx \left(\om_{y0} - \om_{x0} \right) \tau  \]
which behaves as
\[ \varphi_1(2\tau) \rarw 0, \;\;\; {\rm when} \;\;\; \om_{y0} \rarw \om_{x0} \]
i.e. as the betatron tunes approach each other. 

It appears that none of the factors $\Phi_{jy}$ will have a vanishing phase, which would require that all have to be evaluated.
\beqrs
\Phi_{2y}(J_x, J_y) & = & \half  {\rm Im}\left(  \int \int d\phi_x d\phi_y
\exp\left[i\{\phi_x - \phi_y - \Dl\phi_x  - c_{00}\}\right] \exp\left[ i\{ z_{xx}J_x\sin 2 \phi_{x, - \Dl\phi_x} -    z_{xy}J_y\sin 2 \phi_{y, - \Dl\phi_y}\} \right] \right)  \\
&  = & 2\pi^2 {\rm Im}\left( \exp[- i(\Dl\phi_x+ c_{00})]  J_{-1/2}(z_{xx}J_x)\exp[i \Dl\phi_x]  J_{1/2}(-z_{xy}J_y) \exp[-i\Dl\phi_y] \right) \non \\
&  = & 2\pi^2{\rm Im}\left( \exp[ i(-\Dl\phi_y - c_{00})] J_{-1/2}(z_{xx}J_x) J_{-1/2}(-z_{xy}J_y)  \right)
\eeqrs
The phase factor is
\beqrs
\varphi_2(t) & = & -\Dl \phi_y - c_{00}  =  -[\om_{y0} + w_{xy}J_x + w_{yy}J_y](t - \tau) - [\om_{x0} + w_{xx}J_x + w_{xy}J_y]\tau - \half q_x \\
& = & -\om_{y0}(t-\tau) + \om_{x0}\tau + w_{xx}J_x \tau + w_{xy}[J_x(t - \tau) + J_y \tau]  + w_{yy}J_y (t - \tau)  - \half q_x 
\eeqrs
At the time of the x echo,
\[ \varphi_2(2\tau) = -\left(\om_{y0} + \om_{x0} - w_{xx}J_x + w_{xy}[J_x+  J_y ]  + w_{yy}J_y \right) \tau  - \half q_x \]
I do not see conditions under which $\varphi_2 \rarw 0$, so this term $\Phi_{2y}$   could be dropped.

\beqr
\Phi_{3y}(J_x, J_y) & = & \half {\rm Re}\left(  \int \int d\phi_x d\phi_y
\exp\left[i\{\phi_y - \phi_x + \Dl \phi_x + z_{xx}J_x\sin 2 \phi_{x, - \Dl\phi_x} - z_{xy} J_y \sin 2 \phi_{y, - \Dl\phi_y} - c_{00} \}\right]\right)  \\
& = & 2\pi^2 {\rm Re}\left( \exp[ i(\Dl\phi_x - c_{00})]  J_{1/2}(z_{xx}J_x)\exp[- i\Dl\phi_x]  J_{-1/2}(-z_{xy}J_y) \exp[i\Dl\phi_y] \right) \non \\
& = &  2\pi^2 {\rm Re}\left( \exp[ i(\Dl\phi_y - c_{00})]  J_{1/2}(z_{xx}J_x)  J_{-1/2}(-z_{xy}J_y) \right) \non \\
\eeqr

\beqr
\Phi_{4y}(J_x, J_y) & = & \half {\rm Re}\left(  \int \int d\phi_x d\phi_y
\exp\left[i\{3\phi_x + \phi_y  - 3\Dl \phi_x  + z_{xx}J_x\sin 2 \phi_{x, - \Dl\phi_x} 
  - z_{xy} J_y \sin 2 \phi_{y, - \Dl\phi_y} - c_{00} \}\right]\right)  \non \\
& = & 2\pi^2 {\rm Re}\left( \exp[ i(-3\Dl\phi_x - c_{00})]  J_{3/2}(z_{xx}J_x)\exp[3 i\Dl\phi_x]  J_{1/2}(-z_{xy}J_y) \exp[i\Dl\phi_y] \right)
\non \\
& = & 2\pi^2 {\rm Re}\left( \exp[i\varphi_1(t)]  J_{3/2}(z_{xx}J_x)  J_{1/2}(-z_{xy}J_y) \right)
\eeqr

\beqr
\Phi_{5y}(J_x, J_y) & = & \half {\rm Re}\left(  \int \int d\phi_x d\phi_y
\exp\left[i\{-\phi_x -\phi_y + \Dl \phi_x  + z_{xx}J_x\sin 2 \phi_{x, - \Dl\phi_x}
- z_{xy}J_y  \sin 2 \phi_{y, - \Dl\phi_y} - c_{00} \}\right]\right)  \non \\
& = & 2\pi^2 {\rm Re}\left( \exp[ i(\Dl\phi_x - c_{00})]  J_{-1/2}(z_{xx}J_x)\exp[- i\Dl\phi_x]  J_{-1/2}(-z_{xy}J_y) \exp[-i\Dl\phi_y] \right)
\non \\
& = & 2\pi^2 {\rm Re}\left( \exp[ i\varphi_2(t)]J_{-1/2}(z_{xx}J_x)J_{-1/2}(-z_{xy}J_y)  \right)
\eeqr
The phase factor is the same as in $\Phi_{2y}$, hence $\Phi_{5y}$ could also be dropped.

\beqr
\Phi_{6y}(J_x, J_y) & = & \half {\rm Re}\left(  \int \int d\phi_x d\phi_y
\exp\left[i\{ 3\phi_x - \phi_y - 3\Dl \phi_x   + z_{xx}J_x\sin 2 \phi_{x, - \Dl\phi_x} 
  - z_{xy}J_y  \sin 2 \phi_{y, - \Dl\phi_y} - c_{00} \}\right]\right)  \non \\
& = & 2\pi^2 {\rm Re}\left( \exp[ i(-3\Dl\phi_x - c_{00})]  J_{3/2}(z_{xx}J_x)\exp[3 i\Dl\phi_x]  J_{-1/2}(-z_{xy}J_y) \exp[-i\Dl\phi_y] \right)
\non \\
& = & 2\pi^2 {\rm Re}\left( \exp[ i\varphi_2(t)]J_{3/2}(z_{xx}J_x)J_{-1/2}(-z_{xy}J_y) \right)
\eeqr
$\Phi_{6y}$ can also be dropped.

Dropping $\Phi_{2y}, $
and substituting into Eq.(\ref{eq: ymom_1}), we have
\beqr
\lan y(t) \ran & = & -\fr{2 \bt_K \theta_x }{(2\pi)^2 \eps_x^2 \eps_y} \sqrt{\fr{\bt_y }{\bt_x}} \int d J_x dJ_y \;   \sqrt{J_x  J_y}
\exp[-\frac{J_x}{\eps_x} - \frac{J_y}{\eps_y}]  \left\{ \Phi_{1y} - \Phi_{3y} + \Phi_{4y} \right\}  \non \\
& = &  -\fr{\bt_K \theta_x }{ \eps_x^2 \eps_y} \sqrt{\fr{\bt_y }{\bt_x}} \int d J_x dJ_y \;   \sqrt{J_x  J_y}
\exp[-\frac{J_x}{\eps_x} - \frac{J_y}{\eps_y}]  \non \\
&  & \times \left\{ {\rm Im}\left( \exp[ i \varphi_1(t)] J_{-1/2}(z_{xx}J_x) J_{-1/2}(-z_{xy}J_y)  \right)
- {\rm Re}\left( \exp[ i \varphi_1(t)]  J_{1/2}(z_{xx}J_x)  J_{-1/2}(-z_{xy}J_y) \right) \non \right. \\
&  &  \left. + {\rm Re}\left( \exp[i\varphi_1(t)]  J_{3/2}(z_{xx}J_x)  J_{1/2}(-z_{xy}J_y) \right) \right\}
\eeqr
At this point it is not clear to me that the contributions from $\Phi_{3y}, \Phi_{4y}$ can be dropped, so I need to evaluate all three
terms. First, write
\beqr
\varphi_1(t) & = &  \om_{y0}(t-\tau) - \om_{x0}\tau - w_{xx}J_x \tau + w_{xy}[J_x(t - \tau) - J_y \tau]  + w_{yy}J_y (t - \tau)  - \half q_x
\non \\
& = & \om_{y0}(t-\tau) - \om_{x0}\tau - \half q_x - [w_{xx}\tau - w_{xy}(t - \tau)]J_x  - [w_{xy}\tau - w_{yy}(t - \tau)] J_y
\eeqr
Introduce the scaled variables
\beqr
u_x & = & J_x/\eps_x, \;\;\; u_y  =  J_y/\eps_y, \;\;\;  \xi_{xy, 1}(t) = [w_{xx}\tau - w_{xy}(t - \tau)] \eps_x  \non \\
\xi_{xy,2} & = & [w_{xy}\tau - w_{yy}(t - \tau)]\eps_y, \;\;\; \Phi_{xy}(t) = \om_{y0}(t-\tau) - \om_{x0}\tau \non \\
Q_x & = & z_{xx}\eps_x \simeq q_x w_{xx}\eps_x \tau, \;\;\; Q_y  =  z_{xy}\eps_y\simeq q_y w_{xy}\eps_y \tau   \non \\
\varphi_1(t) & = & \Phi_{xy} - \half q_x - \xi_{xy,1}u_x -  \xi_{xy,2}u_y
\eeqr

The three integrals are of the form
\beqr
I_1 & = & {\rm Im}\left[ \left( \int du_x \sqrt{u_x}\exp[ -u_x -i \xi_{xy,1}u_x] J_{-1/2}(Q_{x}u_x) \right)
\left( \int du_y \sqrt{u_y}\exp[ -u_y - i \xi_{xy,2}u_y] J_{-1/2}(-Q_{y}u_y) \right)\right] \non \\
& = & {\rm Im}\left[ \left( \int du_x \sqrt{u_x}\exp[ - a_{xy,1}u_x] J_{-1/2}(Q_{x}u_x) \right)
\left( \int du_y \sqrt{u_y}\exp[ -a_{xy,2}u_y] J_{-1/2}(-Q_{y}u_y) \right) \right]  \\
I_3 & = & {\rm Re}\left[ \left( \int du_x \sqrt{u_x}\exp[ - a_{xy,1}u_x] J_{1/2}(Q_{x}u_x) \right)
\left( \int du_y \sqrt{u_y}\exp[ -a_{xy,2}u_y] J_{-1/2}(-Q_{y}u_y) \right) \right] \\
I_4 & = & {\rm Re}\left[ \left( \int du_x \sqrt{u_x}\exp[ - a_{xy,1}u_x] J_{3/2}(Q_{x}u_x) \right)
\left( \int du_y \sqrt{u_y}\exp[ -a_{xy,2}u_y] J_{-1/2}(-Q_{y}u_y) \right) \right] \\
a_{xy,1} & = &  1 + i \xi_{xy,1},  \;\;\;\;  a_{xy,2} =  1 + i \xi_{xy,2}  \non
\eeqr
The $u_y$ integrations are the same in all cases.

Mathematica yields the following
\beqrs
\int_0^{\infty} du \; \sqrt{u}\exp[-a u] J_{-1/2}(Q u) & = & \sqrt{\fr{2}{\pi}}\fr{a}{\sqrt{Q}(a^2 + Q^2)}  \\
\int_0^{\infty} du \; \sqrt{u}\exp[-a u] J_{-1/2}(-Q u) & = & -i \sqrt{\fr{2}{\pi}}\fr{a}{\sqrt{Q}(a^2 + Q^2)} \\
\int_0^{\infty} du \; \sqrt{u}\exp[-a u] J_{1/2}(Q u) & = & \sqrt{\fr{2}{\pi}}\fr{\sqrt{Q}}{(a^2 + Q^2)}  \\
\int_0^{\infty} du \; \sqrt{u}\exp[-a u] J_{3/2}(Q u) & = & \sqrt{\fr{2}{\pi}}
\fr{(a^2 + Q^2){\rm ArcTan}[Q/a] - a Q}{Q^{3/2}(a^2 + Q^2)}  
\eeqrs
Hence
\beqrs
I_1 & = & -\fr{2}{\pi} {\rm Im}\left[ \fr{a_{xy,1}}{\sqrt{Q_x}(a_{xy,1}^2 + Q_x^2)} \fr{ia_{xy,2}}{\sqrt{Q_y}(a_{xy,2}^2 + Q_y^2)}
  \right] \\
& = & -\fr{2}{\pi} \fr{1}{\sqrt{Q_x Q_y}}{\rm Im}\left[ \fr{a_{xy,1}}{(a_{xy,1}^2 + Q_x^2)} \fr{ia_{xy,2}}{(a_{xy,2}^2 + Q_y^2)} \right]
\eeqrs
Writing
\beqrs
a_{xy,1}^2 + Q_x^2 & = & (1 + i \xi_{xy,1})^2 + Q_x^2 \equiv A_{xy,1} \exp[i 2 \Theta_{xy,1} ] \\
\Rarw A_{xy, 1} & = & [(1 - \xi_{xy,1}^2 + Q_x^2)^2 + 4 \xi_{xy,1}^2 ]^{1/2}, \;\;\; \Theta_{xy,1} =
{\rm Arctan}[\fr{\xi_{xy,1}}{(1 - \xi_{xy,1}^2 + Q_x^2)} ] \\
a_{xy,2}^2 + Q_y^2 & = & (1 + i \xi_{xy,2})^2 + Q_y^2 \equiv A_{xy,2} \exp[i 2 \Theta_{xy,2} ] \\
\Rarw A_{xy, 1} & = & [(1 - \xi_{xy,2}^2 + Q_y^2)^2 + 4 \xi_{xy,2}^2 ]^{1/2}, \;\;\; \Theta_{xy,2} =
      {\rm Arctan}[\fr{\xi_{xy,2}}{(1 - \xi_{xy,2}^2 + Q_y^2)}]
      \eeqrs

\subsubsection{Second order moment $\lan x^2 \ran$ }
 
Since $x^2 = 2 \bt_x J_x \cos^2\phi_x$, we have
\beq
\lan x^2(t) \ran = \half (2\bt_x) \int dJ_x dJ_y d\phi_x d\phi_y J_x [1 + \cos 2\phi_x] \psi_5(J_x, \phi_x, J_y, \phi_y, t)
\eeq
With the notations of the previous subsections, we can write the DF from Eq.(\ref{eq: psi5_2D}) as
\beqrs
\psi_5 & = & -\fr{ \bt_K \theta_x }{(2\pi)^2 \eps_x^2 \eps_y} \sqrt{\fr{2 J_x}{\bt_x}} [1- q_x \sin 2\phi_{x,-\Dl \phi_x}]
 \exp[-\frac{J_x[1- q_x \sin 2\phi_{x,-\Dl \phi_x}]}{\eps_x} - \frac{J_y[1+ q_y \sin 2\phi_{y,-\Dl \phi_y}]}{\eps_y}]
 (1 - \half q_x \sin 2\phi_{x,-\Dl \phi_x})  \\
& & \times
\sin \left( \phi_{x,-\Dl\phi_x} - \left[ \om_{x0} + w_{xx}J_x(1 - q_x \sin 2 \phi_{x, - \Dl\phi_x})   +
  w_{xy}(1 +  q_y \sin 2\phi_{y, - \Dl\phi_y} ) J_y \right] \tau - \half q_x \right) \non \\
& \approx  & -\fr{ \bt_K \theta_x }{(2\pi)^2 \eps_x^2 \eps_y} \sqrt{\fr{2 J_x}{\bt_x}}
 \exp[-\frac{J_x[1- q_x \sin 2\phi_{x,-\Dl \phi_x}]}{\eps_x} - \frac{J_y[1+ q_y \sin 2\phi_{y,-\Dl \phi_y}]}{\eps_y}]  \\
&  &  \times      [1- \fr{3}{2} q_x \sin 2\phi_{x,-\Dl \phi_x}]  \sin(\phi_0) \\
 \phi_0 & = &  \phi_{x,-\Dl\phi_x} + c_{xx}\sin 2 \phi_{x, - \Dl\phi_x} - c_{xy}\sin 2 \phi_{y, - \Dl\phi_y} - c_{00} 
 \eeqrs
 Combining the trigonometric terms,
 \beqrs
     [1- \fr{3}{2} q_x \sin 2\phi_{x,-\Dl \phi_x}]  \sin(\phi_0)  & = & \sin\phi_0 - \fr{3}{4}q_x[\cos(\phi_0 - 2\phi_{x,-\Dl \phi_x}) -
       \cos(\phi_0 + 2\phi_{x,-\Dl \phi_x}) ] \\
     & = & \sin\phi_0 - \fr{3}{4}q_x[\cos(\phi_{0-}) - \cos(\phi_{0+})]  \\
\phi_{0-} & \equiv & \phi_0 - 2\phi_{x,-\Dl \phi_x} = -\phi_x + \Dl\phi_x + c_{xx}\sin 2 \phi_{x, - \Dl\phi_x} - c_{xy}\sin 2 \phi_{y, - \Dl\phi_y} - c_{00} ) \\
\phi_{0+} & \equiv & \phi_0 + 2\phi_{x,-\Dl \phi_x} = 3\phi_x - \Dl\phi_x + c_{xx}\sin 2 \phi_{x, - \Dl\phi_x} - c_{xy}\sin 2 \phi_{y, - \Dl\phi_y} - c_{00} )
\eeqrs

The second dipole moment in 2D is now
\beqr
\lan x^2(t) \ran & = & -\fr{\sqrt{2\bt_x} \bt_K \theta_x }{(2\pi)^2 \eps_x^2 \eps_y} \int d J_x dJ_y \;\;\;  J_x^{3/2}
\exp[-\frac{J_x}{\eps_x} - \frac{J_y}{\eps_y}]  \non \\
&  & \times d\phi_x d\phi_y  \exp[\frac{ q_x  J_x\sin 2\phi_{x,-\Dl \phi_x}}{\eps_x} - \frac{q_y  J_y \sin 2\phi_{y,-\Dl \phi_y}}{\eps_y}] \non \\
& & \times  \left\{ (1 + \cos 2\phi_x)\left[\sin\phi_0 - \fr{3}{4}q_x[\cos(\phi_{0-}) - \cos(\phi_{0+}) ] \right] \right\}
\eeqr
Consider the two contributions separately,
\beqrs
 \lan x^2(t) \ran & = & -\fr{\sqrt{2\bt_x} \bt_K \theta_x }{(2\pi)^2 \eps_x^2 \eps_y} \int d J_x dJ_y \;\;\;  J_x^{3/2}
 \exp[-\frac{J_x}{\eps_x} - \frac{J_y}{\eps_y}]  \left( I  + II \right)  \\
 I & = &  \int \int
d\phi_x d\phi_y  \exp[\frac{ q_x  J_x\sin 2\phi_{x,-\Dl \phi_x}}{\eps_x} - \frac{q_y  J_y \sin 2\phi_{y,-\Dl \phi_y}}{\eps_y}] \non \\
& & \times  \left[\sin\phi_0 - \fr{3}{4}q_x[\cos(\phi_{0-}) - \cos(\phi_{0+}) ] \right]  \\
& = &  I_1  + I_2 + I_3 \\
II  & = &  \int \int d\phi_x d\phi_y  \exp[\frac{ q_x  J_x\sin 2\phi_{x,-\Dl \phi_x}}{\eps_x} - \frac{q_y  J_y \sin 2\phi_{y,-\Dl \phi_y}}{\eps_y}] \non \\
& & \times  \cos 2\phi_x \left[\sin\phi_0 - \fr{3}{4}q_x[\cos(\phi_{0-}) - \cos(\phi_{0+}) ] \right] 
\eeqrs
Here
\beqrs
I_1 & = &  \int \int
d\phi_x d\phi_y  \exp[\frac{ q_x  J_x\sin 2\phi_{x,-\Dl \phi_x}}{\eps_x} - \frac{q_y  J_y \sin 2\phi_{y,-\Dl \phi_y}}{\eps_y}] \non  \sin\phi_0 \\
&  = & {\rm Im}\left\{\int \int
d\phi_x d\phi_y  \exp[i\left( \phi_0 - i\frac{ q_x  J_x\sin 2\phi_{x,-\Dl \phi_x}}{\eps_x} + i \frac{q_y  J_y \sin 2\phi_{y,-\Dl \phi_y}}{\eps_y} \right)  \right\} \non \\
 & = & {\rm Im}\left\{\exp[-i(\Dl\phi_x - c_{00})] \int \int
d\phi_x d\phi_y  \exp[i\left( \phi_{x} + \fr{z_{xx}\sin 2\phi_{x,-\Dl \phi_x}}{\eps_x} - \fr{z_{xy} \sin 2\phi_{y,-\Dl \phi_y}}{\eps_y} \right)  \right\}  
\eeqrs

\clearpage

\clearpage

\section{Spectral Analysis} \label{sec: spectrum}

\bit
\item Spectrum with a single kick, and linear analysis
\item Spectrum with a single kick, nonlinear analysis
\item Extracting information about detuning from echo spectrum 
\item How else is the echo spectrum useful?

For example: in the presence of energy spread and chromaticity, how would the echo spectrum be affected?
How can energy spread affect echo amplitudes besides the additional tune spread from chromaticity?

Can the echo spectrum reveal something about impedances? How are impedances measured from coherent tune shifts? This likely requires the analysis of echoes in the presence of a resistive wall wake. The impedance will affect the decoherence, shortening the decoherence time. How will it affect the echo amplitude?

\item With the nonlinear theory, does the echo spectrum get affected by the initial amplitude? If so, could the echo spectrum be affected by the presence of, and therefore detect, nearby resonances. 

\eit

In the absence of diffusion, the time variation of the echo amplitude is 
determined by the factor 
\beqr
 A_F & = & \frac{\xi(3 - \xi^2) \cos\Phi + (1 - 3\xi^2)\sin\Phi}{(1 + \xi^2)^3} \\
\Phi & = & \om_{\bt}(t - 2 \tau), \;\;\; \xi = \om_{rev}\mu(t - 2\tau)
\eeqr
So, the spectrum is determined by the two parameters $\om_{\bt}, \mu$. 
The complete echo amplitude is given by $\bt_K\theta q \om' \tau \eps A_F$.

The above amplitude factor can be rewritten as
\beqr
A_F(t) & = & \frac{1}{(1 + \xi^2)^{3/2}}\sin(\Phi + \chi) \\
\tan\chi &  =  & \frac{\xi(3 - \xi^2)}{(1-3\xi^2)}
\nonumber 
\eeqr
Taking the Fourier transform,
\beqrs
\tilde{A}_F(\om) & = & \int_{-\infty}^{\infty} dt \;  e^{i\om t} A_F(t) \\
 & = & \frac{1}{2i}\int_{-\infty}^{\infty} dt \; e^{i\om t} \frac{1}{(1 + \xi^2)^{3/2}}
\left[e^{i(\Phi+\chi)}-e^{-i(\Phi+\chi)} \right] 
\eeqrs
The first term contributes to the negative frequency spectrum while the second contributes
to the positive frequency part. Considering only the second term
\beqrs
\tilde{A}_F(\om > 0)  & = & -\frac{1}{2i}\int_{-\infty}^{\infty} dt \; e^{i\om t} 
\frac{1}{(1 + \xi^2)^{3/2}}e^{-i(\Phi+\chi)} \\ 
& = & -\frac{1}{2i} e^{i 2\om_{\bt}\tau} \int_{-\infty}^{\infty} dt \; e^{i(\om - \om_{\bt})t} 
\frac{1}{(1 + \xi^2)^{3/2}}e^{-i\chi}
\eeqrs
This can be evaluated by a contour integration method, see Appendix A. 
The result is
\beqr
\tilde{A}_F(\om) & = &  -\frac{\pi}{6}\frac{e^{i2(\om - \om_{\bt})\tau}}{\mu\om_{rev}}
\dl^3 e^{-\dl}, \;\;\;\; \dl = \frac{\om - \om_{\bt}}{\mu\om_{rev}} \ge 0 \\
 & = & 0 , \;\;\;\;\;  \dl < 0
\eeqr
From this it follows that the spectrum has a peak at $\dl = 3$ or at a tune
given by
\beq
\nu_{peak} = \nu_{\bt} + 3\mu
\eeq
Thus the peak of the echo spectrum is shifted from the nominal tune $\nu_{\bt}$ by three
times the 
detuning parameter to one side and there are no frequencies below (if $\mu > 0$) the tune 
or above (if $\mu < 0$) the nominal tune.

The full width at half maximum of the echo spectrum is
\beq
\dl_{FWHM} = 4.12, \;\;\; \Dl\nu_{Echo,FWHM} = 4.12\mu
\eeq

\bfig
\centering
\includegraphics[scale=1.0]{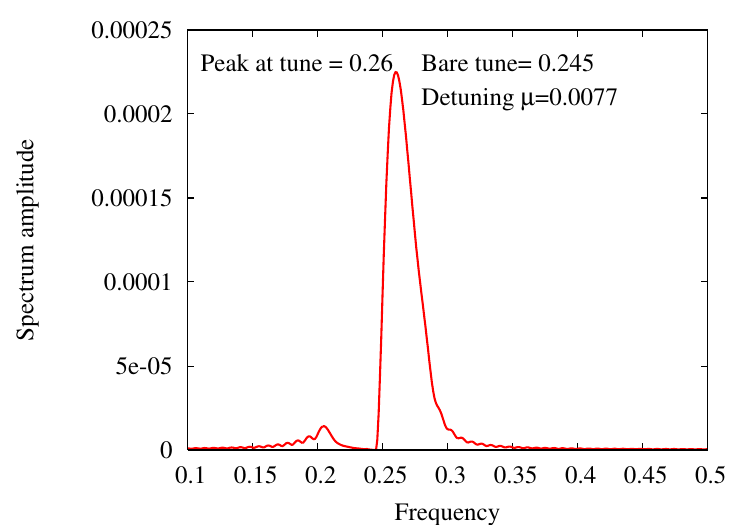}
\caption{Spectrum of the theoretical echo pulse without diffusion}
\label{fig: spectrum_theory}
\efig
Fig. \ref{fig: spectrum_theory} shows a numerical calculation of the echo spectrum,
This spectrum is obtained with parameters $\nu_{\bt} \equiv \om_{\bt}/\om_{rev} = 0.245$ 
and the detuning parameter is $\mu = 0.0077$. The theoretical peak is expected at
$\nu_{\bt}+3\mu= 0.268$ while the numerical calculation shows the peak at 0.260.

\subsection{FFT from simulation}

Do an FFT of the echo pulse. How does the spectrum change (in particular the dominant frequency)
as the quad kick is increased upto saturation and beyond?

Li on Jan 2, 2017

I have double-checked my simulation and results, and it seems that they do match your theory 
predictions. The broad shape of the peaks was due to the small resolution of the horizontal axis. 
Viewed using a range of 0 to 0.5, the peaks appear much more defined. Furthermore, using 
mu ~ -0.0012, the location of the peaks matched the predicted 0.241 (bare tune = 0.245). However, 
there appears to be no changes to the FFT peak in the saturation regime (see attachment).

\clearpage

\section{Longitudinal Echoes}

Bunched beam longitudinal echoes have been observed at the AGS (1998) and at HERA (2002).

Action Items: Theory of Bunched Beam Echoes
\bit
\item Completely linear theory in both phase and voltage kicks
\item Linear in phase, nonlinear in voltage
\item Nonlinear in phase and voltage
\eit

\subsection{Longitudinal action angle variables for small amplitudes}
The  longitudinal action, angle $(J_s, \varphi)$ variables for small amplitude motion as in SY Lee's book,
pages 234-235 , Eq.(3.73)
\beq
\phi - \phi_s = \sqrt{\fr{2 h \eta J_s}{Q_s}}\cos \varphi, \;\;\;  \dl = - \sqrt{\fr{2 Q_s J_s}{h \eta}}\sin \varphi
\eeq
where the zero amplitude synchrotron tune $Q_s$ is
\beq
Q_s = \sqrt{\fr{h e V_0 |\eta \cos \phi_s|}{2\pi \bt^2 E}} \equiv \nu_s \sqrt{|\cos\phi_s|}, \;\;\;\;
\nu_s = \sqrt{\fr{h e V_0 |\eta |}{2\pi \bt^2 E}}
\eeq
Introduce the variables
\beq
  a  =  \half \fr{h\eta}{Q_s}, \;\;\; b = \half \fr{Q_s}{h\eta}= \fr{1}{4a},  \;\;\;  \theta = \phi - \phi_s, 
\eeq
The action and angle variables are
\beqr
J_s & = &  \half \fr{h\eta}{Q_s}[ \dl^2 + (\fr{Q_s}{h\eta})^2 (\phi - \phi_s)^2 ], \;\;\;\; \tan\varphi = - \fr{h\eta}{Q_s}
\fr{\dl}{\phi - \phi_s} \\
J_s & = & a \dl^2  + b \theta^2, \;\;\;  \varphi = - {\rm Arctan}[2a \fr{\dl}{\theta}]  \\
\theta & = &  2 \sqrt{a J_s}\cos\varphi, \;\;\;\;  \dl = - 2\sqrt{b J_s}\cos\varphi
\eeqr
The averaged Hamiltonian Eq.(3.75) and equations of motion are
\beqr
H & = & \om_{rev}Q_s J_s - \fr{\om_{rev}h \eta}{16}(1 + \fr{5}{3}\tan^2\phi_s)J_s^2  + \ldots \\
\fr{d\varphi}{dt} & = & \om_{rev} \left[Q_s -  \fr{h \eta}{8}(1 + \fr{5}{3}\tan^2\phi_s)J_s \right] \\
\fr{d J_s}{dt} & = & 0
\eeqr
The action dependent tune and frequency are
\beq
Q_s(J_s) = Q_s -  \fr{h \eta}{8}(1 + \fr{5}{3}\tan^2\phi_s)J_s, \;\;\;\; \om(J_s) = \om_{rev} Q_s(J_s
\eeq

\subsection{Linear theory in both phase and voltage kicks}

Using the phase space variables $(\phi, \dl)$, the equations of motion (SY Lee's book, Eqs (3.35)-(3.36),, pg 224), the
equations of motion and Hamiltonian are
\beqr
\fr{d\phi}{dt} & = & h \om_{rev}\eta \dl, \;\;\; \fr{d\dl}{dt} = \fr{\om_{rev}}{2\pi} \fr{e V_0}{\bt^2 E}(\sin\phi - \sin\phi_s) \\
H & = & \half h \om_{rev}\eta \dl^2 + \fr{\om_{rev}}{2\pi} \fr{e V_0}{\bt^2 E}[\cos\phi - \cos\phi_s + (\phi - \phi_s)\sin\phi_s]
\eeqr
$h$ is the harmonic number
We assume that the initial distribution is determined entirely by the longitudinal action $J_s(\phi, \dl)$, so that
\beq
\psi(\phi, \dl) = \psi_0(J_s)
\eeq

At time $t=0$, kick the phase by $\Dl \phi_k$ so that after the kick, the variables are
\[ \phi(t=0+) = \phi + \Dl \phi_k, \;\;\;\;  \dl(t=0+) = \dl \]
and the distribution function is
\beq
\psi_1(\phi, \dl) = \psi_0(J_s(\phi - \Dl \phi_k, \dl)
\eeq
Linearizing in the kick, \[ \psi_1(\phi, \dl) = \psi_0(J_s) -  \psi_0'(J_s) \Dl \phi_k \]
The variables after the kick evolve as $J_s = {\rm const}, \phi(t) = \phi(0) + h \om_{rev}\eta \int \dl(t) \; dt $.
These equations are not convenient to use.

In action angle variables, we have 
\beqr
\psi_1(J_s, \varphi) & = & \psi_0(J_s(\phi - \Dl \phi_k)), \\
J_s(\phi - \Dl \phi_k) & = & \half \fr{h\eta}{Q_s}[ \dl^2 + (\fr{Q_s}{h\eta})^2 (\phi - \Dl \phi_k - \phi_s)^2 ] \non \\
  & = &  J_s - \fr{Q_s}{h\eta} (\phi - \phi_s)\Dl\phi_k + \half \fr{Q_s}{h\eta} (\Dl\phi_k)^2  \non \\
  & = &  J_s - \sqrt{\fr{2 Q_s J_s}{h \eta}}\cos \varphi\Dl\phi_k + \half \fr{Q_s}{h\eta} (\Dl\phi_k)^2 
\eeqr
After the kick, the action angle variables evolve as $J_{s}(t) = J_{s}$, $\varphi(t) = \varphi + \om(J_s)t $, hence the
distribution function is
\beqr
\psi_2(J_s, \varphi, t) & = & \psi_1(J_s, \varphi - \om(J_s)t)  \non \\
& = &  \psi_0(J_s - \sqrt{\fr{2 Q_s J_s}{h \eta}}\cos (\varphi - \om(J_s)t)\Dl\phi_k + \half \fr{Q_s}{h\eta} (\Dl\phi_k)^2)  \\
& \approx & \psi_0(J_s) - \psi_0'(J_s)[ \sqrt{\fr{2 Q_s J_s}{h \eta}}\cos (\varphi - \om(J_s)t)\Dl\phi_k ]
\eeqr
In the last expression, I have dropped the term in $(\Dl \phi_k)^2$.

The beam current monitor measures the zeroth moment as
\beqr
I_2(\phi, t) &  = & \int \psi_2(\dl, \phi) \; d\dl \non \\
& = & I_0(\phi) - \Dl\phi_k \sqrt{\fr{2 Q_s }{h \eta}}\int \psi_0'(J_s) [ \sqrt{J_s}\cos (\varphi - \om(J_s)t)  ]d\dl
\eeqr
The time dependent part of the current is given by the second term above, so the 1st term is dropped.

Now I will assume that
\beq
\psi_0(J_s) = \fr{1}{2\pi J_{s, 0}}\exp[- J_s/J_{s, 0}]
\eeq
Note: This results in a Gaussian in both $\phi$ and $\dl$. This should be OK in $\dl$ but not so much in $\phi$.

The change in current is
\beq
I_2(\phi, t) = \fr{\Dl\phi_k }{2\pi J_{s, 0}^2} \sqrt{\fr{2 Q_s }{h \eta}}\int \exp[- J_s/J_{s, 0}]
\left[ \sqrt{J_s} \cos (\varphi - \om(J_s)t)  \right] d\dl 
\eeq
Write
\[ \om(J_s) = \om_s + \om_s' J_s , \;\;\;  \om_s' = -  \fr{h \om_{rev} \eta}{8}(1 + \fr{5}{3}\tan^2\phi_s)
\]
Use
\[ \sqrt{J_s} \cos \varphi = \sqrt{\fr{Q_s}{2 h \eta}} \theta =   \sqrt{b}\theta, \;\;\;
\sqrt{J_s} \sin \varphi = - \sqrt{\fr{h \eta}{2 Q_s}}\dl = - \sqrt{a}\dl
\]
Expand
\beqrs
\sqrt{J_s} \cos (\varphi - \om(J_s)t) & = &  \sqrt{J_s} \cos \varphi \cos \om(J_s)t + \sqrt{J_s} \sin \varphi \sin \om(J_s)t  \\
 & = & \sqrt{b}\theta \cos[ \om_s t + \om_s' (a \dl^2 + b \theta^2)t] - \sqrt{a}\dl \sin[ \om_s t + \om_s' (a \dl^2 + b \theta^2)t]
\eeqrs
Hence
\beqr
I_2(\phi, t) & = & \fr{\Dl\phi_k }{2\pi J_{s, 0}^2} \sqrt{4b}\int_{-\infty}^{\infty} \exp[- J_s/J_{s, 0}]  \non \\
&   & \times  \left[ \sqrt{b}\theta \cos[ \om_s t + \om_s' (a \dl^2 + b \theta^2)t] - \sqrt{a}\dl \sin[ \om_s t + \om_s' (a \dl^2 + b \theta^2)t]   \right] \;  d\dl \non \\
& = & \fr{\Dl\phi_k }{2\pi J_{s, 0}^2} 2 b \theta \int_{-\infty}^{\infty} \exp[- J_s/J_{s, 0}]
\left ( \cos[ \om_s t + \om_s' (a \dl^2 + b \theta^2)t] \right) \; d\dl
\eeqr
where in the last step we used the fact that the second integrand is an odd function and the integral vanishes. Using
\[ \int_{-\infty}^{\infty} \exp[- p x^2] \; dx = \sqrt{\fr{\pi}{p}} \]
we have
\beqr
I_2(\phi, t) & = & \fr{b \theta \Dl\phi_k }{\pi J_{s, 0}^2}\exp[- \fr{b \theta^2}{J_{s, 0}}  ]  {\rm Re}\left( 
\exp[i( \om_s t + b\om_s'  \theta^2t)] \int_{-\infty}^{\infty}\exp[ a(- 1/J_{s, 0} + i \om_s' t)\dl^2] \; d\dl  \right) \non \\
& = & \fr{b \theta \Dl\phi_k }{\pi J_{s, 0}^2}  \sqrt{\pi} \exp[- \fr{b \theta^2}{ J_{s, 0}}] \sqrt{\fr{J_{s, 0}}{a}} {\rm Re}\left( 
\exp[i( \om_s  + b\om_s'  \theta^2)t] \sqrt{\fr{1}{[1 - i \om_s' J_{s, 0} t]}} \right)
\eeqr
We can write
\beqrs
\sqrt{\fr{1}{[1 - i \om_s' J_{s, 0} t]}} & = & \sqrt{\fr{[1 + i \om_s' J_{s, 0} t]}{1 + (\om_s'  J_{s, 0} t)^2}} \\
1 + i \om_s' J_{s, 0} t] &= & A_1(t) \exp[i\Phi_1(t)]  \\
\Rarw \sqrt{\fr{1}{[1 - i \om_s' J_{s, 0} t]}} & = & \sqrt{\fr{\exp[i\Phi_1(t)]}{A_1(t)}}
\eeqrs
where
\beq
A_1(t)  =  \sqrt{1 + (\om_s'  J_{s, 0} t)^2}, \;\;\;\;  \Phi_1(t) = {\rm Arctan}[\om_s' J_{s, 0} t]  \label{eq: A1Phi1}
\eeq
which leads to
\beqrs
    {\rm Re}\left( \exp[i( \om_s  + b\om_s'  \theta^2)t] \sqrt{\fr{1}{[1 - i \om_s' J_{s, 0} t]}} \right)
    = \fr{1}{\sqrt{A_1(t)}}\cos[( \om_s  + b\om_s'  \theta^2)t + \half \Phi_1(t) ]
\eeqrs
    Hence
\beqr
I_2(\phi, t) & = & \fr{b \theta \Dl\phi_k }{\sqrt{\pi} \sqrt{a}J_{s, 0}^{3/2}} \exp[- \fr{b \theta^2}{ J_{s, 0}}]
\fr{1}{\sqrt{A_1(t)}}\cos[( \om_s  + b\om_s'  \theta^2)t + \half \Phi_1(t) ] \non \\
& = & \fr{2  \theta \Dl\phi_k }{\sqrt{\pi} }(\fr{b}{J_{s, 0}})^{3/2} \exp[- \fr{b \theta^2}{ J_{s, 0}}]
\fr{1}{\sqrt{A_1(t)}}\cos[( \om_s  + b\om_s'  \theta^2)t + \half \Phi_1(t) ] 
\eeqr
The amplitude of the current modulation is
\beq
I_2(\phi, t)^{amp} = \fr{2  \theta \Dl\phi_k }{\sqrt{\pi} }(\fr{b}{J_{s, 0}})^{3/2} \exp[- \fr{b \theta^2}{ J_{s, 0}}]
\fr{1}{\sqrt{A_1(t)}}
\eeq
As a function of $\theta = \phi - \phi_s$, it starts from zero at $\theta=0$, reaches a maximum at $\theta = \sqrt{J_{s,0}/(2b)}$
or at
\[ \phi_{max} - \phi_s = \sqrt{\fr{h \eta J_{s, 0}}{Q_s}} \]
The modulation effectively vanishes for
\[ \sqrt{b/J_{s, 0}} \theta_{large} \ge \pi , \;\;\;  \Rarw  \phi_{large} - \phi_s \ge  \sqrt{\fr{2 h\eta J_{s, 0}}{Q_s}} \pi \]
This suggests that if the bunch length $\sg_{\phi} \le \phi_{max}-\phi_s$, the maximum of the modulation may be outside the
bunch. Preferably, one would want $\sg_{\phi} \ge \phi_{large}-\phi_s$. 
  These features can be seen in Fig. \ref{fig: I2amp}.
  \bfig
  \centering
\includegraphics[scale=0.75]{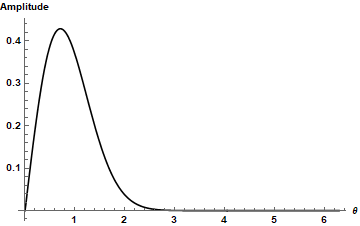}
  \caption{The function $\theta \exp[- \theta^2]$ as a function of $\theta$.}
  \label{fig: I2amp}
\efig

CHECK: Can the current $I_2$ be calculated without the Taylor expansion and linearizing in $\Dl \theta_k$ ? It is possible in the
transverse case. see Chao's notes.

\noi Applying the voltage kick at time $\tau$

At time $\tau$, the distribution function is
\beqr
\psi_3 & = & \psi_2(t = \tau) =  \psi_0(J_s - 2\sqrt{b J_s}\cos (\varphi - \om(J_s) \tau )\Dl\phi_k + \half \fr{Q_s}{h\eta} (\Dl\phi_k)^2)  \\
& \approx & \psi_0(J_s) - 2\psi_0'(J_s)\Dl\phi_k \sqrt{b J_s }\cos (\varphi - \om(J_s) \tau)    \label{eq: psi3}
\eeqr

The kick  changes the energy variable to
\beqr
\dl(\tau_+) & = & \dl(\tau)  + \fr{e V_k}{\bt^2 E}[ \sin(h_k \phi) -  \sin(h_k \phi_s) ] \equiv \dl(\tau) + \Dl \dl_k  \\
\Dl \dl_k & \equiv & \fr{e V_k}{\bt^2 E}[ \sin(h_k \phi) -   -  \sin(h_k \phi_s) ]= q_k[ \sin(h_k \phi)  -  \sin(h_k \phi_s) ], \;\;\; q_k \equiv \fr{e V_k}{\bt^2 E}
\eeqr
where $(V_k, h_k)$ are the voltage and harmonic number of the kick.  If needed, I could linearize the $\Dl \dl_k$ parameter
for small amplitudes about $\phi_s$. Write $\phi = \phi_s + \theta$, where $\theta \ll 1 $ then
\beqrs
\sin \phi & = & \sin\phi_s \cos\theta + \sin\theta \cos\phi_s \approx \sin\phi_s + \cos\phi_s \theta  \\
\Rarw \Dl \dl_k & \approx & (q_k h_k \cos\phi_s ) \theta \equiv r_k \theta, \;\;\;\; r_k = q_k h_k \cos\phi_s
\eeqrs

The distribution function after the kick is
\beqr
\psi_4(\phi, \dl) &  = & \psi_3(\phi, \dl - \Dl\dl_k)  \approx   \psi_3(J_s, \varphi) - \fr{\del \psi_3}{\del \dl} \Dl \dl_k \non \\
& = &  \psi_3(J_s, \varphi) - [\fr{\del \psi_3}{\del d J_s}\fr{\del J_s}{\del \dl} + \fr{\del \psi_3}{\del d\varphi}\fr{\del \varphi}{\del \dl}  ] \Dl \dl_k 
\eeqr
From the definitions we have
\beqrs
\fr{\del J_s}{\del \dl} & = & 2 a \dl = - 4 a \sqrt{b J_s} \sin \varphi =  - \sqrt{\fr{ J_s}{b}} \sin \varphi  \\
\sec^2\varphi \fr{\del \varphi}{\del \dl} & = & - 2a \fr{1}{\theta} , \;\;\\;
\Rarw \fr{\del \varphi}{\del \dl} = - 2a \fr{1}{2\sqrt{a}}\cos\varphi = -\sqrt{a}\cos\varphi 
\eeqrs
From Eq.(\ref{eq: psi3}) 
\beqrs
\fr{\del \psi_3}{\del d J_s} & = & \psi_0'(J_s) -  2\sqrt{b} \Dl\phi_k\fr{\del }{\del d J_s}\left[\psi_0'(J_s) \sqrt{ J_s}\right]
\cos (\varphi - \om(J_s) \tau) \\
&  &  -  2\sqrt{b} \Dl\phi_k\psi_0'(J_s) \om'(J_s) \tau \sqrt{ J_s}]\sin (\varphi - \om(J_s) \tau) \\
\fr{\del \psi_3}{\del d \varphi} & = & 2\sqrt{b} \Dl\phi_k\psi_0'(J_s) \om(J_s) \tau \sqrt{ J_s}]\cos (\varphi - \om(J_s) \tau)
\eeqrs
For sufficiently long times $\tau$ so that
\[ | \om'(J_s) \tau J_{s, 0}| \gg 1 \]
the third term in $\del \psi_3/\del J_s$ will dominate all the other terms (the same argument as in Chao). Keeping only this term
\beqrs
\psi_4(J_s, \varphi) & \approx & - \Dl \dl_k \fr{\del \psi_3}{\del d J_s}\fr{\del J_s}{\del \dl}
\approx - 2\sqrt{b} \Dl\phi_k \Dl \dl_k  \psi_0'(J_s) \om'(J_s) \tau \sqrt{ J_s}]\sin (\varphi - \om(J_s) \tau) 
\sqrt{\fr{ J_s}{b}} \sin \varphi   \\
& = & - 2  \Dl\phi_k \Dl \dl_k   \om'(J_s) \tau   J_s \psi_0'(J_s) \sin (\varphi - \om(J_s) \tau)
\eeqrs
Using the linearized form of the kick $\Dl \dl_k = r_k \theta = 2 r_k \sqrt{a J_s} \cos \varphi$ to finally obtain
\beq
\psi_4(J_s, \varphi) \approx 2 \sqrt{a}\Dl\phi_k r_k \om'(J_s) \tau   J_s^{3/2} \psi_0'(J_s) \sin 2 \varphi \sin (\varphi - \om(J_s) \tau)
\eeq
This closely resembles Eq.(26) in Chao's notes for the transverse case.

At time $t > \tau$ after the quad like kick, the distribution function is
\beqr
\psi_5(J_s, \varphi, t) & = &  \psi_4(J_s, \varphi - \om(J_s)(t - \tau))   \non \\
& = & 2 \sqrt{a}\Dl\phi_k r_k \om'(J_s) \tau   J_s^{3/2} \psi_0'(J_s) \sin [2 \varphi - 2 \om(J_s)(t - \tau)] \sin [\varphi - \om(J_s) t]
\eeqr

The zeroth moment or beam current $I_5$ is
\beqrs
I_5(\phi, t > \tau) & = & \int \psi_5(J_s, \varphi, t) \;  d\dl
\eeqrs
We have on expanding (as done for $I_2$)
\[ \sqrt{J_s} \sin (\varphi - \om(J_s)t) =  - \sqrt{a}\dl \cos[ \om(J_s) t] - \sqrt{b}\theta \sin[ \om(J_s ) t 
 \]
 Using $\psi_0'(J_s) = -\exp[-J_s/J_{s, 0}]/(2\pi J_{s,0}^2)$
 \beqrs
 I_5(\phi, t > \tau) & = &  \fr{1}{2\pi J_{s,0}^2 }2 \sqrt{a}\Dl\phi_k r_k \om' \tau  \int d\dl \; 
 J_s \exp[-\fr{J_s}{J_{s, 0}}] \sin [2 \varphi - 2 \om(J_s)(t - \tau)]  \non \\
 & & \left\{  \sqrt{a}\dl \cos[ \om(J_s) t ] - \sqrt{b}\theta \sin[ \om(J_s) t ] \right\}
 \eeqrs
 Expand
 \beqrs
 J_s \sin [2 \varphi - 2 \om(J_s)(t - \tau)] & = & J_s[ \sin 2\varphi \cos 2 \om(J_s)(t - \tau) - \cos 2\varphi \sin 2 \om(J_s)(t - \tau) ] \\
& = & 2 (\sqrt{J_s}\sin\varphi)(\sqrt{J_s}\cos\varphi)\cos 2 \om(J_s)(t - \tau) - (2 (\sqrt{J_s}\cos\varphi)^2 - J_s)
 \sin 2 \om(J_s)(t - \tau) \\
 & = & - 2\sqrt{ab}\theta\dl \cos 2 \om(J_s)(t - \tau) - (2 b\theta^2 - J_s) \sin 2 \om(J_s)(t - \tau)
 \eeqrs
We have $2\sqrt{ab} = 1$.  Hence
 \beqrs
 \mbox{} &  &   J_s \sin [2 \varphi - 2 \om(J_s)(t - \tau)] \left\{ \sqrt{a}\dl \cos[ \om(J_s) t ] - \sqrt{b}\theta \sin[ \om(J_s) t ] \right\} \\
 & = & \left[- \theta\dl \cos 2 \om(J_s)(t - \tau) - (2 b\theta^2 - J_s) \sin 2 \om(J_s)(t - \tau) \right]
 \left\{  \sqrt{a}\dl \cos[ \om(J_s) t ] - \sqrt{b}\theta \sin[ \om(J_s) t ] \right\}
 \eeqrs
 In the integrand for $I_5$, we can drop all the odd functions of $\dl$. Note that $\om(J_s)$ is an even function of $\dl$.
 Hence
\beqrs
 I_5(\phi, t > \tau) & = &  \fr{1}{2\pi J_{s,0}^2 }2 \sqrt{a}\Dl\phi_k r_k \om' \tau  \int d\dl \; 
 \exp[-\fr{J_s}{J_{s, 0}}]\left\{
   \sqrt{b}\theta (2 b\theta^2 - J_s) \sin 2 \om(J_s)(t - \tau) \sin[ \om(J_s) t ]   \right. \\
& &  \left. - \sqrt{a}\theta\dl^2 \cos 2 \om(J_s)(t - \tau)\cos[ \om(J_s) t ]  \right\}
\eeqrs
Use
\beqrs
\sin 2 \om(J_s)(t - \tau) \sin \om(J_s) t & = & \half \left( \cos[\om(J_s)(t - 2\tau)] - \cos[\om(J_s)(3t - 2 \tau)] \right) \\
\cos 2 \om(J_s)(t - \tau) \cos \om(J_s) t & = & \half \left( \cos[\om(J_s)(t - 2\tau)] + \cos[\om(J_s)(3t - 2 \tau)] \right)
\eeqrs
Since $ J_s = a\dl^2 + b\theta^2 $
\[ 2 b\theta^2 - J_s = b\theta^2 - a \dl^2 \]
Let
\[ \tau_1 = t - 2\tau, \;\;\; \tau_2 = 3t - 2\tau \]
then
\beqrs
\cos[\om(J_s)(t - 2\tau)] & = & \cos[ (\om_s+ b \om_s'\theta^2)\tau_1 + a\om_s' \dl^2 \tau_1]
\equiv \cos[\om_+\tau_1 + a\om_s' \dl^2 \tau_1] \\
\cos[\om(J_s)(3t - 2\tau)] & = & \cos[ (\om_s+ b \om_s'\theta^2)\tau_2 + a\om_s' \dl^2 \tau_2]
\equiv \cos[\om_+\tau_2 + a\om_s' \dl^2 \tau_2] \\
\om_+ & = &  \om_s+ b \om_s'\theta^2
\eeqrs

the integration terms are
\beqrs
T_1 & = &  b \sqrt{b}\theta^3  \int d\dl \;  \exp[-\fr{J_s}{J_{s, 0}}] \sin 2 \om(J_s)(t - \tau) \sin[ \om(J_s) t ]  \\
& = & \half b \sqrt{b}\theta^3  \int d\dl \;  \exp[-\fr{J_s}{J_{s, 0}}]  
\left( \cos[\om_+\tau_1 + a\om_s' \dl^2 \tau_1] -  \cos[\om_+\tau_2 + a\om_s' \dl^2 \tau_2]  \right) \\
& = & \half b \sqrt{b}\theta^3  \exp[- b\theta^2/J_{s, 0}] \int d\dl \;  \exp[-\fr{a \dl^2}{J_{s, 0}}]  
{\rm Re}\left( \exp[i \om_+\tau_1]\exp[ i a\om_s' \dl^2 \tau_1] -  \exp[i \om_+\tau_2]\exp[i a\om_s' \dl^2 \tau_2]  \right) \\
T_2 & =   &  - a \sqrt{b}\theta   \int d\dl \;  \dl^2 \exp[-\fr{J_s}{J_{s, 0}}] \sin 2 \om(J_s)(t - \tau) \sin[ \om(J_s) t ]  \\
& = & - \half a \sqrt{b}\theta   \int d\dl \;  \dl^2 \exp[-\fr{J_s}{J_{s, 0}}]
\left( \cos[\om_+\tau_1 + a\om_s' \dl^2 \tau_1] -  \cos[\om_+\tau_2 + a\om_s' \dl^2 \tau_2]  \right) \\
T_3 & = & - \sqrt{a}\theta \int d\dl \;  \dl^2  \exp[-\fr{J_s}{J_{s, 0}}] \cos 2 \om(J_s)(t - \tau)\cos[ \om(J_s) t ]  \\
& = & - \half \sqrt{a} \theta \int d\dl \;  \dl^2  \exp[-\fr{J_s}{J_{s, 0}}]
\left( \cos[\om_+\tau_1 + a\om_s' \dl^2 \tau_1] +  \cos[\om_+\tau_2 + a\om_s' \dl^2 \tau_2]  \right) \\
T_2 + T_3 & = &  -\half \theta \sqrt{a} \left[ (\sqrt{ab} + 1) \int d\dl \;  \dl^2  \exp[-\fr{J_s}{J_{s, 0}}] \cos[\om_+\tau_1 + a\om_s' \dl^2 \tau_1] \right. \\
  & & \left.  + (-\sqrt{ab} + 1) \int d\dl \;  \dl^2  \exp[-\fr{J_s}{J_{s, 0}}] \cos[\om_+\tau_1 + a\om_s' \dl^2 \tau_2] \right]  \\
& = & -\qrtr \theta \sqrt{a}  \exp[- b\theta^2/J_{s, 0}] \\
&  &  \times  \int d\dl \;  \dl^2 \exp[-\fr{a \dl^2}{J_{s, 0}}]
 {\rm Re}\left(3 \exp[i \om_+\tau_1]\exp[ i a\om_s' \dl^2 \tau_1] + \exp[i \om_+\tau_2]\exp[i a\om_s' \dl^2 \tau_2]  \right)
\eeqrs
where I used $\sqrt{ab} = 1/2$.

There are 2 different integrals
\beqrs
In_1 & = & \int_{-\infty}^{\infty} d\dl \; \exp[ - (A+ i B)\dl^2] = \sqrt{ \fr{\pi}{A+i B}} \\
In_2 & = & \int_{-\infty}^{\infty} d\dl \; \dl^2 \exp[ - (A+ i B)\dl^2] = \sqrt{\fr{\pi}{4 (A+iB)^3}}
\eeqrs

Hence
\beqrs
T_1 & = &  \half b \sqrt{\pi b}\theta^3  \exp[- b\theta^2/J_{s, 0}]  \\
&  & {\rm Re}\left( \exp[i \om_+\tau_1] \sqrt{\fr{J_{s,0}}{a(1 - i \om_s' \tau_1]}} -
\exp[i \om_+\tau_2] \sqrt{\fr{J_{s,0}}{a(1 - i \om_s' \tau_2]}} \right)
\eeqrs
while
\beqrs
T_2 + T_3 & = &   -\fr{1}{8} \theta \sqrt{ \pi a}  \exp[- b\theta^2/J_{s, 0}] \\
&  & {\rm Re} \left( 3 \exp[i \om_+\tau_1]\sqrt{(\fr{J_{s,0}}{a(1 - i \om_s' \tau_1]})^3} +
\exp[i \om_+\tau_2]\sqrt{(\fr{J_{s,0}}{a(1 - i \om_s' \tau_2]})^3}  \right) 
\eeqrs

We had earlier obtained
\[ \sqrt{\fr{1}{(1 - i \om_s' \tau_1]}} = \fr{1}{\sqrt{A_1(\tau_1)}} \exp[\half i \Phi_1(\tau_1) ] \]
 Hence
 \beqr
 T_1 + T_2 + T_3 & = &  \half b \sqrt{\pi b}\sqrt{\fr{J_{s,0}}{a} } \theta^3  \exp[- b\theta^2/J_{s, 0}]  \non \\
 &  & \left( \fr{1}{\sqrt{A_1(\tau_1)}}\cos [ \om_+\tau_1 + \half \Phi_1(\tau_1)] + \fr{1}{\sqrt{A_1(\tau_2)}}
 \cos [ \om_+\tau_2 + \half \Phi_1(\tau_2)]  \right)  \non \\
 &  &  -\fr{1}{8} \theta \sqrt{ \pi a  }\sqrt{(\fr{J_{s,0}}{a})^3}  \exp[- b\theta^2/J_{s, 0}]  \non \\
 &  & \left( \fr{3}{\sqrt{A_1(\tau_1)^3}}\cos [ \om_+\tau_1 + \fr{3}{2} \Phi_1(\tau_1)] +
 \fr{1}{\sqrt{A_1(\tau_2)^3}}\cos [ \om_+\tau_2 + \fr{3}{2} \Phi_1(\tau_2)]  \right)  \non \\
 \eeqr

 \section{Conclusions}

 In this paper, we have provided the theoretical foundation for several aspects of beam echoes. Numerical validation of some of these results are available in the references listed below. Similar validation of the newer results will follow in
 forthcoming publications. There are very many open areas that need a full theoretical formulation, these are left to the imagination of the reader.

\noi {\bf Acknowledgments} \\
I thanks all the undergraduate students who worked with me over the years on various aspects pf beam echoes.
In chronological order they are: Yuan-Shen Li, Alex Gross, Dhruv Desai, Annika Gabriel, Minani  Alexis and
Ben Luke.

\clearpage

\section{Appendix A: Bessel functions}

Power series expansion
\beq
J_{\al}(z) = (\frac{z}{2})^{\al} \sum_k \fr{(-1)^k}{k! \Gm(k + \al + 1)} (\fr{z}{2})^{2k}
\eeq
for real index $\al$. 
It follows that for integer index $n$
\beqr
J_{-n}(z) & = &  (-1)^n J_n(z) \\
J_{\al}(-z) & = &  (-1)^{\al} J_{\al}(z)
\eeqr
They obey the recurrence relation
\beq
J_{n-1}(z) + J_{n+1}(z) = \fr{2 n}{z}J_n(z)
\eeq
Hence
\[ J_0(z) + J_2(z) = \fr{2}{z}J_1(z), \;\;\;\; J_3(z) = \fr{4}{z}J_2(z) - J_1(z)  \]

Integrations
\beqr
H_{1, 0}(a, b) & = & \int dz\; z \exp[- a z] J_0(b z) =  \fr{a}{(a^2 + b^2)^{3/2}} \\
H_{1, 1}(a, b) & = & \int dz\; z \exp[- a z] J_1(b z)  =    \frac{b}{(a^2 + b^2)^{3/2}} \\
H_{1, 2}(a, b) & = &  \int dz\; z \exp[- a z] J_2(b z)  =  \fr{2 (a^2 + b^2)^{3/2} - a(2a^2 +3 b^2)}{b^2 (a^2 + b^2)^{3/2}}  \label{eq: J2int} \\
H_{1,  3}(a, b) & = &  \int dz\; z \exp[- a z] J_3(b z)  =   \fr{8a^4+12a^2 b^2 + 3b^4 - 8 a(a^2+b^2)^{3/2}}
{b^3 (a^2 + b^2)^{3/2}}   \\
H_{2, 0}(a, b) & = &  \int dz\; z^2 \exp[- a z] J_0(b z)  =  \frac{(2 a^2 - b^2)}{(a^2 + b^2)^{5/2}}  \\
H_{2, 1}(a, b) & = &  \int dz\; z^2 \exp[- a z] J_1(b z)  =   \fr{3 a b}{(a^2 + b^2)^{5/2}}  \\
H_{2, 2}(a, b) & = &  \int dz\; z^2 \exp[- a z] J_2(b z)  =    \fr{3  b^2 }{(a^2 + b^2)^{5/2}} \\
H_{2, 3}(a, b) & = &  \int dz\; z^2 \exp[- a z] J_3(b z)  =  \fr{8 (a^2 + b^2)^{5/2} - a(8a^4 + 20a^2 b^2 + 15 b^4)}{b^3(a^2  + b^2)^{5/2}}
\eeqr

The last two integrals should vanish when $b\rarw 0$, since in this limit $J_1(b z) = 0 = J_2(bz)$.
As a check, expanding the right hand size of Eq.(\ref{eq: J2int}) in a power series in $v = b/a$, we have
\beqrs
\fr{2 a^2(\sqrt{a^2 + b^2} -a) + b^2(2\sqrt{a^2 + b^2} - 3a)}{b^2 (a^2 + b^2)^{3/2}} & = & 
\fr{1}{a^2} \fr{2(\sqrt{1+v^2}-1) + v^2(\sqrt{1+v^2}-3)}{v^2[1+v^2]^{3/2}}
\\
& = &  \fr{1}{a^2} \left[ \fr{3}{4}v^2 + O(v^4)\right]
\eeqrs
which does vanish when $v = b/a \rarw 0$.

\section{Appendix B: Useful Identities}

\beq
 {\rm Arctan}[x] = \half \ln \left[\fr{1 - i x}{1 + i x}\right] 
\eeq

\end{document}